\documentclass[11pt, a4paper]{article}
\usepackage[T1]{fontenc}
\usepackage{a4wide, url}
\usepackage[english]{babel}
\usepackage{graphicx}
\usepackage{array}
\usepackage{natbib}
\usepackage[textfont=sc,labelfont=bf]{caption}
\usepackage[blocks]{authblk}
\usepackage{setspace}
\usepackage{multirow}
\usepackage{amsmath}
\usepackage{amsfonts}
\usepackage{amssymb}
\usepackage{pgf} 
\usepackage{bm}
\usepackage{paralist}
\usepackage{enumitem,kantlipsum}
\usepackage{xr}
\DeclareMathAlphabet\mathbfcal{OMS}{cmsy}{b}{n}

\newcommand{\mbf}{\mathbf}
\newcommand{\beq}{\begin{equation}}
\newcommand{\eeq}{\end{equation}}
\newcommand{\bea}{\begin{eqnarray}}
\newcommand{\eea}{\end{eqnarray}}
\newcommand{\bit}{\begin{itemize}}
\newcommand{\eit}{\end{itemize}}
\newcommand{\ben}{\begin{enumerate}} 
\newcommand{\een}{\end{enumerate}}
\newcommand{\bpm}{\begin{pmatrix}}
\newcommand{\epm}{\end{pmatrix}}
\newcommand{\bbm}{\begin{bmatrix}}
\newcommand{\ebm}{\end{bmatrix}}
\newcommand{\eps}{\varepsilon}

\renewcommand{\l}{\left}
\renewcommand{\r}{\right}

\newcommand{\E}[0]{\mathsf{E}}

\newcommand{\Cov}[0]{\mathsf{Cov}}

\newcommand{\nn}{\nonumber}
\newcommand{\wh}{\widehat}
\newcommand{\wt}{\check}

\newtheorem{ass}{Assumption}

\newtheorem{prop}{Proposition}
\newtheorem{rem}{Remark}
\newtheorem{lem}{Lemma}

\begin{document}
\title{{\Large{ \bf  Large-Dimensional   Dynamic Factor Models: Estimation of Impulse-Response Functions with $I(1)$  Cointegrated Factors}}}

\author{{\normalsize Matteo {\sc Barigozzi}${}^{1}$\hskip 1cm \normalsize Marco {\sc Lippi}${}^{2}$\hskip 1cm \normalsize Matteo {\sc Luciani}}${}^{3}$}

\date{\small{\today}}

\maketitle

\vskip-.5cm
\begin{abstract}   
\noindent We study a large-dimensional Dynamic Factor Model where: (i)~the vector of factors $\mathbf F_t$ is $I(1)$  and driven by a number of shocks that is smaller than the dimension of $\mathbf F_t$; and, (ii)~the idiosyncratic components are either  $I(1)$ or  $I(0)$. Under~(i), the factors $\mathbf F_t$ are cointegrated and can be modeled as a Vector Error Correction Model (VECM). Under (i) and (ii), we  provide  consistent estimators, as both the cross-sectional size $n$ and the time dimension $T$ go to infinity, for the factors, the loadings, the shocks, the coefficients of the VECM and therefore the Impulse-Response Functions (IRF) of the observed variables to the shocks.~Furthermore: possible deterministic linear trends are fully accounted for, and the case of an unrestricted VAR in the levels $\mathbf F_t$, instead of a VECM, is also studied. The finite-sample properties the proposed estimators are explored by means of a MonteCarlo exercise. Finally, we revisit two distinct and widely studied empirical applications. By correctly modeling the long-run dynamics of the factors, our results partly overturn those obtained by recent literature. Specifically, we find that: (i) oil price shocks have just a temporary effect on US real activity; and, (ii) in response to a positive news shock, the economy first experiences a significant boom, and then a milder recession.


\medskip

\vspace{3mm}



\end{abstract}

\thispagestyle{empty}

 \footnotetext[1]{matteo.barigozzi@unibo.it -- Universit\`a di Bologna, Italy.} 

\footnotetext[2]{marco.lippi@eief.it -- Einaudi Institute for Economics and Finance, Roma, Italy.}

\footnotetext[3]{matteo.luciani@frb.gov -- Federal Reserve Board of Governors, Washington DC, USA.\\

\noindent Special thanks go to Paolo Paruolo and Lorenzo Trapani for helpful comments.~This paper has benefited also from discussions with Antonio Conti, Domenico Giannone, Dietmar Bauer, and all participants to the 39th Annual NBER  Summer Institute.~Part of this  paper  was  written  while  Matteo  Luciani  was \textit{charg\'e de recherches} F.R.S.-F.N.R.S., and he gratefully acknowledges their financial support.~Of course, any errors are our responsibility.\\

\noindent Disclaimer: the views expressed in this paper are those of the authors and do not necessarily reflect those of the Board of Governors or the Federal Reserve System.
 }

\newpage
\section{Introduction}\label{introduction}

Since the early 2000s large-dimensional Dynamic Factor Models (DFM)  have become increasingly popular in the econometric and  macroeconomic literature, and they are nowadays commonly used by policy institutions. They have been extensively used in policy analysis based on impulse-response functions (IRF) \citep{GRS04,FGLR09,eickmeier09, fornigambettiJME, BCL, fornigambettisala, juvenalpetrella2015, smokinggun, dahlhaus2017},  in forecasting \citep{stockwatson02JASA,FHLR05,Nowcasting,LucioADFM,FGLS18}, and in the construction of both business cycle indicators and inflation indexes \citep{inflationcoin,eurocoin}.

Starting with a large dataset of macroeconomic variables, DFMs are based on the idea that  all the variables in the dataset are driven by a small number of  common  shocks, their residual dynamics being explained by idiosyncratic components. The common shocks, which are {\it pervasive}, i.e., they affect all the variables in the dataset, are interpreted as the macroeconomic shocks. The idiosyncratic components, which are specific to one or a few variables, are interpreted as  (a)~local or sectoral shocks, or (b)~measurement errors; hence they are of little interest in macroeconomic analysis.

Formally, each variable in the $n$-dimensional dataset $x_{it},\ i=1,\ldots,n$, is decomposed into the sum of two unobservable components: the common component $\chi_{it}$, and the idiosyncratic component $\xi_{it}$ \citep{FHLR00,fornilippi01,stockwatson02JASA,bai03}. Moreover, the common components are linear combinations of an $r$-dimensional vector of  {\it common factors} $\mathbf F_t=(F_{1t}\ \cdots\ F_{rt})'$,
\begin{align}
x_{it} &=\chi_{it}+ \xi_{it}, \label{pennetta} \\
\chi_{it}&=\lambda_{i1}F_{1t}+\lambda_{i2}F_{2t}+\cdots + \lambda_{ir}F_{rt}=\bm\lambda_i '\mbf F_t, \label{pennetta2} 
\end{align}
where $\bm  \lambda _i=(\lambda_{i1}\ \cdots\ \lambda_{ir})'$. 

Most of the variables contained in macroeconomic datasets are non-stationary; hence, the factors, and, possibly, also the idiosyncratic components, are non-stationary. When the factors are non-stationary, it holds that
\begin{equation} \label{ostapenko} \Delta \mathbf F_t=\mathbf C(L) \mathbf u_t,\end{equation} 
where $\mathbf C(L)$ is an $r\times q$ square-summable matrix in the lag operator, and $\mathbf u_t=(u_{1t} \ \cdots\ u_{qt})'$ is a $q$-dimensional  orthonormal  white-noise vector of {\it common shocks}.

The goal of this  paper is to estimate the IRFs of the common components $\chi_{it}$, and therefore of the variables $x_{it}$, to the common shocks $\mathbf u_t$ in the  non-stationary DFM defined by \eqref{pennetta}--\eqref{ostapenko}, i.e., to estimate $\bm\lambda_i ' \frac{\mbf C(L)}{1-L}$. Specifically, we consider the general case in which: (i) the factors are $I(1)$, singular, and cointegrated, (ii) the idiosyncratic components are either $I(1)$ or $I(0)$, and (iii)  the presence of deterministic linear trends is explicitly taken into account. As we discuss in Section \ref{sec:model}, all these are relevant features in macroeconomic datasets.

The common practice in the applied DFM literature consists in taking first differences of the non-stationary variables $x_{it}$, thus obtaining a stationary dataset $\Delta x_{it}$ with stationary factors $\Delta \mathbf F_t$, and then applying principal components to $\Delta x_{it}$, which yields consistent estimates of $\Delta \mbf F_t$ and the loadings $\bm\lambda_i$. An estimate of $\mathbf C(L) $ and $\mathbf u_t$ is then obtained by estimating a VAR for $\Delta \mathbf F_t$, see e.g., \citet{FGLR09}. Finally, all the identification techniques, based on macroeconomic theory, that are used in Structural VAR analysis (SVAR) can be applied also in the DFM setting with no modification to obtain structural shocks and IRFs---see for example \citet{FGLR09}, \citet{baiwang2015}, and \citet{stockwatson16}.

However, it is well known that if the factors are cointegrated, then a VAR for $\Delta \mathbf F_t$ is not an admissible representation. Rather, we should write a Vector Error Correction Model (VECM) for $\mbf F_t$, i.e., a VAR for $\mbf F_t$ with $r-c$ unit roots, where $c$ is the cointegration rank of $\mbf F_t$. Therefore, in order to obtain consistent estimates of the IRFs we need to consider estimation of a DFM with $I(1)$  cointegrated factors.

The crucial question then is: are the factors likely to be cointegrated? The answer is ``yes,''  and there are two main reasons why this is the case. Firstly, as predicted by macroeconomic theory, some of the macroeconomic shocks $\mbf u_t$ permanently affect the economy (e.g., technological shocks), while some others (such as monetary policy shocks or oil price shocks) have only transitory effects. In other words, in \eqref{ostapenko} the matrix $\mbf C(1)$ is likely to have reduced rank, which is equivalent to saying that the common factors are cointegrated.

Secondly, \citet{BLL} show that if $\mathbf F_t$ is a singular stochastic vector---i.e., $r$, the dimension of $\mathbf F_t$, is greater than $q$, the dimension of $\mathbf u_t$---then the common factors are cointegrated with cointegration rank $c=r-q+d$, where $0\leq d<q$, so that the cointegration rank is at least $r-q$.  Moreover, under the assumption that the entries of $\mathbf C(L)$ are rational functions of $L$, $\mathbf F_t$  has the VECM representation:
\begin{equation}\label{poverawilliams} 
\mathbf  G(L) \Delta \mathbf F_t + \bm \alpha \bm \beta' \mathbf F_{t-1} = \mathbf h +\mathbf K \mathbf u_t,
\end{equation}
where $\bm \alpha $ and $\bm \beta $ are both $r\times c$ and full rank, $\mathbf K$ is $r\times q$, and $\mathbf G(L)$ is a {\it finite-degree} matrix polynomial.  Therefore, it is legitimate to ask: are the factors likely to be singular? Once again, the answer is ``yes.'' Indeed, as pointed out in several papers, e.g., \cite{baing07}, \cite{FGLR09}, and \cite{stockwatson16}, equation \eqref{pennetta2} is just a convenient \textit{static} representation derived from a ``deeper''  set of  \textit{dynamic} equations linking the common components $\chi_{it}$ to the common shocks $\mathbf u_t$. Moreover, singularity of $\mathbf F_t$  is  strongly supported by empirical evidence, see,  e.g., \cite{GRS04}, \cite{amengualwatson07}, \cite{fornigambettiJME}, \cite{smokinggun} for US macroeconomic databases, \cite{BCL} for the euro area.

So far, the literature has proved consistency (and derived the rate of convergence) for an estimator of the IRFs for DFMs when either the variables are stationary or can be transformed to stationarity by differencing, i.e., when the factors are not cointegrated \citep{FGLR09}. However, the literature has not studied estimation of IRFs when the factors are cointegrated, which, as argued above, is a relevant empirical case in macroeconomics. Our paper fills this gap by proposing two estimators.

\smallskip \noindent
A. Having estimated the loadings $\bm\lambda_i$ and the factors $\mbf F_t$, the first estimator is obtained by fitting a VECM as in \eqref{poverawilliams} on the estimated factors. We show that, as $n,T\to\infty$ our estimator of the IRFs is consistent and converges with a rate that not only depends on $n$ and $T$, but also on the number of idiosyncratic components that are $I(1)$, and on the number of variables for which a deterministic trend is present.

\smallskip \noindent
B. As an alternative to the estimator of the IRFs based on the VECM, we prove consistency of the IRFs obtained by means of an unrestricted VAR in the levels for the estimated factors. Like in the standard VAR analysis, this approach is consistent at each given lag but it does not provide consistent estimates of the long-run features of the IRFs, see  also \citet{phillips98}.  This result is corroborated by a numerical exercise in which the VECM and the unrestricted VAR performances are close  at short horizons, whereas at long horizons, the VECM performs better.

\smallskip
Both our estimator of the loadings, which is based on principal component analysis on differenced data, and our estimator of the factors are closely related to those proposed by \citet{baing04}. However, our estimator of the factors, although asymptotically equivalent to the one proposed by \citet{baing04}, has important finite sample differences owing to a different estimation of the trend slope. A numerical comparison shows that our estimator of the factors tends to perform better than the one proposed by \citet{baing04} for estimation of IRFs.\footnote{Note that since we allow for the idiosyncratic components to be $I(1)$, the approach of estimating the factors by principal components in levels, as in \citet{bai04}, is not valid.} 

Our results  can be applied, with minor modifications, also to a Factor Augmented VAR (FAVAR) (\citealp{BBE05,baing06}) with $I(1)$ variables. Indeed, FAVARs are equivalent to a restricted version of DFMs \citep{stockwatson16}.

The potential advantages of our proposed approach are illustrated by means of two empirical applications. In the first application, we study the effects of oil price shocks on the US economy. We compare the IRFs estimated with a non-stationary DFM, as proposed in this paper, with those obtained by \citet{stockwatson16} with a stationary DFM, and we show that once we account for cointegration in the common factors, the estimate of the long-run effects of an oil price shock changes dramatically. Indeed, while \citet{stockwatson16} estimate that oil price shocks have persistent effects on the US economy, we find that the effects of an oil price shock vanish after five to eight years, a finding consistent with the idea that only technological shocks are capable of having a permanent effect on the real side of the economy. 

In the second empirical application, we study the effects of news shocks on the US economy. To do so, we compare the IRFs estimated with a non-stationary FAVAR, where the factors are either extracted as proposed in this paper, or as proposed by \citet{fornigambettisala}, i.e., under the  assumption that all the idiosyncratic components are $I(0)$.  The IRFs obtained with our approach partly overturn the results in \citet{fornigambettisala} in that we find that in response to a positive news shock, hours worked respond positively, and the economy experiences a significant boom, and then a milder recession.

Lastly, let us mention  that our non-stationary DFM has recently  been used by \citet{LuciaAssetPrice} to study the response of asset prices to monetary policy shocks. When  estimated using a standard SVAR, the response is very slow and not statistically significant. However, by using our non-stationary DFM, \citet{LuciaAssetPrice} find strong and quick asset price reactions, both on euro area and US data. 

The  paper is organized as follows.~In Section \ref{sec:model} we present the model and its assumptions.~Section \ref{sec:est}  establishes consistency and rates for our estimators of the IRFs. In Section \ref{sec:nfactors} we propose an information criterion to  determine the number of permanent shocks $q-d$, which allows us to infer the cointegration rank of the factors. In Section \ref{sec:sim}, by means of a MonteCarlo simulation exercise, we study the finite sample properties of our estimators.~Finally, in Section \ref{sec:emp} we apply our methodology to a US quarterly macroeconomic dataset and in two separate exercises we study the  impact of oil price  and  of news shocks. In Section \ref{sec:end} we conclude. The proofs of our main results  are in Appendix \ref{app:prop}. A complementary appendix contains the proofs of all lemmas, details on identification of the IRFs, a comparison with FAVARs, and additional numerical results.


\section{The non-stationary Dynamic Factor model} \label{sec:model}

\subsection{$I(1)$ vectors and cointegration}\label{sec:dueuno}
Throughout the paper, we will adopt the following definitions for $I(0)$, $I(1)$, and cointegrated stochastic vectors.  They are standard and hold both for non-singular vectors, as in  all textbooks (see, e.g., \citealp{Johansen95}, Ch. 3), and  for singular vectors.

\begin{compactenum}
\item[ (I)]  Consider an $r\times q$ matrix $\mathbf A(L) = \mathbf A_0+\mathbf A_1 L+\cdots,$ with the assumption 
that the series  $\sum_{j=0}^\infty \mathbf A_jz^j$ converges for all complex number $z$ such that $|z|<1+\delta$ for some $\delta >0$. This  condition is fulfilled  when the entries of 
$\mathbf A(L)$ are rational functions of $L$  with no poles inside or on the unit circle (the VARMA case).
Given the $r$-dimensional  stationary stochastic vector
$$\mathbf z_t= \mathbf A(L) \mathbf v_t ,$$
where  $\mathbf v_t$  is a $q$-dimensional white noise, $q\leq r$,  we say that 
$\mathbf z_t$ is $I(0)$ if $\mathbf A(1) \neq \mathbf 0$.

\item[(II)] The  $r$ dimensional  stochastic vector $\mathbf z_t$ is $I(1)$ if $\Delta\mathbf  z_t$ is $I(0)$.

\item [(III)] The $r$-dimensional $I(1)$ vector  $\mathbf z_t$ is cointegrated of order $c$, $0<c<r$, if (1)~there exist linearly independent $r$-dimensional 
vectors $\bm\beta_k$, $k=1,\ldots,c$, such that $\bm\beta_k' \mathbf z_t$ is stationary, (2)~if $\bm \gamma'  \mathbf z_t$ is stationary
then $\bm \gamma $ is a linear combination of the vectors $\bm \beta_k$. 

\end{compactenum}

Some important properties for our model follow from these definitions.
\begin{rem}\upshape{\label{ghiniarann1}
\begin{compactenum}[(a)]
 
 \item[] 
 
\item[(a)] \vskip -.1cm Some of  the coordinates of  an $I(1)$ vector can be stationary.

\item [(b)] If one of the coordinates of the $I(1)$ vector $\mathbf z_t$  is stationary, then 
$\mathbf z_t$ is cointegrated.

\item[(c)]  The cointegration rank of $\mathbf z_t$ is equal to $r$ minus the rank of $\mathbf A(1)$.

\item [(d)] It easy to see that 
 $\mathbf z_t$ is cointegrated with cointegration rank $c$ if and only if $\mathbf z_t$ can be linearly transformed into a vector whose first $c$ coordinates are stationary and the remaining $r-c$ are $I(1)$.  For,
let $\mathbf z_t$ be cointegrated of order $c$ with cointegration vectors $\bm \beta_k$, $k=1,\ldots,c$.  Let $\bm \beta=(\bm \beta_1 \ \cdots\ \bm \beta _c) $   and  $\mathbf B = (\bm \beta \ \ \bm \beta_\perp)$, where $\bm \beta_\perp$ is an $r\times (r-c)$ matrix whose columns are linearly independent and orthogonal to the columns of $\bm \beta$.
Then,   the first $c$ coordinates of $\mathbf  z^*_t=\mathbf B' \mathbf z_t$ are stationary while  the remaining 
$r-c$ are~$I(1)$.   


\item [(e)]   Note that if  $\mathbf z_t$ is $I(1)$ and  $r>q$, then obviously  
$\mathbf z_t$ is cointegrated with cointegration rank at least $r-q$, that is, $c= (r-q)+d$ with $0\leq d <q$.
\end{compactenum}
}
\end{rem}

\subsection{Assumptions on common and idiosyncratic components}\label{sec:duedue}

Define $\mathbf x_t=(x_{1t}\ \cdots\ x_{nt})'$,   $\bm  \chi_t=(\chi_{1t}\ \cdots\ \chi_{nt})'$,   $\bm \xi_t=(\xi_{1t}\ \cdots\ \xi_{nt})'$,   $\bm \Lambda=(\bm  \lambda_1\ \cdots\ \bm \lambda_n)'$. Then, the non-stationary DFM that we consider in this paper and given in equations \eqref{pennetta} and \eqref{ostapenko} become:
\begin{align}
 \mathbf x_t & = \bm  \chi_{t}+\bm \xi_{t}=\bm \Lambda \mathbf F_t +\bm  \xi_t,\label{macron}\\ 
 \Delta\mathbf F_t& = \mathbf C(L) \mathbf u_t.\label{eq:model2vector}
\end{align}
Firstly, we suppose that $\mbf F_t$ has two equivalent representations: an ARIMA and a VECM. Specifically, we assume the following.

\begin{ass}\label{ASS:common}\textbf{(Common shocks and common factors)}

\begin{compactenum}[(a)]

\item [(a)] $\mathbf u_t=(u_{1t}\ \cdots\ u_{qt})'$ is a strong orthonormal $q$-dimensional vector white noise, i.e.,  $\E[\mbf u_t]=\mbf 0_q$, 
$\E[\mbf u_t\mbf u_t']=\mbf I_q$, and  $\mbf u_t$ and $\mbf u_{t-k}$ are independent for any $k\neq 0$, moreover $\E [u_{jt}^{4} ]\leq M_1$, for some positive real $M_1$ independent of $j$. \smallskip

\item [(b)] The  $r$-dimensional stochastic vector $\mathbf F_t$ is $I(1)$ and  has the ARIMA representation
\beq\label{eq:model2} 
\mbf S(L)\Delta\mbf F_t = \mbf Q(L)\mbf u_t,
\eeq
where: (i)~$\mathbf S(L)$ is an $r\times r$ finite-degree  matrix polynomial with $\det(\mbf S(z))\neq 0$ for $|z|\le 1$; (ii)~$\mathbf S(0)=\mathbf I_r$;
(iii)~$\mathbf  Q(L)$ is a finite-degree $r\times q$ matrix polynomial, $\mathbf Q(1)\neq \mbf 0$; (iv)~${\mathrm {rk}}(\mathbf Q(0))=q$.    
Note that, defining $d=q-{\rm rk}(\mathbf  Q(1))$, so that  $0\le d<q$, the cointegration rank of $\mathbf F_t$ is $c= r- \mathrm {rk} (\mathbf Q(1))= (r-q)+d$, see Remark \ref{ghiniarann1}, (c).

\item [(c)] 
The vector  $\mathbf F_t$ has the VECM representation
\begin{equation}\label{faziocoglione}
\mbf G(L) \Delta \mbf F_t+\bm \alpha \bm \beta' \mbf F_{t-1} = \mbf h +   \mbf K \mbf u_t,
\end{equation}
where: (A)~$\bm \alpha$ and $\bm\beta$ are full rank $r\times c$ matrices; (B)~$\mathbf K=\mathbf Q(0)$; (C)~$\mathbf h$ is a constant vector; (D)~$\mathbf G(L)$ is a {\it finite-degree} matrix polynomial with $\mathbf G(0)=\mathbf I_r$.\smallskip

 \item [(d)]  $\mbox {rk}(\E[\Delta\mathbf  F_t\Delta\mathbf F_t' ])=r$ and $\E[\Delta F_{it}^2]>\E[\Delta F_{jt}^2]>0$, for any $i,j=1,\ldots ,r$ with $i<j$.
 \item [(e)] The number of common shocks and factors $q$ and $r$ are finite integers independent of $n$.
 \end{compactenum}
\end{ass}

Condition (a) is stronger than the usual assumption made in a stationary setting, in which $\mbf u_t$ is just required to be white noise, and it is equivalent to Assumption B in \citet{baing04}. Condition (b) implies that $\mbf C(L)= \mbf S(L)^{-1}\mbf Q(L)$ in \eqref{eq:model2vector}, and therefore that the vector $\mathbf F_t$ has  rational spectral density. 
Regarding  (c), by combining  the Granger Representation Theorem \citep{englegranger1987} with recent results on singular stochastic vectors, see \cite{andersondeistler08},  \cite{BLL} prove that a VECM representation like  \eqref{faziocoglione}, \textit{with a finite degree $\mbf G(L)$},  holds generically, i.e., except for a negligible subset in the parameter space, under the assumptions that  $\mathbf F_t$ is singular with    rational spectral density.    This is the motivation for  \textit{assuming}  here the existence of representation \eqref{faziocoglione}. 



\begin{rem}\label{rem:permanent}
\upshape{As a consequence of Assumption \ref{ASS:common} (b), in \eqref{eq:model2vector} we have ${\rm rk}(\mathbf  C(1))=q-d$; hence we can write $\mbf C(1)=\bm\psi\bm\eta'$, where $\bm\psi$ is $r\times q-d$ and $\bm\eta$ is $q\times q-d$ and both have full-rank. Therefore, by defining ${\bm\eta}_{\perp}$ as the $q\times d$ matrix whose columns are independent and orthogonal to the columns of $\bm\eta$, we can always transform $\mbf u_t$ as $\mathbf v_t=  (\mathbf v_{1t}'\  \mathbf v_{2t}')'= (\bm\eta\ \ \bm\eta_\perp)' \mathbf u_t$, where $\mathbf v_{1t}$ has dimension $q-d$ while $\mathbf v_{2t} $ has dimension $d$, such that the $q-d$ shocks in $\mathbf v_{1t}$ have a permanent effect on $\mathbf F_t$, whereas  the $d$ shocks in $\mathbf v_{2t}$ have a transitory effect. Thus the number of permanent shocks is $r$ minus the cointegration rank (since $q-d=r-c$), as in the non-singular case, while the number of transitory shocks $d$ is the complement to $q$, not $r$, as though $r-q$ transitory shocks had a zero coefficient. }
 \end{rem}

We then make the following assumptions on the factor loadings.

\begin{ass} \label{asm:factor}\textbf{(Loadings)} 
\begin{inparaenum}[(a)]
\item As $n\to\infty,$  $n^{-1}\bm\Lambda'\bm\Lambda\to\mbf I_r$;
\item $\Vert \bm\lambda_{i}\Vert \le C$, for some positive real $C$ independent of $i$.
 \end{inparaenum}
\end{ass}

Condition (a) implies that  the $r$ factors are not redundant, i.e., no representation with a number of factors smaller than $r$ is possible. In particular, note that Assumptions \ref{ASS:common} (d) and \ref{asm:factor} (a) are common identifying assumptions imposed in stationary factor models, see, e.g., \cite{stockwatson02JASA}.\footnote{Equivalently, we could assume $\E[\Delta\mathbf  F_t\Delta\mathbf F_t' ]=\mbf I_r$ and $n^{-1}\bm\Lambda'\bm\Lambda\to\mbf V$, as $n\to\infty$, with $\mbf V$ positive definite and with distinct eigenvalues, see, e.g., \citet{fan13}.} The following remark shows that this choice has no implication for IRF estimation. 

\begin{rem}\label{ghiniarann}\upshape{
In  model \eqref{macron} the 
factors $\mathbf F_t$ are not identified. For, given the non  singular $r\times r$ matrix
$\mathbf H$, 
\begin{equation}\label{serveprima}\mathbf x_t = \left [ \bm\Lambda \mathbf H\right ]\left [\mathbf H^{-1} \mathbf F_t\right ]+\bm \xi_t= \bm \Lambda^* \mathbf F^*_t+\bm \xi_t.\end{equation}
Using $\mathbf F^*_t$ implies changes in the matrices in \eqref{eq:model2vector}, \eqref{eq:model2}, and \eqref{faziocoglione} and the loadings that are easy to compute: 
$$\begin{aligned}&\bm \Lambda^*=\bm \Lambda\mathbf H,\ \  \mathbf S^*(L) = \mathbf H^{-1} \mathbf S(L) \mathbf H, \ \ \mathbf Q^*(L) = \mathbf H^{-1} \mathbf Q(L),\ \  \mathbf C^*(L)= \mathbf H^{-1} \mathbf C(L) , \\ &\mathbf G^*(L) = \mathbf H^{-1}\mathbf G(L) \mathbf H,\ \ \bm \alpha^*= \mathbf H^{-1} \bm \alpha,  \ \ \bm \beta ^* = \mathbf H'\bm \beta, 
\ \ \mathbf K^*=\mathbf H^ {-1} \mathbf K.\end{aligned}$$
Note that 
$ \bm \Lambda^* \mathbf C^*(L)=\bm \Lambda \mathbf C(L),$
so that the raw IRFs
of the $x$'s with respect to $\mathbf u_t$, corresponding to the factors $\mathbf F^*_t$ and  to the factors $\mathbf F_t$ are equal. As a consequence, identification of the IRFs based on any economic criterion is independent of the particular factors used, i.e., of the identifying assumptions imposed on $\mbf F_t$ and $\bm\Lambda$. In this respect, although Assumptions \ref{ASS:common} (d) and \ref{asm:factor} (a) might seem restrictive, they are innocuous and are particularly convenient in proving consistency of the estimated factors up to a sign. The theory developed in the next section can be adapted to allow for other identifying constraints. 

Furthermore, because the factors $\mathbf F_t$  are identified up to a linear transformation and in view of Remark \ref{ghiniarann1} (d),  the question of whether some of the factors are stationary  while the remaining ones are $I(1)$ is perfectly equivalent to the question of whether  and ``how much'' the factors are cointegrated, see \citet{bai04}. In other words, the case of $I(0)$ factors is implicitly considered under condition (c), whereas we do not consider in this paper the case of $I(2)$ variables.}
\end{rem}

Regarding the idiosyncratic components we assume the following.

\begin{ass}\label{ASS:idio}\textbf{(Idiosyncratic components)}
For any $i\in\mathbb N$,
\beq
\label{eq:model3} 
(1-\rho_iL)\xi_{it} = d_i(L)\eps_{it},
\eeq
where 
\begin{compactenum}[(a)]
\item $\bm  \eps_t=(\eps_{1t}\  \cdots\ \eps_{nt})'$ is a strong $n$-dimensional vector white noise, i.e., $\E[\bm\eps_t]=\mbf 0_n$,  $\E[\bm\eps_t\bm\eps_t']=\bm\Gamma^\eps_0$, and   $\bm\eps_t$ and $\bm\eps_{t-k}$ are independent for any $k\neq 0$, moreover  $\E[|\eps_{it}|^{\kappa_1}|\eps_{jt}|^{\kappa_2}]\leq M_2$, for some positive real $M_2$ independent of $i$ and $j$ and any $\kappa_1+\kappa_2=4$;
\item $\bm\Gamma^\eps_0$ is positive definite and such that $\max_{j=1,\ldots, n}\sum_{i=1}^n|\E[\eps_{it}\eps_{jt}]|\leq M_3$, for some positive real $M_3$ independent of $n$;
\item $d_i(L)=\sum_{k=0}^\infty d_{ik}$, with $\sum _{k=0}^\infty k |d_{ik}|\leq M_4,$ for some positive real $M_4$ independent of $i$; 
\item $|\rho_i|\leq 1$,  so that $I(1)$ idiosyncratic components are allowed;
\item  $u_{jt}$ and $\eps_{is}$ are independent for any $j=1,\ldots, q$, $i\in\mathbb N$, and $t,s\in\mathbb Z$.

\end{compactenum}
\end{ass}

Condition (a) is similar to Assumption C(i) in \citet{baing04} but is less stringent since we here require only 4$^\textrm{th}$ order finite moments as compared to finite 8$^\textrm{th}$ order moments. Condition (b) allows for contemporaneous  cross-sectional dependence of the idiosyncratic shocks, $\bm \eps_t$. In particular, we  require a mild form of sparsity as proposed by \citet{fan13} and often found empirically, see, e.g., \citet{boivinng06}, \citet{baing08JoE}, and \citet{LucioADFM} in a stationary setting. As a consequence, the components of $\Delta\bm\xi_t$ are also allowed to be both cross-sectionally and serially correlated. 

Condition (c) in Assumption \ref{ASS:idio} implies square summability of the matrix polynomials in \eqref{eq:model3} so that $\xi_{it}$ is non-stationary if and only if $\rho_i=1$. Assuming that $|\rho_i|<1$, that is, all idiosyncratic components are stationary, implies that  any 
$p$-dimensional vector $(x_{i_1,t} \cdots x_{i_p,t})'$, with $p\geq q-d+1$, would be cointegrated---for example, if $q=3$ and $d=0$ then all  $4$-dimensional sub-vectors of $\mathbf x_t$ are cointegrated ($3$-dimensional  if $d=1$). 
Moreover, when applying the test proposed in \citet{baing04} on the US macroeconomic time series analyzed in Section \ref{sec:emp}, and typically analyzed in the empirical DFM literature, we found that the unit root hypothesis is not rejected for nearly half of the estimated idiosyncratic components. Finally, condition (e) is in agreement with the economic interpretation of the model, in which common and idiosyncratic shocks are two independent sources of variation. 

 It can be shown that Assumptions \ref{ASS:common} through \ref{ASS:idio} imply that the $r$ largest eigenvalues of the covariance matrix of $\Delta \mbf x_{t}$ diverge linearly in $n$, while the remaining $n-r$ stay bounded (see Lemma \ref{lem:evalcov} in the complementary appendix for a proof). This result allows us to estimate the number of factors $r$,
while analogous results on the eigenvalues of the spectral density matrix of $\Delta \mbf x_t$,  allow the estimation of $q$ and 
the cointegration rank $c$ of the factors $\mathbf F_t$, see Section \ref{sec:nfactors} for details. 

We conclude with the following assumption, which has the consequence that $\bm \chi^{}_0=\mathbf 0_n$, $\bm \xi^{}_0=\mathbf 0_n$, and $\mathbf x^{}_0=\mathbf 0_n$, a requirement commonly made in unit root analysis.

\begin{ass}\label{initcond}  
For all $i\in \mathbb N$ and $t\leq 0$, $\mathbf u_t=\mathbf 0_q$, and $\varepsilon_{it}=0$.
\end{ass}

In practice, when dealing with macroeconomic time series, deterministic linear trends can also be present; hence we typically do not observe $\mbf x_t$, but  the $n$-dimensional vector $\mbf y_{t}=(y_{1t}\cdots y_{nt})'$, such that
\beq\label{eq:modeltrend}
y_{it} =  a_i +  b_i t +  x_{it},
\eeq
where $a_i,b_i\in\mathbb R$, and $x_{it}$ satisfies Assumptions \ref{ASS:common} through \ref{ASS:idio}. 

For series belonging to the real side of the economy, e.g., GDP, $b_i$ is likely to be strongly significant; however, for nominal series, e.g., inflation, $b_i$ is likely to be not significantly different from zero. Indeed, when considering the US macroeconomic time series analyzed in Section \ref{sec:emp}, we reject the null-hypothesis $b_i=0$ for only about half of the series (see Appendix \ref{app:testB} for details on the adopted testing procedure). Consequently, we introduce the following assumption that poses an asymptotic limit to the number of series with a deterministic linear trend.

\begin{ass}\label{ASS:trend}
Let $n_{b}$ be the number of variables among $y_{1t},\ldots,  y_{nt}$ for which $b_i\ne 0$, then, $n_{b}= O(n^{\eta})$ for some  $\eta\in[0,1)$.
\end{ass}


%


\section{Estimation}\label{sec:est} 

The object of interest of this paper is the true IRF of $x_{it}$, for $i=1,\ldots n$, to the shock $u_{jt}$, for $j=1,\ldots, q$, which we denote as (see also \eqref{macron} and \eqref{eq:model2vector})
\beq\label{trueIRF}
\phi_{ij}(L) = \bm\lambda_i'\l[\frac{\mbf c_j(L)}{1-L}\r],
\eeq
where ${\bm\lambda}_i'$ is the $i$-th row of ${\bm\Lambda}$, ${\mbf c}_j(L)$ is the $j$-th column of ${\mbf C}(L)$, and the notation used is convenient and makes sense, provided that we do not forget that such IRF is not square summable. Note that in view of \eqref{eq:modeltrend} the IRF in \eqref{trueIRF} has to be interpreted as a deviation from the deterministic linear trend.

We follow a procedure similar to \cite{FGLR09} in the stationary setting: (i) we estimate  the loadings, the common factors,  their  VECM dynamics and the  raw (non-identified) IRFs,  (ii)~we identify the structural common shocks and IRFs by imposing a set of restrictions based on economic logic. We now describe in detail these steps and study the asymptotic behavior of all our estimators for both $n$ and $T$ tending to infinity. 

Note that, in practice, the number of common factors $r$, of common shocks $q$, and of the cointegration relations $c=r-q+d$ is unknown, and in Section \ref{sec:nfactors}, we show that these quantities can be consistently estimated with probability tending to one, as $n,T\to\infty$. Therefore, throughout this section, we can assume that $r$, $q$, and $c$ are known.

Hereafter, we denote estimated quantities with a hat, like in $\mathbf {\widehat {\bm\Lambda}}$, without explicit notation for their dependence  on both $n$ and $T$. We also denote the spectral norm of a  matrix $\mbf B$ by  $\Vert\mbf B\Vert=(\mu_1^{\mbf B'\mbf B})^{1/2}$, where $\mu_1^{\mbf B'\mbf B}$ is the largest eigenvalue of $\mbf B'\mbf B$.

\subsection{Loadings and common factors}\label{sec:detrend}
Assume to observe the $n$-dimensional vector $\mbf y_{t}=(y_{1t}\cdots y_{nt})'$ satisfying \eqref{eq:modeltrend} over the period  $t=1, \ldots, T$, then
the model for $\Delta y_{it}= y_{it}- y_{it-1}$ with $t=2,\ldots, T$, reads
\beq\label{eq:moddiff}
\Delta  y_{it} = b_i+\Delta x_{it}= b_i+ \bm \lambda_i' \Delta \mathbf  F_t+\Delta \xi_{it}.
\eeq
We first present and discuss our approach to estimation of loadings and common factors, and in Lemma  \ref{lem:load} below, we prove their asymptotic properties. Then, in Remark \ref{rem:baing} below, we compare our estimators with those in \citet{baing04}.

The loadings estimator is computed by principal component analysis on the differenced data. Let $\wh{\bm\Gamma}_0$ be the $n\times n$ sample covariance matrix of $\Delta\mbf y_t=(\Delta y_{1t}\cdots \Delta_{nt})'$ and let $\mathbf {\widehat W}$ be the $n\times r$ matrix with the right normalized eigenvectors of  $\bm{\widehat \Gamma}_0$, corresponding to the first
$r$ eigenvalues, on the columns. Our estimator of the loadings matrix $\bm\Lambda$ is given by
\beq\label{parlacondudu_load}
\bm{\widehat \Lambda}=\sqrt n \,\mathbf {\widehat W}.
\eeq

In order to estimate the common factors, we explicitly introduce an estimator of the slope coefficients $b_i$. Consider the set $\mathcal I_b$ of values of $i$ such that $b_i\ne 0$, then for any $i\in\mathcal I_b$, we de-trend $y_{it}$ by least squares regression on a constant and a linear trend, giving the estimator
\beq\label{eq:biols}
\wh{b}_i = 
\frac{\sum_{t=1}^T(t-\frac {T+1} 2)(y_{it}-\bar y_i)}{\sum_{t=1}^T (t-\frac {T+1} 2)^2},
\eeq
where $\bar y_i$ is the sample mean of $y_{it}$. If instead $i\in\mathcal I_b^c$, we set $\widehat b_i=0$. In practice ${\mathcal I}_b$  is unknown and in Appendix \ref{app:testB} we introduce a test for the null-hypothesis that $b_i=0$ for all $i=1,\ldots,n$. In particular, we show that as $n,T\to\infty$ the probability of type I and type II errors of our testing procedure tends to zero, hence hereafter, we can assume that $\mathcal I_b$ is known.



By defining $\wh{x}_{it}=y_{it}-\wh{b}_it$, our estimator of  the common factors is given by projecting $\wh{\mbf x}_t=(\wh{x}_{1t}\cdots \wh{x}_{nt})'$ onto the estimated loadings:
\beq\label{parlacondudu}
\wh{\mbf F}_t=\frac{1}{n}\wh{\bm\Lambda}'\wh{\mbf x}_t=\frac 1n\sum_{i=1}^n \wh{\bm\lambda}_i\wh{x}_{it}.
\eeq
Consistency of this procedure is proved in the following Lemma.

\begin{lem}\label{lem:load}  
Let Assumptions \ref{ASS:common} through \ref{initcond} hold. Then, there exists an $r\times r$ diagonal matrix $\mathbf J$ with entries $\pm 1$, depending on $n$ and $T$, such that, as $n,T\to \infty$,
\begin{inparaenum}
\item [(i)] for all $i$, $\Vert\wh{\bm\lambda}_i'-\bm\lambda_i'\mbf J\Vert = O_p(\max(n^{-1/2},T^{-1/2}))$.
\end{inparaenum}
If also Assumption \ref{ASS:trend} holds, then:
\begin{inparaenum}
\item [(ii)] for all $i\in\mathcal I_b$, $\vert \wh b_i-b_i\vert= O_p(T^{-1/2})$;
\item [(iii)] given $t$, $T^{-1/2}\Vert\wh{\mbf F}_t-\mbf J\mbf F_t\Vert= O_p(\max(n^{-1/2},T^{-1/2}, n^{-(1-\eta)}))$.
\end{inparaenum}\end{lem}

Notice that, since for different values of $n$ and $T$ we get different estimators of the loadings $\wh{\bm\lambda}_i$ and the factors $\widehat{\mathbf F}_t$, then  in general also the matrix $\mathbf J$ depends on $n$ and $T$. However, in light of Remark \ref{ghiniarann} above and as shown in the proofs of Propositions \ref{vecm} and \ref{var} below, such indeterminacy poses no problem for consistency of estimated IRFs.

The result on the loadings estimator which is obtained from the differenced data, is derived in a way that is similar to the approach used by \citet{stockwatson02JASA}, \citet{FGLR09}, and \citet{fan13}. The result on the factors estimator is new and the next remark provides an intuition for it.

\begin{rem}\upshape{
An immediate consequence of Lemma \ref{lem:load} is that if all  series have a deterministic linear trend, i.e., $\eta=1$, then $\wh{\mbf F}_t$ is not a consistent estimator of the common factors $\mbf F_t$. Indeed, first note that, since $\wh{x}_{it}=y_{it}-\wh{b}_it$, because of \eqref{eq:modeltrend} we can re-write \eqref{parlacondudu} as 
\beq\label{fat1}
\wh{\mbf F}_t = \frac 1 n\sum_{i=1}^n\wh{\bm\lambda}_i x_{it} + \frac 1n\sum_{i=1}^n \wh{\bm\lambda}_i a_i + 
\frac 1 {n}\sum_{i\in\mathcal I_b} \wh{\bm\lambda}_i \big(b_i-\wh b_i\big)t.
\eeq
Then, since $x_{it}=\bm\lambda_i'\mbf F_t+\xi_{it}$, from \eqref{fat1} it follows that the factors estimation error is
\begin{align}\label{fat2}
\frac 1{\sqrt T}\big(\wh{\mbf F}_t-\mbf J\mbf F_t\big)&=\frac 1 {n\sqrt T} \sum_{i=1}^n{\bm\lambda}_i\xi_{it}+ \frac 1 {n\sqrt T}\sum_{i=1}^n {\bm\lambda}_i a_i + 
\frac 1 {n\sqrt T}\sum_{i\in\mathcal I_b} {\bm\lambda}_i \big(b_i-\wh b_i\big)t +o_p(1),
\end{align}
where the last term on the right hand side is the loadings estimation error (see part (i) of Lemma \ref{lem:load} above). Now, while the first term on the right-hand-side of \eqref{fat2} is $O_p(n^{-1/2})$ and the second term is $O_p(T^{-1/2})$, the third term due to the linear deterministic trends will not vanish unless $\eta<1$. As already discussed above, the assumption $\eta<1$ is realistic for a typical macroeconomic dataset. In an extensive numerical analysis conducted in Section \ref{sec:sim} and the complementary appendix, we show that our estimators perform well even for values of $\eta$ close to one. 
}
\end{rem}


In \citet{baing04} principal component analysis on differenced data $\Delta \mbf y_{t}$ is used to compute both the loadings estimator and an estimator $\Delta\widetilde{\mbf F}_t$ of the differenced factors. An estimator $\widetilde {\mbf F}_t$ of $\mbf F_t$ is then computed as $\widetilde {\mbf F}_t=\sum_{s=2}^t \Delta\widetilde {\mbf F}_s$. In the next Remark, we compare the two approaches.


\begin{rem}\label{rem:baing} \upshape{
First, from Lemmas 1 and 2 in \citet{baing04} it follows that $\Delta\widetilde {\mbf F}_s$ is a consistent estimator of $\mbf J(\Delta {\mbf F}_s-\overline{\Delta \mbf F})$, where $\overline{\Delta \mbf F}$ is the sample mean of $\Delta {\mbf F}_s$, and, therefore, 
 $T^{-1/2}\Vert\widetilde {\mbf F}_t-\mbf J {\mbf F}_t+\mbf J {\mbf F}_1+\mbf J (\mbf F_T-\mbf F_1)(t-1)/(T-1)\Vert=o_p(1)$, as $n,T\to\infty$. So $\widetilde {\mbf F}_t$ is a consistent estimator of  ${\mbf F}_t$ only up to a location shift. Although, this result is enough for the purposes of testing for unit roots, as in \citet{baing04}, it is not enough for the purposes of the present paper. 

Second, because $\Delta\widetilde{\mbf F}_t$ is estimated by principal components that require each $\Delta y_{it}$ to be centered, $\widetilde{\mbf F}_t$ is estimated as if the data where de-trended by using $\overline {\Delta y_i}=(T-1)^{-1} \sum_{t=2}^T \Delta y_{it}$ as an estimator of the slope. More precisely, since  $\Delta\widetilde{\mathbf F}_s=n^{-1}\sum_{i=1}^n\widehat{\bm\lambda}_i(\Delta  y_{is} -\overline {\Delta  y_i})$, from \eqref{eq:moddiff} we immediately have
\begin{align}
\widetilde {\mathbf F}_t &=\frac 1n\sum_{i=1}^n\widehat{\bm\lambda}_i x_{it} -\frac 1 n\sum_{i=1}^n\wh{\bm\lambda}_i x_{i1} +  
\frac 1 n \sum_{i\in\mathcal I_b} \wh{\bm\lambda}_i \big(b_i-\overline{\Delta y}_i\big)(t-1).\nn
\end{align}
By comparing this expression with the one obtained for $\wh{\mbf F}_t$ in \eqref{fat1}, we see that, because of the two different de-trending procedures, the two estimators differ just by a constant term and a term linear in $t$.  Then, it is clear that also $\widetilde {\mathbf F}_t$ is a consistent estimator if and only if $\eta<1$.

Third, although $\widetilde {\mbf F}_t$ and $\wh{\mbf F}_t$  are asymptotically equivalent (both $\widehat b_i$  and $\overline{\Delta y}_i$ are $\sqrt T$-consistent estimators of $b_i$), there is an important finite sample difference. Indeed, since the principal components $\Delta\widetilde{\mbf F}_t$ have zero sample mean by construction, we always have $\widetilde{\mbf F}_1=\widetilde{\mbf F}_T$, thus fixing the estimator at $T$ equal to the initial condition which can be arbitrarily specified.\footnote{Note that we can also write
\begin{align}
\widetilde {\mathbf F}_t &=  \frac 1 n\sum_{i=1}^n\widehat{\bm\lambda}_i\sum_{s=2}^t(\Delta  y_{is} -\overline {\Delta  y_i}) = 
\frac 1 n\sum_{i=1}^n\widehat{\bm\lambda}_i\bigg[  y_{it}- y_{i1} -  \frac {(t-1)}{(T-1)}( y_{iT}-y_{i1})  \bigg],\nn
\end{align}
then $\widetilde {\mathbf F}_1=\mbf 0_r$ and   $\widetilde {\mathbf F}_T=\mbf 0_r$.
} 
Instead, when using our approach based on $\wh{b}_i$, since in general $\wh{x}_{i1}\ne \wh{x}_{iT}$, from \eqref{parlacondudu} we also have that in general $\wh{\mbf F}_{1}\neq \wh{\mbf F}_{T}$. A numerical comparison of the finite sample properties of the two methods, which is shown in Section \ref{sec:sim} and the complementary appendix, suggests that our estimation method is to be preferred. 

}
 \end{rem}

We conclude with the following remark on the role of the intercept term $a_i$.

\begin{rem}\label{rem:const} \upshape{Although in \eqref{eq:modeltrend} we have not assumed $a_i$ to be zero, we have not included any estimator of the intercept when deriving $\wh{\mbf F}_t$ in \eqref{parlacondudu}. Indeed, no consistent estimator of $a_i$ is available in the present setting. Nevertheless, the results in Lemma \ref{lem:load} hold irrespectively of the choice of such estimator, and therefore, without loss of generality, we can always set $\wh{a}_i= 0$ for all $i$.\footnote{Equivalently, we could set $\wh a_i$ equal to any generic value and then in \eqref{parlacondudu} use $\wh{x}_{it}=y_{it}-\wh b_i t-\wh a_i$ for estimating $\wh{\mbf F}_t$.} The same comment applies to the factor estimator by \citet{baing04}, where usually the condition $\widetilde {\mbf F}_1=\mbf 0_r$ is imposed. Note that by Assumption \ref{initcond}, we have $a_i=y_{i0}$, which is not observed, therefore, for simplicity, we let also $a_i=0$ in the following.}\footnote{Note that if this were not the case, then we could weaken Assumption \ref{initcond} to allow for $\E[\mbf F_{t}]=\mbf c$ with $\mbf c=(c_1\cdots c_r)'$ with $c_j\ne 0$ for some $j=1,\ldots, r$, such that $a_i=\bm\lambda_i'\mbf c$. In this case, we would need to estimate both the VECM in \eqref{VECMhatF} and the VAR in \eqref{eq:fattoriVARinlivelli} including also a constant term.}
\end{rem}

\subsection{IRFs when estimating a VECM for the common factors}\label{sec:vecmvar}

We now turn to estimation of the VECM  in \eqref{faziocoglione},   with $c=r-q+d$ cointegration relations,   see  Assumption \ref{ASS:common}:  
\begin{equation}\label{VECMhatF}
{\Delta\mbf F}_t = \bm \alpha \bm \beta' {\mbf F}_{t-1} + \sum_{k=1}^p \mbf G_k{\Delta\mbf F}_{t-k}+  \mbf w_t, \quad  \mbf w_t=\mbf K\mbf u_t.
\end{equation}
As a consequence of Assumption \ref{initcond} we set $\mbf h=\mbf 0$.
  
Different estimators for the cointegration vector, $\bm\beta$, are possible.~As suggested by the asymptotic and numerical studies in \citet{phillips91} and \citet{gonzalo94}, we opt for the estimation approach proposed by \citet{Johansen95}.~Although typically derived 
from the maximization of a Gaussian likelihood, this estimator is nothing else but the solution of an eigen-problem naturally associated to a reduced rank regression model, where no specific assumption about the distribution of the errors is necessary in order to establish consistency, see, e.g., \cite{velu1986}.


We briefly review estimation of the VECM in \eqref{VECMhatF} when  using the estimated factors $\wh{\mbf F}_t$, instead of the unobserved $\mbf F_t$, and when setting  $p=1$, for simplicity.\footnote{We refer to \citet[Chapter 6]{Johansen95} for a detailed description of the estimators in the case $p>1$.} Denote as $\wh{\mbf e}_{0t}$ and $\wh{\mbf e}_{1t}$ the residuals of the least squares regressions of $\Delta\wh{\mbf F}_t$ and of $\wh{\mbf  F}_{t-1}$ on $\Delta\wh{\mbf F}_{t-1}$, respectively, and define the matrices $\wh{\mbf S}_{ij}=T^{-1}\sum_{t=1}^T\wh{\mbf e}_{it}\wh{\mbf e}_{jt}'$. Let $\wh{\mu}_j$ be the $j$-th largest eigenvalue of the matrix $(\wh{\mbf S}_{11}-\wh{\mbf S}_{10}\wh{\mbf S}_{00}^{-1}\wh{\mbf S}_{01})$. Then, following \citet{Johansen95}, the estimator of the $c$ cointegration vectors, $\wh{\bm\beta}_1,\ldots, \wh{\bm\beta}_c$, are  such that, for any $j=1,\ldots c$, they solve
$(\wh{\mbf S}_{11}-\wh{\mbf S}_{10}\wh{\mbf S}_{00}^{-1}\wh{\mbf S}_{01})\wh{\bm\beta}_j = \wh{\mu}_j\wh{\bm\beta}_j.$
The vectors $\wh{\bm\beta}_j$ are then the $c$ columns of the estimated matrix $\wh{\bm\beta}$. The other parameters of the VECM, $\bm\alpha$ and $\mbf G_1$, are estimated in a second step as the least squares estimators of the regression
\[
\Delta\wh{\mbf F}_t = \bm\alpha (\wh{\bm\beta}'\wh{\mbf F}_{t-1})+ \mbf G_1 \Delta\wh{\mbf F}_{t-1} + \mbf w_t.
\]
From this regression, we also obtain the vector of residuals $\wh{\mbf w}_t$, which is an estimator of $\mbf w_t$. Denote the $r\times r$ sample covariance matrix of $\wh{\mbf w}_t$ as $\wh{\bm\Gamma}_0^w$. Let $\wh{\mbf W}^w$ be the $r\times q$ matrix with the right normalized eigenvectors of $\wh{\bm\Gamma}_0^w$, corresponding to the first $q$ eigenvalues, on the columns, and let $\wh{\mbf M}^w$ be the $q\times q$ diagonal matrix of those eigenvalues. Then, the estimators of $\mbf K$ and the common shocks $\mbf u_t$ are given by $\wh{\mbf K}=\wh{\mbf W}^w(\wh{\mbf M}^w)^{1/2}$ and $\wh{\mbf u}_t=(\wh{\mbf M}^w)^{-1/2}\wh{\mbf W}^{w'}\wh{\mbf w}_t$, respectively.

A VECM($p$) with cointegration rank $c$ can also be written as a VAR($p+1$) with $r-c$ unit roots. Therefore, after estimating \eqref{VECMhatF}, we have the estimated matrix polynomial $\widehat{\mbf A}^{\mbox{\tiny{VECM}}}(L)= \mbf I_r - \sum_{k=1}^{p+1}\widehat{\mbf A}^{\mbox{\tiny{VECM}}}_k L^k$,
with coefficients given by 
\begin{align}
\widehat{\mbf A}^{\mbox{\tiny{VECM}}}_1 &= \widehat{\mbf G}_1-\widehat{\bm\alpha}\widehat{\bm\beta}'+\mbf I_r,\nn\\ 
\widehat{\mbf A}^{\mbox{\tiny{VECM}}}_{k} &= \widehat{\mbf G}_{k}-\widehat{\mbf G}_{k-1},\; \mbox{ for } k=2,\ldots, p,\; \mbox{ and }\; \widehat{\mbf A}^{\mbox{\tiny{VECM}}}_{p+1} = -\widehat{\mbf G}_{p},\label{eq:AVECMinv}
\end{align}
such that ${\rm rk}(\widehat{\mbf A}^{\mbox{\tiny{VECM}}}(1))={\rm rk}(\widehat{\bm\alpha}\widehat{\bm\beta}')=c$. Then, for $i=1,\ldots,n$ and  $j=1,\ldots, q$, the raw (non-identified) IRFs estimator is defined as
\beq\label{eq:IRF1_NOTID}
\widetilde{\phi}^{\mbox{\tiny{VECM}}}_{ij}(L)=\wh{\bm\lambda}_i' \l[\wh{\mbf A}^{\mbox{\tiny{VECM}}}(L)\r]^{-1}\wh{\mbf k}_j,
\eeq
where $\wh{\bm\lambda}_i'$ is the $i$-th row of $\wh{\bm\Lambda}$, $\wh{\mbf k}_j$ is the $j$-th column of $\wh{\mbf K}$. 

As we show in Proposition \ref{vecm} below, $\wh{\mbf K}$ is a consistent estimator of $\mbf K$ only up to right multiplication by an orthogonal $q\times q$ transformation $\mbf R$. Therefore, the IRFs in \eqref{eq:IRF1_NOTID} are in general not identified unless we also estimate $\mbf R$ and
economic theory tells us that the choice of the identifying transformation can be determined by the economic meaning attached to the common shocks, $\mbf u_t$. 
In general, for a given set of identifying restrictions, ${\mbf R}$ depends on the other parameters of the model, that is, it is determined by a mapping ${\mbf R}\equiv {\mbf R}({\bm\Lambda},{\mbf A}(L),{\mbf K})$. In the typical case of just- or under-identifying restrictions, to estimate $\mbf R$ we just have to consider the $q$ rows of the raw estimated IRFs, denoted as $\widetilde{\bm\Phi}_{[q]}(L)$, corresponding to the economic variables which are relevant for identification of the shocks, and then we define the estimator $\wh{\mbf R}$ such that $\widetilde{\bm\Phi}_{[q]}(L)\wh{\mbf R}$ satisfies our desired restrictions. In this case, due to orthogonality, an estimator $\wh{\mbf R}$ is obtained by solving a linear system of  $q(q - 1)/2$ equations with $q(q - 1)/2$ unknowns, which depends on $\widetilde{\bm\Phi}_{[q]}(L)$ and therefore on $\wh{\bm\Lambda}$, $\wh{\mbf A}^{\mbox{\tiny{VECM}}}(L)$, and $\wh{\mbf K}$. Among the most common identifying restrictions considered in the literature there are the zero impact restrictions (imposed on $\wh{\bm\Phi}_{[q]}(0)$) and the long-run restrictions (imposed on $\wh{\bm\Phi}_{[q]}(1)$), see Section \ref{sec:emp} for two examples. 


The estimated and identified IRFs are then defined by combining the estimated parameters and the identification restrictions. In particular, for $i=1,\ldots,n$ and  $j=1,\ldots, q$, the dynamic reaction of the $i$-th variable to the $j$-th common shock is estimated as
\beq\label{eq:IRF1}
\wh{\phi}_{ij}^{\mbox{\tiny{VECM}}}(L) = \wh{\bm\lambda}_i'\l[\wh{\mbf A}^{\mbox{\tiny{VECM}}}(L)\r]^{-1}\wh{\mbf K}\,\wh{\mbf r}_j,
\eeq
where $\wh{\bm\lambda}_i'$ is the $i$-th row of $\wh{\bm\Lambda}$, $\wh{\mbf r}_j$ is the $j$-th column of $\wh{\mbf R}$.  

Consistent estimation of \eqref{eq:IRF1} in presence of estimated factors, is possible under the following additional assumption.

\begin{ass}\label{asm:rates} ${}$
\begin{compactenum} [(a)]
\item Let $n_{1}$ be the number of $I(1)$ variables among $\xi_{1t},\ \ldots, \ \xi_{nt}$. Then,  
$n_{1}= O(n^{\delta})$ for some  $\delta\in[0,1)$;
\item let ${\cal I}_{0}$ and  ${\cal I}_{1}$  be the sets $\{i\leq n,\ \hbox{such that $\xi_{it}$ is $I(0)$}\}$  and 
$\{i\leq n,\ \hbox{such that $\xi_{it}$ is $I(1)$}\}$, respectively, then, $n^{-\gamma}\sum_{i\in {\cal I}_0}\sum_{j\in  {\cal I}_1}|\E[\eps_{it}\eps_{jt}]|\leq M_9$,
for some  $\gamma<\delta$ and some positive real $M_9$ independent of $n$.
\end{compactenum}
\end{ass}

Under condition (a), we put an asymptotic  limit to the number of $I(1)$ idiosyncratic components, i.e.,  those $\xi_{it}$ such that $\rho_i=1$, see Assumption \ref{ASS:idio} (d). Their number $n_{1}$  can grow to infinity but more slowly than the number of the $I(0)$ components. As already discussed, this assumption seems realistic in typical macroeconomic datasets. Moreover, the numerical results in Section \ref{sec:sim} and the complementary appendix show that our estimators perform well even for values of $\delta$ close to one. Finally, with reference to the partitioning of the vector of idiosyncratic components into $I(1)$ and $I(0)$ coordinates, condition (b) limits the dependence between the two blocks more than the dependence within each block, which is in turn controlled by Lemma \ref{rem:idio}.\footnote{We could, in principle, consider any $\gamma<1$, in which case the rates of convergence of Proposition \ref{vecm} below would also depend on $\gamma$. However, since the main message of those results would be qualitatively unaffected, we impose, for simplicity, $\gamma<\delta$.}

We then have consistency of the estimated VECM parameters and the IRFs. For simplicity, we assume that the degree of $\wh{\mbf A}^{\mbox{\tiny{VECM}}}(L)$ in \eqref{eq:IRF1} is $p=1$, the generalization to any degree, $p>1$, being straightforward. 

\begin{prop}\label{vecm}{\textbf{(Consistency of Impulse-Response Functions based on VECM)}}\\
Define $\vartheta_{nT,\delta,\eta}=\max\left({T^{1/2} n^{-(1-(\delta+\eta)/2)}},{T^{1/2} n^{-(1-\eta)}}, n^{-(1-\delta)/2}, n^{-(1-\eta)/2},{T^{-1/2}}\right)$. Let Assumptions \ref{ASS:common} through \ref{asm:rates} hold and assume $T^{1/2} /n\to 0$, as $n,T\to\infty$. Then, there exists a $c\times c$ orthogonal matrix $\mbf Q$ depending on $n$ and $T$, such that, as $n,T\to\infty$, 
 \begin{inparaenum}[(i)]
\item  $\Vert\wh{\bm\beta}-\mbf J\bm\beta\mbf Q\Vert=O_p(T^{-1/2}\vartheta_{nT,\delta,\eta})$; 
\item  $\Vert\wh{\bm\alpha}-\mbf J\bm\alpha\mbf Q\Vert=O_p(\vartheta_{nT,\delta,\eta})$;
\item  $\Vert\wh{\mbf G}_1-\mbf J\mbf G_1\mbf J\Vert=O_p(\vartheta_{nT,\delta,\eta})$;
\end{inparaenum}
where $\mbf J$ is defined in Lemma \ref{lem:load}.

If we further assume that there exists an integer $\bar n$ such that $\mbf K'\mbf K$ has distinct eigenvalues for $n>\bar n$, then there exists a $q\times q$ orthogonal matrix $\mbf R$, depending on $n$ and $T$, such that, as $n,T\to\infty$, 
\begin{inparaenum}[(i)] 
\item  [(iv)] $\Vert\wh{\mbf K}-\mbf J\mbf K\mbf R'\Vert=O_p(\vartheta_{nT,\delta,\eta})$;
\item  [(v)] given $t$, $\Vert\wh{\mbf u}_t-\mbf R\mbf u_t\Vert=O_p(\vartheta_{nT,\delta,\eta})$.
\end{inparaenum}

Denote as $\phi_{ijk}$ the $k$-th coefficient of the polynomial $\phi_{ij}(L)$ in \eqref{trueIRF} and as $\wh{\phi}_{ijk}^{\mbox{\tiny{VECM}}}$ the $k$-th coefficient of the polynomial $\wh{\phi}_{ij}^{\mbox{\tiny{VECM}}}(L)$ in \eqref{eq:IRF1}. Then, as $n,T\to\infty$, 
 \begin{inparaenum}[(i)]
\item [(vi)] given $i,j$ and $k$, $|\wh{\phi}_{ijk}^{\mbox{\tiny{VECM}}}-\phi_{ijk}|= O_p(\vartheta_{nT,\delta,\eta})$; 
\item [(vii)]  given $i$ and $j$, $\lim_{k\to \infty}\vert \wh{\phi}_{ijk}^{\mbox{\tiny{VECM}}}- {\phi}_{ijk}\vert=O_p(\vartheta_{nT,\delta,\eta})$.
\end{inparaenum}
\end{prop}

The rate of convergence in Proposition \ref{vecm} is determined by $\vartheta_{nT,\delta,\eta}$ and we can distinguish two cases depending on the ratio $\delta/\eta$ being greater or smaller than one or in other words depending on whether the number of series with $I(1)$ idiosyncratic components dominates over the number of those with linear trends or vice versa. First, consider the case $\delta/\eta\ge 1$, then, we have 
\beq\label{eq:thetanTdelta}
\vartheta_{nT,\delta,\eta} = \l\{\begin{array}{lcl}
T^{1/2}n^{-(1-(\delta+\eta)/2)} & \mbox{if} & T^{1/(2-\delta-\eta)}<n\le T^{1/(1-\eta)},\\
n^{-(1-\delta)/2}& \mbox{if} & T^{1/(1-\eta)}\le n \le T^{1/(1-\delta)},\\
T^{-1/2}& \mbox{if} &n\ge T^{1/(1-\delta)},
\end{array}
\r.
\eeq
while, when $\delta/\eta< 1$ we have\footnote{If $\delta=\eta$ then \eqref{eq:thetanTdelta2} coincides with \eqref{eq:thetanTdelta}.}
\beq\label{eq:thetanTdelta2}
\vartheta_{nT,\delta,\eta} = \l\{\begin{array}{lcl}
T^{1/2}n^{-(1-\eta)} & \mbox{if} & T^{1/(2-2\eta)}<n\le T^{1/(1-\eta)},\\
T^{-1/2}& \mbox{if} &n\ge T^{1/(1-\eta)}.
\end{array}
\r.
\eeq

The conditions $\delta<1$ and $\eta<1$, required in Assumptions \ref{ASS:trend} (a) and \ref{asm:rates} (a), are then necessary for consistency. As already mentioned above, both conditions are realistic in typical macroeconomic datasets. The condition $\vartheta_{nT,\delta,\eta}\to 0$, as $n,T\to\infty$, is instead sufficient to guarantee consistency, and it implies that at least we must have $T^{1/2}/n\to 0$ (when $\delta=\eta=0$), a typical constraint when considering estimation of factor augmented regressions in a stationary setting, see, e.g., \citet{baing06}. However, when $\delta>0$ and/or $\eta>0$, we need $n$ to grow faster than $\sqrt T$ in order to have consistency and, in particular, if $T^{1/(1-\max(\delta,\eta))}/n\to 0$, then the classical $\sqrt T$-consistency, in principle, can still be achieved.

The rates in \eqref{eq:thetanTdelta} and \eqref{eq:thetanTdelta2} are the consequence of our two-step estimation procedure: when estimating a VECM using the estimated factors, the estimated coefficients have an error which grows with $T$, however, since the estimated factors are cross-sectional averages of the $x$'s (see also \eqref{parlacondudu}), we can keep such error under control by allowing for an increasingly large cross-sectional dimension, $n$. The following remarks provide some more intuition about the role of $\delta$ and $\eta$ in the results in Proposition \ref{vecm}.


\begin{rem}\label{rem:vecmnew}\upshape{
The estimation error of the Error Correction term in the VECM must account for the deviation of the estimated cointegration relations $\wh{\bm\beta}'\wh{\mbf F}_t$ from the stationary process ${\bm\beta}'{\mbf F}_t$. Specifically, $\wh{\bm\beta}'\wh{\mbf F}_t$ contains two non-stationary sources of error. The first one is due to the idiosyncratic components and is proportional to their weighted average ${(n\sqrt T)^{-1}}\sum_{t=1}^T\sum_{i=1}^n\bm\lambda_i\xi_{it}$. While in the stationary factor model literature this is typically controlled by means of conditions on the cross-sectional dependence of idiosyncratic components like our Assumption \ref{ASS:idio} (b), in the present setting, stronger requirements also on the number of $I(1)$ idiosyncratic components are needed. In particular, under our assumptions, this error term has variance of order $T^2 n^{-4+2\delta}$. 

The second source of error is due to the de-trending procedure discussed in Section \ref{sec:detrend} and  is proportional to ${(n\sqrt T)^{-1}}\sum_{t=1}^T\sum_{i=1}^n\bm\lambda_i(\wh{b}_i- b_i)t$ (see \eqref{fat2} above). Although these errors are strongly cross-sectionally dependent, they are still controllable because the estimator $\wh{b}_i$ of the slope is consistent. In particular, under our assumptions, this error term has variance of order $T^2n^{-4+4\eta}$. 

Summing up, both errors are of the same magnitude with respect of $T$, but with respect to $n$, the second one is larger. Therefore, $\delta$ and $\eta$ have different roles in determining consistency, with $\eta$ being more relevant.
}
\end{rem}

\begin{rem}\label{rem:vecm2}{\upshape
Due to the factor estimation error, we do not have, in general, the classical $T$-consistency for the estimated cointegration vector $\wh{\bm\beta}$. Still, $\wh{\bm\beta}$ converges to the true value, $\bm\beta$, at a faster rate with respect to the rate of consistency of the other estimated VECM parameters. This is enough to consistently apply the two-step VECM estimation as in \citet{Johansen95}. 
}
\end{rem}

\begin{rem}\label{rem:vecm4}{\upshape The estimated VECM parameters approach the true parameters only up to three transformations $\mbf J$, $\mbf Q$, and $\mbf R$. The matrix $\mbf J$ reflects the fact that the factors are identified ones only up to a sign (see Lemma \ref{lem:load}), while the matrix $\mbf Q$ represents the usual indeterminacy in the identification of the cointegration relations. Consistently with Remark \ref{ghiniarann}, these matrices have no role in the estimation of the IRFs. 
%
The matrix $\mbf R$ represents indeterminacy in the identification of the matrix $\mbf K$, and, as discussed above, an estimator $\wh{\mbf R}$ can be estimated by means of economic restrictions imposed on the non-identified IRFs. Consistency of $\wh{\mbf R}$ when considering just- or under-identifying restrictions for which  the map ${\mbf R}\equiv {\mbf R}({\bm\Lambda},{\mbf A}(L),{\mbf K})$ is analytic, is straightforward \citep{FGLR09}. The case of over-identifying restrictions can be treated in a similar way \citep{han18}. Last, note that the requirement of asymptotically distinct eigenvalues of $\mbf K'\mbf K$, which restricts $\mbf R$ to be an orthogonal matrix, is a common requirement in the literature, see, e.g., Assumption 7 in \citet{FGLR09}.}
\end{rem}
%

%

\subsection{IRFs when estimating a VAR in levels for the common factors}\label{tretre}

In presence of non-singular cointegrated vectors, several papers have addressed the issue of whether and when 
a VECM or an unrestricted VAR for the levels should be used for estimation. \citet{simsstockwatson} show that the parameters of a cointegrated VAR are consistently estimated using an unrestricted VAR in the levels. On the other hand, \citet{phillips98} shows that if the variables are cointegrated, then the long-run features of the IRFs are consistently estimated only if the unit roots are explicitly taken into account, that is, within a VECM specification, see also \citet{paruolo97a}. This result is confirmed numerically in \citet{BLL} also for the singular case, $r>q$.

Nevertheless, since by estimating an unrestricted VAR it is still possible to estimate short-run IRFs consistently without the need to determine the number of unit roots, and therefore without having to estimate the cointegration relations, this approach has become very popular in empirical research \citep{simsstockwatson}. For this reason, here we also study the properties of IRFs when we consider least squares estimation of an unrestricted VAR($p$) model in levels for the common factors:
\beq\label{eq:fattoriVARinlivelli}
\mbf F_t = \sum_{k=1}^p \mbf A_k \mbf F_{t-k} +  \mbf w_t, \quad  \mbf w_t=\mbf K\mbf u_t.
\eeq
Denote by $\wh{\mbf A}_k^{\mbox{\tiny{VAR}}}$ the least squares estimators of the coefficient matrices, obtained using $\wh{\mbf F}_t$, and by $\wh{\mbf K}$ and $\wh{\mbf u}_t$, the estimators of $\mbf K$ and $\mbf u_t$, which are obtained as in the VECM case but this time starting from the sample covariance of the VAR residuals. However, as before, $\mbf K$ can be identified only up to right multiplication by an orthogonal matrix $\mbf R$ and an estimator $\wh{\mbf R}$ can be obtained by imposing appropriate economic restrictions.

By letting $\wh{\mbf A}^{\mbox{\tiny{VAR}}}(L)=\mbf I_r-\sum_{k=1}^p\wh{\mbf A}_k^{\mbox{\upshape{\tiny VAR}}}L^k$, for $i=1,\ldots,n$ and $j=1,\ldots, q$, the estimated and identified IRF of the $i$-th variable to the $j$-th shock is defined as
\beq\label{eq:IRF1var}
\wh{\phi}_{ij}^{\mbox{\tiny{VAR}}}(L) = \wh{\bm\lambda}_i'\l[\wh{\mbf A}^{\mbox{\tiny{VAR}}}(L)\r]^{-1}\wh{\mbf K}\,\wh{\mbf r}_j,
\eeq
where $\wh{\bm\lambda}_i'$ is the $i$-th row of $\wh{\bm\Lambda}$, $\wh{\mbf r}_j$ is the $j$-th column of $\wh{\mbf R}$. 

Consistency of these estimators is given in the following Lemma. For simplicity, we assume that the degree of $\wh{\mbf A}^{\mbox{\tiny{VAR}}}(L)$ in \eqref{eq:IRF1var} is $p=1$.  Generalization to any degree, $p>1$, is straightforward. 

\begin{prop}\label{var}{\textbf{(Consistency of Impulse-Response Functions based on VAR)}}\\
Define $\zeta_{nT,\eta}=\max\left(n^{-(1-\eta)},n^{-1/2},T^{-1/2}\right)$. Let Assumptions \ref{ASS:common} through \ref{ASS:trend} hold.  Then, as $n,T\to\infty$,
\begin{inparaenum}[(i)]
\item $\Vert\wh{\mbf A}_1^{\mbox{\upshape{\tiny VAR}}} -\mbf J\mbf A_1\mbf J\Vert=O_p(\zeta_{nT,\eta})$;
\end{inparaenum}
where $\mbf J$ is defined in Lemma \ref{lem:load}.

If we further assume that there exists an integer $\bar n$ such that $\mbf K'\mbf K$ has distinct eigenvalues for $n>\bar n$, then there exists a $q\times q$ orthogonal matrix $\mbf R$,  depending on $n$ and $T$, such that, as $n,T\to\infty$,
\begin{inparaenum}
\item [(ii)] $\Vert\wh{\mbf K}-\mbf J\mbf K\mbf R'\Vert=O_p(\zeta_{nT,\eta})$;
\item [(iii)] given $t$, $\Vert\wh{\mbf u}_t-\mbf R\mbf u_t\Vert=O_p(\zeta_{nT,\eta})$.
\end{inparaenum}

Denote as $\phi_{ijk}$ the $k$-th coefficients of the polynomial $\phi_{ij}(L)$ in \eqref{trueIRF} and as $\wh{\phi}_{ijk}^{\mbox{\tiny{VAR}}}$ the $k$-th coefficient of the polynomial $\wh{\phi}_{ij}^{\mbox{\tiny{VAR}}}(L)$ in \eqref{eq:IRF1var}. Then, as $n,T\to\infty$, 
\begin{inparaenum}
\item [(iv)] given $i,j$ and $k$, $|\wh{\phi}_{ijk}^{\mbox{\tiny{VAR}}}-\phi_{ijk}|= O_p(\zeta_{nT,\eta})$;
\item [(v)] given $i$ and $j$, $\lim_{k\to \infty}\vert\wh{\phi}_{ijk}^{\mbox{\tiny{VECM}}}- {\phi}_{ijk}\vert=O_p(1)$.
\end{inparaenum}
\end{prop}

From this result, we see that using an unrestricted VAR in levels for the estimated factors has both advantages and disadvantages compared to using a VECM. On the one hand, consistency of IRFs can be achieved with a possibly faster convergence rate and without having to require stationarity of some idiosyncratic components or any constraint on the relative rates of divergence of $n$ and $T$. This is possible since the cointegration matrix $\bm\beta$ need not be estimated. Note, however, that the presence of deterministic linear trends affects the rate of convergence also in this case. On the other hand, the long-run IRFs $\wh{\phi}_{ij}^{\mbox{\tiny{VAR}}}(1)$ are inconsistent, a result which is the direct consequence of the fact that we are not correctly modeling the cointegration among the factors. These two contrasting aspects pose a trade-off for the empirical researcher between (i) estimation of a model which is misspecified but simpler to estimate, which however is valid in the short- medium-run only (VAR), or (ii) estimation of the correctly specified model, which requires estimating more parameters but is consistent at all lags (VECM). These facts are confirmed in Sections \ref{sec:sim} and \ref{sec:emp} when comparing the two approaches on simulated and real data.


We conclude by comparing our approach with FAVARs.

\begin{rem}\label{marcellino}\upshape{In FAVAR models IRFs are estimated from a VAR including some exogenously observed variables, say $z_{it}$, and some latent factors extracted from other observed variables $w_{it}$ (\citealp{BBE05}). As observed by \citet[Section 5.2]{stockwatson16}, such an approach is equivalent to a DFM for $w_{it}$ and $z_{it}$, where both variables are driven by the same common shocks, but the latter has zero idiosyncratic component and unit factor loadings (see Section \ref{app:favar} in the complementary appendix for details). As a consequence, the results of Proposition \ref{var} are directly applicable to IRF estimation in non-stationary FAVAR models. For similar reasons, the results of Proposition \ref{vecm} can be applied to IRF analysis when considering cointegration between the factors and some observed variables, i.e., in the case of a Factor Augmented VECM (FAVECM), see also Section \ref{emp:FGS} below for an application.\footnote{The FAVECM has not to be confused with the FECM proposed by \citet{marcellino3}, where the factors and all the observed variables are assumed to be cointegrated since  the idiosyncratic components are assumed to be $I(0)$.} 
}
\end{rem}
 

\section{Determining the number of factors and shocks}\label{sec:nfactors} 

In the previous section, we made the assumption that $r$, $q$, and $d$ are known. Of course, this is not the case in practice, and we need a method to determine them. Hereafter, for simplicity of notation, we define $\tau=q-d$ the number of common permanent shocks, such that the cointegration rank is $c=r-q+d=r-\tau$.

In light of the results in Lemma \ref{lem:evalcov}, we can determine $r$ by using existing methods based on the behavior of the eigenvalues of the covariance of the variables $\Delta x_{it}$. A non-exhaustive list of possible approaches includes the contributions by \citet{baing02}, \citet{onatski09}, \citet{ABC10}, and \citet{ahnhorenstein13}. 

In order to determine $q$ and $\tau$, we can instead study the spectral density matrix of $\Delta x_{it}$, $\Delta\chi_{it}$ and $\Delta\xi_{it}$, which are defined by
\beq\label{eq:spettro}
\bm\Sigma^{\Delta x}(\theta) = \bm\Sigma^{\Delta\chi}(\theta)+ \bm\Sigma^{\Delta\xi}(\theta)= \frac 1 {2\pi}\bm\Lambda\mbf C(e^{-i\theta}){\mbf C'(e^{i\theta})}\bm\Lambda'+\bm\Sigma^{\Delta\xi}(\theta), \quad \theta\in[-\pi,\pi].
\eeq
It can be shown that Assumptions \ref{ASS:common} through \ref{ASS:idio} imply that the $q$ largest eigenvalues of $\bm\Sigma^{\Delta x}(\theta)$ diverge linearly in $n$, while the remaining $n-q$ stay bounded. This is true at all frequencies but at frequency $\theta=0$, where only the $\tau$ largest eigenvalues of $\bm\Sigma^{\Delta x}(0)$ diverge linearly in $n$ (see Lemma \ref{rem:idiospect} in the complementary appendix for a proof). 
 
The values of $q$ and $\tau$ can, therefore, be determined by analyzing the behavior of the eigenvalues of the spectral density matrix.
In particular, let $\wh{\bm\Gamma}_k$ be the $n\times n$ sample lag $k$ autocovariance matrix of the differenced data $\Delta\mbf y_t$ and consider the lag-window estimator of the spectral density matrix of $\Delta\mbf y_t$:
\[
\wh{\bm\Sigma}^{\Delta y}(\theta) =\frac 1{2\pi} \sum_{k=-B_T}^{B_T} \wh{\bm\Gamma}_k e^{-ik\theta}w(B_T^{-1} k)
\]
where $B_T$ is a suitable bandwidth, and $w(\cdot)$ is a positive even weight function. Let $\wh{\nu}_j(\theta)$ be the  eigenvalues of $\wh{\bm\Sigma}^{\Delta y}(\theta)$. Then, \citet{hallinliska07} define the estimator for $q$ as (see also \citealp{onatski10}, for a similar approach):\footnote{Other methods for determining $q$, not discussed in this paper, are proposed by \citet{amengualwatson07} and \citet{baing07}. Both require knowing $r$ before determining $q$.} 
\begin{align}
\wh{q} &= \arg\!\!\!\!\!\!\!\!\min_{k=0,\ldots,q_{\max}} \bigg[\log\bigg(\frac1 {n(2B_T+1)}\sum_{h=-B_T}^{B_T}\sum_{j=k+1}^{n}\wh{\nu}_j(\theta_h)\bigg)+k s(n,T)\bigg],\label{eq:cucappello}
\end{align}
where $s(n,T)$ is some suitable penalty function, and $q_{\max}$ is a given maximum number of common shocks such that $q< q_{\max}\le n$. Similarly, we introduce the following information criterion for determining $\tau$, based on the behavior of the eigenvalues of the spectral density matrix at zero-frequency:\footnote{Alternative approaches, not discussed in this paper, are: (i) the unit root test for factors by \citet{baing04}, (ii) panel cointegration tests \citep[see, e.g., ][]{GUW15}, and (iii) the classical cointegration tests (see, e.g., \citealp{phillipsouliaris}, and \citealp{Johansen95}).  However, the tests in (i) and (ii)  are designed only for the non-singular case, $r = q$. Likewise, the tests in (iii), which were designed for observed variables,  should be applied to the estimated factors, thus potentially suffering from a pre-estimation error.}
\begin{align}
\wh{\tau} &= \arg\!\!\!\!\!\!\!\!\min_{k=0,\ldots,\tau_{\max}} \bigg[\log\bigg(\frac1 n\sum_{j=k+1}^{n}\wh{\nu}_j(0)\bigg)+k p(n,T)\bigg],\label{eq:taucappello}
\end{align}
where $p(n,T)$ is some suitable penalty functions, and $\tau_{\max}$ is a given maximum number of common trends such that $\tau<\tau_{\max}\le n$. 
We then have the following sufficient conditions for consistently determining $q$ and $\tau$ by means of \eqref{eq:cucappello} and \eqref{eq:taucappello}, respectively (for $\wh q\,$ see also \citealp{hallinliska07}).

\begin{prop}[Number of common permanent shocks]\label{prop:tau} 
Let $\rho_T=(B_T\log B_T T^{-1})^{-1/2}$ and assume that
\begin{inparaenum}[(i)]
\item as $T\to\infty$, $\rho_T\to \infty$ and $\rho_T/T\to 0$;
\item as $n,T\to\infty$, $s(n,T)\to 0$ and $(n\rho_T^{-1}) s(n,T)\to \infty$;
\item as $n,T\to\infty$, $p(n,T)\to 0$ and $(n\rho_T^{-1}) p(n,T)\to \infty$.
\end{inparaenum}
Then, under Assumptions \ref{ASS:common} through \ref{ASS:trend}, as  $n,T\to\infty$, $\mathsf P(\wh{q}=q)\to 1$ and $\mathsf P(\wh{\tau}=\tau)\to 1$.
\end{prop}


Finally, since by definition we have $\tau=r-c$, by virtue of Proposition \ref{prop:tau}, once we determine $\tau$, $q$, and $r$, we immediately have the estimated cointegration rank $\wh c=\wh r-\wh \tau$ and also an estimate of the number of transitory shocks $d$ given by $\wh d=\wh q-\wh\tau$.

\section{Simulations}\label{sec:sim}
The goal of this section is to study the finite sample properties of the IRFs estimators presented in the previous sections. 
We simulate data, from the non-stationary DFM with $r =4$ common factors, $q=3$ common shocks, and $\tau=1$ common permanent shock, thus the cointegration rank of the common factors is $c=r-\tau=3$. More precisely, for any $i=1,\ldots, n$, and $t=1,\ldots, T$ and for given values of $n$ and $T$, each time series follows the data generating process:
\begin{align*}
y_{it} &=b_it + \bm \lambda_i' \mbf F_t + \xi_{it}, \qquad \mbf A(L)\mbf F_t = \mbf K \mbf R \mbf u_t, \qquad \rho_{i}(L)\xi_{it}  = \eps_{it},
\end{align*}
where $\bm \lambda_i$ is $r\times 1$, $\mbf A(L)$ is an $r\times r$ polynomial matrix of degree 2, $\mbf K$ is $r \times q$, and $\mbf R$  is $q \times q$. Details on the way these parameters and the shocks are generated follow.

Starting with the common component, for any $i$ the loadings vector $\bm\lambda_i$ is such that its entries $\lambda_{ij}$ are generated from a $\mathcal{N}(1, 1)$ distribution independently across $i$ and $j=1,\ldots, r$, and for any $t$, the vector of common shocks $\mbf u_t$ is simulated from a $\mathcal{N}(\mbf 0, \mbf I_q)$ distribution, independently across $t$. Then, to generate $\mbf A(L)$ we exploit a particular Smith-McMillan factorization \citep[][]{watson94} according to which $\mbf A(L)=\mathbfcal{U}(L) \mathbfcal{M}(L) \mathbfcal{V}(L)$, where 
$\mathbfcal{M}(L)= \mbox{diag} \left( (1-L)\mbf I_{\tau}, \mbf I_c\right)$, $\mathbfcal{V}(L)=\mbf I_r$, and $\mathbfcal{U}(L)=(\mbf I_r-\mathbfcal{U}_1 L)$, where the diagonal elements of $\mathbfcal{U}_1$ are drawn from a uniform distribution on $[0.5,0.8]$, the off-diagonal elements from a uniform distribution on $[0,0.3]$, and $\mathbfcal{U}_1$ is then rescaled to ensure that its largest eigenvalue is $0.6$. In this way, $\mbf F_t$ follows a VAR(2) with $\tau$ unit roots, or, equivalently, a VECM(1) with $c$ cointegration relations. Finally, the matrix $\mbf K$ is generated as in \citet{baing07}: let $\tilde{\mbf K}$ be a $r \times r$ diagonal matrix of rank $q$ with entries drawn from a uniform distribution on $[.8,1.2]$, and let $\check{\mbf K}$ be a $r \times r$ orthogonal matrix, then, $\mbf K$ is equal to the first $q$ columns of the matrix $\check{\mbf K} \tilde{\mbf K}^{\frac 1 2}$. At each MonteCarlo replication, we draw $\bm\lambda_i$, $\mbf A(L)$, $\mbf u_t$, $\mbf K$, thus simulating the common components $\chi_{it}=\bm\lambda_i'\mbf F_t$ and the IRFs coefficients $\phi_{ijk}$. We then choose $\mbf R$ such that the following restrictions hold for the zero-lag simulated IRFs: $\phi_{12,0}=\phi_{13,0}=\phi_{23,0}=0$.

\begin{table}[t!]\caption{MonteCarlo Simulations - Impulse-Response Functions} \label{table:vecm}
\centering
\textsc{\small Mean Squared Errors - VECM} \\\medskip

\scriptsize
\begin{tabular*}{\textwidth}{@{}@{\extracolsep{\fill}}ccccc|ccccccccc@{}}\hline\hline
&    $T$    &    $n$    &    $\delta$   &   $n_1$   &    $k=0$  &    $k=1$  &    $k=4$  &    $k=8$  &    $k=12$ &    $k=16$     &    $k=20$ &    $k=100$    &    \\\hline
&  100 &  50 & 0.50 &   7 & 0.22 & 0.21 & 0.35 & 0.44 & 0.47 & 0.48 & 0.48 & 0.49 &  \\       
 & 100 &  50 & 0.50 &   7 & 0.11 & 0.11 & 0.20 & 0.26 & 0.28 & 0.29 & 0.30 & 0.31 &  \\       
 & 100 &  50 & 0.75 &  19 & 0.14 & 0.14 & 0.27 & 0.35 & 0.40 & 0.42 & 0.44 & 0.47 &  \\       
 & 100 &  50 & 0.85 &  28 & 0.16 & 0.16 & 0.29 & 0.41 & 0.47 & 0.51 & 0.53 & 0.57 &  \\       
 & 100 &  50 & 0.95 &  41 & 0.15 & 0.17 & 0.31 & 0.43 & 0.50 & 0.54 & 0.57 & 0.61 &  \\       
 & 100 &  50 & 1.00 &  50 & 0.15 & 0.18 & 0.33 & 0.46 & 0.54 & 0.58 & 0.60 & 0.64 &  \\ \hline
 & 100 &  75 & 0.50 &   9 & 0.09 & 0.10 & 0.18 & 0.22 & 0.23 & 0.24 & 0.24 & 0.24 &  \\       
 & 100 &  75 & 0.75 &  25 & 0.11 & 0.12 & 0.22 & 0.29 & 0.32 & 0.34 & 0.35 & 0.37 &  \\       
 & 100 &  75 & 0.85 &  39 & 0.11 & 0.12 & 0.22 & 0.32 & 0.37 & 0.41 & 0.42 & 0.45 &  \\       
 & 100 &  75 & 0.95 &  60 & 0.09 & 0.11 & 0.24 & 0.36 & 0.43 & 0.47 & 0.50 & 0.53 &  \\       
 & 100 &  75 & 1.00 &  75 & 0.09 & 0.12 & 0.25 & 0.38 & 0.46 & 0.51 & 0.53 & 0.57 &  \\ \hline
 & 100 & 100 & 0.50 &  10 & 0.09 & 0.10 & 0.17 & 0.21 & 0.22 & 0.22 & 0.22 & 0.23 &  \\       
 & 100 & 100 & 0.75 &  32 & 0.09 & 0.10 & 0.20 & 0.27 & 0.30 & 0.32 & 0.33 & 0.34 &  \\       
 & 100 & 100 & 0.85 &  50 & 0.10 & 0.11 & 0.22 & 0.31 & 0.36 & 0.39 & 0.41 & 0.44 &  \\       
 & 100 & 100 & 0.95 &  79 & 0.09 & 0.11 & 0.22 & 0.33 & 0.41 & 0.45 & 0.47 & 0.51 &  \\       
 & 100 & 100 & 1.00 & 100 & 0.08 & 0.11 & 0.22 & 0.34 & 0.41 & 0.45 & 0.46 & 0.49 &  \\ \hline
 & 200 & 200 & 0.50 &  14 & 0.04 & 0.04 & 0.07 & 0.09 & 0.09 & 0.10 & 0.10 & 0.10 &  \\       
 & 200 & 200 & 0.75 &  53 & 0.03 & 0.04 & 0.07 & 0.10 & 0.11 & 0.12 & 0.13 & 0.15 &  \\       
 & 200 & 200 & 0.85 &  90 & 0.03 & 0.04 & 0.08 & 0.11 & 0.14 & 0.17 & 0.18 & 0.24 &  \\       
 & 200 & 200 & 0.95 & 153 & 0.03 & 0.04 & 0.09 & 0.15 & 0.19 & 0.23 & 0.26 & 0.38 &  \\       
 & 200 & 200 & 1.00 & 200 & 0.03 & 0.04 & 0.10 & 0.16 & 0.21 & 0.25 & 0.28 & 0.40 &  \\ \hline
 & 300 & 300 & 0.50 &  17 & 0.02 & 0.02 & 0.04 & 0.05 & 0.06 & 0.06 & 0.06 & 0.06 &  \\       
 & 300 & 300 & 0.75 &  72 & 0.02 & 0.02 & 0.05 & 0.06 & 0.07 & 0.08 & 0.08 & 0.11 &  \\       
 & 300 & 300 & 0.85 & 128 & 0.02 & 0.03 & 0.05 & 0.07 & 0.09 & 0.10 & 0.12 & 0.18 &  \\       
 & 300 & 300 & 0.95 & 226 & 0.02 & 0.03 & 0.06 & 0.09 & 0.12 & 0.15 & 0.18 & 0.33 &  \\       
 & 300 & 300 & 1.00 & 300 & 0.02 & 0.03 & 0.06 & 0.10 & 0.13 & 0.16 & 0.19 & 0.36 &  \\ \hline
\end{tabular*}
\begin{tabular}{@{}p{\textwidth}}
\scriptsize  
MSE for the estimated IRFs by fitting a VECM on $\wh{\mbf F}_t$ as in \eqref{VECMhatF}. $T$ is the number of observations, $n$ is the number of variables, and $n_1=\lceil n^\delta\rceil $ is the number of $I(1)$ idiosyncratic components. In these simulations there are $n_b=\lceil n^\eta\rceil$ variables with a deterministic linear trend, with $\eta=\delta$ or equivalently $n_b=n_1$.
\end{tabular}
\end{table}

Turning to the idiosyncratic components, the vector of idiosyncratic shocks $\bm\varepsilon_t=(\eps_{1t}\cdots \eps_{nt})'$ is simulated from a $\mathcal{N}(\mbf 0, \bm\Gamma^{\varepsilon})$ distribution, independently across $t$, and with the $(i,j)$th entry of $\bm\Gamma^{\varepsilon}$ given by $\Cov(\eps_{it},\eps_{jt}) = 0.5^{|i-j|}$. Therefore, we allow for cross-correlation among the idiosyncratic shocks. Note that the amount of cross-correlation that we allow for is higher than most simulation exercises available in the literature \citep[e.g.,][]{baing04}. For each MonteCarlo replication, we allow for $n_1=\lceil n^\delta\rceil$ randomly selected idiosyncratic components with a unit root. In particular, each $\xi_{it}$ follows an AR(2) where the first root of the polynomial $\rho_i(L)$ is either 0 or 1 depending on whether $\xi_{it}\sim I(0)$ or $I(1)$, while the second root is drawn from a uniform distribution on ${[0,0.5]}$. Moreover, for each MonteCarlo replication, we allow for $n_b=\lceil n^\eta\rceil$  randomly selected variables with a deterministic linear trend having slope $b_i$ drawn from a uniform distribution on ${[0.3,0.5]}$. In all replications, the first variable $y_{1t}$ is simulated with a deterministic linear trend and an $I(1)$ idiosyncratic component. Finally, each idiosyncratic component $\xi_{it}$ is rescaled so that $\Delta\xi_{it}$ accounts for 40\% of the variance of the corresponding $\Delta x_{it}$. 

For each MonteCarlo replication, the DFM is estimated as explained in Section \ref{sec:est}. Specifically, the factors are estimated as in \eqref{parlacondudu}, while the IRFs are estimated either by fitting a VECM on $\Delta\wh{\mbf F}_t$ as in \eqref{VECMhatF}, or a VAR on $\wh{\mbf F}_t$ as in \eqref{eq:fattoriVARinlivelli}. The numbers $r$, $q$, and $\tau$ are assumed to be known. Furthermore, we assume to know which are the $n_b$ variables with a linear trend, which is therefore removed by mean of least squares regression as indicated in \eqref{eq:biols}.

All results are based on 2000 MonteCarlo replications, and we consider Mean Squared Errors (MSE) of estimated IRFs averaged across all series, all shocks and all replications. We consider different cross-sectional and sample sizes ($n$ and $T$), different numbers of $I(1)$ idiosyncratic components ($n_1$), and of variables with deterministic trend ($n_b$), and for simplicity, we let $n_1=n_b$.\footnote{In the complementary appendix, we provide also results for the IRF of the first series to the first shock only, $\phi_{11}(L)$, and for other values of $n$, $T$, $n_1$ and $n_b$.} 

\begin{table}[t!]\caption{MonteCarlo Simulations - Impulse-Response Functions} \label{table:var}
\centering
\textsc{\small Mean Squared Errors - Unrestricted VAR in Levels} \\\medskip

\scriptsize
\begin{tabular*}{\textwidth}{@{}@{\extracolsep{\fill}}ccccc|ccccccccc@{}}\hline\hline
&    $T$    &    $n$    &    $\delta$   &   $n_1$   &    $k=0$  &    $k=1$  &    $k=4$  &    $k=8$  &    $k=12$ &    $k=16$     &    $k=20$ &    $k=100$    &    \\\hline
 & 100 &  50 & 0.50 &   7 & 0.11 & 0.11 & 0.19 & 0.29 & 0.38 & 0.45 & 0.51 & 0.74 &  \\       
 & 100 &  50 & 0.75 &  19 & 0.14 & 0.14 & 0.25 & 0.36 & 0.45 & 0.52 & 0.57 & 0.75 &  \\       
 & 100 &  50 & 0.85 &  28 & 0.15 & 0.15 & 0.28 & 0.40 & 0.50 & 0.58 & 0.64 & 0.77 &  \\       
 & 100 &  50 & 0.95 &  41 & 0.14 & 0.16 & 0.29 & 0.41 & 0.51 & 0.58 & 0.63 & 0.75 &  \\       
 & 100 &  50 & 1.00 &  50 & 0.15 & 0.17 & 0.31 & 0.43 & 0.53 & 0.59 & 0.64 & 0.77 &  \\ \hline
 & 100 &  75 & 0.50 &   9 & 0.09 & 0.10 & 0.18 & 0.27 & 0.36 & 0.43 & 0.49 & 0.76 &  \\       
 & 100 &  75 & 0.75 &  25 & 0.11 & 0.12 & 0.21 & 0.32 & 0.42 & 0.50 & 0.56 & 0.76 &  \\       
 & 100 &  75 & 0.85 &  39 & 0.11 & 0.12 & 0.22 & 0.34 & 0.45 & 0.53 & 0.59 & 0.76 &  \\       
 & 100 &  75 & 0.95 &  60 & 0.09 & 0.11 & 0.23 & 0.37 & 0.48 & 0.56 & 0.62 & 0.76 &  \\       
 & 100 &  75 & 1.00 &  75 & 0.09 & 0.12 & 0.24 & 0.39 & 0.50 & 0.58 & 0.64 & 0.75 &  \\ \hline
 & 100 & 100 & 0.50 &  10 & 0.09 & 0.10 & 0.17 & 0.26 & 0.35 & 0.42 & 0.48 & 0.75 &  \\       
 & 100 & 100 & 0.75 &  32 & 0.09 & 0.10 & 0.20 & 0.31 & 0.41 & 0.49 & 0.55 & 0.77 &  \\       
 & 100 & 100 & 0.85 &  50 & 0.10 & 0.11 & 0.21 & 0.33 & 0.43 & 0.51 & 0.57 & 0.75 &  \\       
 & 100 & 100 & 0.95 &  79 & 0.09 & 0.10 & 0.21 & 0.35 & 0.47 & 0.57 & 0.63 & 0.76 &  \\       
 & 100 & 100 & 1.00 & 100 & 0.08 & 0.11 & 0.21 & 0.36 & 0.46 & 0.54 & 0.60 & 0.73 &  \\ \hline
 & 200 & 200 & 0.50 &  14 & 0.04 & 0.04 & 0.07 & 0.12 & 0.17 & 0.21 & 0.26 & 0.68 &  \\       
 & 200 & 200 & 0.75 &  53 & 0.03 & 0.04 & 0.08 & 0.13 & 0.18 & 0.24 & 0.30 & 0.71 &  \\       
 & 200 & 200 & 0.85 &  90 & 0.03 & 0.04 & 0.08 & 0.14 & 0.20 & 0.26 & 0.32 & 0.72 &  \\       
 & 200 & 200 & 0.95 & 153 & 0.03 & 0.04 & 0.09 & 0.16 & 0.23 & 0.30 & 0.37 & 0.74 &  \\       
 & 200 & 200 & 1.00 & 200 & 0.03 & 0.04 & 0.10 & 0.17 & 0.24 & 0.31 & 0.37 & 0.72 &  \\ \hline
 & 300 & 300 & 0.50 &  17 & 0.02 & 0.02 & 0.04 & 0.07 & 0.10 & 0.13 & 0.16 & 0.58 &  \\       
 & 300 & 300 & 0.75 &  72 & 0.02 & 0.02 & 0.05 & 0.08 & 0.11 & 0.15 & 0.18 & 0.61 &  \\       
 & 300 & 300 & 0.85 & 128 & 0.02 & 0.03 & 0.05 & 0.09 & 0.13 & 0.17 & 0.21 & 0.67 &  \\       
 & 300 & 300 & 0.95 & 226 & 0.02 & 0.03 & 0.06 & 0.10 & 0.15 & 0.19 & 0.24 & 0.69 &  \\       
 & 300 & 300 & 1.00 & 300 & 0.02 & 0.03 & 0.07 & 0.10 & 0.15 & 0.20 & 0.25 & 0.69 &  \\ \hline
\end{tabular*}
\begin{tabular}{@{}p{\textwidth}}
\scriptsize  
\scriptsize  
MSE for the estimated IRFs by fitting an unrestricted VAR on $\wh{\mbf F}_t$ as in \eqref{eq:fattoriVARinlivelli}. $T$ is the number of observations, $n$ is the number of variables, and $n_1=\lceil n^\delta\rceil$ is the number of $I(1)$ idiosyncratic components.  In these simulations there are $n_b=\lceil n^\eta\rceil$ variables with a deterministic linear trend, with $\eta=\delta$ or equivalently $n_b=n_1$.
\end{tabular}
\end{table}

Table \ref{table:vecm} shows MSEs for the estimated IRFs when using a VECM. In agreement with the predictions of Proposition \ref{vecm}, four main features emerge: (i) the MSEs decrease monotonically as $n$ and $T$ grow; (ii) the MSEs are larger at longer horizons, and also, as $n$ and $T$ get larger,  at long horizons they decrease less than at short horizons; (iii) the MSEs are inversely related to the number of non-stationary idiosyncratic components, and for given $n$ and $T$ at long horizons are smaller for smaller values of $\delta$;  (iv) the MSEs are quite substantial when $n=50$ regardless of the horizon and of $\delta$, thus indicating that a large number of variables is needed to estimate the model sufficiently well.

Table \ref{table:var} shows MSE for the estimated IRFs when using an unrestricted VAR in levels. At short horizons, the MSEs are comparable to those of the VECM case, whereas, at long horizons, the MSEs are larger than in the VECM case. This result is in accordance with Proposition \ref{var} according to which the long-run IRFs estimated by fitting an unrestricted VAR in levels on the estimated factors are not consistent.

\begin{table}[t!]\caption{MonteCarlo Simulations - Impulse-Response Functions} \label{table:varD}
\centering
\textsc{\small Mean Squared Errors relative to VAR in Differences - VECM} \\\medskip

\scriptsize
\begin{tabular*}{\textwidth}{@{}@{\extracolsep{\fill}}ccccc|ccccccccc@{}}\hline\hline
&    $T$    &    $n$    &    $\delta$   &   $n_1$   &    $k=0$  &    $k=1$  &    $k=4$  &    $k=8$  &    $k=12$ &    $k=16$     &    $k=20$ &    $k=100$    &    \\\hline
& 100 &  50 & 0.50 &   7 & 1.07 & 0.66 & 0.41 & 0.44 & 0.47 & 0.49 & 0.50 & 0.52 &  \\       
 & 100 &  50 & 0.75 &  19 & 0.93 & 0.67 & 0.49 & 0.55 & 0.61 & 0.65 & 0.67 & 0.71 &  \\       
 & 100 &  50 & 0.85 &  28 & 1.23 & 0.78 & 0.54 & 0.64 & 0.72 & 0.77 & 0.81 & 0.87 &  \\       
 & 100 &  50 & 0.95 &  41 & 0.94 & 0.71 & 0.55 & 0.65 & 0.75 & 0.81 & 0.85 & 0.91 &  \\       
 & 100 &  50 & 1.00 &  50 & 1.07 & 0.77 & 0.61 & 0.72 & 0.83 & 0.89 & 0.93 & 0.99 &  \\ \hline
 & 100 &  75 & 0.50 &   9 & 0.95 & 0.60 & 0.36 & 0.38 & 0.39 & 0.40 & 0.41 & 0.41 &  \\       
 & 100 &  75 & 0.75 &  25 & 1.03 & 0.69 & 0.44 & 0.49 & 0.54 & 0.57 & 0.59 & 0.62 &  \\       
 & 100 &  75 & 0.85 &  39 & 1.04 & 0.68 & 0.45 & 0.55 & 0.63 & 0.69 & 0.72 & 0.76 &  \\       
 & 100 &  75 & 0.95 &  60 & 1.04 & 0.68 & 0.49 & 0.63 & 0.74 & 0.81 & 0.85 & 0.92 &  \\       
 & 100 &  75 & 1.00 &  75 & 1.03 & 0.70 & 0.52 & 0.67 & 0.79 & 0.87 & 0.91 & 0.97 &  \\ \hline
 & 100 & 100 & 0.50 &  10 & 0.99 & 0.62 & 0.36 & 0.37 & 0.38 & 0.38 & 0.38 & 0.39 &  \\       
 & 100 & 100 & 0.75 &  32 & 0.96 & 0.61 & 0.40 & 0.45 & 0.49 & 0.52 & 0.54 & 0.56 &  \\       
 & 100 & 100 & 0.85 &  50 & 1.01 & 0.66 & 0.45 & 0.54 & 0.61 & 0.66 & 0.69 & 0.75 &  \\       
 & 100 & 100 & 0.95 &  79 & 1.02 & 0.66 & 0.45 & 0.58 & 0.70 & 0.77 & 0.81 & 0.87 &  \\       
 & 100 & 100 & 1.00 & 100 & 1.02 & 0.68 & 0.47 & 0.61 & 0.72 & 0.78 & 0.82 & 0.86 &  \\ \hline
 & 200 & 200 & 0.50 &  14 & 0.94 & 0.43 & 0.19 & 0.19 & 0.20 & 0.20 & 0.20 & 0.20 &  \\       
 & 200 & 200 & 0.75 &  53 & 0.94 & 0.45 & 0.20 & 0.22 & 0.24 & 0.26 & 0.28 & 0.32 &  \\       
 & 200 & 200 & 0.85 &  90 & 0.95 & 0.45 & 0.21 & 0.25 & 0.30 & 0.35 & 0.39 & 0.51 &  \\       
 & 200 & 200 & 0.95 & 153 & 0.95 & 0.48 & 0.24 & 0.31 & 0.40 & 0.48 & 0.54 & 0.78 &  \\       
 & 200 & 200 & 1.00 & 200 & 0.95 & 0.48 & 0.26 & 0.34 & 0.44 & 0.52 & 0.59 & 0.84 &  \\ \hline
 & 300 & 300 & 0.50 &  17 & 0.90 & 0.32 & 0.12 & 0.12 & 0.13 & 0.13 & 0.13 & 0.13 &  \\       
 & 300 & 300 & 0.75 &  72 & 0.91 & 0.33 & 0.13 & 0.14 & 0.15 & 0.17 & 0.18 & 0.23 &  \\       
 & 300 & 300 & 0.85 & 128 & 0.91 & 0.36 & 0.15 & 0.17 & 0.20 & 0.23 & 0.25 & 0.40 &  \\       
 & 300 & 300 & 0.95 & 226 & 0.91 & 0.39 & 0.18 & 0.21 & 0.27 & 0.33 & 0.39 & 0.73 &  \\       
 & 300 & 300 & 1.00 & 300 & 0.92 & 0.40 & 0.18 & 0.22 & 0.28 & 0.35 & 0.41 & 0.79 &  \\ \hline
\end{tabular*}
\begin{tabular}{@{}p{\textwidth}}
\scriptsize  
\scriptsize  
Ratio between the MSE for the estimated IRFs obtained by fitting a VECM on $\wh{\mbf F}_t$ as in \eqref{VECMhatF}, and the MSE for the estimated and cumulated IRFs obtained by estimating a VAR on $\Delta\widetilde{\mbf F}_t$ as in \citet{FGLR09}. Values smaller than one indicate a better performance of our method.
$T$ is the number of observations, $n$ is the number of variables, and $n_1=\lceil n^\delta\rceil$ is the number of $I(1)$ idiosyncratic components. In these simulations there are $n_b=\lceil n^\eta\rceil$ variables with a deterministic linear trend, with $\eta=\delta$ or equivalently $n_b=n_1$.
\end{tabular}
\end{table}

\begin{table}[t!]\caption{MonteCarlo Simulations - Impulse-Response Functions} \label{table:vecmII}
\centering
\textsc{\small Mean Squared Errors relative to \citet{baing04} - VECM} \\\medskip

\scriptsize
\begin{tabular*}{\textwidth}{@{}@{\extracolsep{\fill}}ccccc|ccccccccc@{}}\hline\hline
&    $T$    &    $n$    &    $\delta$   &   $n_1$   &    $k=0$  &    $k=1$  &    $k=4$  &    $k=8$  &    $k=12$ &    $k=16$     &    $k=20$ &    $k=100$    &    \\\hline
& 100 &  50 & 0.50 &   7 & 0.97 & 0.94 & 0.87 & 0.85 & 0.84 & 0.83 & 0.83 & 0.81 &  \\       
 & 100 &  50 & 0.75 &  19 & 0.93 & 0.91 & 0.89 & 0.90 & 0.90 & 0.90 & 0.89 & 0.86 &  \\       
 & 100 &  50 & 0.85 &  28 & 1.04 & 0.96 & 0.89 & 0.93 & 0.95 & 0.96 & 0.96 & 0.91 &  \\       
 & 100 &  50 & 0.95 &  41 & 0.85 & 0.84 & 0.89 & 0.94 & 0.96 & 0.96 & 0.94 & 0.88 &  \\       
 & 100 &  50 & 1.00 &  50 & 1.08 & 0.99 & 0.99 & 1.03 & 1.03 & 1.01 & 0.99 & 0.90 &  \\ \hline
 & 100 &  75 & 0.50 &   9 & 0.97 & 0.91 & 0.86 & 0.82 & 0.79 & 0.77 & 0.76 & 0.75 &  \\       
 & 100 &  75 & 0.75 &  25 & 0.99 & 0.99 & 0.95 & 0.93 & 0.92 & 0.89 & 0.88 & 0.83 &  \\       
 & 100 &  75 & 0.85 &  39 & 1.03 & 0.97 & 0.90 & 0.93 & 0.94 & 0.93 & 0.91 & 0.85 &  \\       
 & 100 &  75 & 0.95 &  60 & 1.01 & 0.94 & 0.93 & 1.01 & 1.03 & 1.02 & 1.00 & 0.92 &  \\       
 & 100 &  75 & 1.00 &  75 & 1.02 & 0.95 & 0.96 & 1.03 & 1.03 & 1.01 & 0.99 & 0.89 &  \\ \hline
 & 100 & 100 & 0.50 &  10 & 1.00 & 0.95 & 0.85 & 0.79 & 0.75 & 0.73 & 0.72 & 0.70 &  \\       
 & 100 & 100 & 0.75 &  32 & 0.96 & 0.94 & 0.89 & 0.85 & 0.83 & 0.82 & 0.81 & 0.78 &  \\       
 & 100 & 100 & 0.85 &  50 & 1.04 & 0.98 & 0.94 & 0.96 & 0.95 & 0.93 & 0.92 & 0.86 &  \\       
 & 100 & 100 & 0.95 &  79 & 1.03 & 0.95 & 0.93 & 0.99 & 1.00 & 0.99 & 0.96 & 0.88 &  \\       
 & 100 & 100 & 1.00 & 100 & 1.00 & 0.94 & 0.93 & 0.99 & 0.99 & 0.97 & 0.94 & 0.85 &  \\ \hline
 & 200 & 200 & 0.50 &  14 & 0.98 & 0.91 & 0.80 & 0.78 & 0.74 & 0.71 & 0.70 & 0.66 &  \\       
 & 200 & 200 & 0.75 &  53 & 0.99 & 0.90 & 0.77 & 0.77 & 0.75 & 0.73 & 0.71 & 0.65 &  \\       
 & 200 & 200 & 0.85 &  90 & 0.99 & 0.88 & 0.80 & 0.86 & 0.88 & 0.88 & 0.88 & 0.79 &  \\       
 & 200 & 200 & 0.95 & 153 & 0.99 & 0.89 & 0.82 & 0.93 & 0.99 & 1.01 & 1.02 & 0.89 &  \\       
 & 200 & 200 & 1.00 & 200 & 0.98 & 0.87 & 0.82 & 0.95 & 1.02 & 1.05 & 1.05 & 0.90 &  \\ \hline
 & 300 & 300 & 0.50 &  17 & 0.98 & 0.87 & 0.75 & 0.75 & 0.72 & 0.70 & 0.68 & 0.63 &  \\       
 & 300 & 300 & 0.75 &  72 & 0.98 & 0.85 & 0.73 & 0.76 & 0.76 & 0.75 & 0.73 & 0.65 &  \\       
 & 300 & 300 & 0.85 & 128 & 0.98 & 0.85 & 0.75 & 0.80 & 0.83 & 0.84 & 0.85 & 0.78 &  \\       
 & 300 & 300 & 0.95 & 226 & 0.98 & 0.85 & 0.78 & 0.91 & 0.99 & 1.04 & 1.06 & 0.95 &  \\       
 & 300 & 300 & 1.00 & 300 & 0.98 & 0.83 & 0.75 & 0.90 & 1.00 & 1.05 & 1.08 & 0.96 &  \\ \hline
\end{tabular*}
\begin{tabular}{@{}p{\textwidth}}
\scriptsize  
Ratio between the MSE for the estimated IRFs obtained by fitting a VECM on $\wh{\mbf F}_t$ as in \eqref{VECMhatF}, and the MSE for the estimated IRFs obtained by fitting a VECM on the common factors estimated as in \citet{baing04}. Values smaller than one indicate a better performance of our method.
$T$ is the number of observations, $n$ is the number of variables, and $n_1=\lceil n^\delta\rceil$ is the number of $I(1)$ idiosyncratic components. In these simulations there are $n_b=\lceil n^\eta\rceil$ variables with a deterministic linear trend, with $\eta=\delta$ or equivalently $n_b=n_1$.
\end{tabular}
\end{table}

\begin{table}[t!]\caption{MonteCarlo Simulations - Impulse-Response Functions} \label{table:varII}
\centering
\textsc{\small Mean Squared Errors relative to \citet{baing04} - Unrestricted VAR in Levels} \\\medskip

\scriptsize
\begin{tabular*}{\textwidth}{@{}@{\extracolsep{\fill}}ccccc|ccccccccc@{}}\hline\hline
&    $T$    &    $n$    &    $\delta$   &   $n_1$   &    $k=0$  &    $k=1$  &    $k=4$  &    $k=8$  &    $k=12$ &    $k=16$     &    $k=20$ &    $k=100$    &    \\\hline
 & 100 &  50 & 0.50 &   7 & 0.98 & 0.97 & 0.94 & 0.94 & 0.95 & 0.96 & 0.96 & 1.00 &  \\       
 & 100 &  50 & 0.75 &  19 & 0.92 & 0.94 & 0.95 & 0.99 & 1.01 & 1.01 & 1.02 & 1.00 &  \\       
 & 100 &  50 & 0.85 &  28 & 1.06 & 1.01 & 0.96 & 0.99 & 1.02 & 1.03 & 1.03 & 1.00 &  \\       
 & 100 &  50 & 0.95 &  41 & 0.83 & 0.83 & 0.90 & 0.99 & 1.02 & 1.02 & 1.02 & 1.00 &  \\       
 & 100 &  50 & 1.00 &  50 & 1.02 & 0.96 & 0.97 & 1.01 & 1.02 & 1.02 & 1.01 & 1.01 &  \\ \hline
 & 100 &  75 & 0.50 &   9 & 1.01 & 0.97 & 0.94 & 0.94 & 0.95 & 0.96 & 0.97 & 0.99 &  \\       
 & 100 &  75 & 0.75 &  25 & 1.05 & 1.04 & 0.98 & 0.98 & 1.00 & 1.00 & 1.00 & 0.99 &  \\       
 & 100 &  75 & 0.85 &  39 & 0.97 & 0.95 & 0.95 & 1.02 & 1.04 & 1.05 & 1.05 & 1.00 &  \\       
 & 100 &  75 & 0.95 &  60 & 1.01 & 0.97 & 0.98 & 1.04 & 1.05 & 1.06 & 1.05 & 1.01 &  \\       
 & 100 &  75 & 1.00 &  75 & 1.01 & 0.97 & 0.98 & 1.04 & 1.06 & 1.06 & 1.04 & 1.00 &  \\ \hline
 & 100 & 100 & 0.50 &  10 & 1.00 & 0.97 & 0.94 & 0.94 & 0.95 & 0.95 & 0.96 & 0.99 &  \\       
 & 100 & 100 & 0.75 &  32 & 0.98 & 0.97 & 0.95 & 0.97 & 0.99 & 1.00 & 1.01 & 1.00 &  \\       
 & 100 & 100 & 0.85 &  50 & 1.01 & 1.00 & 0.98 & 1.01 & 1.03 & 1.04 & 1.04 & 1.00 &  \\       
 & 100 & 100 & 0.95 &  79 & 1.02 & 0.98 & 0.98 & 1.04 & 1.07 & 1.07 & 1.06 & 1.00 &  \\       
 & 100 & 100 & 1.00 & 100 & 0.99 & 0.96 & 0.97 & 1.04 & 1.06 & 1.06 & 1.04 & 1.00 &  \\ \hline
 & 200 & 200 & 0.50 &  14 & 0.99 & 0.95 & 0.90 & 0.92 & 0.94 & 0.95 & 0.96 & 0.99 &  \\       
 & 200 & 200 & 0.75 &  53 & 1.00 & 0.96 & 0.91 & 0.94 & 0.97 & 0.99 & 1.00 & 1.00 &  \\       
 & 200 & 200 & 0.85 &  90 & 0.99 & 0.94 & 0.90 & 0.96 & 1.01 & 1.04 & 1.06 & 1.01 &  \\       
 & 200 & 200 & 0.95 & 153 & 1.00 & 0.95 & 0.93 & 1.03 & 1.09 & 1.11 & 1.12 & 0.99 &  \\       
 & 200 & 200 & 1.00 & 200 & 0.99 & 0.93 & 0.92 & 1.02 & 1.08 & 1.10 & 1.11 & 0.98 &  \\ \hline
 & 300 & 300 & 0.50 &  17 & 0.99 & 0.92 & 0.86 & 0.89 & 0.90 & 0.92 & 0.93 & 0.99 &  \\       
 & 300 & 300 & 0.75 &  72 & 0.99 & 0.92 & 0.87 & 0.93 & 0.96 & 0.98 & 1.00 & 1.01 &  \\       
 & 300 & 300 & 0.85 & 128 & 0.99 & 0.93 & 0.89 & 0.96 & 1.00 & 1.04 & 1.06 & 1.03 &  \\       
 & 300 & 300 & 0.95 & 226 & 0.99 & 0.93 & 0.89 & 0.98 & 1.05 & 1.09 & 1.11 & 1.03 &  \\       
 & 300 & 300 & 1.00 & 300 & 0.99 & 0.90 & 0.86 & 0.99 & 1.07 & 1.12 & 1.15 & 1.01 &  \\ \hline
\end{tabular*}
\begin{tabular}{@{}p{\textwidth}}
\scriptsize  
Ratio between the MSE for the estimated IRFs by fitting an unrestricted VAR on $\wh{\mbf F}_t$ as in \eqref{eq:fattoriVARinlivelli}, and the MSE for the estimated IRFs obtained by fitting an unrestricted VAR in levels on the common factors estimated as in \citet{baing04}. Values smaller than one indicate a better performance of our method. $T$ is the number of observations, $n$ is the number of variables, and $n_1=\lceil n^\delta\rceil$ is the number of $I(1)$ idiosyncratic components.  In these simulations there are $n_b=\lceil n^\eta\rceil$ variables with a deterministic linear trend, with $\eta=\delta$ or equivalently $n_b=n_1$.
\end{tabular}
\end{table}

In Table \ref{table:varD}, we show the MSEs of the VECM approach relative to the stationary approach where the factors, $\Delta\widetilde{\mbf F}_t$ are estimated by principal component analysis on differenced data, as in \citet{baing02}, and the IRFs are computed from a VAR on $\Delta\widetilde{\mbf F}_t$, as in \citet{FGLR09}. This approach is equivalent to saying that we are imposing the existence of $q$ unit roots when estimating the model, as opposed to the $\tau$ assumed in generating the factors. Results clearly show that this approach produces worse estimators of the IRFs than our approach (values less than one in the table).

Tables \ref{table:vecmII} and \ref{table:varII} present the MSEs relative to the case in which the factors are estimated as suggested by \citet{baing04}.  As explained in Section \ref{sec:detrend}, the difference between the \citet{baing04} procedure and ours depends on the way we de-trend data. One main conclusion can be drawn from these tables: while at short horizons, the two approaches are essentially equivalent in terms of MSE, at longer horizons, our procedure performs better (values less than one in the tables), and this is true both for the VECM case and for the unrestricted VAR in levels case.

To conclude, we use  the same data generating process considered above to study the performance of the information criterion \eqref{eq:taucappello}, proposed in Section \ref{sec:nfactors} for determining $\tau$. Table \ref{tab:trend} shows the percentage of times in which we estimate the number of common permanent shocks $\tau=1$ correctly. For the sake of comparison, we also report results for the information criterion \eqref{eq:cucappello}, proposed by \citet{hallinliska07}, for estimating $q=3$. 
Results show that for $n\ge 100$  our criterion works fairly well by giving the correct answer more than 90\% of the times, in most of the configurations of the parameters considered.\footnote{Other results are in the complementary appendix. Note also that the actual implementation of these criteria requires a procedure of fine-tuning of the penalty. Indeed, for any constant $c>0$, the functions $c\, s(n,T)$ and $c\, p(n,T)$ are also admissible penalties, and, therefore, a whole range of values of $c$ has to be explored, see \citet{hallinliska07} for details.}

\begin{table}[t!]\caption{MonteCarlo Simulations - Number of Common Shocks}\label{tab:trend}
\centering 
\textsc{\small Percentages of Correct Answers}\\[.3cm]
\scriptsize
\begin{tabular*}{\textwidth}{@{}@{\extracolsep{\fill}}ccccccc|cccccc@{}}\hline\hline
&   $T$ &   $n$ &   $n_1$   &   $\wh{\tau}=\tau$    &   $\wh{q}=q$  &   &   $T$ &   $n$ &   $n_1$   &   $\wh{\tau}=\tau$    &   $\wh{q}=q$  &   
\\  \hline
&   100 &   50  &   7   &   93.3    &   60.6    &   &   100 &   100 &   10  &   82.2    &   96.4    &   \\
&   100 &   50  &   19  &   98.4    &   61.0    &   &   100 &   100 &   32  &   96.6    &   95.9    &   \\
&   100 &   50  &   28  &   98.1    &   64.2    &   &   100 &   100 &   50  &   99.3    &   95.9    &   \\
&   100 &   50  &   41  &   97.0    &   71.1    &   &   100 &   100 &   79  &   99.5    &   98.0    &   \\
&   100 &   50  &   50  &   96.3    &   84.4    &   &   100 &   100 &   100 &   99.0    &   99.3    &   \\\hline
&   100 &   75  &   9   &   89.1    &   86.5    &   &   200 &   200 &   14  &   70.5    &   100.0   &   \\
&   100 &   75  &   25  &   98.2    &   87.3    &   &   200 &   200 &   53  &   93.0    &   100.0   &   \\
&   100 &   75  &   39  &   99.3    &   86.7    &   &   200 &   200 &   90  &   98.5    &   100.0   &   \\
&   100 &   75  &   60  &   99.0    &   92.4    &   &   200 &   200 &   153 &   99.9    &   100.0   &   \\
&   100 &   75  &   75  &   98.2    &   95.8    &   &   200 &   200 &   200 &   100.0   &   100.0   &   \\\hline
\end{tabular*}
\begin{tabular}{@{}p{\textwidth}}
\scriptsize  Percentage of cases in which the information criteria \eqref{eq:cucappello} and \eqref{eq:taucappello} returned the correct number of all common shocks ($\wh{q}=q$) and of common permanent shocks ($\wh{\tau}=\tau$). $T$ is the number of observations, $n$ is the number of variables, and $n_1=\lceil n^\delta\rceil$ is the number of $I(1)$ idiosyncratic components. In these simulations there are $n_b=\lceil n^\eta\rceil$ variables with a deterministic linear trend, with $\eta=\delta$ or equivalently $n_b=n_1$.
\end{tabular}
\end{table}


\section{Empirical applications}\label{sec:emp}
In this section, we evaluate the practical usefulness of our methodology by considering two different empirical applications. In the first one, we estimate the effects of an oil price shock on the US economy by means of our non-stationary DFM, and we compare our results with those in \citet{stockwatson16}, who instead use a stationary DFM. In the second one, we estimate the effects of news shocks on the US business cycle by means of our non-stationary DFM, and we compare our results with those in \citet{fornigambettisala}, who instead use a FAVAR with factors extracted from the variables in levels as in \citet{bai04}.

\subsection{Application 1: the effect of oil price shocks}
Quantifying the effects of unexpected oil price changes on the US economy has been a question of particular interest ever since the oil price shocks of the 1970s. Starting with the seminal paper of \citet{hamilton83}, the majority of the papers has addressed this issue using SVAR models \citep[e.g.,][among others]{barskyK02,Kilian08,Kilian09,BG07}; however, a number of them have used DFMs \citep[e.g.,][]{aastveit14,anetal2014,juvenalpetrella2015}. The main conclusion of this literature is that oil price shocks have a significant effect. 

\citet{stockwatson16} (henceforth SW) consider a panel of 207 quarterly US macroeconomic time series from 1985:Q1 to 2014:Q4 to estimate the effects of an oil price shock on the US economy. In particular, they use a stationary DFM, where all non-stationary variables are differentiated, and the IRFs are estimated by cumulating the IRFs obtained from a VAR on the differenced factors. Specifically, SW identify the oil price shock by assuming that it is the only shock that has a contemporaneous effect on the oil price, which corresponds to a classical Choleski identification with the oil price ordered first, see Section \ref{app:ident} in the complementary appendix for technical details. This is a common and widely used assumption based on the idea that unexpected changes to the oil price are predetermined with respect to the US economy, see \citet{kilianvega2011} for a discussion.\footnote{Under this identification scheme, an oil price shock is an unpredicted and unpredictable change in the oil price. An alternative and very popular identification scheme consists in disentangling oil supply shocks from oil demand shocks, see, e.g., 
\citet{Kilian08,Kilian09}, \citet{BauHam}, and \citet{CaldaraCavalloIacoviello}.}

Using the same dataset and identification strategy as in SW, we estimate the effects of an oil price shock using our proposed non-stationary DFM.\footnote{Of the 207 series analyzed, the test by \citet{baing04} suggests that at about 90 series have an $I(1)$ idiosyncratic component, while our test in Appendix \ref{app:testB} suggests that about 100 series have a linear deterministic trend.} There are two main differences between our approach and the one used by SW. First, since we estimate either a VECM or a VAR in levels for the estimated factors, the IRFs are not cumulated, and, therefore, do not possess the undesirable property that all shocks have generically long-run effects on the levels of the variables, a property that is typical of stationary DFMs and that is at odds with macroeconomic theory. Second, we consider a singular autoregressive representation of the factors, as indicated by the analyzed data. In particular, the \citet{baing02} information criterion indicates that $r=8$, and the \citet{amengualwatson07} and \citet{hallinliska07}
information criteria indicate $q=3$ common shocks (see also Table 2(c) in SW). Note that, while SW set $r=8$, they do not impose singularity. 

\begin{figure}[t!]\caption{Impulse-Response Functions to an Oil Price Shock}\label{fig:IRFoil}
\centering

\smallskip
\setlength{\tabcolsep}{.01\textwidth}
\begin{tabular}{@{}ccc|ccc@{}}\hline\hline
&\textsc{\small Panel a: VECM} & &
&\textsc{\small Panel b: VAR}\\\hline 
&&&&\\[-8pt]
\rotatebox{90}{\small \hskip 30pt  Oil price} & 
\includegraphics[width=.4\textwidth]{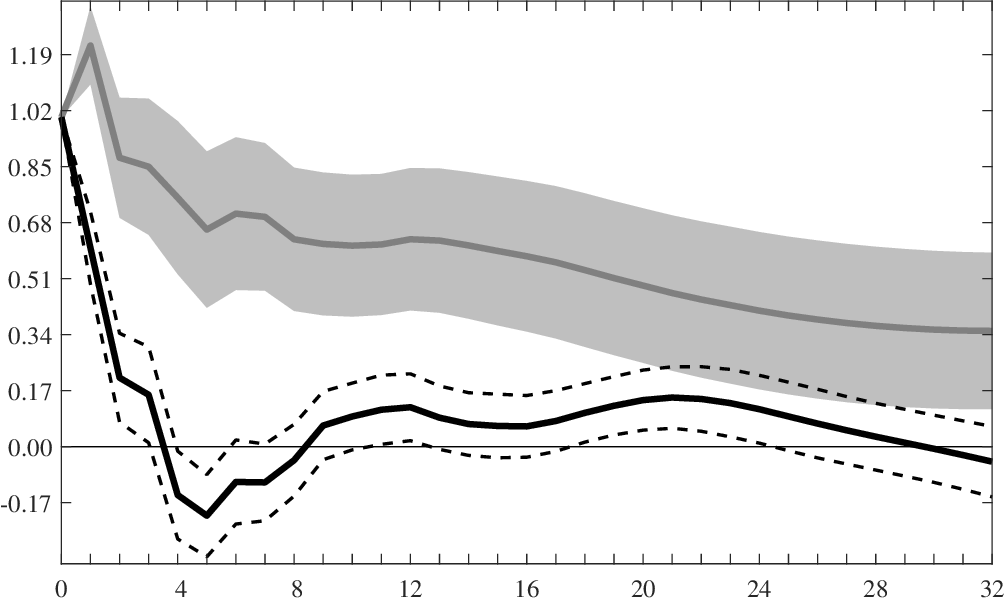} & & &
\includegraphics[width=.4\textwidth]{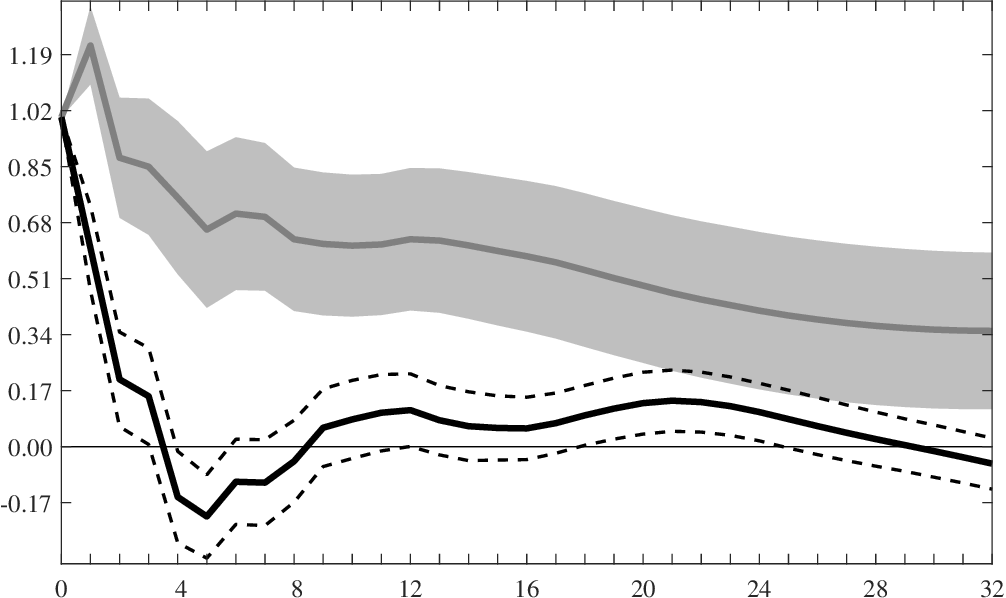}\\[2pt]
\rotatebox{90}{\small \hskip 40pt  GDP}  & 
\includegraphics[width=.4\textwidth]{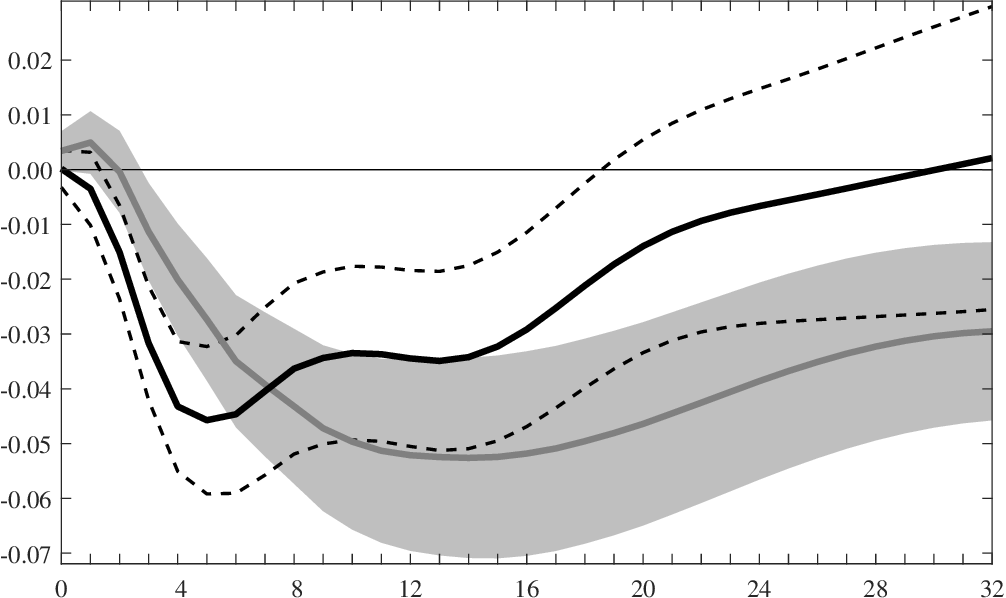} & & &
\includegraphics[width=.4\textwidth]{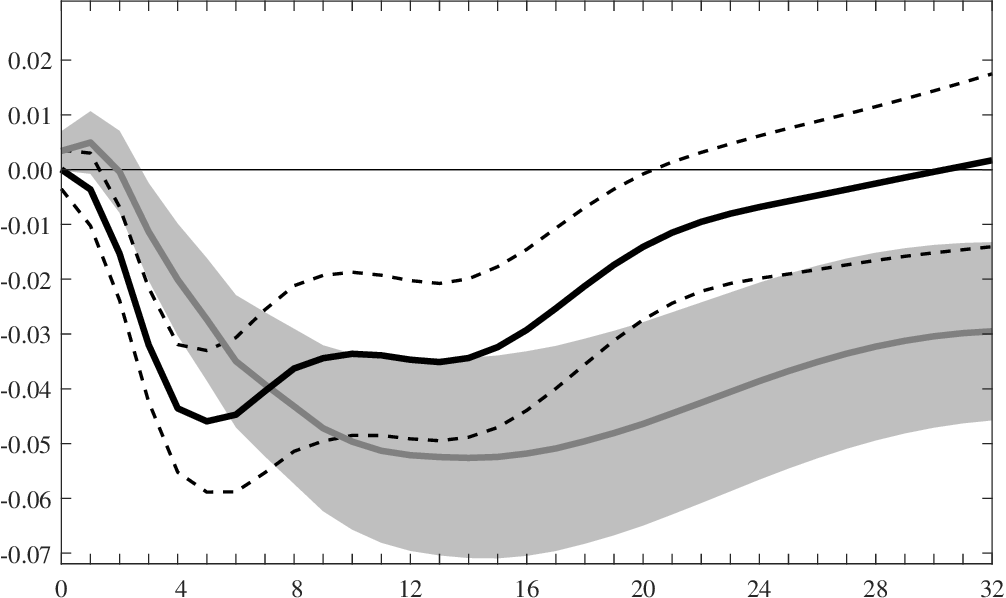} \\[2pt]
\rotatebox{90}{\small \hskip 25pt \footnotesize Consumption} &
\includegraphics[width=.4\textwidth]{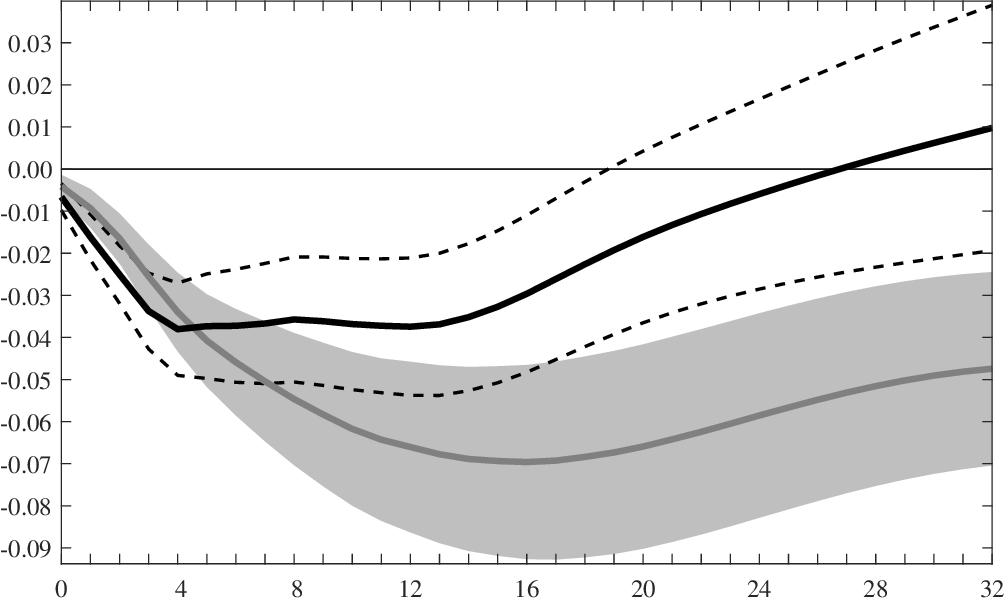} & & &
\includegraphics[width=.4\textwidth]{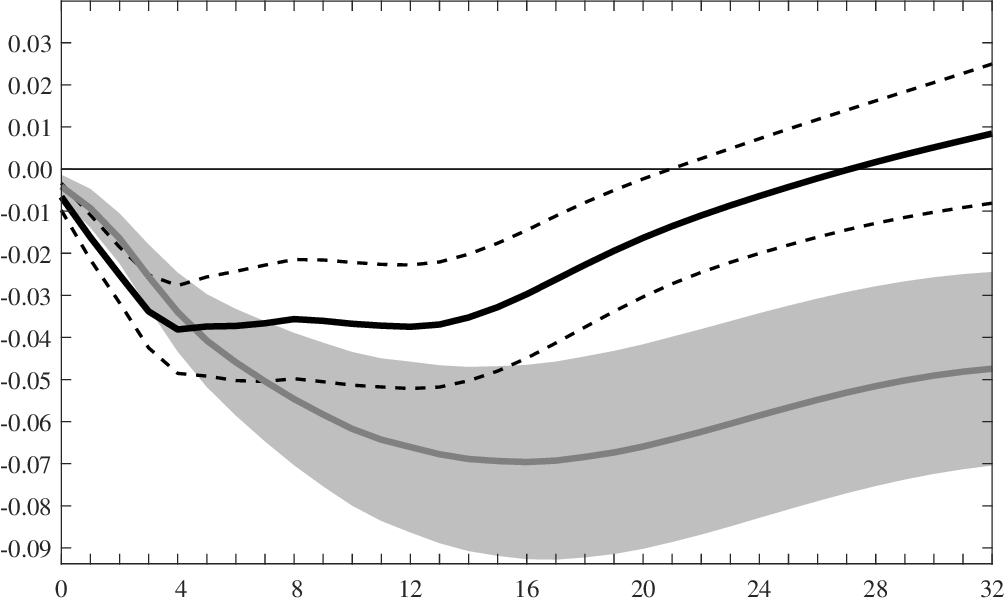} \\[2pt]
\rotatebox{90}{\small \hskip 20pt Fixed Investment} & 
\includegraphics[width=.4\textwidth]{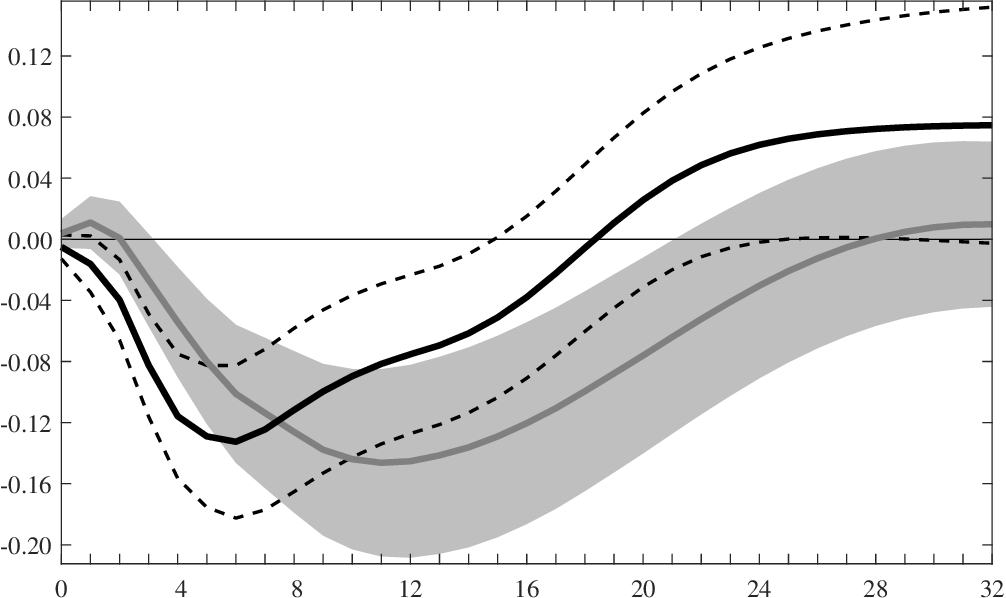} & & &
\includegraphics[width=.4\textwidth]{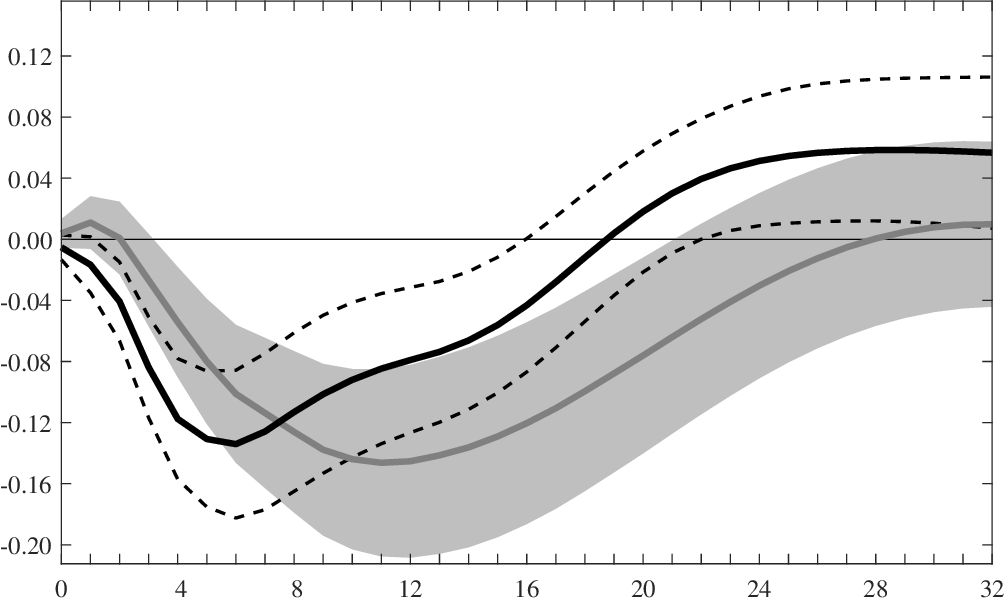}  \\[2pt]\hline
\end{tabular}
\begin{tabular}{@{}p{\textwidth}}
\scriptsize In each plot, the thick gray line is the IRF estimated by SW with a stationary DFM, while the shaded area is the 68\% bootstrap confidence band. The thick black line is the IRF estimated with the non-stationary DFM, while the dotted line delimit the 68\% bootstrap confidence band. The x-axis are quarters after the shocks, the y-axis are percentage points for all variables, but for Global Commodity Demand for the y-axis is standard deviations.
\end{tabular}
\end{figure}

Figure \ref{fig:IRFoil} compares the IRFs estimated by SW (gray lines), with those estimated with our method (black lines) either estimating a VECM (Panel A) or a VAR in levels (Panel B) for the factors---the VECM is estimated with $c=7$ cointegration relations as determined via the information criterion given in Section \ref{sec:nfactors}. Two crucial differences emerge: first, while SW estimate that an oil price shock has a persistent effect on the oil price---after a shock that increases the oil price by one percentage point, the oil price is estimated to be permanently higher by about 0.4 percentage points---our model estimates that the oil price returns to its initial level about a year after the shock.\footnote{In our model, the oil price is the refiners' acquisition cost (RAC). This is a common practice in the literature \citep[e.g.,][]{ConflittiLuciani}, and using another of the oil price indicators in the SW dataset instead of RAC has virtually no consequences on the results.} 
Second, while SW estimate that an oil price shock has a permanent effect on real activity (i.e., GDP, consumption, and investments), our model estimate that the effects of an oil price shock wipe out in about five-to-eight years, which is consistent with the idea that only technological shocks are capable of having a permanent effect on the real side of the economy. 

Summing up, our results partly overturn those in SW and those in the literature applying the same identification technique (e.g.,  \citealp{BG07}). According to this literature, an oil price shock has a permanent effect on real activity; according to our result, an oil price shock has only a temporary effect on real activity. In particular,  our results differ from those of SW because they cumulate
the IRFs obtained from a VAR estimated on the differenced factors. In contrast, our approach has a built-in error correction mechanism which disciplines the long-run behavior of the estimated IRFs.


Finally, as we can see by comparing Panel (A) and Panel (B) in Figure \ref{fig:IRFoil}, there are no significant differences between the IRFs estimated by fitting a VECM or an unrestricted VAR on the levels of the factors. This is not surprising because, as we showed in Section \ref{sec:emp}, the methods estimate the short to medium-run IRFs consistently.

\subsection{Application 2: the effect of news shocks}\label{emp:FGS}
Starting with the seminal paper of \citet{BeaudryPortier}, there has been a renewed interest in the idea that expectations about future fundamentals can be a driver of the business cycle. According to this theory, news  about future productivity (a.k.a. news shocks) can generate a boom today and a bust in the future, if the realized productivity improvement is less than expected. In their paper, \citet{BeaudryPortier} by estimating a small size VECM  find that a positive news shock has a positive impact on stock prices, output, consumption, investment, and hours worked. These results generated lots of interest because they are theoretically controversial. Indeed, in a neoclassical setting, in response to a positive news shock, hours worked should decrease---the wealth effect coming from higher stock prices induces households to consume more, and work less (i.e., desire more leisure)---and (at least initially) output and investment should decrease as well.  In subsequent analyses, \citet{BarskySims}, who use a small size SVAR in levels, overturn some of the results in \citet{BeaudryPortier}. Specifically, they find that in response to a positive news shock, shock  output and investment initially decline; after that, the response of output and investment tracks the path of Total Factor Productivity (TFP), rather than anticipate it. Furthermore, \citet{BarskySims} estimate a negative response of hours worked to a positive news shock.\footnote{There is a large number of papers that have analyzed the effects of news shock on the US economy, both by means of VAR models \citep[e.g.,][]{BeaudryPortier2,KurmannOtrok} and by means of estimated DSGE models \citep[e.g.,][]{BlanchardHullierLorenzoni,SGUribe}. The overall conclusion is that the effects of news shocks on the US economy are sizable.} 

\citet{fornigambettisala} (FGS hereafter) estimate the IRF to a news shock from a panel of 107 US quarterly macroeconomic time series, covering the period 1960:Q1 to 2010:Q4. In particular, they estimate a FAVAR with two observed factors (TFP and stock prices) and three latent factors extracted from principal components in levels, thus implicitly assuming all idiosyncratic components to be stationary. The news shock is identified by imposing that (\textit{i}) it does not move TFP on impact, and (\textit{ii}) it has maximal impact on TFP at the 60 quarters (15 years) horizon, see Section \ref{app:ident} in the complementary appendix for technical details. 

Using the same dataset and identification strategy as in FGS, we estimate the effects of a news shock using our IRFs estimator in a FAVAR setting, as discussed in Remark \ref{marcellino} in Section \ref{tretre}, see also Section \ref{app:FAVARapp} in the complementary appendix for technical details.\footnote{Of the 107 series analyzed, the test by \citet{baing04} suggests that at about 50 series have an $I(1)$ idiosyncratic component, while our test in Appendix \ref{app:testB} suggests that about 60 series have a linear deterministic trend.} There are two main differences between our approach and the one used by FGS. First, we estimate the factors from differenced data properly de-trended, as explained in Section \ref{sec:detrend}. In this way, we avoid the risk of detecting spurious factors due to the possible presence of $I(1)$ idiosyncratic components and/or deterministic linear trends, see \citet{onatskiwang19}. Second, in addition to the FAVAR, we also consider IRFs obtained from a FAVECM with four cointegration relations (as suggested by the criteria in Section \ref{sec:nfactors}), which also account for cointegration between TFP, stock prices, and the three common factors.

\begin{figure}[t!]\caption{Impulse-Response Functions to a News Shock}\label{fig:IRFnews} \medskip
\centering

\setlength{\tabcolsep}{.002\textwidth}
\begin{tabular}{@{}ccc@{}}\hline\hline
&&\\[-8pt]
\multicolumn{3}{l}{\textsc{\small Panel a: VECM}}\\\hline 
&&\\[-8pt] 
\footnotesize TFP & \footnotesize Stock prices& \footnotesize Hours\\
\includegraphics[width=.33\textwidth]{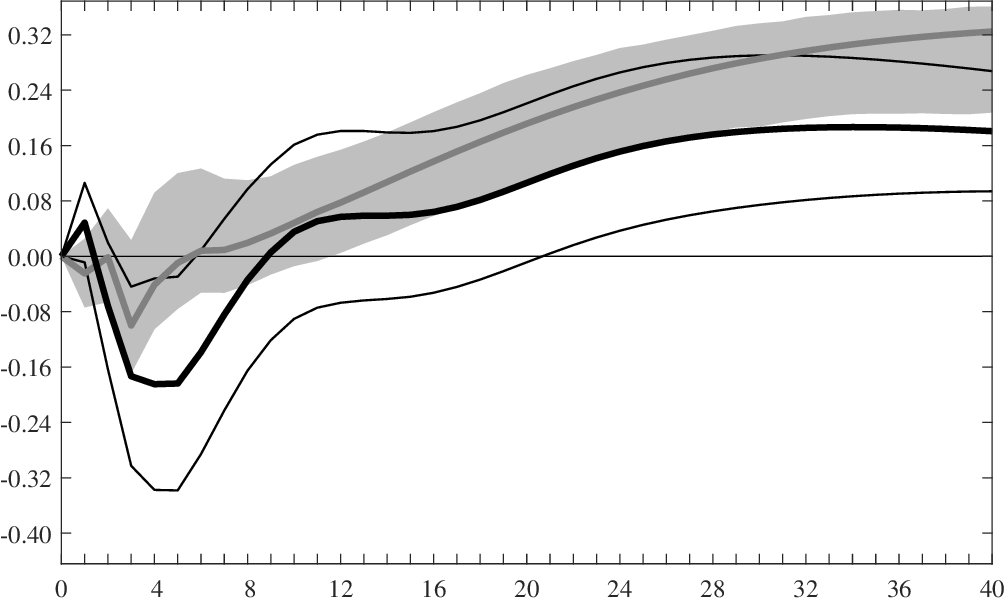} &
\includegraphics[width=.33\textwidth]{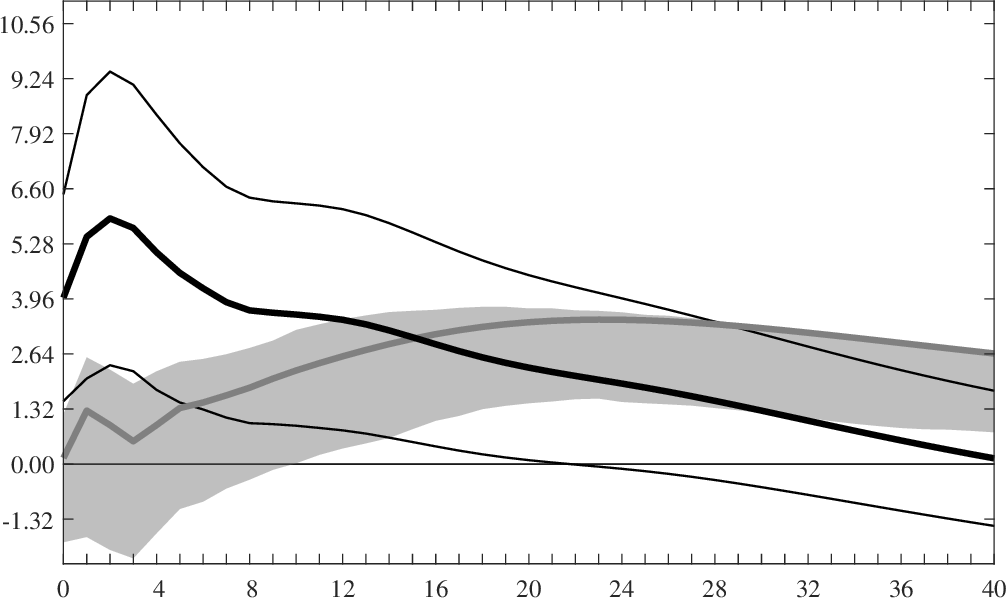} & 
\includegraphics[width=.33\textwidth]{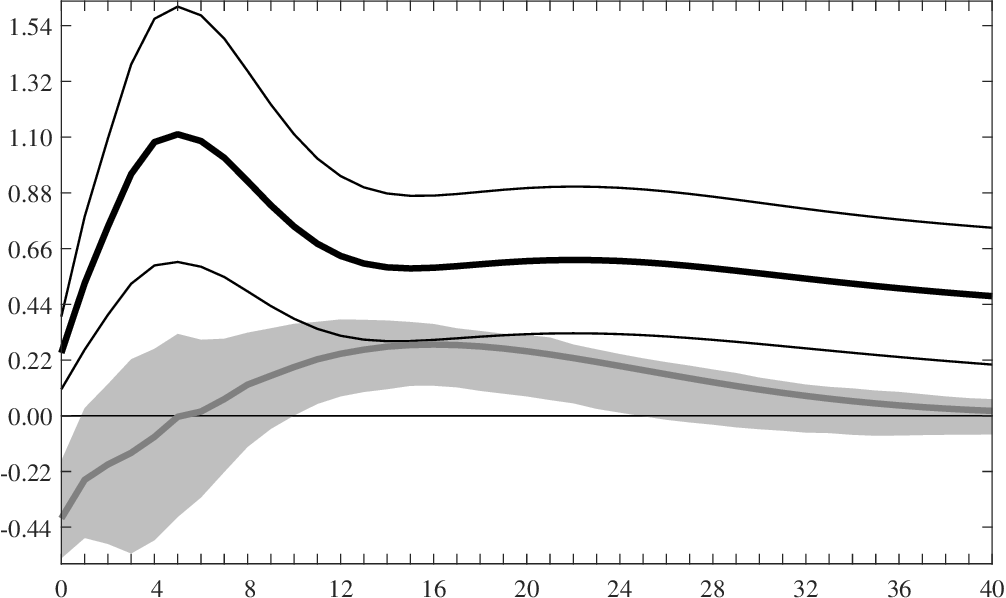} \\
\footnotesize Consumption  & \footnotesize Output & \footnotesize Investment \\
\includegraphics[width=.33\textwidth]{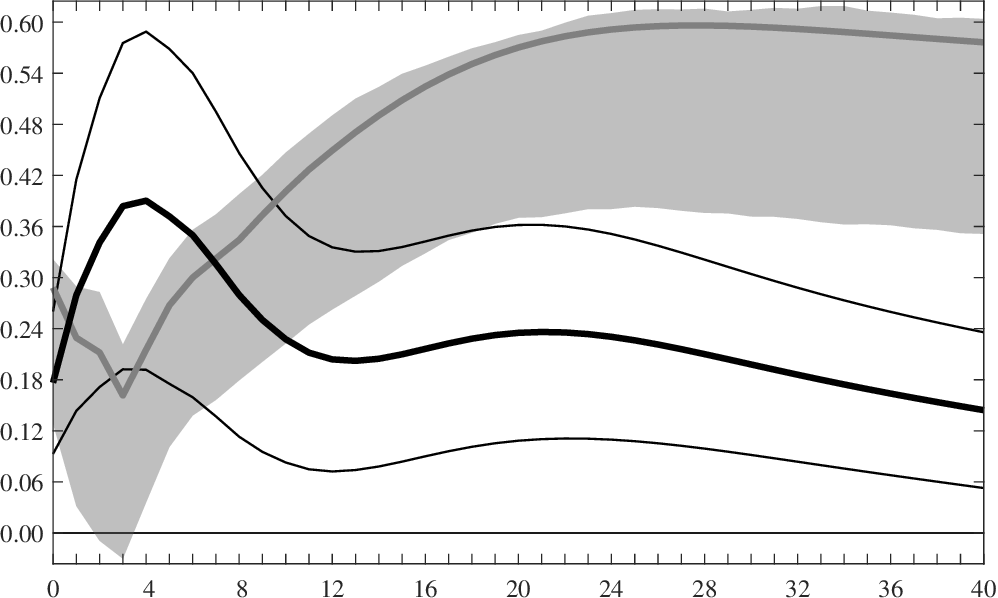} &
\includegraphics[width=.33\textwidth]{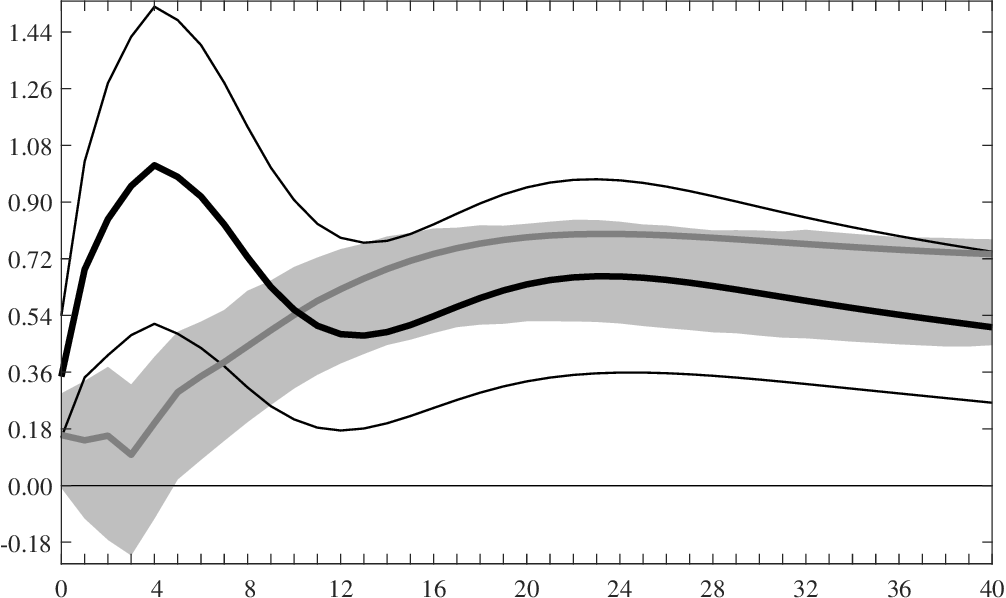} &
\includegraphics[width=.33\textwidth]{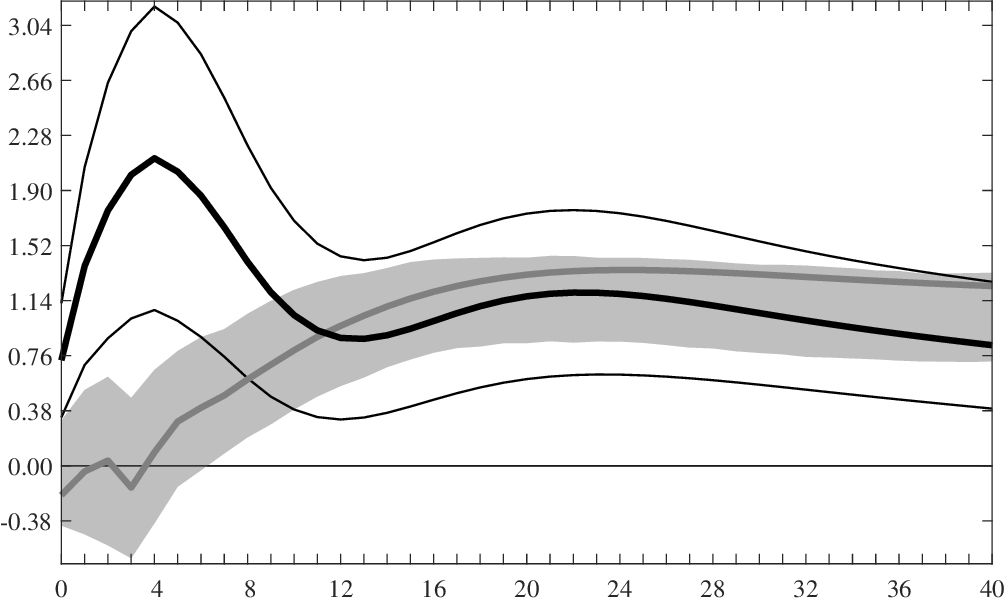} \\\hline
&&\\[-8pt]
\multicolumn{3}{l}{\textsc{\small Panel b: VAR}}\\\hline 
&&\\[-8pt]
\footnotesize TFP & \footnotesize Stock prices& \footnotesize Hours\\
\includegraphics[width=.33\textwidth]{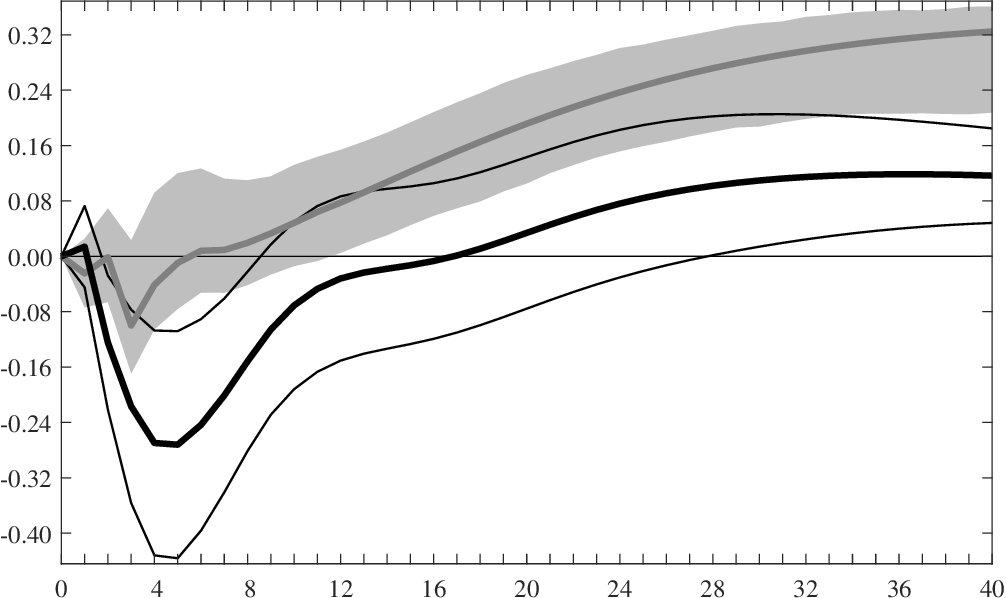} &
\includegraphics[width=.33\textwidth]{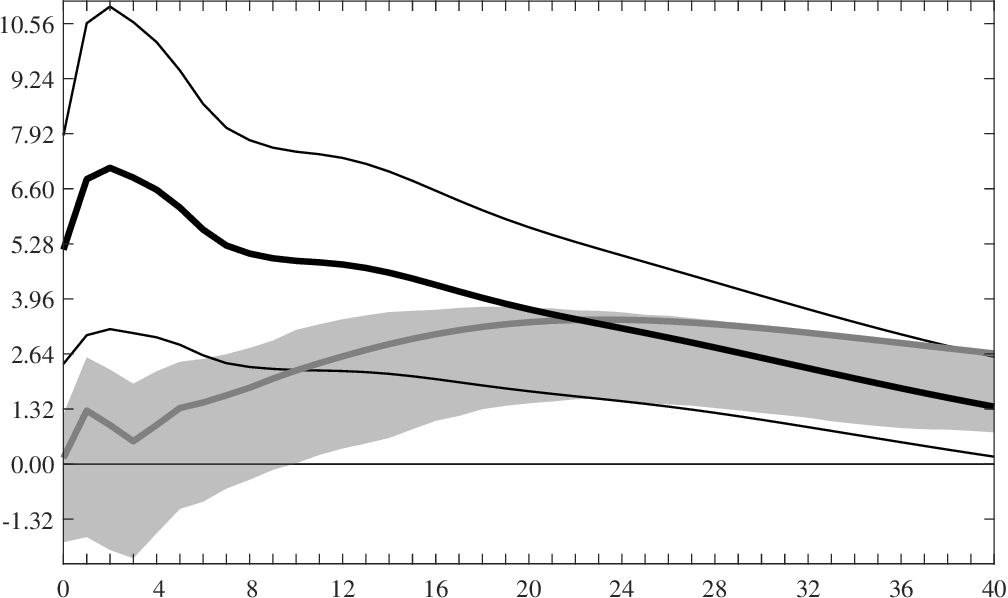} & 
\includegraphics[width=.33\textwidth]{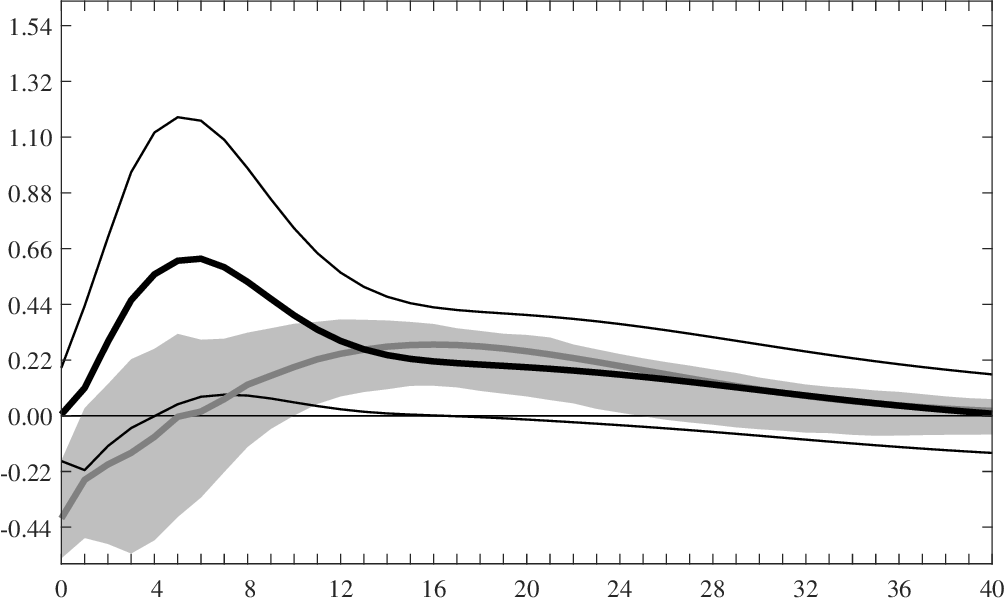} \\
\footnotesize Consumption  & \footnotesize Output & \footnotesize Investment \\
\includegraphics[width=.33\textwidth]{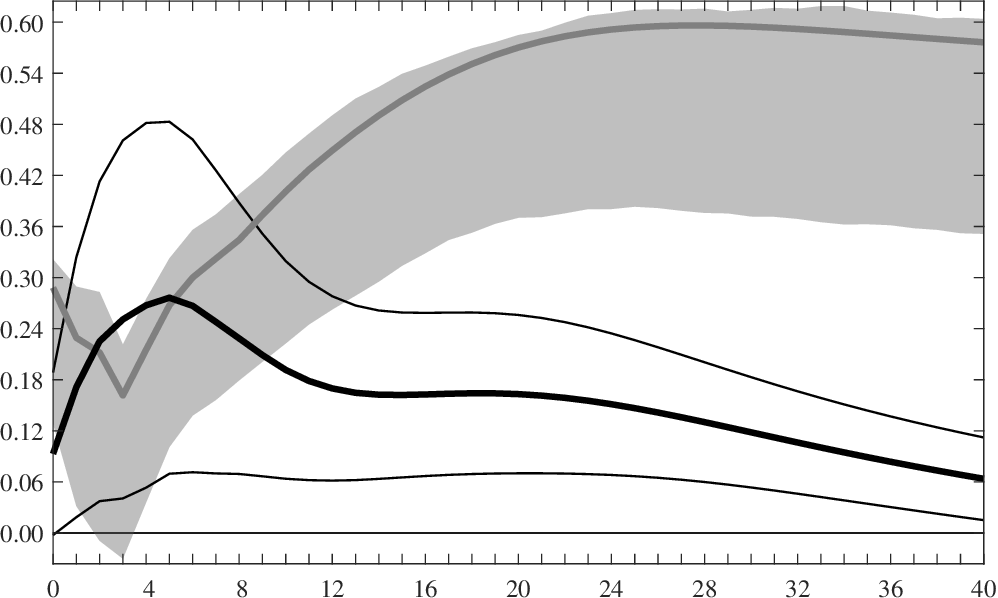} &
\includegraphics[width=.33\textwidth]{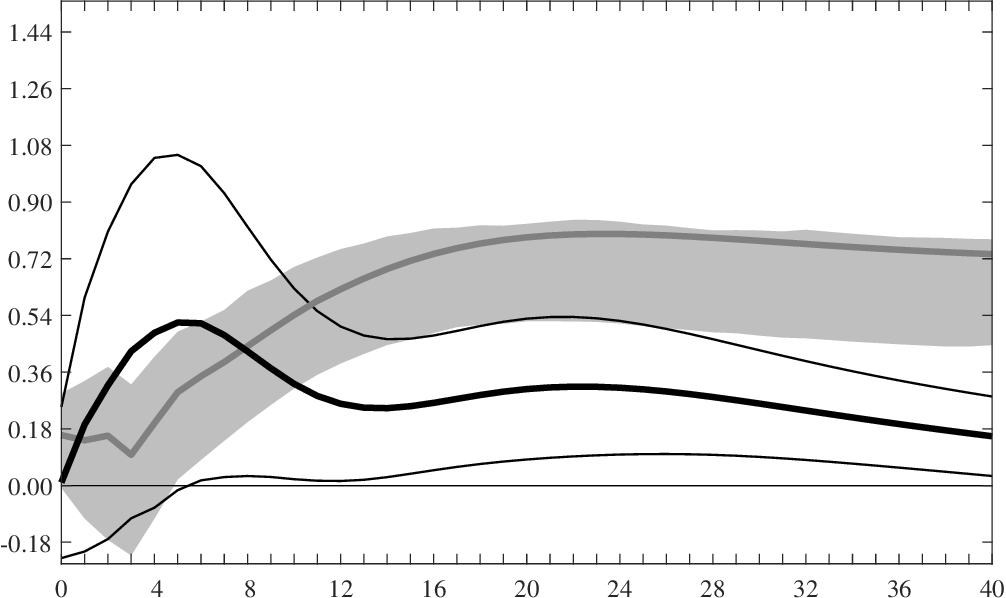} &
\includegraphics[width=.33\textwidth]{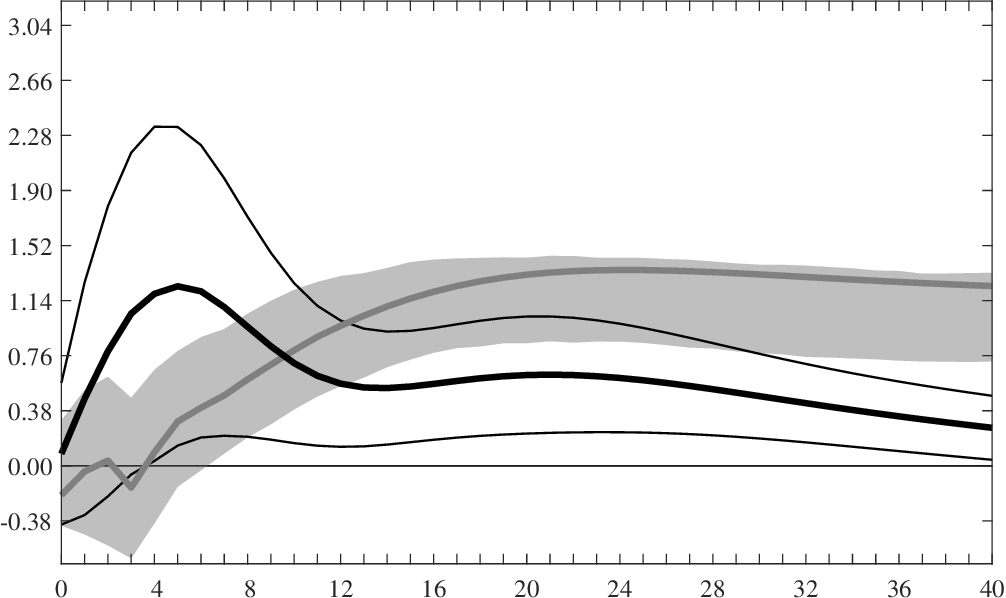} \\\hline
\end{tabular}
\begin{tabular}{@{}p{\textwidth}}
\scriptsize In each plot, the thick gray line is the IRF estimated by FGS, while the shaded area is the 68\% bootstrap confidence band. The thick black line is the IRF estimated with our model, while the dotted line delimit the 68\% bootstrap confidence band. The x-axis are quarters after the shocks, the y-axis are percentage points.
\end{tabular}
\end{figure}

Figure \ref{fig:IRFnews} compares the IRFs to a news shock estimated by FGS (gray lines), with those estimated with our method (black lines), either estimating a FAVECM (Panel A) or a FAVAR in levels (Panel B) for the factors. Three main differences emerge. First, we estimate that hours worked respond positively to a news shock. Second, we estimate that investment and GDP increase on impact together with consumption, and their response leads the response of TFP. Third, as we can see from the hump-shaped response of consumption investment and output, our model predicts that in response to a positive news shock the economy first experiences a significant temporary boom, and then a temporary milder recession. In other words, our results are more in line with those of \citet{BeaudryPortier} and partly overturn those in \citet{BarskySims} and \citet{fornigambettisala}. 

A comment related only to our methodology is also in order: as we can see by comparing Panel (A) and Panel (B) in Figure \ref{fig:IRFnews}, there are some differences between the IRFs estimated by fitting a FAVECM or a FAVAR. These differences emerge since the news shock is identified by imposing a restriction at 60 quarters horizon, and, as we discussed in Section \ref{tretre}, the long-run IRFs estimated with a VAR in levels are not consistently estimated in the long run. Therefore, we recommend for this application to use the FAVECM approach.


\section{Conclusions}\label{sec:end}
In this paper, we introduce a non-stationary Dynamic Factor Model (DFM) for large datasets, and we propose an estimator for the impulse response functions (IRFs). The natural use of this class of models in a macroeconomic context motivates the main assumptions upon which the present theory is built.

%

%
Estimation of IRFs is obtained with a two-step estimator based on principal components, and on a VECM---or an unrestricted VAR in levels---for the latent $I$(1) common factors. We prove consistency of the IRFs estimator when both the cross-sectional dimension $n$ and the sample size $T$ of the dataset grow to infinity.~Furthermore, we also propose an information criterion to determine the number of common permanent shocks in a large dimensional setting. 

A numerical study shows the validity and usefulness of our approach. Results show that if the short run is the focus, both the VECM and the unrestricted VAR in levels perform equally well, while if the long run is the focus, the VECM must be preferred. 

In two empirical applications, we find that: (i) oil price shocks have just a temporary effect on US real activity; and, (ii) in response to a positive news shock, the economy first experiences a significant  boom, and then a  milder recession. Our results partly overturn those obtained by \citet{stockwatson16} and \citet{fornigambettisala}, respectively, and show the importance of correctly accounting for the presence of cointegration in the common factors when estimating the IRFs.

In conclusion, compared to the stationary model commonly used in the literature, the non-stationary model proposed in this paper, which accounts for cointegration in the common factors, a feature that both economic  and econometric theory suggest to be extremely likely, offers a more realistic representation of the data. Moreover, our approach has the advantage that it does not require to transform the variables to stationarity. Our empirical analysis shows that when estimating IRFs, cointegration matters, hence data should not be transformed. However, we have not investigated whether it matters when the goal is not estimating IRFs---for example, \citet{ngmcracken} show that the cost of over differencing the data when forecasting is negligible. This is  an empirical question and is part of our future research.

\small
\singlespacing
{{
\setlength{\bibsep}{.2cm}
\bibliographystyle{chicago}
\bibliography{BLL_biblio}
}}



  \setcounter{section}{0}%
\setcounter{subsection}{-1}
\setcounter{equation}{0}
\setcounter{lem}{0}
\renewcommand{\thelem}{A\arabic{lem}}
\renewcommand{\theequation}{A\arabic{equation}}

\appendix
\section{Technical appendix}\label{app:prop}
\small
\subsection*{Norms} 
For any $m\times p$ matrix $\mbf B$ with generic element $b_{ij}$, we denote its spectral norm as $\Vert\mbf B\Vert=(\mu_1^{\mbf B'\mbf B})^{1/2}$, where $\mu_1^{\mbf B'\mbf B}$ is the largest eigenvalue of $\mbf B'\mbf B$, 
the Frobenius norm as $\Vert\mbf B\Vert_F=(\mbox{tr}(\mbf B'\mbf B))^{1/2}=(\sum_i\sum_j b_{ij}^2)^{1/2}$, 
and the column and row norm as $\Vert\mbf B\Vert_1=\max_{j}\sum_{i}|b_{ij}|$ and $\Vert\mbf B\Vert_{\infty}=\max_{i}\sum_{j}|b_{ij}|$, respectively. Throughout we also make use of Weyl's inequality, for two $n\times n$ symmetric matrices $\mbf A$ and $\mbf B$, with eigenvalues $\mu_j^{A}$ and $\mu_j^B$:
\beq\label{eq:weyl}
|\mu_j^{A}-\mu_j^B|\leq \Vert\mbf A-\mbf B\Vert, \qquad j=1,\ldots, n.
\eeq

\subsection{Proof of Proposition \ref{vecm}}
First let us introduce some useful notation. Throughout define $\wt{\mbf F}_t=\mbf J\mbf F_t$ and $\wt{\bm\beta}=\mbf J\bm\beta$, where $\mbf J$ is an $r\times r$ diagonal matrix with entries $\pm 1$ defined in Lemma \ref{lem:load}, and note that  $\wt{\bm\beta}'\wt{\mbf F}_t=\bm\beta'{\mbf F}_t$. Thus $\wt{\bm\beta}$ is the matrix of cointegration vectors of $\wt{\mbf F}_t$ and we denote its orthogonal complement as $\wt{\bm\beta}_{\perp}$, such that $\wt{\bm\beta}_{\perp}'\wt{\bm\beta}=\mbf 0_{r-c\times c}$. Define the matrices 
\bea
&&\wh{\mbf M}_{00}=\frac 1 T\sum_{t=1}^T\Delta\wh{\mbf F}_t\Delta\wh{\mbf F}_t',
\quad\wh{\mbf M}_{01}=\frac 1 T\sum_{t=1}^T\Delta\wh{\mbf F}_t\wh{\mbf F}_{t-1}',
\quad\wh{\mbf M}_{02}=\frac 1 T\sum_{t=1}^T\Delta\wh{\mbf F}_t\Delta\wh{\mbf F}_{t-1}',\nn\\
&&\wh{\mbf M}_{11}=\frac 1 T\sum_{t=1}^T\wh{\mbf F}_t\wh{\mbf F}_t',
\quad \wh{\mbf M}_{21}=\frac 1 T\sum_{t=1}^T\Delta\wh{\mbf F}_{t-1}'\wh{\mbf F}_{t-1},
\quad \wh{\mbf M}_{22}=\frac 1 T\sum_{t=1}^T\Delta\wh{\mbf F}_{t-1}\Delta\wh{\mbf F}_{t-1}',\nn\\
&&\wh{\mbf S}_{00}=\wh{\mbf M}_{00}-\wh{\mbf M}_{02}\wh{\mbf M}_{22}^{-1}\wh{\mbf M}_{20},\;\;\;
\wh{\mbf S}_{01}=\wh{\mbf M}_{01}-\wh{\mbf M}_{02}\wh{\mbf M}_{22}^{-1}\wh{\mbf M}_{21},\;\;\;
\wh{\mbf S}_{11}=\wh{\mbf M}_{11}-\wh{\mbf M}_{12}\wh{\mbf M}_{22}^{-1}\wh{\mbf M}_{21},\nn
\eea
and denote by $\mbf M_{ij}$ and $\mbf S_{ij}$, for $i,j=0,1,2$, the analogous ones but computed by using $\wt{\mbf F}_t$. Finally, define the conditional covariance matrices
\[
\wt{\bm\Omega}_{00}=\E[\Delta\wt{\mbf F}_t\Delta\wt{\mbf F}_t'|\Delta\wt{\mbf F}_{t-1}],
\;\; \wt{\bm\Omega}_{\wt{\beta}\wt{\beta}}=\E[\wt{\bm\beta}'\wt{\mbf F}_{t-1}\wt{\mbf F}_{t-1}'\wt{\bm\beta}|\Delta\wt{\mbf F}_{t-1}],
\;\;\check{\bm\Omega}_{0\wt{\beta}}=\E[\Delta\wt{\mbf F}_t\wt{\mbf F}_{t-1}'\wt{\bm\beta}|\Delta\wt{\mbf F}_{t-1}],
\;\; \check{\bm\Omega}_{\wt{\beta}0}=\check{\bm\Omega}_{0\wt{\beta}}'.\nn
\]

Let us start from part (i). Notice that if we denote the residuals of the regression of $\Delta\wh{\mbf F}_t$ and of $\wh{\mbf F}_{t-1}$ on $\Delta\wh{\mbf F}_{t-1}$ as $\wh{\mbf e}_{0t}$ and $\wh{\mbf e}_{1t}$, respectively then $\wh{\mbf S}_{ij}=T^{-1}\sum_{t=1}^T\wh{\mbf e}_{it}\wh{\mbf e}_{jt}'$, with $i,j=0,1$. Consider the generalized eigenvalues problem
\beq\label{eq:evecj1}
\det\big(\wh{\mu}_j \wh{\mbf S}_{11}-\wh{\mbf S}_{10}\wh{\mbf S}_{00}^{-1}\wh{\mbf S}_{01}\big)=0,\quad j=1,\ldots ,r.
\eeq
If $\wh{\mbf U}$ are the normalized eigenvectors of $\wh{\mbf S}_{11}^{-1/2}\wh{\mbf S}_{10}\wh{\mbf S}_{00}^{-1}\wh{\mbf S}_{01}\wh{\mbf S}_{11}^{-1/2}$, then $\wh{\mbf P}=\wh{\mbf S}_{11}^{-1/2}\wh{\mbf U}$ are eigenvectors of  $\wh{\mbf S}_{11}-\wh{\mbf S}_{10}\wh{\mbf S}_{00}^{-1}\wh{\mbf S}_{01}$ with eigenvalues $\wh{\mu}_j$.~Then, the estimator $\wh{\bm\beta}$ proposed by \citet{Johansen95} is given by the $c$ columns of $\wh{\mbf P}$ corresponding to the $c$ largest eigenvalues.

Analogously define $\wh{\mbf U}^0$ as the normalized eigenvectors of ${\mbf S}_{11}^{-1/2}{\mbf S}_{10}{\mbf S}_{00}^{-1}{\mbf S}_{01}{\mbf S}_{11}^{-1/2}$ and define  $\wh{\mbf P}^0={\mbf S}_{11}^{-1/2}\wh{\mbf U}^0$. Then the estimator $\wh{\bm\beta}^0$ that we would obtain if estimating a VECM on $\check{\mbf F}_t$, is the matrix of the $c$ columns of $\wh{\mbf P}^0$, corresponding to the $c$ largest eigenvalues $\wh{\mu}_j^0$ of ${\mbf S}_{11}-{\mbf S}_{10}{\mbf S}_{00}^{-1}{\mbf S}_{01}$, and such that
\beq\label{eq:evecj2}
\det\big(\wh{\mu}_j^0 \mbf S_{11}-\mbf S_{10}\mbf S_{00}^{-1}\mbf S_{01}\big)=0,\quad j=1,\ldots ,r.
\eeq
Notice that by definition the two estimators $\wh{\bm\beta}$ and $\wh{\bm\beta}^0$ are normalized in such a way that $\wh{\bm\beta}'\wh{\mbf S}_{11}\wh{\bm\beta}=\mbf I_c$ and $\wh{\bm\beta}^{0'}{\mbf S}_{11}\wh{\bm\beta}^0=\mbf I_c$.

Consider then the $r\times r$ matrix $\mbf A_T= \big(\check{\bm\beta}\;\; (T^{-1/2}{\wt{\bm\beta}_{\perp*}})\big),$
where $\wt{\bm\beta}_{\perp*}^{}=\wt{\bm\beta}_{\perp}^{}(\wt{\bm\beta}_{\perp}'\wt{\bm\beta}_{\perp}^{})^{-1}$, and consider the equations
\begin{align}
\det\big[ \mbf A_T'\big(\wh{\mu}_j\wh{\mbf S}_{11}-\wh{\mbf S}_{10}\wh{\mbf S}_{00}^{-1}\wh{\mbf S}_{01}\big)\mbf A_T\big]=0,\quad j=1,\ldots ,r,\label{eq:evecj3}\\
\det\big[ \mbf A_T'\big(\wh{\mu}_j^0\mbf S_{11}-\mbf S_{10}\mbf S_{00}^{-1}\mbf S_{01}\big)\mbf A_T\big]=0,\quad j=1,\ldots ,r.\label{eq:evecj4}
\end{align}
Clearly \eqref{eq:evecj3} has the same solutions as \eqref{eq:evecj1}, but its eigenvectors are now given by $\mbf A_T^{-1}\wh{\mbf P}$ and those corresponding to the largest $c$ eigenvalues are $\mbf A_T^{-1}\wh{\bm\beta}$. Analogously for \eqref{eq:evecj4} we have the eigenvectors $\mbf A_T^{-1}\wh{\mbf P}^0$ and the $c$ largest are given by $\mbf A_T^{-1}\wh{\bm\beta}^0$.  Moreover, 
\begin{align}
&\mbf A_T'\big(\wh{\mbf S}_{11}-\wh{\mbf S}_{10}\wh{\mbf S}_{00}^{-1}\wh{\mbf S}_{01}\big)\mbf A_T-\mbf A_T'\big({\mbf S}_{11}-{\mbf S}_{10}{\mbf S}_{00}^{-1}{\mbf S}_{01}\big)\mbf A_T=\nn\\
=& \l\{\l[\begin{array}{cc}
\wt{\bm\beta}'\wh{\mbf S}_{11}\wt{\bm\beta} & T^{-1/2}\wt{\bm\beta}'\wh{\mbf S}_{11}\wt{\bm\beta}_{\perp*}\\
 T^{-1/2}\wt{\bm\beta}_{\perp*}'\wh{\mbf S}_{11}\wt{\bm\beta} & T^{-1}\wt{\bm\beta}_{\perp*}'\wh{\mbf S}_{11}\wt{\bm\beta}_{\perp*}
\end{array}
\r]-
 \l[\begin{array}{cc}
\wt{\bm\beta}'{\mbf S}_{11}\wt{\bm\beta} & T^{-1/2}\wt{\bm\beta}'{\mbf S}_{11}\wt{\bm\beta}_{\perp*}\\
 T^{-1/2}\wt{\bm\beta}_{\perp*}'{\mbf S}_{11}\wt{\bm\beta} & T^{-1}\wt{\bm\beta}_{\perp*}'{\mbf S}_{11}\wt{\bm\beta}_{\perp*}
\end{array}
\r]
\r\}\nn\\
&-\l\{
\l[\begin{array}{cc}
\wt{\bm\beta}'\wh{\mbf S}_{10}\wh{\mbf S}_{00}^{-1}\wh{\mbf S}_{01}\wt{\bm\beta}&T^{-1/2}\wt{\bm\beta}'\wh{\mbf S}_{10}\wh{\mbf S}_{00}^{-1}\wh{\mbf S}_{01}\wt{\bm\beta}_{\perp*}\\
T^{-1/2}\wt{\bm\beta}_{\perp*}'\wh{\mbf S}_{10}\wh{\mbf S}_{00}^{-1}\wh{\mbf S}_{01}\wt{\bm\beta}&T^{-1}\wt{\bm\beta}_{\perp*}'\wh{\mbf S}_{10}\wh{\mbf S}_{00}^{-1}\wh{\mbf S}_{01}\wt{\bm\beta}_{\perp*}
\end{array}\r]
\r.\nn\\
&\l.
\;\;\;-\l[\begin{array}{cc}
\wt{\bm\beta}'{\mbf S}_{10}{\mbf S}_{00}^{-1}{\mbf S}_{01}\wt{\bm\beta}&T^{-1/2}\wt{\bm\beta}'{\mbf S}_{10}{\mbf S}_{00}^{-1}{\mbf S}_{01}\wt{\bm\beta}_{\perp*}\\
T^{-1/2}\wt{\bm\beta}_{\perp*}'{\mbf S}_{10}{\mbf S}_{00}^{-1}{\mbf S}_{01}\wt{\bm\beta}&T^{-1}\wt{\bm\beta}_{\perp*}'{\mbf S}_{10}{\mbf S}_{00}^{-1}{\mbf S}_{01}\wt{\bm\beta}_{\perp*}
\end{array}\r]\r\}= O_p(\vartheta_{nT,\delta,\eta}).\label{eq:evecj5}
\end{align}

This result is proved by using Lemma \ref{lem:shat}(ii), \ref{lem:shat}(iii) and  \ref{lem:shat}(vi) for the first term on the rhs, and by using Lemma \ref{lem:shat}(i), \ref{lem:shat}(iv) and \ref{lem:shat}(v) for the second term. Thus, from \eqref{eq:evecj5}, for any $j=1,\ldots, r$, from Weyl's inequality \eqref{eq:weyl}, we have
\beq\label{eq:evhat}
\big|\wh{\mu}_j-\wh{\mu}_j^0\big|\leq \big\Vert\mbf A_T'\big(\wh{\mbf S}_{11}-\wh{\mbf S}_{10}\wh{\mbf S}_{00}^{-1}\wh{\mbf S}_{01}\big)\mbf A_T-\mbf A_T'\big({\mbf S}_{11}-{\mbf S}_{10}{\mbf S}_{00}^{-1}{\mbf S}_{01}\big)\mbf A_T\big\Vert = O_p(\vartheta_{nT,\delta,\eta}).
\eeq
 Then, because of Lemmas \ref{rem:DFF}(ii) and \ref{lem:shat2}, and from \eqref{eq:evecj5}, \eqref{eq:evhat}, and Slutsky's theorem, as $n,T\to\infty$, we have  \citep[see also  Lemma 13.1 in][]{Johansen95}
\begin{align}
\det&\bigg[\mbf A_T' \Big(\wh{\mu}_j \wh{\mbf S}_{11}-\wh{\mbf S}_{10}\wh{\mbf S}_{00}^{-1}\wh{\mbf S}_{01}\Big)\mbf A_T\bigg]=\det\bigg[\mbf A_T' \Big(\wh{\mu}^0_j {\mbf S}_{11}-{\mbf S}_{10}{\mbf S}_{00}^{-1}{\mbf S}_{01}\Big)\mbf A_T\bigg]+O_p(\vartheta_{nT,\delta,\eta})\label{eq:evecj6}\\
\stackrel{d}{\to}&\det\bigg(\wh{\mu}^0_j\wt{\bm\Omega}_{\wt{\beta}\wt{\beta}}-\wt{\bm\Omega}_{\wt{\beta}0}\wt{\bm\Omega}_{00}^{-1}\wt{\bm\Omega}_{0\wt{\beta}}\bigg)\;\det \bigg[\wh{\mu}^0_j\wt{\bm\beta}_{\perp*}'\mbf C(1)
\bigg(\int_0^1\mbf W_q(\tau)\mbf W_q'(\tau)\mbox {\upshape d} \tau\bigg)
\mbf C'(1)\wt{\bm\beta}_{\perp*}^{}\bigg].\nn
\end{align}
where $\mbf W_q(\cdot)$ is a $q$-dimensional Brownian motion with covariance $\mbf I_q$. 
The first term on the rhs of \eqref{eq:evecj6} has only $c$ solutions different from zero (the matrix is positive definite) while the remaining $r-c$ solutions come from the second term and are all zero since ${\rm rk}(\mbf C(1)) = r-c=q-d$. Therefore, as $n,T\to\infty$ both $\mbf A_T^{-1}\wh{\mbf P}$ and $\mbf A_T^{-1}\wh{\mbf P}^0$ span a space of dimension $c$ given by their first $c$ eigenvectors, which by definition are given by $\mbf A_T^{-1}\wh{\bm \beta}$ and $\mbf A_T^{-1}\wh{\bm \beta}^0$, respectively. As a consequence, there exist a positive real $D_1$ such that  $\wh{\mu}_j^0>D_1$ for $j=1,\ldots, c$. From \eqref{eq:evecj5} and Theorem 2 in \citet{yu15}, there exists an orthogonal $c\times c$ matrix $\mbf O_c$ such that 
\begin{align}\label{eq:AVAV2}
\big\Vert\mbf A_T^{-1}&\wh{\bm \beta}\mbf O_c-\mbf A_T^{-1}\wh{\bm \beta}^0\big\Vert\leq 
\frac{2^{3/2}\sqrt c\big\Vert\mbf A_T'\big(\wh{\mbf S}_{11}-\wh{\mbf S}_{10}\wh{\mbf S}_{00}^{-1}\wh{\mbf S}_{01}\big)\mbf A_T-\mbf A_T'\big({\mbf S}_{11}-{\mbf S}_{10}{\mbf S}_{00}^{-1}{\mbf S}_{01}\big)\mbf A_T\big\Vert}{\wh{\mu}_c^0}\nn\\
&\leq \frac{2^{3/2}\sqrt c\big\Vert\mbf A_T'\big(\wh{\mbf S}_{11}-\wh{\mbf S}_{10}\wh{\mbf S}_{00}^{-1}\wh{\mbf S}_{01}\big)\mbf A_T-\mbf A_T'\big({\mbf S}_{11}-{\mbf S}_{10}{\mbf S}_{00}^{-1}{\mbf S}_{01}\big)\mbf A_T\big\Vert} {D_1} =O_p(\vartheta_{nT,\delta,\eta}).
\end{align}
%
Define the transformed estimators
\begin{align}
\widetilde{\bm\beta} = \wh{\bm\beta}\mbf O_c(\check{\bm\beta}_*'\wh{\bm\beta}\mbf O_c)^{-1}
, \qquad\widetilde{\bm\beta}^0 =  \wh{\bm\beta}^0(\check{\bm\beta}_*'\wh{\bm\beta}^0)^{-1}.\label{eq:newbeta}
\end{align}
From Lemma 13.1 in \citet{Johansen95}, we have (recall that $\check{\bm\beta}'_{\perp}\check{\bm\beta}=\mbf 0_{r-c\times c}$)
\beq\label{eq:opuno}
\mbf A_T^{-1} \widetilde{\bm\beta}^0 =\mbf A_T^{-1}\big(\check{\bm\beta}+\check{\bm\beta}_{\perp*}^{}\check{\bm\beta}_{\perp}'\widetilde{\bm\beta}^0\big)= \l(\begin{array}{c}
\mbf I_c\\
\sqrt T \check{\bm\beta}'_{\perp}\widetilde{\bm\beta}^0
\end{array}
\r)=
\l(\begin{array}{c}
\mbf I_c\\
\sqrt T \check{\bm\beta}'_{\perp}(\widetilde{\bm\beta}^0-\check{\bm\beta})
\end{array}
\r)
=\l(\begin{array}{c}
\mbf I_c\\
o_p(1)
\end{array}
\r),
\eeq
since $\mbf A_T^{-1} \widetilde{\bm\beta}^0$ spans a space of dimension $c$. In the same way, we have
\beq
\mbf A_T^{-1} \widetilde{\bm\beta} = \l(\begin{array}{c}
\mbf I_c\\
\sqrt T\check{\bm\beta}'_{\perp} \widetilde{\bm\beta}
\end{array}
\r) = \l(\begin{array}{c}
\mbf I_c\\
\sqrt T\check{\bm\beta}'_{\perp} (\widetilde{\bm\beta}-\wt{\bm\beta})
\end{array}
\r)=\l(\begin{array}{c}
\mbf I_c\\
\sqrt T\check{\bm\beta}'_{\perp} (\widetilde{\bm\beta}^0-\wt{\bm\beta})+\sqrt T\check{\bm\beta}'_{\perp} (\widetilde{\bm\beta}-\widetilde{\bm\beta}_0)
\end{array}
\r).\label{eq:opuno2}
\eeq
Now since $\mbox{span}(\mbf A_T^{-1} \widetilde{\bm\beta})=\mbox{span}(\mbf A_T^{-1} \wh{\bm\beta})$, also \eqref{eq:opuno2} spans a space of dimension $c$. Then, since also $\mbox{span}(\mbf A_T^{-1} \widetilde{\bm\beta}^0)=\mbox{span}(\mbf A_T^{-1} \wh{\bm\beta}^0)$, by comparing \eqref{eq:opuno} and \eqref{eq:opuno2}, and using \eqref{eq:AVAV2} and \eqref{eq:newbeta}, we have
\beq\label{eq:opuno3}
\big\Vert\sqrt T\check{\bm\beta}'_{\perp} (\widetilde{\bm\beta}-\widetilde{\bm\beta}_0)\big\Vert = \big\Vert\mbf A_T^{-1} \widetilde{\bm\beta}-\mbf A_T^{-1} \widetilde{\bm\beta}^0\big\Vert=O_p(\vartheta_{nT,\delta,\eta}).
\eeq
Therefore, given that $\Vert\check{\bm\beta}'_{\perp} \Vert=O(1)$, from \eqref{eq:opuno} and \eqref{eq:opuno3}, we have
\beq
\big\Vert\widetilde{\bm\beta}-\wt{\bm\beta}\big\Vert\leq \big\Vert\widetilde{\bm\beta}^0-\wt{\bm\beta}\big\Vert+\big\Vert\widetilde{\bm\beta}^0-\widetilde{\bm\beta}\big\Vert=o_p\l(\frac{1}{\sqrt T}\r)+O_p\l(\frac{\vartheta_{nT,\delta,\eta}}{\sqrt T}\r).\label{eq:betatildecons}
\eeq
Finally, from \eqref{eq:newbeta}, we can always define a $c\times c$ orthogonal matrix $\mbf Q$, which depends on $\mbf O_c$, and such that $\widetilde{\bm\beta}\mbf Q=\widehat{\bm\beta}$ \citep[see also pp.179-180 in][for a discussion about the choice of the identification matrix $\mbf Q$]{Johansen95}. Therefore, we have
\[
\big\Vert\wh{\bm\beta}-\wt{\bm\beta}\mbf Q\big\Vert=\big\Vert\wh{\bm\beta}-\mbf J{\bm\beta}\mbf Q\big\Vert = O_p\l(\frac{\vartheta_{nT,\delta,\eta}}{\sqrt T}\r),
\]
which completes the proof of part (i).\smallskip

Once we have $\wh{\bm\beta}$, the other parameters are estimated by linear regression as
\beq\label{eq:alphaG}
\wh{\bm\alpha}=\wh{\mbf S}_{01}\wh{\bm\beta}\big(\wh{\bm\beta}'\wh{\mbf S}_{11}\wh{\bm\beta}\big)^{-1}, \qquad \wh{\mbf G}_1=\big(\wh{\mbf M}_{02}-\wh{\bm\alpha}\wh{\bm\beta}'\wh{\mbf M}_{12}\big)\wh{\mbf M}_{22}^{-1}.
\eeq
For part (ii), first notice that, by definition from a VECM for $\mbf F_t$, since $\mbf u_t$ is independent because of Assumption \ref{ASS:common}(a), we have 
$
\bm\alpha = \E[\Delta \mbf F_t\mbf F_{t-1}'\bm\beta| \Delta \mbf F_{t-1}]\big(\E[\bm\beta' \mbf F_t\mbf F_{t-1}'\bm\beta| \Delta \mbf F_{t-1}]\big)^{-1}.
$
Therefore, since conditioning on $\Delta \mbf F_{t-1}$ is equivalent to conditioning on $\mbf J\Delta \mbf F_{t-1}=\Delta\wt{\mbf F}_{t-1}$ and since $\bm\beta'\mbf F_t = \wt{\bm\beta}'\wt{\mbf F}_t$,  we immediately have
\begin{align}
\check{\bm\alpha} =\mbf J\bm\alpha =& \mbf J \E[\Delta \mbf F_t\check{\mbf F}_{t-1}'\wt{\bm\beta}| \Delta \check{\mbf F}_{t-1}]\big(\E[\wt{\bm\beta}' \wt{\mbf F}_t\wt{\mbf F}_{t-1}'\wt{\bm\beta}| \Delta \wt{\mbf F}_{t-1}]\big)^{-1}\nn\\
=&\E[\Delta \check{\mbf F}_t\check{\mbf F}_{t-1}'\wt{\bm\beta}| \Delta \check{\mbf F}_{t-1}]\big(\E[\wt{\bm\beta}' \wt{\mbf F}_t\wt{\mbf F}_{t-1}'\wt{\bm\beta}| \Delta \wt{\mbf F}_{t-1}]\big)^{-1}= \wt{\bm\Omega}_{0\wt{\beta}}^{}\wt{\bm\Omega}_{\wt{\beta}\wt{\beta}}^{-1}.\nn
\end{align}
Then, 
\begin{align}
\big\Vert \wh{\mbf S}_{01}\wh{\bm\beta} - \wt{\bm\Omega}_{0\wt{\beta}}^{}\mbf Q\big\Vert \leq& \big\Vert \wh{\mbf S}_{01}(\wh{\bm\beta} -\wt{\bm\beta}\mbf Q)\big\Vert + \big\Vert\wh{\mbf S}_{01}\wt{\bm\beta}\mbf Q-\mbf S_{01}\wt{\bm\beta}\mbf Q\big\Vert + \big\Vert\mbf S_{01}\wt{\bm\beta}\mbf Q-\wt{\bm\Omega}_{0\wt{\beta}}\mbf Q\big\Vert = O_p(\vartheta_{nT,\delta,\eta}),\label{eq:alpha1}
\end{align}
using part (i) and the fact that $\Vert\wh{\mbf S}_{01}\Vert=O_p(T^{1/2})$ for the first term on the rhs, Lemma \ref{lem:shat}(iv) for the second term, and   
Lemma \ref{lem:shat2}(iii) for the third term. Analogously we have
\begin{align}
\big\Vert \wh{\bm\beta}' \wh{\mbf S}_{11}\wh{\bm\beta} - \mbf Q'\wt{\bm\Omega}_{\wt\beta\wt{\beta}}^{}\mbf Q\big\Vert \leq& \big\Vert(\wh{\bm\beta}' -\mbf Q'\wt{\bm\beta}') \wh{\mbf S}_{11}(\wh{\bm\beta} -\wt{\bm\beta}\mbf Q)\big\Vert + \big\Vert\mbf Q'\wt{\bm\beta}'\wh{\mbf S}_{11}\wt{\bm\beta}\mbf Q-\mbf Q'\wt{\bm\beta}'\mbf S_{11}\wt{\bm\beta}\mbf Q\big\Vert\nn\\
& + \big\Vert\mbf Q'\wt{\bm\beta}'\mbf S_{11}\wt{\bm\beta}\mbf Q-\mbf Q'\wt{\bm\Omega}_{\wt{\beta}\wt{\beta}}\mbf Q\big\Vert = O_p(\vartheta_{nT,\delta,\eta}),\label{eq:alpha2}
\end{align}
using part (i) and the fact that $\Vert\wh{\mbf S}_{11}\Vert=O_p(T)$ for the first term, Lemma \ref{lem:shat}(ii) for the second term, and   
Lemma \ref{lem:shat2}(ii) for the third term. Therefore, from \eqref{eq:alphaG}, \eqref{eq:alpha1}, and \eqref{eq:alpha2}, and since $\mbf Q$ is orthogonal, we have
\begin{align}
\big\Vert\wh{\bm\alpha}-\wt{\bm\alpha}\mbf Q\big\Vert=\big\Vert\wh{\bm\alpha}-\mbf J{\bm\alpha}\mbf Q\big\Vert = O_p(\vartheta_{nT,\delta,\eta}),\nn
\end{align}
which proves part (ii).\smallskip

For part (iii), notice that, by definition, we have:
\beq
\check{\mbf G}_1 = \mbf H\mbf G_1\mbf H' = \big(\bm\Gamma_1^{\Delta\wt{F}}-\check{\bm\alpha}\E[\wt{\bm\beta}'\wt{\mbf F}_{t-1}\Delta\wt{\mbf F}_{t-1}']\big)(\bm\Gamma_0^{\Delta\wt{F}})^{-1}.\label{eq:GG1}
\eeq
Then, from \eqref{eq:alphaG}, 
\begin{align}
\big\Vert\wh{\mbf G}_1-\check{\mbf G}_1\big\Vert\leq &\big\Vert\big(\wh{\mbf M}_{02}-\wh{\bm\alpha}\wh{\bm\beta}'\wh{\mbf M}_{12}\big)\wh{\mbf M}_{22}^{-1}-\big(\wh{\mbf M}_{02}-\wt{\bm\alpha}\wt{\bm\beta}'\wh{\mbf M}_{12}\big)\wh{\mbf M}_{22}^{-1}\big\Vert\nn\\
&+ \big\Vert\big(\wh{\mbf M}_{02}-\wt{\bm\alpha}\wt{\bm\beta}'\wh{\mbf M}_{12}\big)\wh{\mbf M}_{22}^{-1}-\big({\mbf M}_{02}-\wt{\bm\alpha}\wt{\bm\beta}'{\mbf M}_{12}\big){\mbf M}_{22}^{-1}\big\Vert\nn\\
&+\big\Vert\big({\mbf M}_{02}-\wt{\bm\alpha}\wt{\bm\beta}'{\mbf M}_{12}\big){\mbf M}_{22}^{-1}-\big(\bm\Gamma_1^{\Delta\wt{F}}-\check{\bm\alpha} \E[\wt{\bm\beta}'\wt{\mbf F}_{t-1}\Delta\wt{\mbf F}_{t-1}']\big)(\bm\Gamma_0^{\Delta\wt{F}})^{-1}\big\Vert= O_p(\vartheta_{nT,\delta,\eta}),\nn
\end{align}
since the first term on the rhs is $O_p(\vartheta_{nT,\delta,\eta})$ by parts (i) and (ii) and since $\check{\bm\alpha}\mbf Q\mbf Q'\check{\bm\beta}'=\check{\bm\alpha}\check{\bm\beta}'$, the second term is $O_p(\vartheta_{nT,\delta,\eta})$ by Lemma \ref{lem:mhat}(iii), \ref{lem:mhat}(iv) and \ref{lem:mhat}(vii), and the third term is $O_p(T^{-1/2})$ by Lemma \ref{rem:DFF}(i) and \ref{rem:DFF}(vi) and Chebychev's inequality. This, together with \eqref{eq:GG1}, proves part (iii).\smallskip

For part (iv), first consider the VECM residuals $\wh{\mbf w}_t = \Delta\wh{\mbf F}_t-\wh{\bm\alpha}\wh{\bm\beta}'\wh{\mbf F}_{t-1}-\wh{\mbf G}_1\Delta\wh{\mbf F}_{t-1}$ and notice that their sample covariance is also written as 
\begin{align}
\wh{\bm\Gamma}_0^w =& \frac 1T \sum_{t=1}^T \wh{\mbf w}_t \wh{\mbf w}_t'=\frac 1 T\sum_{t=1}^T(\Delta\wh{\mbf F}_t-\wh{\bm\alpha}\wh{\bm\beta}'\wh{\mbf F}_{t-1}-\wh{\mbf G}_1\Delta\wh{\mbf F}_{t-1})(\Delta\wh{\mbf F}_t-\wh{\bm\alpha}\wh{\bm\beta}'\wh{\mbf F}_{t-1}-\wh{\mbf G}_1\Delta\wh{\mbf F}_{t-1})'\nn\\
=& \wh{\mbf M}_{00} + \wh{\bm\alpha} \wh{\bm\beta}'\wh{\mbf M}_{11} \wh{\bm\beta}\wh{\bm\alpha}' + \wh{\mbf G}_1\wh{\mbf M}_{22}\wh{\mbf G}_1'-\wh{\mbf M}_{01}\wh{\bm\beta}\wh{\bm\alpha}'-\wh{\bm\alpha} \wh{\bm\beta}'\wh{\mbf M}_{12}\wh{\mbf G}_1'-\wh{\bm\alpha} \wh{\bm\beta}'\wh{\mbf M}_{10} -\wh{\mbf G}_1\wh{\mbf M}_{20}-\wh{\mbf G}_1\wh{\mbf M}_{21}\wh{\bm\beta}\wh{\bm\alpha}'.\nn
\end{align}
Then from parts (i), (ii) and (iii), Lemma \ref{lem:mhat}(ii) through \ref{lem:mhat}(vi) and \ref{lem:mhat}(ix), and Lemma \ref{rem:DFF}(i) and \ref{rem:DFF}(vi), we can prove that
\beq\label{eq:K1}
\big\Vert\wh{\bm\Gamma}_0^w-\mbf J{\bm\Gamma}_0^w\mbf J\big\Vert=O_p(\vartheta_{nT,\delta,\eta}),
\eeq 
where ${\bm\Gamma}_0^w = \E\big[\mbf w_t\mbf w_t'\big] = \E\big[(\Delta{\mbf F}_t-{\bm\alpha}{\bm\beta}'{\mbf F}_{t-1}-{\mbf G}_1\Delta{\mbf F}_{t-1})(\Delta{\mbf F}_t-{\bm\alpha}{\bm\beta}'{\mbf F}_{t-1}-{\mbf G}_1\Delta{\mbf F}_{t-1})'\big]$. 

By \eqref{VECMhatF}, we have $\mbf w_t = \mbf K\mbf u_t$, therefore, since the shocks $\mbf u_t$ are orthonormal by Assumption \ref{ASS:common}(a), we have $\bm{\Gamma}_0^w=\mbf K\mbf K'$. Denote as $\mu_j^w$, $j=1,\ldots,q$, the $q$ largest eigenvalues of $\bm{\Gamma}_0^w$, which are also the $q$ eigenvalues of $\mbf K'\mbf K$ and are asymptotically distinct by the assumption made in the statement of Proposition \ref{vecm}. Then, since $\mbf K=\mbf Q(0)=\mbf C(0)$, from Assumption \ref{ASS:common} and the model given in \eqref{eq:model2vector}, we have ${\rm rk}(\mbf K)= q$, and therefore there exist positive reals $\underline D_j, \overline D_j$ and an integer $\bar n$, such that $\underline D_j>\overline D_{j+1}$ for $j=1,\ldots, q-1$, and  $\underline D_j\le \mu_j^w\le \overline D_j$, for $n>\bar n$ and $j=1,\ldots, q$. 

Denote as $\mu_j^w$ the eigenvalues of $\bm{\Gamma}_0^w$, which are also the eigenvalues of $\mbf K'\mbf K$. Then, Denote by $\bm w_j^w$ the eigenvector corresponding to $\mu_j^w$ and define as $\mbf M^w$ the $q\times q$  diagonal matrix with entries $\mu_j^w$ and as $\mbf W^w=(\bm w_1^w\cdots \bm w_q^w)$ the corresponding $r\times q$ matrix of normalized eigenvectors. For any $q\times q$ invertible matrix $\mbf P$, we can always write  $\mbf w_t = \big[\mbf K\mbf P\big]\big[\mbf P^{-1}\mbf u_t\big] =\mbf H\mbf v_t$. In particular, let us choose $\mbf P$ to be such that (recall the identity $\mbf w_t =\mbf W^w\mbf W^{w'}\mbf w_t$)
\beq\label{identKU}
\mbf v_t=\mbf P^{-1}\mbf u_t=(\mbf M^{w})^{-1/2}\mbf W^{w'} \mbf w_t, \qquad \mbf H=\mbf K\mbf P = \mbf W^w(\mbf M^{w})^{1/2}.
\eeq
Then, for this choice of $\mbf P$, we have $\bm\Gamma^v_0= \E[\mbf v_t\mbf v_t'] = (\mbf P)^{-1}(\mbf P)^{-1'}=\mbf I_q$, and therefore $\mbf P$ must be orthogonal, i.e., $\mbf P^{-1}=\mbf P'$.

Now, consider the estimators: $\wh{\mbf K}=\wh{\mbf W}^w(\wh{\mbf M}^{w})^{1/2}$ and $\wh{\mbf u}_t=(\wh{\mbf M}^{w})^{-1/2}\wh{\mbf W}^{w'}\wh{\mbf w}_t$, where $\wh{\mbf W}^w=(\wh{\bm w}_1^w\cdots \wh{\bm w}_q^w)$ is the $r\times q$ matrix of the first $q$ normalized eigenvectors of $\wh{\bm\Gamma}^w_0$ and $\wh{\mbf M}^{w}$ is the $q\times q$ diagonal matrix of the corresponding eigenvalues $\wh{\mu}_j^w$. 
Then, since $\underline D_j>\overline D_{j+1}$ for $j=1,\ldots, q-1$, by Corollary 1 in \citet{yu15} and because of \eqref{eq:K1}, for $j=1,\ldots, q$, we have (note that $\mbf J\mbf W^w$ are eigenvectors of $\mbf J\bm\Gamma_0^w\mbf J$ with eigenvalues $\mu_j^w$)
\begin{align}
\big\Vert \wh{\bm w}_j^w-\mbf J\bm w_j^w s_j^w\big\Vert&\le \frac{2^{3/2}\big\Vert\wh{\bm\Gamma}_0^w-\mbf J{\bm\Gamma}_0^w\mbf J\big\Vert}{\min((\mu_{j-1}^w-\mu_{j}^w),(\mu_j^w-\mu_{j+1}^w))}\le \frac{2^{3/2}\big\Vert\wh{\bm\Gamma}_0^w-\mbf J{\bm\Gamma}_0^w\mbf J\big\Vert}{\min((\underline D_{j-1}-\overline D_j),(\underline D_j-\overline D_{j+1}))}=O_p(\vartheta_{nT,\delta,\eta}),\label{dkqw}
\end{align}
where $s_j^w=\mbox{sign}(\wh{\bm w}_j^{w'}\mbf J\bm w_j^w)$ and we define $\mu_0^w=\infty$. Define as $\mbf J^w$ the $q\times q$ diagonal matrix with entries $s_j^w$, then from \eqref{dkqw}, we have
\beq
\big\Vert\wh{\mbf W}^w-\mbf J\mbf W^w \mbf J^w\big\Vert\le \sqrt{\sum_{j=1}^q\big\Vert \wh{\bm w}_j^w-\mbf J\bm w_j^w s_j^w\big\Vert^2}=O_p(\vartheta_{nT,\delta,\eta}).\label{dkqw2}
\eeq
Now, let us consider the estimated eigenvalues. From, \eqref{eq:K1} and using Weyl's inequality \eqref{eq:weyl}, we have
\beq\label{muqmuhatq}
\big|\wh{\mu}^{w}_j-{\mu}^{w}_j\big|\leq \big\Vert\wh{\bm\Gamma}_0^w-\mbf J{\bm\Gamma}_0^w\mbf J\big\Vert=O_p(\vartheta_{nT,\delta,\eta}),\qquad j=1,\ldots ,q,
\eeq
which implies
\beq\label{eq:K2sqrt}
\big|(\wh{\mu}^{w}_j)^{1/2}-({\mu}^{w}_j)^{1/2}\big|\le \frac{\big|\wh{\mu}^{w}_j-{\mu}^{w}_j\big|}{2({\mu}^{w}_j)^{1/2}}\le \frac{\big|\wh{\mu}^{w}_j-{\mu}^{w}_j\big|} {2 D^{1/2}} =O_p(\vartheta_{nT,\delta,\eta}),\qquad j=1,\ldots ,q.
\eeq
Therefore, from \eqref{eq:K2sqrt}, we have
\beq\label{eq:K2sqrt2}
\Vert(\wh{\mbf M}^w)^{1/2}-(\mbf M^w)^{1/2}\Vert\le \sqrt{\sum_{j=1}^q \big((\wh{\mu}^{w}_j)^{1/2}-({\mu}^{w}_j)^{1/2}\big)^2}=O_p(\vartheta_{nT,\delta,\eta}),
\eeq
Let us define the orthogonal matrix $\mbf R=\mbf J^w\mbf P'$, then, using \eqref{identKU}, \eqref{dkqw2}, and \eqref{eq:K2sqrt2}, we have (notice that $\mbf P \mbf J^w=\mbf P(\mbf M^{w})^{-1/2}\mbf J^w (\mbf M^{w})^{1/2}$ and $\mbf H(\mbf M^{w})^{-1/2}=\mbf W^w$)
\begin{align}
\big\Vert \wh{\mbf K} -\mbf J\mbf K\mbf R'\big\Vert &= \big\Vert\wh{\mbf W}^w(\wh{\mbf M}^w)^{1/2} -\mbf J\mbf K\mbf P\mbf J^{w}\big\Vert 
=\big\Vert\wh{\mbf W}^w(\wh{\mbf M}^w)^{1/2} -\mbf J\mbf K\mbf P(\mbf M^{w})^{-1/2}\mbf J^w (\mbf M^{w})^{1/2}\big\Vert \nn\\
&=\big\Vert\wh{\mbf W}^w(\wh{\mbf M}^w)^{1/2} -\mbf J\mbf H(\mbf M^{w})^{-1/2}\mbf J^w (\mbf M^{w})^{1/2}\big\Vert =
\big\Vert\wh{\mbf W}^w(\wh{\mbf M}^w)^{1/2} -\mbf J\mbf W^w\mbf J^w (\mbf M^{w})^{1/2}\big\Vert \nn\\
&\le \big\Vert\wh{\mbf W}^w -\mbf J\mbf W^w\mbf J^w \big\Vert \,\Vert (\mbf M^{w})^{1/2}\Vert
+\big\Vert(\wh{\mbf M}^w)^{1/2} - (\mbf M^{w})^{1/2}\big\Vert + o_p(\vartheta_{nT,\delta,\eta})=O_p(\vartheta_{nT,\delta,\eta}),\nn
\end{align}
because $\Vert (\mbf M^{w})^{1/2}\Vert=(\mu_1^w)^{1/2}\le (\overline D_1)^{1/2}$ for $n>\bar n$, and $\Vert\mbf W^w\Vert=\Vert \mbf J\Vert=\Vert \mbf J^w\Vert=1$. This proves part (iv).\smallskip 

For part(v), first notice that, given $t$, we have
\begin{align}\label{consistvecmerror}
\big\Vert\wh{\mbf w}_t - \mbf J\mbf w_t\big\Vert =& \big\Vert (\Delta\wh{\mbf F}_t-\wh{\bm\alpha}\wh{\bm\beta}'\wh{\mbf F}_{t-1}-\wh{\mbf G}_1\Delta\wh{\mbf F}_{t-1})-(\mbf J\Delta{\mbf F}_t-\mbf J{\bm\alpha}\mbf Q\mbf Q'{\bm\beta}'\mbf J\mbf J{\mbf F}_{t-1}-\mbf J{\mbf G}_1\mbf J\mbf J\Delta{\mbf F}_{t-1})\big\Vert\nn\\
\leq& \big\Vert\Delta\wh{\mbf F}_t-\mbf J\Delta{\mbf F}_t\big\Vert 
+\big\Vert\Delta\wh{\mbf F}_t-\mbf J\Delta{\mbf F}_t\big\Vert \, \Vert \mbf G_1\Vert 
+ \big\Vert\wh{\mbf G}_1-\mbf J{\mbf G}_1\mbf J\big\Vert\, \Vert \Delta\wh{\mbf F}_{t-1}\Vert\nn\\
&+\big\Vert\wh{\bm\alpha}-\mbf J{\bm\alpha}\mbf Q\big\Vert \,\Vert\mbf Q'{\bm\beta}'{\mbf F}_{t-1}\Vert
+\big\Vert\wh{\bm\beta}'-\mbf Q'{\bm\beta}'\mbf J\big\Vert\,\Vert{\mbf F}_{t-1}\Vert \, \Vert\bm\alpha\mbf Q\Vert \nn\\
&+\big\Vert\bm\beta'\mbf J(\wh{\mbf F}_{t-1}-\mbf J\mbf F_{t-1})\big\Vert\, \Vert\mbf Q\Vert\,\Vert\bm\alpha\mbf Q\Vert+o_p(\vartheta_{nT,\delta,\eta})= O_p(\vartheta_{nT,\delta,\eta}).
\end{align}
Indeed, for the first and second term on the rhs of \eqref{consistvecmerror} by taking differences and multiplying by $\sqrt T$ in Lemma \ref{lem:load}(iii) and using \eqref{UNNAMED_L} in the proof of Lemma \ref{lem:mhat}, we immediately have that $\Vert\Delta\wh{\mbf F}_t-\mbf J\Delta{\mbf F}_t\Vert =O_p(T^{-1/2})$, for the third, fourth and fifth terms we can use parts (iii), (ii), and  (i), respectively, and  $\Vert\mbf F_{t-1}\Vert=O_p(\sqrt T)$ by Lemma \ref{lem:main}(ii) and $\Vert \mbf Q'\bm\beta'\mbf F_{t-1}\Vert=O_p(1)$ by Lemma \ref{lem:BN}(ii), and obviously $\Vert \mbf G_1\Vert=O(1)$, $\Vert\bm\alpha\mbf Q\Vert = O(1)$, $\Vert\mbf Q\Vert= O(1)$, and $\Vert \mbf J\Vert=1$. While, for the last term on the rhs of \eqref{consistvecmerror}, using the same approach as in the proof of Lemma \ref{lem:mhat}(ix) (see \eqref{eq:mhatc2}), we have
\beq
\big\Vert\bm\beta'\mbf J(\wh{\mbf F}_{t-1}-\mbf J\mbf F_{t-1})\big\Vert = \big\Vert\check{\bm\beta}'(\wh{\mbf F}_{t-1}-\mbf J{\mbf F}_{t-1})\big\Vert = O_p(\vartheta_{nT,\delta,\eta}).\nn
\eeq
Second, since ${\rm rk}(\bm\Gamma_0^w)=q$ then $\mu_q^w>0$ for any $n\in\mathbb N$ and therefore $\mbf M^w$ is always invertible. Moreover,
since, for $n>\bar n$, $\wh{\mu}_q^w\ge \underline D_q+O_p(\vartheta_{nT,\delta,\eta})$ because of \eqref{muqmuhatq}, then the inverse of $\wh{\mbf M}^w$ exists with probability tending to one as $n,T\to\infty$. Then, from \eqref{eq:K2sqrt}, we have
\begin{align}
\Vert(\wh{\mbf M}^w)^{-1/2}-(\mbf M^w)^{-1/2}\Vert&\leq \Vert(\wh{\mbf M}^w)^{-1/2}-(\mbf M^w)^{-1/2}\Vert_F \le \sum_{j=1}^q\l\vert\frac {(\mu_j^w)^{1/2}-(\wh{\mu}^{w}_j)^{1/2}}{(\wh{\mu}_j^w\mu_j^w)^{1/2}}\r\vert\nn\\
&\leq \frac{q \max_{j=1,\ldots, q} \vert(\wh{\mu}^{w}_j)^{1/2}-({\mu}^{w}_j)^{1/2}\vert}{\underline D_q+O_p(\vartheta_{nT,\delta,\eta})}=O_p(\vartheta_{nT,\delta,\eta}).\label{eq:K2sqrtinverse}
\end{align}
Then, from \eqref{identKU}, \eqref{dkqw2}, \eqref{consistvecmerror}, and \eqref{eq:K2sqrt}, since $\Vert \mbf w_t\Vert=O_p(1)$, 
$\Vert(\mbf M^w)^{-1/2}\Vert=(\mu_q^w)^{-1/2}\le (\underline D_q)^{-1/2}$ for $n>\bar n$, and $\Vert\mbf W^w\Vert=\Vert \mbf J\Vert=\Vert \mbf J^w\Vert=1$, we have 
\begin{align}
\big\Vert \wh{\mbf u}_t&-\mbf R\mbf u_t\big\Vert =\big\Vert (\wh{\mbf M}^{w})^{-1/2}\wh{\mbf W}^{w'}\wh{\mbf w}_t-\mbf J^w\mbf P'\mbf u_t
\big\Vert
=\big\Vert (\wh{\mbf M}^{w})^{-1/2}\wh{\mbf W}^{w'}\wh{\mbf w}_t-(\mbf M^w)^{-1/2}\mbf J^w(\mbf M^w)^{1/2}\mbf P'\mbf u_t
\big\Vert\nn\\
=&\big\Vert (\wh{\mbf M}^{w})^{-1/2}\wh{\mbf W}^{w'}\wh{\mbf w}_t-(\mbf M^w)^{-1/2}\mbf J^w(\mbf M^w)^{1/2}\mbf v_t
\big\Vert=\big\Vert (\wh{\mbf M}^{w})^{-1/2}\wh{\mbf W}^{w'}\wh{\mbf w}_t-(\mbf M^w)^{-1/2}\mbf J^w\mbf W^{w'}\mbf W^w(\mbf M^w)^{1/2}\mbf v_t
\big\Vert\nn\\
=&\big\Vert (\wh{\mbf M}^{w})^{-1/2}\wh{\mbf W}^{w'}\wh{\mbf w}_t-(\mbf M^w)^{-1/2}\mbf J^w\mbf W^{w'}\mbf H\mbf v_t
\big\Vert=\big\Vert (\wh{\mbf M}^{w})^{-1/2}\wh{\mbf W}^{w'}\wh{\mbf w}_t-(\mbf M^w)^{-1/2}\mbf J^w\mbf W^{w'}\mbf J\mbf J\mbf w_t
\big\Vert\nn\\
\leq& \big\Vert \wh{\mbf W}^{w'}-\mbf J^w\mbf W^{w'} \mbf J\big\Vert\, \Vert \mbf w_t\Vert\,\Vert (\mbf M^w)^{-1/2}\Vert+ \big\Vert \wh{\mbf w}_t-\mbf J\mbf w_t\big\Vert\,\,\Vert (\mbf M^w)^{-1/2}\Vert +\big\Vert(\wh{\mbf M}^w)^{-1/2} - (\mbf M^{w})^{-1/2}\big\Vert\,\Vert\mbf w_t\Vert+ o_p(\vartheta_{nT,\delta,\eta})\nn\\
=&O_p(\vartheta_{nT,\delta,\eta}),\nn
\end{align}
and this proves part (v).\smallskip

For part (vi) consider an estimator of $\mbf R$, given by $\wh{\mbf R}$ obtained by imposing suitable restrictions on the raw IRFs \eqref{eq:IRF1_NOTID}, then the true IRF and the identified estimated IRF of $x_{it}$ to $u_{jt}$ at lag $k$ are given by (see also the definitions in \eqref{trueIRF} and \eqref{eq:IRF1})
\begin{align}
&\phi_{ijk} = \bm\lambda_i' \mbf B_k \mbf k_j = \big[\bm\lambda_i'\mbf J\big]\big[\mbf J\mbf B_k\mbf J\big]\big[\mbf J\mbf k_j\big],\qquad \widehat{\phi}_{ijk}^{\mbox{\tiny VECM}} = \wh{\bm\lambda}_i'\wh{\mbf B}_k\wh{\mbf K}\wh{\mbf r}_j,\nn
\end{align}
where $\mbf B_k$ is the $k$-th coefficient of $(1-L)^{-1}\mbf C(L)$, $\wh{\mbf B}_k$ is the $k$-th coefficient of $[\wh{\mbf A}^{\mbox{\tiny VECM}}(L)]^{-1}$, $\mbf k_j$ is the $j$-th column of $\mbf K$, and $\wh{\mbf r}_j$ is the $j$-th column of $\wh{\mbf R}$. 

The estimated VECM with $p=1$ can always be written as a VAR(2) with estimated matrix polynomial, $\widehat{\mbf A}^{\mbox{\tiny{VECM}}}(L)=\mbf I_r-\widehat{\mbf A}^{\mbox{\tiny{VECM}}}_1L-\widehat{\mbf A}^{\mbox{\tiny{VECM}}}_2L^2$, where $\widehat{\mbf A}^{\mbox{\tiny{VECM}}}_1 = \widehat{\mbf G}_1+\widehat{\bm\alpha}\widehat{\bm\beta}'+\mbf I_r$, and $\widehat{\mbf A}^{\mbox{\tiny{VECM}}}_2 = -\widehat{\mbf G}_{1}$. 
Then, from parts (i), (ii) and (iii), we have, for $k=1,2$, 
\beq\label{eq:AHATVECMlemma}
\big\Vert\wh{\mbf A}_k^{\mbox{\tiny VECM}}-\mbf J\mbf A_k\mbf J\big\Vert=O_p(\vartheta_{nT,\delta,\eta}).
\eeq
Define the infinite matrix polynomial
$
\wh{\mbf B}(L) = \big[\wh{\mbf A}^{\mbox{\tiny VECM}}(L)\big]^{-1} = (\mbf I_r-\wh{\mbf A}_1^{\mbox{\tiny VECM}}L-\wh{\mbf A}_2^{\mbox{\tiny VECM}}L^2)^{-1}=\sum_{k=0}^{\infty} \wh{\mbf B}_kL^k,
$
such that $\wh{\mbf B}(0)=\mbf I_r$, $\wh{\mbf B}_1=\wh{\mbf A}_1^{\mbox{\tiny VECM}}$, $\wh{\mbf B}_2=(\wh{\mbf A}_1^{\mbox{\tiny VECM}}\wh{\mbf B}_1+\wh{\mbf A}_2^{\mbox{\tiny VECM}})$, $\wh{\mbf B}_3=(\wh{\mbf A}_1^{\mbox{\tiny VECM}}\wh{\mbf B}_2+\wh{\mbf A}_2^{\mbox{\tiny VECM}}\wh{\mbf B}_1)$, and so on. Then, from \eqref{eq:AHATVECMlemma}, we have, for a given $k$,  
\beq\label{eq:bhat}
\big\Vert\wh{\mbf B}_k-\mbf J\mbf B_k\mbf J\big\Vert=O_p(\vartheta_{nT,\delta,\eta}).
\eeq
The estimator $\wh{\mbf R}$ is in general a function of $\wh{\bm\lambda}_i$, $\wh{\mbf B}(L)$, and $\wh{\mbf K}$, and for regular identification schemes, such that this mapping is analytical, using Lemma \ref{lem:load}(i), part (iv), and \eqref{eq:bhat}, we have 
\beq\label{eq:rhat}
\big\Vert\wh{\mbf R}-\mbf R\big\Vert=O_p(\vartheta_{nT,\delta,\eta}).
\eeq
Moreover, from part (iv) and \eqref{eq:rhat}, and since for any matrix $\mbf A$, 
$\Vert\mbf A\Vert_F\le \sqrt {{\rm rk}(\mbf A)}\, \Vert\mbf A\Vert$, we have
\beq\label{eq:rhatcol}
\big\Vert \wh{\mbf K}\wh{\mbf r}_j-\mbf J \mbf k_j \big\Vert \le \sqrt {rq}\big\Vert \wh{\mbf K}\wh{\mbf R}-\mbf J \mbf K \big\Vert_{F}\le q\sqrt r\big\Vert \wh{\mbf K}\wh{\mbf R}-\mbf J \mbf K \big\Vert= O_p(\vartheta_{nT,\delta,\eta}).
\eeq
Then, by Lemma \ref{lem:load}(i), part (iv), and using \eqref{eq:bhat} and \eqref{eq:rhatcol}, for given $i,j$ and $k$, we have (note that $\vartheta_{nT,\delta,\eta}\ge \max(T^{-1/2},n^{-1/2})$)
\begin{align}
\big\vert
\widetilde{\phi}_{ijk}^{\mbox{\tiny VECM}}-\phi_{ijk} 
\big\vert \le &
\big\Vert  \wh{\bm\lambda}_i'-\bm\lambda_i'\mbf J\big \Vert \, \Vert\mbf B_k\Vert\, \Vert \mbf K \mbf r_j\Vert + 
\big\Vert\wh{\mbf B}_k-\mbf J\mbf B_k\mbf J\big\Vert\, \Vert\bm\lambda_i\Vert \, \Vert \mbf K\mbf r_j\Vert\nn\\
&+\big\Vert \wh{\mbf K}\wh{\mbf r}_j - \mbf J\mbf k_j\big \Vert\, \Vert\bm\lambda_i\Vert \,  \Vert\mbf B_k\Vert + o_p(\vartheta_{nT,\delta,\eta})= O_p(\vartheta_{nT,\delta,\eta}),
\end{align}
because $\Vert\mbf B_k\Vert=O(1)$, $\Vert \mbf K \mbf r_j\Vert=O(1)$, and by Assumption \ref{asm:factor}(b) $\Vert\bm\lambda_i\Vert=O(1)$. This proves part (vi) while part (vii) can be proved as in Theorem 2.9 by \citet{phillips98}. This completes the proof. \hfill $\Box$

\subsection{Proof of Proposition \ref{var}}
Define the $r\times r$ transformation $\bm{\mathcal D}=(\bm\beta\;\bm\beta_{\perp})'$, where $\bm\beta$ is the $r\times c$ cointegration vector of $\mbf F_t$, and $\bm\beta_{\perp}$ is such that ${\bm\beta}_{\perp}'{\bm\beta}=\mbf 0_{r-c\times r}$. Then, the vector process $\mbf Z_t =\bm{\mathcal D}\mbf F_t$, is partitioned into an $I(0)$ vector $\mbf Z_{0t}=\bm\beta'\mbf F_t$ and an $I(1)$ vector $\mbf Z_{1t}=\bm\beta_{\perp}'\mbf F_t$. The vectors $\mbf Z_{0t}$ and $\mbf Z_{1t}$ are orthogonal. 

Now consider the models for $\mbf F_t$, $\mbf Z_{0t}$, and $\mbf Z_{1t}$:
\begin{align}
\mbf F_{t} = \mbf A_{1}\mbf F_{t-1}+\mbf w_{t},\quad\mbf Z_{0t} = \mbf Q_{0}\mbf F_{t-1}+\bm\beta'\mbf w_{t},\quad\mbf Z_{1t} = \mbf Q_{1}\mbf F_{t-1}+\bm\beta_{\perp}'\mbf w_{t},\nn
\end{align}
where $\mbf Q_0$ is $c\times r$ and $\mbf Q_1$ is $r-c\times r$, and $\mbf w_t=\mbf K\mbf u_t$. Denote the ordinary least squares estimators of the above models, when using $\mbf F_t$, as $\wh{\mbf A}_1^{1\mbox{\tiny VAR}}$, $\wh{\mbf Q}_0$, and $\wh{\mbf Q}_1$ . Then,  
\begin{align}\label{eq:Q0}
\big\Vert\wh{\mbf Q}_0-\mbf Q_0\big\Vert= \bigg\Vert\bigg(\frac 1T \sum_{t=1}^T \bm\beta'\mbf F_{t-1}\mbf u_{t}'\mbf K' \bm\beta\bigg)\bigg(\frac 1T \sum_{t=1}^T \bm\beta'\mbf F_{t-1}\mbf F_{t-1}'\bm\beta\bigg)^{-1}\bigg\Vert= O_p\l(\frac 1{\sqrt T}\r).
\end{align}
Indeed, the first term on the rhs is $O_p(T^{-1/2})$ from \eqref{eq:fBN} and by independence of $\mbf u_t$ in Assumption \ref{ASS:common}(a), while the second term is $O_p(1)$ by Lemma \ref{rem:DFF}(v). Similarly,
\begin{align}\label{eq:Q1}
\big\Vert\wh{\mbf Q}_1-\mbf Q_1\big\Vert= \bigg\Vert\bigg(\frac 1{T^2} \sum_{t=1}^T \bm\beta'_{\perp}\mbf F_{t-1}\mbf u_{t}'\mbf K' \bm\beta_{\perp}\bigg)\bigg(\frac 1{T^2} \sum_{t=1}^T \bm\beta_{\perp}'\mbf F_{t-1}\mbf F_{t-1}'\bm\beta_{\perp}\bigg)^{-1}\bigg\Vert= O_p\l(\frac 1{T}\r).
\end{align}
Indeed, the first term on the rhs is $O_p(T^{-1})$ from \eqref{eq:fBN} and by independence of $\mbf u_t$ in Assumption \ref{ASS:common}(a), while the second term is $O_p(1)$ by Lemma \ref{rem:DFF}(ii). 
Moreover,
\beq\label{eq:AQQ}
\mbox{vec}\big(\wh{\mbf A}_1^{1\mbox{\tiny VAR}}\big)
= (\bm{\mathcal D}^{-1}\otimes \mbf I_r)
\l(\begin{array}{c}
\mbox{vec}(\wh{\mbf Q}_0')\\
\mbox{vec}(\wh{\mbf Q}_1') 
\end{array}
\r).
\eeq
Analogous formulas to \eqref{eq:Q0}-\eqref{eq:AQQ} are in Theorem 1 by \citet{simsstockwatson} and, by combining them, 
\beq\label{eq:A1VAR}
\big\Vert\wh{\mbf A}_1^{1\mbox{\tiny VAR}} - \mbf A_1\big\Vert = O_p\l(\frac 1 {\sqrt T}\r).
\eeq
Notice that of the $r^2$ parameters in $\mbf A_1$, $cr$ in $\mbf Q_0$ are estimated consistently with rate $O_p(T^{-1/2})$, while $(r-c)r$ in $\mbf Q_1$ with rate $O_p(T^{-1})$.

If we now denote as $\wh{\mbf A}_1^{0\mbox{\tiny VAR}}$ the ordinary least squares estimator for the VAR when using $\mbf J\mbf F_t$, then $\wh{\mbf A}_1^{0\mbox{\tiny VAR}}=\mbf J\wh{\mbf A}_1^{1\mbox{\tiny VAR}}\mbf J$, and from \eqref{eq:A1VAR}
\beq\label{eq:A0VAR}
\big\Vert\wh{\mbf A}_1^{0\mbox{\tiny VAR}} - \mbf J\mbf A_1\mbf J\big\Vert = O_p\l(\frac 1 {\sqrt T}\r).
\eeq
Define 
\beq\label{eq:mhatL}
\wh{\mbf M}_{1L}=\frac 1 T\sum_{t=1}^T \wh{\mbf F}_t\wh{\mbf F}_{t-1}',\qquad \wh{\mbf M}_{LL}=\frac 1 T\sum_{t=1}^T \wh{\mbf F}_{t-1}\wh{\mbf F}_{t-1}'.
\eeq
Then, we can write the VAR estimators as
\beq\label{eq:AVAR}
\wh{\mbf A}_1^{\mbox{\tiny VAR}}=\frac{\wh{\mbf M}_{1L}}T \l(\frac{\wh{\mbf M}_{LL}}T\r)^{-1}, \quad \wh{\mbf A}_1^{0\mbox{\tiny VAR}}=\frac{{\mbf M}_{1L}}T\l(\frac{{\mbf M}_{LL}}T\r)^{-1},
\eeq
where ${\mbf M}_{1L}$ and ${\mbf M}_{LL}$ are defined as in \eqref{eq:mhatL}, but when using $\mbf J\mbf F_t$. 

Because of Lemma \ref{lem:mhat}(i), we have
\begin{align}
&\bigg\Vert\frac {\wh{\mbf M}_{1L}} T-\frac {{\mbf M}_{1L}} T\bigg\Vert = O_p\l(\max\l(\frac1{\sqrt n},\frac 1 {\sqrt T},\frac 1{n^{1-\eta}}\r)\r),\qquad\bigg\Vert\frac {\wh{\mbf M}_{LL}} T-\frac {{\mbf M}_{LL}} T\bigg\Vert= O_p\l(\max\l(\frac1{\sqrt n},\frac 1 {\sqrt T},\frac 1{n^{1-\eta}}\r)\r),\nn
\end{align}
thus
\beq\label{eq:ancoravar}
\big\Vert\wh{\mbf A}_1^{\mbox{\tiny VAR}} - \wh{\mbf A}_1^{0\mbox{\tiny VAR}}\big\Vert = O_p\l(\max\l(\frac1{\sqrt n},\frac 1 {\sqrt T},\frac 1{n^{1-\eta}}\r)\r).
\eeq
By combining \eqref{eq:ancoravar} with \eqref{eq:A0VAR}
\begin{align}
\big\Vert\wh{\mbf A}_1^{\mbox{\tiny VAR}}-\mbf J\mbf A_1\mbf J
\big\Vert
\leq\big\Vert\wh{\mbf A}_1^{\mbox{\tiny VAR}}-\wh{\mbf A}_1^{0\mbox{\tiny VAR}}\big\Vert+\big\Vert\wh{\mbf A}_1^{0\mbox{\tiny VAR}}-\mbf J\mbf A_1\mbf J\big\Vert=O_p\l(\max\l(\frac1{\sqrt n},\frac 1 {\sqrt T},\frac 1{n^{1-\eta}}\r)\r),\label{eq:AAAA}
\end{align}
which completes the proof of part (i).\smallskip

By noticing that, as a consequence of part (i), \eqref{eq:K1} holds also in this case, but with the rate given in \eqref{eq:AAAA}, we prove parts (iii) and (iv) exactly as in Proposition \ref{vecm}(iv) and (v), respectively. \smallskip

For part (v), define
$
\wh{\mbf B}(L) = \big[\wh{\mbf A}^{\mbox{\tiny VAR}}(L)\big]^{-1} = (\mbf I_r-\wh{\mbf A}_1^{\mbox{\tiny VAR}}L)^{-1}=\sum_{k=0}^{\infty} \wh{\mbf B}_kL^k,
$
such that $\wh{\mbf B}_k=(\wh{\mbf A}_1^{\mbox{\tiny VAR}})^k$. Then, from part (i), we have, for a  given $k$, 
\beq\label{eq:bhatvar}
\big\Vert\wh{\mbf B}_k-\mbf J\mbf B_k\mbf J\big\Vert=O_p\l(\max\l(\frac1{\sqrt n},\frac 1{\sqrt T},\frac 1{n^{1-\eta}}\r)\r).
\eeq
The identified estimated IRF of $x_{it}$ to $u_{jt}$ at lag $k$ is given by (see also \eqref{eq:IRF1var})
\beq\label{eq:irfvarest}
\wh{\phi}_{ijk}^{\mbox{\tiny VAR}} = \wh{\bm\lambda}_i'\wh{\mbf B}_k\wh{\mbf K}\wh{\mbf r}_j,
\eeq
where $\wh{\mbf r}_j$ is the $j$-th column of $\wh{\mbf R}$, which is an estimator of the identifying matrix $\mbf R$. Such estimator  is in general a function of $\wh{\bm\lambda}_i$, $\wh{\mbf B}(L)$, and $\wh{\mbf K}$, and for regular identification schemes, such that this mapping is analytical, using Lemma \ref{lem:load}(i), part (i), and \eqref{eq:bhat}, and similarly to the proof of Proposition \ref{vecm}, we can show that (see \eqref{eq:rhatcol})
\beq\label{eq:rhatcolvar}
\big\Vert \wh{\mbf K}\wh{\mbf r}_j-\mbf J \mbf k_j \big\Vert = O_p\l(\max\l(\frac1{\sqrt n},\frac 1{\sqrt T},\frac 1{n^{1-\eta}}\r)\r).
\eeq
Consistency of the identified estimated IRFs \eqref{eq:irfvarest} is then proved in the same way as in the proof of Proposition \ref{vecm}, by using Lemma \ref{lem:load}(i), part (i), and \eqref{eq:bhatvar} and \eqref{eq:rhatcolvar}. This proves part (v), while part (vi) follows from Theorem 2.3 by \citet{phillips98}. This completes the proof. \hfill $\Box$
\subsection{Proof of Proposition \ref{prop:tau}}
For $\wh q$ the proof is in Proposition 2 in \citet{hallinliska07}. For $\wh \tau$ the proof follows similar steps but when fixing $\theta=0$ and combining it with Lemma \ref{rem:idiospect} and consistency of the spectral density estimator $\wh{\bm\Sigma}^{\Delta y}(\theta)$, which is proved in Proposition 6 in \citet{FHLZ15}.\hfill$\Box$

\subsection{Testing for linear trends}\label{app:testB}
For a given $i$ we have the model $y_{it}=a_i+b_i t+x_{it}$ in \eqref{eq:modeltrend}. We want to test the null and the alternative hypothesis 
\[
H_0: b_i=0, \;\mbox{ vs. }\; H_1: b_i\ne 0.
\]
Consider the following statistic:
\[
S_i = \sqrt T\l\{\frac{\frac 1T\sum_{t=2}^T\Delta y_{it}}{\sqrt{\mathcal V_i}}\r\},\;\text{ with }\; \mathcal V_i=\sum_{h=-M_T}^{M_T}  \l(1-\frac{|h|} {M_T}\r) \wh{\gamma}_{ih}^{\Delta y},
\]
where $\wh{\gamma}_{ih}^{\Delta y}$ is the lag-$h$ sample autocovariance of $\Delta y_{it}$. It can be shown that, under $H_0$, as $T\to\infty$,  if $(M_T)^{-1}+M_T T^{-1}\to 0$ then $S_i\stackrel{d}{\to}N(0,1)$ (see, e.g., \citealp{hamilton}, Propositions 7.11 and 8.3). 

The proposed test is based on the following rejection rule:
\[
\text {if } \vert S_i\vert> c_{T} \; \text{ then reject }\;  H_0,
\]
where we let the sequence of critical values to be such that $c_T\to\infty$ and $c_{T}=o(\sqrt T)$, as $T\to\infty$. As a consequence, the probabilities of type I and type II errors for this test are asymptotically zero.

Indeed,  as $T\to\infty$, the probability of type I errors is such that,
\begin{align}
\mathsf P\l(\vert S_i\vert >c_{T}| b_i=0 \r) &= 2\int_{c_T}^\infty \frac 1{\sqrt{2\pi}}e^{-u^2/2}\mathrm d u = K_b\, \frac{e^{-c_T^2/2}}{c_T}(1+o(1))\to 0.\label{eq:sizeB}
\end{align}
for some positive real $K_b$ independent of $i$. Moreover, under $H_1$, when, say, $b_i=b$ for some real $b> 0$, as $T\to\infty$, we have
\begin{align}
\mathsf P\l(\vert S_i\vert >c_{T}| b_i=b\r) 
&\ge \mathsf P\l( \l.S_i-\sqrt T\,  \frac{b}{\sqrt{\mathcal V_i}} > c_{T}-\sqrt T\, \frac{b}{\sqrt{\mathcal V_i}} \r\vert b_i=b\r)\to\mathsf P \l( \mathcal Z_i > -\infty\r)= 1,\label{eq:powerB}
\end{align}
where $\mathcal Z_i\sim N(0,1)$. Thus, the probability of type II errors tends to zero.

Now, let us consider the implications for the multiple testing problem given by the null and the alternative hypothesis 
\[
H_0: b_i=0\; \text{ for all $i$}, \;\mbox{ vs. }\; H_1: b_i\ne 0 \; \text{ for at least one $i$}.
\]
First, without loss of generality assume that $H_1$ holds with $b_1=b> 0$ and $b_i=0$ for all $i= 2,\ldots, n$. Then,
\beq
\mathsf P\l(\bigcup_{i=1}^n \l\{\vert S_i\vert >c_{T}\r\}| b_1=b; b_i=0,\; i=2,\ldots, n\r) \ge \mathsf P\l( \vert S_1\vert >c_{T} | b_1=b\r) \to 1,\label{eq:powerB2}
\eeq
because of \eqref{eq:powerB}. Thus, the probability of type II errors tends to zero independently of $n$. Turning to the probability of type I errors we have
\beq
\mathsf P\l(\bigcup_{i=1}^n \l\{\vert S_i\vert >c_{T}\r\}| b_i=0,\; i=1,\ldots, n\r)\le \sum_{i=1}^n \mathsf P\l( \vert S_i\vert >c_{T} | b_i=0\r) = n K_b\, \frac{e^{-c_T^2/2}}{c_T}(1+o(1)).\label{eq:sizeB2}
\eeq
By choosing $c_T=O(T^\epsilon)$ for any $\epsilon \in(0,1/2)$, then the probability of type I errors in \eqref{eq:sizeB2} tends always to zero as $n,T\to\infty$. However, in order to avoid power losses in \eqref{eq:powerB2}, we opt  for the choice $c_T=O(\log T)$, which still implies a vanishing probability of type I errors, as $n,T\to\infty$, provided that $T/\sqrt n\to 0$. Note that the latter condition is compatible with the condition $\sqrt T/n\to 0$ assumed in Proposition \ref{vecm} and is reasonable for macroeconomic datasets. Therefore, in practice, we run each of the $n$ tests using a critical value $c_{T}=\log T$.

\clearpage
\small
\setcounter{subsection}{-1}
\setcounter{equation}{0}
\setcounter{lem}{0}
\renewcommand{\thesection}{C}
\renewcommand{\thesubsection}{C\arabic{subsection}}
\renewcommand{\thelem}{C\arabic{lem}}
\renewcommand{\theequation}{C\arabic{equation}}

\subsection*{Preliminary definitions and notation} 
\noindent{\textbf{Norms.}}  For any $m\times p$ matrix $\mbf B$ with generic element $b_{ij}$, we denote its spectral norm as $\Vert\mbf B\Vert=(\mu_1^{\mbf B'\mbf B})^{1/2}$, where $\mu_1^{\mbf B'\mbf B}$ is the largest eigenvalue of $\mbf B'\mbf B$, 
the Frobenius norm as $\Vert\mbf B\Vert_F=(\mbox{tr}(\mbf B'\mbf B))^{1/2}=(\sum_i\sum_j b_{ij}^2)^{1/2}$, 
and the column and row norm as $\Vert\mbf B\Vert_1=\max_{j}\sum_{i}|b_{ij}|$ and $\Vert\mbf B\Vert_{\infty}=\max_{i}\sum_{j}|b_{ij}|$, respectively. Throughout we make use of the following properties.
\ben
\item Subadditivity of the norm, for an $m\times p$ matrix $\mbf A$  and a $p\times s$ matrix $\mbf B$: 
\beq\label{eq:subadd}
\Vert\mbf A\mbf B\Vert\leq \Vert\mbf A\Vert\;\Vert\mbf B\Vert.
\eeq
\item Norm inequalities, for an $n\times n$ symmetric matrix $\mbf A$: 
\beq\label{eq:12inf}
\mu_1^A=\Vert\mbf A\Vert\leq \sqrt{\Vert\mbf A\Vert_1\;\Vert\mbf A\Vert_{\infty}}=\Vert\mbf A\Vert_1, \quad \Vert\mbf A\Vert\leq \Vert\mbf A\Vert_F, \quad \Vert\mbf A\Vert_{F}\le \sqrt n \Vert\mbf A\Vert.
\eeq
\item Weyl's inequality, for two $n\times n$ symmetric matrices $\mbf A$ and $\mbf B$, with eigenvalues $\mu_j^{A}$ and $\mu_j^B$:
\beq\label{eq:weyl}
|\mu_j^{A}-\mu_j^B|\leq \Vert\mbf A-\mbf B\Vert, \qquad j=1,\ldots, n.
\eeq 
\een
\noindent{\textbf{Factors' dynamics.}} It is convenient to write the dynamic model of the factors, \eqref{eq:model2vector}, as
\begin{align}\label{eq:model2scalar}
\Delta F_{jt} &= \mbf c'_j(L)\mbf u_t=\sum_{l=1}^q c_{jl}(L) u_{lt}, \quad j=1,\ldots r, 
\end{align}
where $\mbf c_j(L)$ is an $q\times 1$ infinite rational polynomial matrix with entries $c_{jl}(L)$. Due to rationality, there exists a positive real $K_1$ such that
\beq
\sup_{j=1,\ldots, r}\sup_{l=1,\ldots, q}\sum_{k=0}^{\infty} c_{jlk}^2\leq K_1.\label{eq:sqsumMM1}
\eeq
From Assumption \ref{initcond} we also have $F_{jt} = \sum_{s=1}^t \mbf c_j'(L)\mbf u_s$. \\
\smallskip

\noindent{\textbf{Idiosyncratic dynamics.}} Likewise, for the idiosyncratic components it is convenient to write \eqref{eq:model3} as
\beq\label{eq:idioscalar}
\Delta\xi_{it} = \check{d}_i(L)\varepsilon_{it} ,\quad i=1,\ldots ,n,
\eeq
where $\check{d}_i(L)$ are a infinite polynomials defined as $\check{d}_i(L)=(1-L) (1-\rho_iL)^{-1} d_i(L)$ with $d_i(L)$ also infinite polynomials. Because of Assumption \ref{ASS:idio}(c) there exists a positive real $K_2$ such that
\beq
\sup_{i=1,\ldots, n}\sum_{k=0}^{\infty} \check d_{ik}^{\,2}\leq K_2.\label{eq:sqsumMM2}
\eeq
With reference to Assumption \ref{asm:rates}(a) we have $\rho_i=1$ if $i\in\mathcal I_1$ and $|\rho_i|<1$ if $i\in\mathcal I_1^c$. Hence, by Assumptions \ref{initcond}, we have also $\xi_{it} = \sum_{s=1}^t \check{d}_i(L) \varepsilon_{is}$, which is non-stationary if and only if $i\in\mathcal I_1$.\\
\smallskip
%

\noindent{\textbf{Factors' identification.}} 
The following choice of  the factors is very 
convenient and will be adopted 
in the sequel (see also Remark \ref{ghiniarann}). Let $\mathbf W$ be the $n\times r$ matrix  whose columns 
are the right normalised eigenvectors of the variance-covariance matrix of $\Delta\bm  \chi_t$, corresponding to the first $r$ 
eigenvalues $\mu^{\Delta \chi}_j$, $j=1,\ldots,r$. Following \citet{FGLR09} we identify the differenced factors by defining $\Delta\mathbf F_t = \mathbf W'\Delta \bm  \chi_t$.
Now project $\Delta\bm \chi_t$ on $\Delta\mathbf F_t$: $\Delta\bm \chi_t= \bm{\mathcal A}\hskip1pt\Delta \mathbf F_t +\bm{\mathcal R}_t$.
We see that
$\bm{\mathcal A}=\mathbf W$ and that the variance-covariance matrices  of $\Delta \bm \chi_t$ and of $\mathbf W\Delta\mathbf F_t$ are equal, 
 so that  $\bm{\mathcal R}_t=\mathbf 0$ and  the projection becomes
$\Delta\bm \chi _t = \mathbf W \mathbf W'\Delta\bm \chi_t,$
that is $\left (\mathbf I_n -\mathbf W\mathbf W'\right )\Delta\bm\chi_t=\mathbf 0$.
Since, by Assumption \ref{initcond}, $\bm\chi_0=\mathbf 0$, we obtain $\bm\chi_t = \mathbf W  \mathbf W' \bm \chi_t$, for $t>0$, or, in our preferred 
specification,
$\bm \chi_t = [ \sqrt n \mathbf W ]  [ n^{-1/2}\mathbf W' \bm \chi_t]$. We set henceforth, for all $n\in \mathbb N$, 
\beq\label{fattorinis2}
\bm \Lambda= \sqrt n \mathbf W, \ \ \  
\mathbf F_t =\frac{\displaystyle 1}{\displaystyle \sqrt n} \mathbf W'\bm \chi_t=\frac{1}{n}\bm \Lambda' \bm \chi_t.
\eeq
Note that now the factors  $\mathbf F_t$ and the loadings $\bm \lambda_i$, for a given $i$, depend on $n$.\\
\smallskip

\noindent{\textbf{Sample size of differenced data.}} The data in level is assumed to be observed for $t=1,\ldots, T$, thus the sample size is $T$, which implies that the sample size of the data in differences is $(T-1)$. When both levels and differences are present in the same proof we keep the distinction between the two sample sizes, however, in proofs where no confusion can arise we use just $T$ as sample size.\\


\section{Proof of Lemma \ref{lem:load}}

In order to prove part (i), we first prove results on the asymptotic properties of the sample covariance and of its eigenvalues and eigenvectors.\smallskip

\textit{Sample covariance matrix}. From Assumption \ref{ASS:idio}(e) of independent common and idiosyncratic components, we have $\bm\Gamma_0^{\Delta x}=\bm\Gamma_0^{\Delta \chi}+\bm\Gamma_0^{\Delta \xi}$ and therefore from Lemmas \ref{lem:covX} (which holds uniformly over all $i$ and $j$) and \ref{lem:evalcov}(ii) and Assumption \ref{ASS:idio}(e) we have 
\begin{align}
\bigg\Vert\frac{\wh{\bm\Gamma}_0^{\Delta  y}}{n}-\frac{\bm\Gamma_0^{\Delta \chi}}{n}\bigg\Vert &\leq\bigg\Vert\frac{\wh{\bm\Gamma}_0^{\Delta y}}{n}-\frac{\bm\Gamma_0^{\Delta x}}{n}\bigg\Vert + \bigg\Vert\frac{\bm\Gamma_0^{\Delta \xi}}{n}\bigg\Vert \le  \sqrt{\frac 1 {n^2}\sum_{i=1}^n\sum_{j=1}^n \big(\wh{\gamma}_{ij}^{\Delta y} - \gamma_{ij}^{\Delta x}\big)^2}+\frac{\mu_1^{\Delta\xi}}{n}\nn\\
&\leq \,  O_p\l(\frac 1 {\sqrt T}\r)+\frac{M_7}{n} = O_p\l(\max\l(\frac 1 {\sqrt T},\frac1  n\r)\r).\label{eq:gamma123}
\end{align}
Moreover, by denoting as $\bm\epsilon_i$ an $n$-dimensional vector with 1 as $i$-th entry and all other entries equal to zero, again by Lemmas \ref{lem:covX} and \ref{lem:evalcov}(ii), we have 
\begin{align}
\bigg\Vert\frac{\bm\epsilon_i'}{\sqrt n}&\big(\wh{\bm\Gamma}_0^{\Delta y}-\bm\Gamma_0^{\Delta\chi}\big)\bigg\Vert \leq \bigg\Vert\frac{\bm\epsilon_i'}{\sqrt n}\big(\wh{\bm\Gamma}_0^{\Delta y}-\bm\Gamma_0^{\Delta x}\big)\bigg\Vert + \bigg\Vert\frac{\bm\epsilon_i'\bm\Gamma_0^{\Delta\xi}}{\sqrt n}\bigg\Vert
\leq \sqrt{\frac 1 n \sum_{j=1}^n {\big(\wh{\gamma}_{ij}^{\Delta y} - \gamma_{ij}^{\Delta x}\big)}^2} +  \frac{\mu_1^{\Delta\xi}}{\sqrt n}\nn\\
&\leq O_p\l(\frac 1 {\sqrt T}\r)+\frac{M_7}{\sqrt n}
= O_p\l(\max\l(\frac 1 {\sqrt T},\frac1  {\sqrt n}\r)\r),\label{gammaradicen}
\end{align}
which holds for all $i=1,\ldots,n$ since Lemma \ref{lem:covX} holds uniformly over all $i$ and $j$. Moreover, note that for all $i=1,\ldots, n$, it holds that
\beq\label{gammaradicen2}
\bigg\Vert \frac{\bm\epsilon_i'\bm\Gamma_0^{\Delta \chi}}{\sqrt n}\bigg\Vert = \sqrt{\frac 1 n\sum_{j=1}^n\big(\gamma_{ij}^{\Delta \chi}\big)^2}=
\sqrt{\frac 1 n\sum_{j=1}^n\big(\bm\lambda_i'\bm\Gamma^{\Delta F}_0\bm\lambda_j\big)^2}\le r^2 C^2,
\eeq
because of Assumption \ref{asm:factor}(b) of uniformly bounded loadings, i.e. with $C$ that does not depend on $i$.\smallskip

\textit{Sample eigenvalues}. For the eigenvalues $\mu_j^{\Delta\chi}$ of $\bm\Gamma_0^{\Delta\chi}$ and $\wh{\mu}_j^{\Delta y}$ of $\widehat{\bm\Gamma}_0^{\Delta y}$, and using Weyl's inequality \eqref{eq:weyl}, we have
\beq
\bigg|\frac{\wh{\mu}^{\Delta y}_j}{n}-\frac{\mu^{\Delta \chi}_j}{n}\bigg|\leq\bigg\Vert\frac{\wh{\bm\Gamma}_0^{\Delta y}}{n}-\frac{\bm\Gamma_0^{\Delta \chi}}{n}\bigg\Vert= O_p\l(\max\l(\frac 1 {\sqrt T},\frac1  n\r)\r),\qquad j=1,\ldots, r.\label{eq:MGAMMA}
\eeq
From Lemma \ref{lem:evalcov}(i) and \eqref{eq:MGAMMA}, there exists an integer $\bar n$, such that for $n>\bar n$, we have
\beq\label{eq:MINV3}
\frac{\mu_r^{\Delta\chi}}n\geq \underline M_6,\qquad\frac{\wh{\mu}_r^{\Delta y}}n\geq \underline M_6+O_p\l(\max\l(\frac 1 {\sqrt T},\frac1  n\r)\r).
\eeq
Define as $\mbf M^{\Delta\chi}$ and $\wh{\mbf M}^{\Delta y}$ the diagonal $r\times r$ matrices with diagonal elements $\mu_j^{\Delta\chi}$ and $\wh{\mu}_j^{\Delta y}$, respectively.  From \eqref{eq:MINV3}, the matrix $n^{-1}{\mbf M^{\Delta\chi}}$ is invertible for $n>\bar n$ and the inverse of $n^{-1}{\wh{\mbf M}^{\Delta y}}$ exists with probability tending to one as $n,T\to\infty$. Moreover, by Lemma \ref{lem:evalcov}(i), \eqref{eq:MGAMMA}, and \eqref{eq:MINV3}, for $n>\bar n$ we have 
\begin{align}\label{eq:MINV}
&\bigg\Vert\bigg(\frac{\mbf M^{\Delta\chi}}{n}\bigg)^{-1}\bigg\Vert= \frac{n}{\mu_{r}^{\Delta\chi}} \le \frac 1{\underline M_6},
\end{align}
which implies $\Vert(n^{-1}\mbf M^{\Delta\chi})^{-1}\Vert= O_p(1)$.
Then, from \eqref{eq:MGAMMA} and \eqref{eq:MINV3}, we have
\begin{align}
\bigg\Vert\bigg(\frac{\wh{\mbf M}^{\Delta y}}{n}&\bigg)^{-1}-\bigg(\frac{\mbf M^{\Delta \chi}}{n}\bigg)^{-1}\bigg\Vert\leq\bigg\Vert\bigg(\frac{\wh{\mbf M}^{\Delta y}}{n}\bigg)^{-1}-\bigg(\frac{\mbf M^{\Delta \chi}}{n}\bigg)^{-1}\bigg\Vert_F= \sqrt{\sum_{j=1}^r\bigg(\frac{n}{\wh{\mu}_j^{\Delta y}}-\frac{n}{\mu_j^{\Delta \chi}}\bigg)^2}\nn\\
&\leq \sum_{j=1}^r n\bigg|\frac{\wh{\mu}_j^{\Delta y}-\mu_j^{\Delta\chi}}{\wh{\mu}_j^{\Delta y}\mu_j^{\Delta\chi}}\bigg|\leq \frac{r\max_{j=1,\ldots,r}|\wh{\mu}_j^{\Delta y}-\mu_j^{\Delta\chi}|}{n\underline M_6^2+O_p\l(\max\l(\frac n {\sqrt T}, 1\r)\r)}
=O_p\l(\max\l(\frac 1 {\sqrt T}, \frac 1 n\r)\r).\label{eq:MINV4}
\end{align}
Last, from the identification constraint \eqref{fattorinis2}, we have that $\bm\Gamma_0^{\Delta F}$ is diagonal with entries $\E(\Delta F_{jt}^2)= \mu_j^{\Delta\chi}/n$ for $j=1,\ldots,r$, which are finite and bounded away from zero because of Lemma \ref{lem:evalcov}(i). Then, by Assumption \ref{ASS:common}(d) $\bm\Gamma_0^{\Delta\chi}$ has $r$ non-zero distinct eigenvalues. Moreover, \eqref{fattorinis2} implies also that $n^{-1}\bm\Lambda'\bm\Lambda=\mbf I_r$, for any $n\in\mathbb N$. Therefore, under our identification constraints, Lemma \ref{lem:evalcov}(i) and thus \eqref{eq:MINV3} and \eqref{eq:MINV} hold for any $n\in\mathbb N$. As a consequence, from Lemma \ref{lem:evalcov}(i) there exist  positive reals $\underline C_{j} ,\overline C_{j} $, such that $\underline  C_{j}>\overline C_{j+1}$ for $j=1,\ldots, r-1$, and, for any $n\in\mathbb N$, we have
\beq\label{distinct}
\underline C_{j} \le \frac{\mu_{j}^{\Delta\chi}}{n}\le \overline C_{j}, \qquad j=1,\ldots, r.
\eeq
Notice that then $\overline C_{1}\equiv \overline M_6$ and $\underline C_{r}\equiv \underline M_6$, where $\overline M_6$ and $\underline M_6$ are defined in Lemma \ref{lem:evalcov}(i).  \smallskip

\textit{Sample eigenvectors}. Define as $\mbf w_j^{\Delta\chi}$ and $\wh{\mbf w}_j^{\Delta y}$ the $n\times 1$ normalised eigenvectors corresponding to the $j$-th largest eigenvalue of $\bm\Gamma_0^{\Delta\chi}$ and $\widehat{\bm\Gamma}_0^{\Delta y}$, respectively. Define $s_j=\mbox{sign}(\widehat{\mbf w}_j^{\Delta y'}\mbf w_j^{\Delta\chi})$ and notice that $\widehat{\mbf w}_j^{\Delta y'}\mbf w_j^{\Delta\chi}s_j\ge 0$ for all $j=1,\ldots, r$. Then, from Corollary 1 in \citet{yu15}, defining $\mu_{0}^{\Delta \chi}=\infty$, we have
\begin{align}\label{eq:dk1}
\Vert\widehat{\mbf w}_j^{\Delta y}-\mbf w_j^{\Delta\chi}s_j\Vert\leq 
\frac{2^{3/2}\Vert\wh{\bm\Gamma}_0^{\Delta y}-\bm\Gamma_0^{\Delta\chi}\Vert}
{\min\big((\mu_{j-1}^{\Delta \chi}-\mu_{j}^{\Delta\chi}),(\mu_{j}^{\Delta \chi}-\mu_{j+1}^{\Delta\chi})\big)}, \qquad j=1,\ldots, r.
\end{align}
Then, because of \eqref{distinct} for the denominator of \eqref{eq:dk1}, for any $n\in\mathbb N$ we have
\begin{align}
 {\mu_{j-1}^{\Delta \chi}-\mu_j^{\Delta\chi}} \geq n(\underline C_{j-1}-\overline C_{j}) >0,\quad j=2,\ldots, r,\label{DKdenom1}\\
 {\mu_{j}^{\Delta \chi}-\mu_{j+1}^{\Delta\chi}} \geq n (\underline C_{j}-\overline C_{j+1}) >0,\quad j=1,\ldots, r.\label{DKdenom2}
\end{align}
Define $\mbf J$ as the $r\times r$ diagonal matrix with entries $s_j$ and define also the $n\times r$ orthonormal matrices of eigenvectors $\mbf W^{\Delta\chi}=(\mbf w_1^{\Delta\chi} \cdots \mbf w_r^{\Delta\chi})$ and $\wh{\mbf W}^{\Delta y}=(\wh{\mbf w}_1^{\Delta y}\cdots \wh{\mbf w}_r^{\Delta y})$. 
Then, from \eqref{eq:dk1}, \eqref{DKdenom1}, and \eqref{DKdenom2}, we have
\beq\label{eq:what}
\Vert\widehat{\mbf W}^{\Delta y}-\mbf W^{\Delta\chi}\mbf J\Vert\le  \sqrt{\sum_{j=1}^r\Vert\widehat{\mbf w}_j^{\Delta y}-\mbf w_j^{\Delta\chi}s_j\Vert^2}=O_p\l(\max\l(\frac 1 {\sqrt T},\frac1  n\r)\r).
\eeq

We can now prove part (i). The loadings estimator is defined as $\widehat{\bm\Lambda} = n^{1/2} \widehat{\mbf W}^{\Delta y}$ while from \eqref{fattorinis2} we have $\bm\Lambda = n^{1/2}\mbf W^{\Delta\chi}$. Hence, $\wh{\bm\lambda}_i' = n^{1/2}\bm\epsilon_i'\wh{\mbf W}^{\Delta y}$ and $\bm\lambda_i' = n^{1/2} \bm\epsilon_i'\mbf W^{\Delta\chi}$. Then, notice that the columns of $\mbf W^{\Delta\chi}\mbf J$ are also normalised eigenvectors of $\bm\Gamma_0^{\Delta\chi}$, that is $\bm\Gamma_0^{\Delta\chi}\mbf W^{\Delta\chi}\mbf J=\mbf W^{\Delta\chi}\mbf J\mbf M^{\Delta\chi}$.~Therefore, using \eqref{gammaradicen}, \eqref{gammaradicen2}, \eqref{eq:MINV}, \eqref{eq:MINV4}, and \eqref{eq:what}, for all $i=1,\ldots, n$ we have
\begin{align}
\big\Vert\wh{\bm\lambda}_i' &-\bm\lambda_i'\mbf J\big\Vert=\big\Vert\sqrt n\bm\epsilon_i'\wh{\mbf W}^{\Delta y}- \sqrt n \bm\epsilon_i'\mbf W^{\Delta\chi}\mbf J\big\Vert=
\bigg\Vert
\frac {\bm\epsilon_i'} {\sqrt n}\bigg[\wh{\bm\Gamma}_0^{\Delta y}\wh{\mbf W}^{\Delta y}\bigg(\frac{\wh{\mbf M}^{\Delta y}}{n}\bigg)^{-1}
-\bm\Gamma_0^{\Delta \chi}\mbf W^{\Delta \chi}\mbf J\bigg(\frac{\mbf M^{\Delta \chi}}{n}\bigg)^{-1}\bigg]
\bigg\Vert\nn\\
&\leq \bigg\Vert
\frac {\bm\epsilon_i'} {\sqrt n}\big(\wh{\bm\Gamma}_0^{\Delta y}-\bm\Gamma_0^{\Delta \chi}\big)\bigg\Vert\;\bigg\Vert\bigg(\frac{\mbf M^{\Delta\chi}}{n}\bigg)^{-1}\bigg\Vert +
\bigg\Vert
\frac {\bm\epsilon_i'\bm\Gamma_0^{\Delta \chi}} {\sqrt n}\bigg\Vert\;\bigg\Vert\bigg(\frac{\wh{\mbf M}^{\Delta y}}{n}\bigg)^{-1}-\bigg(\frac{\mbf M^{\Delta \chi}}{n}\bigg)^{-1}\bigg\Vert\label{uniflamb}\\ 
&+ \big\Vert
\wh{\mbf W}^{\Delta y}-\mbf W^{\Delta \chi}\mbf J\big\Vert\;
\bigg\Vert
\frac {\bm\epsilon_i'\bm\Gamma_0^{\Delta \chi}} {\sqrt n}\bigg\Vert\;
\bigg\Vert\bigg(\frac{\mbf M^{\Delta\chi}}{n}\bigg)^{-1}\bigg\Vert + o_p\l(\max\l(\frac 1 {\sqrt T},\frac1  {\sqrt n}\r)\r)= O_p\l(\max\l(\frac 1 {\sqrt T},\frac1  {\sqrt n}\r)\r),\nn
\end{align} 
where we also used the fact that $\Vert\mbf W^{\Delta\chi}\Vert=1$. Note in particular that \eqref{uniflamb} holds uniformly over all $i$ because of \eqref{gammaradicen} and \eqref{gammaradicen2})
This proves part (i).\smallskip

Turning to part (ii), for any $i\in\mathcal I_b$, consider $\wh b_i$ defined in \eqref{eq:biols}, then because of \eqref{eq:modeltrend},  
\begin{align}
\E[\vert\wh b_i - b_i \vert^2]&
=\E\l[\l( \frac{\sum_{t=1}^T(t-\frac {T+1} 2)(x_{it}-\bar x_i)}{\sum_{t=1}^T (t-\frac {T+1} 2)^2}\r)^2\r]= \frac{\E\l[\l(\sum_{t=1}^T t x_{it}-\frac {T+1}2\sum_{t=1}^T x_{it}\r)^2\r]}{\l(\frac 1{12}T(T^2-1)\r)^2},\label{eq:blem1}
\end{align}
where $\bar y_i=T^{-1}\sum_{t=1}^T y_{it}$ and $\bar x_i = T^{-1}\sum_{t=1}^T x_{it}$ and therefore $\bar y_i = \bar x_i +a_i + b_i (T+1)/2$. Then, for all $i\in\mathcal I_b$, we have 
\begin{align}\label{eq:blem2}
\E\bigg[&\bigg(\sum_{t=1}^T x_{it}\bigg)^2\bigg] \le 2 \bigg\{\E\bigg[\bigg(\sum_{t=1}^T\bm\lambda_i'\mbf F_t\bigg)^2\bigg]+
\E\bigg[\bigg(\sum_{t=1}^T\xi_{it}\bigg)^2\bigg] \bigg\}\le 2C^2 \E\bigg[\bigg\Vert \sum_{t=1}^T  \mbf F_t\bigg\Vert^2\bigg] + 2 \E\bigg[\bigg(\sum_{t=1}^T  \xi_{it}\bigg)^2\bigg]\nn\\
&\le 2 C^2 \sum_{t=1}^T\sum_{s=1}^T \bigg\{\sum_{j_1,j_2=1}^r \big|\E[F_{j_1t}F_{j_2s} ]\big| + \big|\E[\xi_{it}\xi_{is}]\big|\bigg\} \le 2C^2 T^2 \Big(r \E[\Vert \mbf F_t\Vert^2]+  \E[\xi_{it}^2]\Big) = O(T^3),
\end{align}
because of  Assumption \ref{asm:factor}(b) of uniformly bounded loadings and Lemma \ref{lem:main}(ii) and \ref{lem:main}(iv) (and specifically since $\E[\xi_{it}^2]=O(T)$ holds uniformly over $i$, see also \eqref{eq:main5}) and using Cauchy-Schwarz inequality. Moreover, by the same arguments leading to \eqref{eq:blem2}, we also have
\begin{align}
\E\bigg[&\bigg(\sum_{t=1}^T t x_{it}\bigg)^2\bigg] \le  2 \bigg\{\E\bigg[\bigg(\sum_{t=1}^T t \bm\lambda_i'\mbf F_t\bigg)^2\bigg]+
\E\bigg[\bigg(\sum_{t=1}^T t \xi_{it}\bigg)^2\bigg] \bigg\}\le 2C^2 \E\bigg[\bigg\Vert \sum_{t=1}^T t \mbf F_t\bigg\Vert^2\bigg] + 2 \E\bigg[\bigg(\sum_{t=1}^T t \xi_{it}\bigg)^2\bigg]\nn\\
&\le 4 C^2 \sum_{t=1}^T\sum_{s=1}^t t s \bigg\{\sum_{j_1,j_2=1}^r \big|\E[F_{j_1t}F_{j_2s} ]\big| + \big|\E[\xi_{it}\xi_{is}]\big|\bigg\}\le 4C^2\sum_{t=1}^T \frac{t^2(t+1)}2\Big(r \E[\Vert \mbf F_t\Vert^2]+  \E[\xi_{it}^2]\Big)\nn\\
&= 4 C^2\frac{T(T+1)(T+2)(3T+1)}{24}\Big(r \E[\Vert \mbf F_t\Vert^2]+  \E[\xi_{it}^2]\Big) = O(T^5).\label{eq:blem3}
\end{align} 
From \eqref{eq:blem2} and \eqref{eq:blem3} we have that the numerator in \eqref{eq:blem1} is $O(T^5)$. Therefore, $\E[\vert\wh b_i - b_i \vert^2]=O(T^{-1})$, for all $i\in\mathcal I_b$ and by Chebychev's inequality we prove part (ii). \smallskip

We can now prove part (iii). First, note that by substituting the expressions for $\bm\Lambda$ and $\widehat{\bm\Lambda}$ in \eqref{eq:what}, we have
\beq\label{eq:what2}
\bigg\Vert \frac{\widehat{\bm\Lambda} - \bm\Lambda\mbf J}{\sqrt n}\bigg\Vert =\big\Vert \wh{\mbf{W}}^{\Delta x}-\mbf W^{\Delta\chi}\mbf J\big\Vert=O_p\l(\max\l(\frac 1 {\sqrt T},\frac1  n\r)\r),
\eeq
which implies also that
\beq\label{eq:whatlemma3}
\bigg\Vert \frac{\widehat{\bm\Lambda}'\bm\Lambda}n - \mbf J\bigg\Vert=O_p\l(\max\l(\frac 1 {\sqrt T},\frac1  n\r)\r).
\eeq
Then, let $\wh{\mbf b}=(\wh b_1\cdots \wh b_n)'$, where $\wh b_i$ is given in \eqref{eq:biols} if $i\in\mathcal I_b$, while $\wh b_i=0$ otherwise and define the de-trended data as $\wh{\mbf x}_t=\mbf y_t-\wh{\mbf b} t$. The factors are estimated as $\wh{\mbf F}_t=n^{-1}\wh{\bm\Lambda}'\wh{\mbf x}_t$. Let also ${\mbf b}=( b_1\cdots  b_n)'$ and $\mbf a=(a_1\cdots a_n)'$ such that $\mbf y_t=\mbf a+\mbf b t+\mbf x_t$. Then, for a given $t$ we have
\begin{align}
\frac 1{\sqrt T}\big\Vert \wh{\mbf F}_t&-\mbf J\mbf F_t\big\Vert =\bigg\Vert\frac{\wh{\bm\Lambda}'\wh{\mbf x}_t}{n\sqrt T}-\frac{\mbf J\mbf F_t}{\sqrt T}\bigg\Vert\leq 
\bigg\Vert\frac{\wh{\bm\Lambda}'\bm\Lambda\mbf F_t}{n\sqrt T}-\frac{\mbf J\mbf F_t}{\sqrt T}+ \frac{\wh{\bm\Lambda}'\bm\xi_t}{n\sqrt T} \bigg\Vert
+\bigg\Vert \frac{\wh{\bm\Lambda}'(\mbf b-\wh{\mbf b})t}{n\sqrt T} \bigg\Vert
+\bigg\Vert \frac{\wh{\bm\Lambda}'\mbf a}{n\sqrt T} \bigg\Vert.
\label{scompF}
\end{align}
The first term on the rhs of \eqref{scompF}, is such that 
\begin{align}
\bigg\Vert\frac{\wh{\bm\Lambda}'\bm\Lambda\mbf F_t}{n\sqrt T}-\frac{\mbf J\mbf F_t}{\sqrt T}+ \frac{\wh{\bm\Lambda}'\bm\xi_t}{n\sqrt T} \bigg\Vert&\leq \bigg\Vert\frac{\wh{\bm\Lambda}'\bm\Lambda}{n}-\mbf J\bigg\Vert\;\bigg\Vert\frac{\mbf F_t}{\sqrt T}\bigg\Vert +\bigg\Vert\frac{\wh{\bm\Lambda}-\bm\Lambda\mbf J}{\sqrt n}\bigg\Vert\;\bigg\Vert\frac{\bm\xi_t}{\sqrt {nT}}\bigg\Vert+ \bigg\Vert \frac{\bm\Lambda'\bm\xi_t}{n\sqrt T} \bigg
\Vert\;\Vert\mbf J\Vert\nn\\
&= O_p\l(\max\l(\frac 1 {\sqrt T},\frac1  {n}\r)\r)+O_p\l(\frac1  {\sqrt n}\r),\label{scompF2}
\end{align}
because of \eqref{eq:whatlemma3}, \eqref{eq:what2}, and Lemma \ref{lem:main}(ii), \ref{lem:main}(iv) and \ref{lem:main}(vi) and since obviously $\Vert\mbf J\Vert=1$. 

The second term on the rhs of  \eqref{scompF} is such that 
\begin{align}
\bigg\Vert \frac{\wh{\bm\Lambda}'(\mbf b-\wh{\mbf b})t}{n\sqrt T} \bigg\Vert\le \bigg\Vert\frac{\wh{\bm\Lambda}-\bm\Lambda\mbf J}{\sqrt n}\bigg\Vert\;\bigg\Vert\frac{(\mbf b-\wh{\mbf b})t}{\sqrt {nT}}\bigg\Vert+ \bigg\Vert \frac{\bm\Lambda'(\mbf b-\wh{\mbf b})t}{n\sqrt T} \bigg
\Vert\;\Vert\mbf J\Vert.\label{scompF3}
\end{align}
Now, because of part (ii), we have
\begin{align}
\E\bigg[\bigg\Vert\frac{(\mbf b-\wh{\mbf b})t}{\sqrt {nT}}\bigg\Vert^2\bigg] = \frac{t^2}{nT}\sum_{i\in\mathcal I_b}\E\big[ (b_i-\wh b_i)^2\big]= O\l(\frac{1}{n^{1-\eta}}\r).
\end{align}
since $t\le T$ and by \eqref{eq:what2} the first term on the rhs of \eqref{scompF3} is $o_p(\max(T^{-1/2},n^{-1}))$. For the second term on the rhs of \eqref{scompF3} we have (obviously $\Vert\mbf J\Vert^2=1$)
\begin{align}
\E\bigg[\bigg\Vert \frac{\bm\Lambda'(\mbf b-\wh{\mbf b})t}{n\sqrt T} \bigg
\Vert^2\bigg]
&\le \frac{t^2}{n^2 T}\sum_{j=1}^r\E\bigg[\bigg(\sum_{i\in\mathcal I_b} \lambda_{ij}(b_i-\wh b_i)\bigg)^2\bigg]\le \frac{t^2 C^2}{n^2 T}\sum_{i\in\mathcal I_b}\sum_{j\in\mathcal I_b}\big\vert\E[(b_i-\wh b_i)(b_j-\wh b_j)]\big\vert\nn\\
&\le \frac{T C^2n^\eta}{n^{2}} \sum_{i\in\mathcal I_b}\E[(b_i-\wh{b}_i)^2] = O\l(\frac 1{n^{2(1-\eta)}} \r),
\end{align}
where we used Assumption \ref{asm:factor}(b) of uniformly bounded loadings, Cauchy-Schwarz inequality and part (ii). Therefore, \eqref{scompF3} is $O_p({n^{-(1-\eta)}} )$. 

For the third term on the rhs of \eqref{scompF}, since $\Vert\mbf a\Vert=O(\sqrt n)$,  we have
\begin{align}
\bigg\Vert \frac{\wh{\bm\Lambda}'\mbf a}{n\sqrt T} \bigg\Vert \le  \bigg\Vert\frac{\wh{\bm\Lambda}-\bm\Lambda\mbf J}{\sqrt n}\bigg\Vert\;\bigg\Vert\frac{\mbf a}{\sqrt {nT}}\bigg\Vert+ \bigg\Vert \frac{\bm\Lambda'\mbf a}{n\sqrt T} \bigg
\Vert\;\Vert\mbf J\Vert=O_p\l(\frac 1 {\sqrt T}\r),\label{scompF4}
\end{align}
By substituting \eqref{scompF2}, \eqref{scompF3}, and \eqref{scompF4} into \eqref{scompF} we prove part (iii). This completes the proof. \hfill$\Box$
\newpage
\setcounter{subsection}{-1}
\setcounter{equation}{0}
\setcounter{lem}{0}
\renewcommand{\thesection}{D}
\renewcommand{\thesubsection}{D\arabic{subsection}}
\renewcommand{\thelem}{D\arabic{lem}}
\renewcommand{\theequation}{D\arabic{equation}}

\section{Auxiliary Lemmas}
\begin{lem}\label{rem:idio} Under Assumptions  \ref{ASS:common} through \ref{ASS:idio}, there exists a positive real $M_5$ such that
 $\mu_{1}^{\varepsilon}\leq M_5$ and $n^{-1}\sum_{i=1}^n\sum_{j=1}^n|\E[\eps_{it}\eps_{jt}]|\leq M_5$, for any $n\in\mathbb N$.
\end{lem}

\noindent{\textbf{Proof.}} 
First notice that, from Assumption \ref{ASS:idio}(b), we have 
\[
\frac 1 n\sum_{i,j=1}^n|\E[\eps_{it}\eps_{jt}]|\leq \max_{i=1,\ldots, n}\sum_{j=1}^n |\E[\eps_{it}\eps_{jt}]|=\Vert\bm\Gamma_0^{\eps}\Vert_1\leq M_3.
\]
Thus, from \eqref{eq:12inf}, we have $\mu_1^{\eps} = \big\Vert\bm\Gamma_0^{\eps} \big\Vert\leq \big\Vert\bm\Gamma_0^{\eps}\big\Vert_1\leq M_3$.
By setting $M_5=M_3$, we complete the proof.\hfill$\Box$

\begin{lem}\label{lem:evalcov} Under Assumptions \ref{ASS:common} through \ref{ASS:idio},  there exist positive reals $\underline{M}_6$, $\overline{M}_6$, $M_7$, $\underline{M}_8$, $\overline{M}_8$ and an integer $\bar n$ such that
\smallskip 
\begin{compactenum}[(i)]
\item  $ \underline{M}_6 \leq n^{-1}\mu_{j}^{\Delta\chi}\leq \overline{M}_6$ for any $j=1,\ldots ,r$  and $n>\bar n$;
\item  $\mu_{1}^{\Delta \xi}\leq M_7$, for any $n\in\mathbb N$;
\item  $\underline{M}_8^{\vphantom{\Delta\chi}}\leq n^{-1}\mu_{j}^{\Delta x}\leq \overline{M}_8$ for any $j=1,\ldots ,r$ and $n>\bar n$;
\item  $\mu_{r+1}^{\Delta x}\leq M_7$, for any $n\in\mathbb N$.
\end{compactenum}
\end{lem}
\noindent{\textbf{Proof.}} 
Throughout, let $\bm\Gamma_0^{\Delta F} =\E[\Delta\mbf F_t\Delta\mbf F_t']$, $\bm\Gamma_0^{\Delta\chi}=\E[\Delta\bm\chi_t\Delta\bm\chi_t']$, $\bm\Gamma_0^{\Delta\xi}=\E[\Delta\bm\xi_t\Delta\bm\xi_t']$, and $\bm\Gamma_0^{\Delta x}=\E[\Delta\mbf x_t\Delta\mbf x_t']$. Then, we can write $\bm\Gamma_0^{\Delta F} = \mbf W^{\Delta F}\mbf M^{\Delta F}\mbf W^{\Delta F'}$,
where $\mbf W^{\Delta F}$ is the $r\times r$ matrix of normalised eigenvectors and $\mbf M^{\Delta F}$ the corresponding diagonal matrix of eigenvalues. 
Define a new $n\times r$ loadings matrix $\bm L=\bm\Lambda\mbf W^{\Delta F}(\mbf M^{\Delta F})^{1/2}$. Under Assumption \ref{asm:factor}(a) there exists an integer $\bar n$ such that $n^{-1}\bm\Lambda'\bm\Lambda=\mbf I_r$, for any $n> \bar n$, therefore, for any $n\ge \bar n$,
\beq\label{eq:LLMFn}
\frac{\bm L'\bm L}{n} = \mbf M^{\Delta F}.
\eeq
By Assumption \ref{ASS:common}(d) and square summability of the coefficients given in \eqref{eq:sqsumMM1}, all eigenvalues of $\bm\Gamma_0^{\Delta F}$ are positive and finite, i.e. there exist positive reals $\underline M_6$ and $\overline M_6$ such that
\beq\label{eq:M5}
\underline{M}_6 \leq \mu_{j}^{\Delta F}\leq \overline{M}_6, \quad j=1,\ldots, r.
\eeq
Then, for $n> \bar n$, 
\beq\nn
\frac {\bm\Gamma_0^{\Delta\chi}}{n} = \frac{\bm\Lambda\mbf W^{\Delta F}\mbf M^{\Delta F}\mbf W^{\Delta F'}\bm\Lambda'}{n}=\frac{\bm L\bm L'}{n}.
\eeq
Therefore, the non-zero eigenvalues of $\bm\Gamma_0^{\Delta\chi}$ are the same as those of $\bm L'\bm L$, and from \eqref{eq:LLMFn}, we have  $n^{-1}\mu_j^{\Delta\chi} = \mu_j^{\Delta F}$, for any $n>\bar n$ and any $j=1,\ldots,r$. Part (i) then follows from \eqref{eq:M5}.\smallskip

As for part (ii), we have
\beq\label{eq:muxi}
\mu_1^{\Delta\xi} =\big\Vert\bm\Gamma_0^{\Delta\xi} \big\Vert \leq\sum_{k=0}^{\infty} \big\Vert\check{\mbf D}_k\big\Vert^2\;  \big\Vert\bm\Gamma_0^{\eps} \big\Vert\leq K_2 M_3= M_7,
\eeq
because of square summability of the coefficients, with $K_2$ defined in \eqref{eq:sqsumMM2}, and from Lemma \ref{rem:idio}.
\smallskip

Finally, parts (iii) and (iv) are immediate consequences of Assumption \ref{ASS:idio}(e) of independent common and idiosyncratic shocks, which implies that $\bm\Gamma_0^{\Delta x}=\bm\Gamma_0^{\Delta\chi}+\bm\Gamma_0^{\Delta\xi}$ and of Weyl's inequality \eqref{eq:weyl}. So, because of parts (i) and (ii), there exist positive reals $\underline M_8$ and $\overline M_8$, such that, for $j=1,\ldots, r$, and for any $n> \bar n$,
\begin{align}
\frac{\mu_j^{\Delta x}}n &\leq \frac{\mu_j^{\Delta\chi}}n+ \frac{\mu_1^{\Delta \xi}}n \leq \overline M_6 + \frac{\mu_1^{\Delta \xi}}n \leq \overline M_6+\frac {M_7} n= \overline M_8,\quad \frac{\mu_j^{\Delta x}}n &\geq \frac{\mu_j^{\Delta\chi}}n+ \frac{\mu_n^{\Delta \xi}}n \geq \underline M_6 + \frac{\mu_n^{\Delta \xi}}n= \underline M_8,\nn
\end{align}
This proves part (iii). When $j=r+1$, using parts (i) and (ii), and since $\mbox{rk}(\bm\Gamma_0^{\Delta\chi})=r$, we have $\mu_{r+1}^{\Delta x} \leq \mu_{r+1}^{\Delta\chi}+ \mu_1^{\Delta \xi}=\mu_1^{\Delta \xi}\leq M_7$, thus proving part (iv). This completes the proof. \hfill $\Box$

\begin{lem}\label{lem:covX}
Let the generic $(i,j)$-th element of the covariance matrix $\bm\Gamma_0^{\Delta x}$ of $\Delta\mbf x_t$ be $\gamma_{ij}^{\Delta x}=\E[\Delta x_{it}\Delta x_{jt}]$. Let the generic  $(i,j)$-th element of the sample covariance matrix $\wh{\bm\Gamma}_0^{\Delta y}$ of $\Delta \mbf y_t$ be $\wh{\gamma}_{ij}^{\Delta y}$.
Then, under Assumptions \ref{ASS:common} through \ref{initcond}, as $T\to\infty$, there exists a positive real $C_0$ which does not depend on $i$ and $j$ such that $\E[|\wh{\gamma}_{ij}^{\Delta y}-\gamma_{ij}^{\Delta x}|^2]\le C_0 T^{-1}$.
\end{lem}

\noindent{\textbf{Proof.}} 
First, note that $\gamma_{ij}^{\Delta x} = \bm\lambda_i'\bm\Gamma_0^{\Delta F}\bm\lambda_j+\gamma_{ij}^{\Delta \xi}$, where $\bm\lambda_i'$ is the $i$-th row of $\bm\Lambda$, $\bm\Gamma_0^{\Delta F}=\E[\Delta \mbf F_t\Delta \mbf F_t']$, and $\gamma_{ij}^{\Delta \xi}=\E[\Delta\xi_{it}\Delta\xi_{jt}]$.\smallskip

Start with the sample covariance of the factors, and consider the fourth moments of $\Delta\mbf F_t$. Using \eqref{eq:model2scalar}, we have
\begin{align}
\sum_{t,s=1}^T \E\big[&\Delta F_{it}\Delta F_{jt}\Delta F_{is}\Delta F_{js}\big]=\sum_{t,s=1}^T\sum_{l,l',h,h'=1}^q\sum_{k,k',m,m'=0}^{\infty}\!\!\!\!\E\big[c_{ilk} u_{lt-k}c_{il'k'} u_{l't-k'}c_{jhm} u_{hs-m}c_{jh'm'} u_{h's-m'}\big]\nn\\
\leq& q^4 K_1^4\sum_{t,s=1}^T \E[ u_{lt}  u_{l't}  u_{hs} u_{h's}]=q^4 K_1^4 \bigg(\sum_{t,s=1}^T\E[ u_{lt}^2] \E[u_{hs}^2] + \sum_{t=1}^T\E[ u_{lt}^2 u_{ht}^2]+\sum_{t=1}^T\E[u_{lt}^4]\bigg),\label{mom4A}
\end{align}
because of Assumption \ref{ASS:common}(a) of independence of $\mbf u_t$ and square summability of the coefficients, with $K_1$ defined in \eqref{eq:sqsumMM1}. Similarly, for any $(i,j)$-th element of $\bm\Gamma_0^{\Delta F}$, denoted as $\gamma_{ij}^{\Delta F}$, we have
\begin{align}
(\gamma_{ij}^{\Delta F})^2&=\big(\E\big[\Delta F_{it}\Delta F_{jt}\big]\big)^2=\bigg(\sum_{l,l'=1}^q\sum_{k,k'=0}^{\infty}\E\big[c_{ilk} u_{lt-k}c_{il'k'} u_{l't-k'}\big]\bigg)^2\nn\\
\leq & \,q^4 K_1^4\sum_{t,s=1}^T(\E[u_{lt}u_{l't}]\E[u_{hs}u_{h's}])=q^4K_1^4\bigg(\sum_{t,s=1}^T\E[ u_{lt}^2] \E[u_{hs}^2]+\sum_{t=1}^{T}(\E[u_{lt}^2])^2\bigg).\label{mom4B}
\end{align}
Now, using \eqref{eq:12inf} and combining \eqref{mom4A} and \eqref{mom4B}, we have
\begin{align}
\E\bigg[\bigg\Vert\frac 1 T& \sum_{t=1}^T \Delta\mbf F_t\Delta\mbf F_t' -\bm\Gamma_0^{\Delta F} \bigg\Vert^2\bigg] \leq 
\sum_{i,j=1}^r\frac {1}{T^2}\E\bigg[\sum_{t,s=1}^T\bigg(\Delta F_{it}\Delta F_{jt}-\gamma_{ij}^{\Delta F}\bigg)\bigg(\Delta F_{is}\Delta F_{js}-\gamma_{ij}^{\Delta F}\bigg)\bigg]\nn\\
=&\sum_{i,j=1}^r\frac {1}{T^2}\sum_{t,s=1}^T\Big(\E\big[\Delta F_{it}\Delta F_{jt}\Delta F_{is}\Delta F_{js}\big]-(\gamma_{ij}^{\Delta F})^2\Big)\nn\\
=&\frac{r^2K_1^4q^4}{T^2}\sum_{t=1}^T\E[ u_{lt}^2] \E[u_{ht}^2]+\frac{r^2K_1^4q^4}{T^2}\sum_{t=1}^T\E[u_{lt}^4]-\frac{r^2K_1^4q^4}{T^2}\sum_{t=1}^T(\E[u_{lt}^2])^2 \leq  \frac{r^2K_1^4q^4M_1}{T} ,\label{eq:LLNF}
\end{align}
since $\E[u_{jt}^2]=1$ for any $j=1,\ldots, q$ and because of Assumption \ref{ASS:common}(a) of existence of fourth moments. 

In the same way, for the idiosyncratic component, using \eqref{eq:idioscalar}, for all $i,j=1,\ldots, n$, we have
\begin{align}
\E\bigg[\bigg\vert\frac 1 T& \sum_{t=1}^T \Delta\xi_{it}\Delta\xi_{jt} -\gamma_{ij}^{\Delta \xi} \bigg\vert^2\bigg] \leq 
\frac {1}{T^2}\sum_{t,s=1}^T\Big(\E\big[\Delta \xi_{it}\Delta \xi_{jt}\Delta \xi_{is}\Delta \xi_{js}\big]-(\gamma_{ij}^{\Delta \xi})^2\Big)\nn\\
&\leq\frac{K_2^4}{T^2}\sum_{t=1}^T\E[ \varepsilon_{it}^2\varepsilon_{jt}^2] \leq \frac{K_2^4M_2}{T} ,\label{eq:LLNxi}
\end{align}
where we used Assumption \ref{ASS:idio}(a) of independence of $\bm\varepsilon_t$ and existence of its fourth moments, and square summability of the coefficients, with $K_2$ defined in \eqref{eq:sqsumMM2}.
By combining \eqref{eq:LLNF} and \eqref{eq:LLNxi} and Assumption \ref{asm:factor}(b) of uniformly bounded loadings, as $T\to\infty$, there exists a positive real $C_1$ which does not depend on $i$ and $j$ such that $\E[|\wh{\gamma}_{ij}^{\Delta x}-\gamma_{ij}^{\Delta x}|^2]\le C_1T^{-1}$.

Then for all $i,j=1,\ldots, n$, we have
\begin{align}
\E&\big[\vert\wh{\gamma}_{ij}^{\Delta y}-\wh{\gamma}_{ij}^{\Delta x}\vert^2\big] = \E\bigg[\bigg|\frac 1 T\sum_{t=1}^T\Big(\big(\Delta y_{it}-\Delta \overline{y}_i\big)
\big(\Delta y_{jt}-\Delta \overline{y}_j\big)-\Delta x_{it}\Delta x_{jt}\Big)\bigg|^2\bigg]\nn\\
&\le2\E\bigg[\bigg|\frac 1 T\sum_{t=1}^T\Delta x_{it}\big(b_j-\Delta \overline{y}_j\big)\bigg|^2\bigg]+\E\bigg[\bigg|\frac 1 T\sum_{t=1}^T\big(b_i-\Delta \overline{y}_i\big)\big(b_j-\Delta \overline{y}_j\big)\bigg|^2\bigg]\nn\\
&\le2 \E \bigg[\bigg|\frac 1 T\sum_{t=1}^T\Delta x_{it}\bigg|^2\bigg]\, \E\big[\big|\big(b_i-\Delta \overline{y}_i\big)\big|^2\big]
+\E\big[\big|\big(b_i-\Delta \overline{y}_i\big)\big(b_j-\Delta \overline{y}_j\big)\big|^2\big].\label{eq:trendapp1}
\end{align}
Now, by definition of sample mean we have for all $i=1,\ldots, n$
\begin{align}
\E\big[\big\vert b_i&-\Delta \overline{y}_i\big\vert^2\big] =\E\bigg[ \bigg\vert \frac 1 T \sum_{t=1}^T \Delta x_{it}\bigg\vert^2\bigg]=\frac 1{T^2}\sum_{t,s=1}^T\big\vert\E[\Delta x_{it}\Delta x_{is}]\big\vert\nn\\
\le&\, \frac 1{T^2}\sum_{t,s=1}^T\big\vert\E[\bm\lambda_i'\Delta \mbf F_{t}\bm\lambda_i'\Delta\mbf F_{s}]\big\vert
+\frac 1{T^2}\sum_{t,s=1}^T\big\vert\E[\Delta\xi_{it}\Delta\xi_{is}]\big\vert\nn\\
\le&\,  \frac{C^2}{T^2}\sum_{t,s=1}^T\sum_{j,\ell=1}^r\sum_{k,h=0}^{\infty}\vert c_{jm_1k}\vert\, \vert c_{\ell m_2 h}\vert\sum_{m_1,m_2=1}^q\vert \E[u_{m_1t-k}u_{m_2s-h}]\vert+\frac{1}{T^2}\sum_{t,s=1}^T\sum_{k,h=0}^{\infty}\vert d_{ik}\vert\, \vert d_{i h}\vert\vert \E[\eps_{it-k}\eps_{is-h}]\vert\nn\\
\le&\, \frac {C^2r^2 q K_1^2 }{T} \E[u_{jt}^2]+\frac {K_2^2 }{T}\max_{i=1,\ldots, n}\E[\eps_{it}^2]=O\l(\frac 1 T\r),\label{eq:trendapp2}
\end{align}
because of Assumption \ref{ASS:common}(a) of independence of $\mbf u_t$ and square summability of the coefficients, with $K_1$ defined in \eqref{eq:sqsumMM1} and  since $\E[u_{jt}^2]=1$ for any $j=1,\ldots, q$, and because of Assumption \ref{ASS:idio}(a) of independence of $\bm\varepsilon_t$ and existence of its fourth moments, and square summability of the coefficients, with $K_2$ defined in \eqref{eq:sqsumMM2} and since $\max_{i=1,\ldots, n}\E[\eps_{it}^2]$ is finite by Assumption \ref{ASS:idio}(b). By using \eqref{eq:trendapp2} in \eqref{eq:trendapp1} we have that as $T\to\infty$, there exists a positive real $C_2$ which does not depend on $i$ and $j$ such that $\E[|\wh{\gamma}_{ij}^{\Delta y}-\wh{\gamma}_{ij}^{\Delta x}|^2]\le C_2T^{-1}$. 

Therefore,
 \begin{align}
\E\big[\vert\wh{\gamma}_{ij}^{\Delta y}-{\gamma}_{ij}^{\Delta x}\vert^2\big]\le \E\big[\vert\wh{\gamma}_{ij}^{\Delta y}-\wh{\gamma}_{ij}^{\Delta x}\vert^2\big] + \E\big[\vert\wh{\gamma}_{ij}^{\Delta x}-{\gamma}_{ij}^{\Delta x}\vert^2\big]\le  \frac{C_1+C_2}{T},
\end{align}
by setting $C_0=C_1+C_2$ we complete the proof.\hfill $\Box$\\
\begin{lem}\label{lem:main}
Under Assumptions \ref{ASS:common} through \ref{initcond}, for any $t$ we have\smallskip
\begin{compactenum}[(i)]
\item $\E[\Vert\Delta{\mbf F}_t\Vert^2] =O(1)$;
\item $\E[\Vert T^{-1/2}{\mbf F}_t\Vert^2]=O(1)$;
\item $\E[\Vert n^{-1/2}\Delta\bm\xi_t \Vert^2]=O(1)$;
\item $\E[\Vert (nT)^{-1/2}\bm\xi_t \Vert^2]=O(1)$;
\item $\E[\Vert n^{-1/2}{\bm\Lambda}'\Delta\bm\xi_t \Vert^2]=O(1)$;
\item $\E[\Vert (nT)^{-1/2}{\bm\Lambda}'\bm\xi_t \Vert^2]=O(1)$.
\end{compactenum}
\end{lem}


\noindent{\textbf{Proof.}} For part (i), just notice that, since by Assumption \ref{ASS:common}(b) $\Delta F_{jt}\sim I(0)$ for any $i=1,\ldots ,r$, then they have finite variance. This proves part (i).\smallskip

For part (ii), from \eqref{eq:model2scalar} we have
\begin{align}
\E\bigg[&\bigg\Vert\frac{\mbf F_t}{\sqrt T}\bigg\Vert^2\bigg] = 
\frac 1 T\sum_{j=1}^r\E\big[F_{jt}^2\big]=
\frac 1 T \sum_{j=1}^r\E\bigg[\bigg(\sum_{s=1}^t \sum_{l=1}^q c_{jl}(L) u_{ls}\bigg)^2\bigg]\nn\\
=& \frac 1 T\sum_{j=1}^r\sum_{s,s'=1}^t\sum_{l,l'=1}^q\sum_{k,k'=0}^{\infty} c_{jlk}c_{jl'k'}\E[ u_{ls-k}u_{l's'-k'}]\leq \frac{rq K_1t}T\leq rq K_1,\label{eq:main2}
\end{align}
since $t\leq T$ and where we used the fact $\mbf u_t$ is a white noise because of Assumption \ref{ASS:common}(a) and we used square summability of the coefficients, with $K_1$ defined in \eqref{eq:sqsumMM1}. This proves part (ii). \smallskip

For part (iii), for any $n\in\mathbb N$ and from \eqref{eq:idioscalar}, we have,
\begin{align}
\E\bigg[&\bigg\Vert\frac{\Delta\bm\xi_t}{\sqrt n}\bigg\Vert^2\bigg] =\frac 1 n \sum_{i=1}^n\E\big[\Delta \xi_{it}^2\big]=\frac 1 n\sum_{i=1}^n\E[(\check{d}_i(L)\varepsilon_{it})^2]\nn\\
=&\frac 1 n\sum_{i=1}^n\sum_{k,k'=0}^{\infty} \check{d}_{jk}\check{d}_{ik'}\E[ \eps_{it-k}\eps_{it-k'}]\leq K_2\max_{i=1,\ldots, n}\E[\eps_{it}^2],\label{eq:main4}
\end{align}
where we used Assumption \ref{ASS:idio}(a) of serially uncorrelated $\bm\eps_t$ and square summability of the coefficients, with $K_2$ defined in \eqref{eq:sqsumMM2}. Also because of the existence of fourth moments in Assumption \ref{ASS:idio}(a) the variance of $\eps_{it}$ is finite for any $i$. This proves part (iii).\smallskip

Similarly, for part (iv), for any $n\in\mathbb N$, we have,
\begin{align}
\E\bigg[&\bigg\Vert\frac{\bm \xi_t}{\sqrt {nT}}\bigg\Vert^2\bigg] = \frac 1 {nT}\sum_{i=1}^n\E\big[\xi_{it}^2\big]=\frac 1 {nT} \sum_{i=1}^n\E\bigg[\bigg(\sum_{s=1}^t \check d_{i}(L) \eps_{is}\bigg)^2\bigg]\nn\\
=& \frac 1 {nT} \sum_{i=1}^n \sum_{s,s'=1}^t\sum_{k,k'=0}^{\infty} \check d_{ik}\check d_{ik'}\E[ \eps_{is-k}\eps_{is'-k'}]\leq \frac{K_2t}T\max_{i=1,\ldots n}\E[\eps_{it}^2]\leq K_2\max_{i=1,\ldots,n}\E[\eps_{it}^2],\label{eq:main5}
\end{align}
since $t\leq T$ and where we used the same assumptions as in \eqref{eq:main4}. This proves part (iv).\smallskip
  
As for part (v), for any $n\in\mathbb N$, we have
\begin{align}
\E\bigg[&\bigg\Vert\frac{\bm\Lambda'\Delta\bm\xi_t}{\sqrt n}\bigg\Vert^2\bigg] =\frac 1 n \sum_{j=1}^r\E\bigg[\bigg(\sum_{i=1}^n\lambda_{ij}\Delta \xi_{it}\bigg)^2\bigg]=\frac 1 n \sum_{j=1}^r\sum_{i,l=1}^n\E\big[\lambda_{ij}\Delta \xi_{it}\lambda_{lj}\Delta \xi_{lt}\big]\nn\\
\leq &\frac {rC^2} n \sum_{i,l=1}^n\sum_{k,k'=0}^{\infty}\check d_{ik}\check d_{lk'}\E[\eps_{it-k}\eps_{lt-k'}]\leq \frac {rC^2K_2} n\sum_{i,l=1}^n\big|\E[\eps_{it}\eps_{lt}]\big|\leq rC^2K_2M_3,\label{eq:main6}
\end{align}
where we used the same assumptions as in \eqref{eq:main4}, Assumption \ref{asm:factor}(b) of bounded loadings, and Lemma \ref{rem:idio}. This proves part (v).\smallskip

Similarly for part (vi), for any $n\in\mathbb N$, we have
\begin{align}
\E\bigg[&\bigg\Vert\frac{\bm\Lambda'\bm\xi_t}{\sqrt {nT}}\bigg\Vert^2\bigg] =\frac 1 {nT} \sum_{j=1}^r\E\bigg[\bigg(\sum_{i=1}^n\lambda_{ij} \xi_{it}\bigg)^2\bigg]=\frac 1 {nT} \sum_{j=1}^r\sum_{i,l=1}^n \E\big[\lambda_{ij} \xi_{it}\lambda_{lj} \xi_{lt}\big]\nn\\
\leq &\frac {rC^2} {nT} \sum_{i,l=1}^n\sum_{s,s'=1}^t \sum_{k,k'=0}^{\infty}\check d_{ik}\check d_{lk'}\E[\eps_{is-k}\eps_{ls'-k'}]\leq \frac {rC^2K_2t} {nT}\sum_{i,l=1}^n\big|\E[\eps_{it}\eps_{lt}]\big|\leq rC^2K_2M_3,\label{eq:main7}
\end{align}
where we used the same assumptions as in \eqref{eq:main6}. This proves part (vi) and completes the proof.\hfill$\Box$\\
\begin{lem}\label{lem:BN} Under Assumptions \ref{ASS:common} and \ref{initcond}: 
\begin{compactenum}[(i)]
\item $\mbf F_t = \mbf C(1)\sum_{s=1}^{t} \mbf u_{s} + \check{\mbf C}(L)\mbf u_t$, such that $\check{\mbf C}(L)$ is an $r\times q$ infinite rational  polynomial matrix with square summable coefficients; moroever, $\mbf C(1)=\bm\psi\bm\eta'$,
where $\bm\psi$ is $r\times r-c$, $\bm\eta$ is $q\times r-c$, $\mbox{rk}(\bm\psi)=\mbox{rk}(\bm\eta)=r-c=q-d$ and $\bm\beta'\mbf C(1)=\mbf 0_{c\times q}$, where $\bm\beta$ is the $r\times c$ cointegration matrix;
\item $\E[\Vert{\bm\beta}'{\mbf F}_t\Vert^2]=O(1)$ for any $t=1,\ldots, T$.
\end{compactenum}
\end{lem}

\noindent{\textbf{Proof.}}
From  Lemma 2.1 in \citet{phillipssolo}, the Beveridge-Nelson decomposition of $\mbf C(L)$ in \eqref{eq:model2vector} gives
\[
\Delta\mbf F_t = \mbf C(1)\mbf u_t + \check{\mbf C}(L)(\mbf u_t-\mbf u_{t-1}),
\]
where $\check{\mbf C}(L)=\sum_{k=0}^{\infty} \check{\mbf C}_kL^k$ with $\check{\mbf C}_k=-\sum_{h=k+1}^{\infty}\mbf C_h$ and has square summable coefficients because of \eqref{eq:sqsumMM1}. Then,
\beq\label{eq:fBN}
\mbf F_t = \mbf C(1) \sum_{s=1}^t \mbf u_s +{\bm\omega}_t,
\eeq
where $\bm\omega_t=\check{\mbf C}(L)(\mbf u_t-\mbf u_0)=\check{\mbf C}(L)\mbf u_t$, since $\mbf u_t=\mbf 0_q$ when $t\leq 0$ by Assumption \ref{initcond}, and $\bm\omega_t\sim I(0)$, because of square summability of the coefficients of $\check{\mbf C}(L)$. Moreover, from Assumption \ref{ASS:common}(c) of cointegration, we have $\mbf C(1)=\bm\psi\bm\eta'$,  where $\bm\psi$ is $r\times r-c$ and $\bm\eta$ is $q\times r-c$. Since $\bm\beta$ is a cointegrating vector for $\mbf F_t$, we must have $\bm\beta'\mbf F_t\sim I(0)$, which from \eqref{eq:fBN} implies $\bm\beta'\mbf C(1)=\mbf 0_{c\times q}$. This proves part (i).\smallskip

Turning to part (ii), from part (i) and \eqref{eq:fBN}, we have 
$$
\bm\beta'\mbf F_t = \bm\beta'\bm\omega_t= \bm\beta'\check{\mbf C}(L) \mbf u_t.
$$ 
Define $\widetilde{\mbf C}(L)=\bm\beta'\check{\mbf C}(L)$ and notice that it has square summable coefficients because of square summability of the coefficients of $\check{\mbf C}(L)$, then
\begin{align}
\E\big[&\big\Vert\bm\beta'{\mbf F}_t\big\Vert^2\big] 
=\sum_{j=1}^r\E[(\widetilde{\mbf c}_j'(L)\mbf u_t)^2] = \sum_{j=1}^r\E\bigg[\bigg(\sum_{l=1}^q \widetilde c_{jl}(L) u_{lt}\bigg)^2\bigg]\nn\\
=& \sum_{j=1}^r\sum_{l,l'=1}^q\sum_{k,k'=0}^{\infty} \widetilde c_{jlk}\widetilde c_{jl'k'}\E[ u_{lt-k}u_{l't-k'}]\leq rq K_1,\label{eq:main1}
\end{align}
where we used the fact $\mbf u_t$ is a white noise because of Assumption \ref{ASS:common}(a) and we used square summability of the coefficients, with $K_1$ defined in \eqref{eq:sqsumMM1}. 
This proves part (ii) and completes the proof. \hfill$\Box$\\
\begin{lem}\label{rem:DFF}
For $k=0,1$, define $\bm\Gamma_k^{\Delta F}=\E[\Delta\mbf F_t^{}\Delta\mbf F_{t-k}']$ and $\bm\Gamma_k^{\omega}=\E[\bm\omega_t\bm\omega_{t-k}']$, where $\bm\omega_t=\check{\mbf C}(L)\mbf u_t$ is defined in \eqref{eq:fBN}. Define also, $\bm\Gamma_{L}^{\omega}=\bm\Gamma_0^{\omega}+2 \sum_{h=1}^{\infty} \bm\Gamma_h^{\omega}$. Denote as $\mbf W_q(\cdot)$ a $q$-dimensional Brownian motion with covariance $\mbf I_q$ and as $\mbf W_r(\cdot)$ an $r$-dimensional Brownian motion with covariance $\mbf I_r$. Under Assumptions \ref{ASS:common} and \ref{initcond}, as $T\to\infty$, \smallskip
\begin{compactenum}[(i)]
\item $\E[\Vert T^{-1}\sum_{t=k+1}^T\Delta \mbf F_t^{}\Delta \mbf F_{t-k}'-\bm\Gamma_k^{\Delta F}\Vert^2]=O(T^{-1})$, for $k=0,1$;
\item $T^{-2}\sum_{t=1}^T \mbf F_t^{}\mbf F_t'\stackrel{d}{\to}\mbf C(1)\big(\int_0^1\mbf W_q(\tau)\mbf W_q'(\tau)\mbox {\upshape d} \tau\big)\mbf C'(1)$;
\item $T^{-1}\sum_{t=1}^T \mbf F_{t-1}^{}\Delta \mbf F_{t}'\stackrel{d}{\to}\mbf C(1)\big(\int_0^1\mbf W_q(\tau)\mbox {\upshape d}\mbf W_q'(\tau)\big)\mbf C'(1)+(\bm\Gamma_{1}^{\omega}-\bm\Gamma_0^{\omega})$;
\item $T^{-1}\sum_{t=1}^T \mbf F_t^{}\mbf F_t'\bm\beta\stackrel{d}{\to}\mbf C(1)\big(\int_0^1\mbf W_q(\tau)\mbox {\upshape d}\mbf W_r'(\tau)\big)(\bm\Gamma_L^\omega)^{1/2}\bm\beta+\bm\Gamma_0^{\omega}\bm\beta$;
\item $\E[\Vert T^{-1}\sum_{t=1}^T \bm\beta'\mbf F_t^{}\mbf F_t'\bm\beta-\bm\beta'\bm\Gamma_0^{\omega}\bm\beta\Vert^2]=\E[\Vert T^{-1}\sum_{t=1}^T \bm\beta'\mbf F_t^{}\mbf F_t'\bm\beta-\E[\bm\beta'\mbf F_t^{}\mbf F_t'\bm\beta]\Vert^2]=O(T^{-1})$;
\item $\E[\Vert T^{-1}\sum_{t=1}^T \Delta\mbf F_t^{}\mbf F_{t-1}'\bm\beta\,-(\bm\Gamma_1^{\omega}-\bm\Gamma_0^{\omega})\bm\beta\Vert^2]\!=\!\E[\Vert T^{-1}\sum_{t=1}^T \Delta\mbf F_t^{}\mbf F_{t-1}'\bm\beta-\E[\Delta\mbf F_t^{}\mbf F_{t-1}'\bm\beta]\Vert^2]=O(T^{-1})$.
\end{compactenum}
 \end{lem}

\noindent{\textbf{Proof.}}
For part (i), the case $k=0$ is already proved in \eqref{eq:LLNF} in the proof of Lemma \ref{lem:covX}. The proof for the case $k= 1$, is analogous.\smallskip

In order to prove the other statements, notice that $\mbox{rk}(\bm\Gamma_{L}^{\omega})=r$ because of Assumption \ref{ASS:common}(d) and define, for $\tau\in[0,1]$,
\[
\bm{\mathcal X}_{u,T}(\tau) = \frac 1 {\sqrt T}\sum_{s=1}^{\lfloor T\tau \rfloor} \mbf u_s,\qquad \bm{\mathcal X}_{\omega,T}(\tau) = \Big(\bm\Gamma_L^\omega\Big)^{-1/2} \frac 1 {\sqrt T}\sum_{s=1}^{\lfloor T\tau \rfloor} \bm \omega_s.
\]
Then, we can write
\begin{align}
\sum_{s=1}^{t} \mbf u_s &= \sqrt T\, \bm{\mathcal X}_{u,T}\bigg(\frac t T\bigg),\label{parsumu}\\
\mbf u_t &= \sqrt T\, \bigg[\bm{\mathcal X}_{u,T}\bigg(\frac t T\bigg)-\bm{\mathcal X}_{u,T}\bigg(\frac {t-1} T\bigg)\bigg], \label{parsumu2}\\
\bm\omega_t &= \sqrt T\, \Big(\bm\Gamma_{L}^{\omega}\Big)^{1/2}
\bigg[
\bm{\mathcal X}_{\omega,T}\bigg(\frac t T\bigg) -\bm{\mathcal X}_{\omega,T}\bigg(\frac {t-1} T\bigg) 
\bigg].\label{omegaX}
\end{align}
As proved in Corollary 2.2 in \citet{phillipsdurlauf98} (see also Theorem 3.4 in \citealp{phillipssolo}), for any $\tau\in [0,1]$, we have, as $T\to\infty$, 
\beq\label{eq:fclt}
\bm{\mathcal X}_{u,T}(\tau) \stackrel{d}{\to} \mbf W_q(\tau),\qquad \bm{\mathcal X}_{\omega,T}(\tau) \stackrel{d}{\to} \mbf W_r(\tau),
\eeq
where $\mbf W_q(\cdot)$ is a $q$-dimensional Brownian motion with covariance $\mbf I_q$ and  $\mbf W_r(\cdot)$ is a $q$-dimensional Brownian motion with covariance $\mbf I_r$.

For part (ii), from Lemma \ref{lem:BN}(i),  we have
\begin{align}
\frac 1 {T^2}\sum_{t=1}^T \mbf F_t\mbf F_t'&=\frac 1{T^2}\sum_{t=1}^T\bigg[\bigg( \mbf C(1) \sum_{s=1}^t\mbf u_s\bigg)\bigg( \mbf C(1) \sum_{s=1}^t\mbf u_s\bigg)'\bigg]\nn\\ 
&+\frac 1{T^2}\sum_{t=1}^T\bigg[\bigg( \mbf C(1) \sum_{s=1}^t\mbf u_s\bigg)\bm\omega_t'+\bm\omega_t\bigg( \mbf C(1) \sum_{s=1}^t\mbf u_s\bigg)'\bigg]+\frac 1{T^2}\sum_{t=1}^T\bm\omega_t\bm\omega_t'.\label{eq:FFT2}
\end{align}
For the first term on the rhs of \eqref{eq:FFT2}, using \eqref{parsumu} and \eqref{eq:fclt}, we have, as $T\to\infty$,
\beq\label{WW12}
\frac 1{T^2}\sum_{t=1}^T\bigg[\bigg( \mbf C(1) \sum_{s=1}^t\mbf u_s\bigg)\bigg( \mbf C(1) \sum_{s=1}^t\mbf u_s\bigg)'\bigg]\stackrel{d}{\to}\mbf C(1)\bigg(\int_0^1\mbf W_q(\tau)\mbf W_q'(\tau)\mbox {\upshape d} \tau\bigg)\mbf C'(1),
\eeq
which is $O_p(1)$, since it has finite covariance, and has rank $r-c$, since $\mbox{rk}(\mbf C(1))=r-c$ because of Assumption \ref{ASS:common}(c). Then, since $\frac{\mbf W_r(\tau)-\mbf W_r(\tau-\mbox {\upshape \scriptsize d} \tau)}{\mbox {\upshape \scriptsize d} \tau}=\frac{\mbox {\upshape \scriptsize d} \mbf W_r(\tau)}{\mbox {\upshape \scriptsize d}\tau}+O(\mbox{d}\tau)$, as $\mbox {\upshape d}\tau\to 0$, using \eqref{omegaX} and \eqref{eq:fclt}, we have, as $T\to\infty$, 
\begin{align}
\frac 1{T}\sum_{t=1}^T \bigg( \mbf C(1) \sum_{s=1}^t\mbf u_s\bigg)\bm\omega_t' \stackrel{d}{\to} 
&\,\mbf C(1)\bigg(\int_0^1\mbf W_q(\tau)\mbox {\upshape d} \mbf W_r'(\tau)\bigg)\Big(\bm\Gamma_{L}^{\omega}\Big)^{1/2},
\label{deriW}
\end{align}
which is $O_p(1)$, since it has finite covariance. Therefore, the second and third term on the rhs of \eqref{eq:FFT2} are $O_p(T^{-1})$. Similarly, the fourth
term on the rhs of \eqref{eq:FFT2} is $O_p(T^{-1})$ since $\Vert\bm\Gamma_0^{\omega}\Vert=O(1)$ and for $k=0,1$, we have
\beq\label{gammaomega0}
\E\bigg[\bigg\Vert\frac 1 T\sum_{t=1}^T\bm\omega_t\bm\omega_{t-k}' - \bm\Gamma_k^{\omega}\bigg\Vert^2\bigg] = O\l(\frac 1{ T}\r),
\eeq
by arguments analogous to those used in proving part (i). By substituting \eqref{WW12}, \eqref{deriW}, and \eqref{gammaomega0} (which implies convergence in probability by Chebychev's inequality) in \eqref{eq:FFT2}, and by Slutsky's theorem, we prove part (ii).\smallskip

For part (iii), from Lemma \ref{lem:BN}(i),  we have
\begin{align}
\frac 1 T\sum_{t=1}^T \mbf F_{t-1}\Delta \mbf F_t'&=\frac 1 T\sum_{t=1}^T\bigg[\bigg(\sum_{s=1}^{t-1}\mbf C(1) \mbf u_s\bigg)\Big(\mbf C(1)\mbf u_t\Big)'\bigg]
+\frac 1 T\sum_{t=1}^T\bigg[\bigg(\sum_{s=1}^{t-1}\mbf C(1) \mbf u_s\bigg)\Delta\bm\omega_{t}'\bigg]\nn\\
&+\frac 1 T\sum_{t=1}^T\bigg[\bm\omega_{t-1}\Big(\mbf C(1)\mbf u_t\Big)'\bigg]+\frac 1 T\sum_{t=1}^T\bm\omega_{t-1}\Delta\bm\omega_{t}'.\label{eq:FFT3}
\end{align}
For the first term on the rhs of \eqref{eq:FFT3}, using \eqref{parsumu}, \eqref{parsumu2}, and \eqref{eq:fclt}, we have, as $T\to\infty$,
\beq\label{WderiW}
\frac 1 T\sum_{t=1}^T\bigg[\bigg(\sum_{s=1}^{t-1}\mbf C(1) \mbf u_s\bigg)\Big(\mbf C(1)\mbf u_t\Big)'\bigg]\stackrel{d}{\to}\mbf C(1)\bigg(\int_0^1\mbf W_q(\tau)\mbox {\upshape d} \mbf W_q'(\tau)\bigg)\mbf C'(1),
\eeq
which is $O_p(1)$, since it has finite covariance, and has rank $r-c$, since $\mbox{rk}(\mbf C(1))=r-c$. For the second term on the rhs of \eqref{eq:FFT3}, since $\Delta\bm\omega_t=\bm\omega_t-\bm\omega_{t-1}$, by following twice the same steps as those leading to \eqref{deriW}, we have
\begin{align}
\frac 1 T\sum_{t=1}^T\bigg[\bigg(\sum_{s=1}^{t-1}\mbf C(1) \mbf u_s\bigg)\Delta\bm\omega_{t}'\bigg]
\stackrel{d}{\to}\mbf 0_{r\times r}.\label{tuttozero}
\end{align}
For the third term on the rhs of \eqref{eq:FFT3} we have 
\beq\label{terzotermine0}
\E\bigg[\bigg\Vert\frac 1 T\sum_{t=1}^T\bigg[\bm\omega_{t-1}\Big(\mbf C(1)\mbf u_t\Big)'\bigg]\bigg\Vert^2\bigg]=O\l(\frac 1{T}\r).
\eeq
by arguments similar to \eqref{gammaomega0} and the fact that $\E[\bm \omega_{t-1}\mbf u_t']=\mbf 0_{r\times r}$, because of orthonormality of $\mbf u_t$ given in Assumption \ref{ASS:common}(a). Last, for the fourth term 
on the rhs of \eqref{eq:FFT3}, we can use \eqref{gammaomega0} to show that
\beq\label{gammaomegah}
\E\bigg[\bigg\Vert \frac 1 T\sum_{t=1}^T\bm\omega_{t-1}\Delta\bm\omega_{t}'-\Big(\bm\Gamma_1^{\omega}-\bm\Gamma_0^{\omega}\Big)\bigg\Vert^2\bigg]= O\l(\frac 1{T}\r).
\eeq
By substituting \eqref{WderiW}, \eqref{tuttozero}, \eqref{terzotermine0} and \eqref{gammaomegah} (both implying convergence in probability by Chebychev's inequality)  in \eqref{eq:FFT3},  and by Slutsky's theorem, we prove part (iii).\smallskip

Turning to part (iv), since $\bm\beta'\mbf F_t = \bm\beta'\bm\omega_t$, from Lemma \ref{lem:BN}(i), we have 
\begin{align}
\frac 1 T\sum_{t=1}^T \mbf F_t\mbf F_t'\bm\beta&
=\mbf C(1) \bigg[\frac 1 T \sum_{t=1}^T\bigg( \sum_{s=1}^t \mbf u_s \bigg)\bm\omega_t'\bigg]\bm\beta+\bigg[\frac 1 T\sum_{t=1}^T\bm\omega_t\bm\omega_t'\bigg]\bm\beta\nn\\
\stackrel{d}{\to}&\,\mbf C(1)\bigg(\int_0^1\mbf W_q(\tau)\mbox {\upshape d}\mbf W_r'(\tau)\bigg)\Big(\bm\Gamma_L^\omega\Big)^{1/2}\bm\beta+\bm\Gamma_0^{\omega}\bm\beta.
\label{eq:fbf}
\end{align}
by analogous arguments as those leading to \eqref{deriW} and using \eqref{gammaomega0} and Slutsky's theorem. This completes the proof of part (iv). \smallskip

Part (v) is proved analogously just by multiplying \eqref{eq:fbf} also on the left by $\bm\beta'$ and then using \eqref{gammaomega0} and the fact that $\bm\beta'\mbf F_t = \bm\beta'\bm\omega_t$ because of  Lemma \ref{lem:BN}(i).\smallskip

Finally, part (vi) is proved by noticing that
\begin{align}
\frac 1 T\sum_{t=1}^T \Delta \mbf F_t\mbf F_{t-1}'\bm\beta&=\bigg(\frac1 T\sum_{t=1}^T\mbf C(1)\mbf u_t\bm\omega_{t-1}' +\frac1 T\sum_{t=1}^T\Delta \bm\omega_t\bm\omega_{t-1}'\bigg)\bm\beta\nn
\end{align}
and using \eqref{terzotermine0} and \eqref{gammaomegah}. This completes the proof. \hfill$\Box$\\
\begin{lem}\label{lem:fidio}
Under Assumptions \ref{ASS:common} through \ref{initcond} and \ref{asm:rates}, as $n,T\to\infty$, \smallskip
\begin{compactenum}
\item [(i)] $\E[\Vert (nT^{2})^{-1} \sum_{t=1}^T \mbf F_t\bm\xi_t'\bm\Lambda\Vert^2]=O(n^{-(2-\delta)})$;
\item [(ii)] $\E[\Vert (\sqrt n T^2)^{-1} \sum_{t=1}^T\mbf F_t\bm\xi_t'\Vert^2 ]=O(n^{-(1-\delta)})$;
\item [(iii)] $\E[\Vert (n^2 T^2)^{-1} \sum_{t=1}^T \bm\Lambda'\bm\xi_t\bm\xi_t'\bm\Lambda\Vert^2]=O(n^{-2(2-\delta)})$;
\item [(iv)] $\E[\Vert (n T^2)^{-1} \sum_{t=1}^T \bm\xi_t\bm\xi_t'\Vert^2]=O(n^{-2(1-\delta)})$;
\item [(v)] $\E[\Vert (nT)^{-1} \sum_{t=1}^T \Delta\mbf F_t\bm\xi_t'\bm\Lambda\Vert^2]=O(T n^{-(2-\delta)})$;
\item [(vi)] $\E[\Vert (\sqrt nT)^{-1} \sum_{t=1}^T \Delta\mbf F_t\bm\xi_t'\Vert^2]=O(T n^{-(1-\delta)})$;
\item [(vii)] $\E[\Vert (n^2T)^{-1} \sum_{t=1}^T \bm\Lambda'\Delta\bm\xi_t\bm\xi_t'\bm\Lambda\Vert^2]=O(T n^{-2(2-\delta)})$;
\item [(viii)] $\E[\Vert (nT)^{-1} \sum_{t=1}^T \Delta\bm\xi_t\bm\xi_t'\Vert^2]=O(T n^{-2(1-\delta)})$.
\item [(ix)] $\E[\Vert (n^{3/2} T^2)^{-1} \sum_{t=1}^T \bm\xi_t\bm\xi_t'\bm\Lambda\Vert^2]=O(n^{-(3-2\delta)})$.
\end{compactenum}
\end{lem}

\noindent{\textbf{Proof.}} Start with part (i):
\begin{align}
\E&\bigg[\bigg\Vert \frac 1{nT^{2}} \sum_{t=1}^T \mbf F_t\bm\xi_t'\bm\Lambda\bigg\Vert^2\bigg] = \frac1{n^2T^4}\sum_{j_1,j_2=1}^r \E\bigg[\bigg(\sum_{t=1}^T F_{j_1t}\sum_{i=1}^n\lambda_{ij_2} \xi_{it}\bigg)^2\bigg]\le\frac{C^2 r}{n^2T^4} \sum_{t,s=1}^T\sum_{j=1}^r \sum_{i_1,i_2=1}^n \Big\vert\E\big[F_{jt} F_{js} \xi_{i_1t}\xi_{i_2s}\big]\Big\vert\nn\\
&\le \frac{C^2 r}{n^2T^4}\sum_{t,s=1}^T\sum_{j=1}^r\Big\vert\E\big[F_{jt} F_{js}\big]\Big\vert\bigg\{
\sum_{i_1,i_2\in\mathcal I_1^c}\Big\vert\E\big[\xi_{i_1t}\xi_{i_2s}\big]\Big\vert + 3 \sum_{i_1,i_2\in\mathcal I_1}\Big\vert\E\big[\xi_{i_1t}\xi_{i_2s}\big]\Big\vert \bigg\}\nn\\
&\le \frac{C^2 r}{n^2T^4}\sum_{t,s=1}^T\sum_{j=1}^r\E\big[F_{jt}^2\big]\bigg\{
\sum_{i_1,i_2\in\mathcal I_1^c}\Big\vert\E\big[\xi_{i_1t}\xi_{i_2s}\big]\Big\vert + 3 \sum_{i_1,i_2\in\mathcal I_1}\Big\vert\E\big[\xi_{i_1t}\xi_{i_2s}\big]\Big\vert \bigg\}\nn\\
&\le \frac{C^2 r}{n^2 T^4} \sum_{t,s=1}^T\sum_{j=1}^r\E\big[F_{jt}^2\big] K_2^2\Big\{  \sum_{i_1,i_2\in\mathcal I_1^c}\Big\vert\E\big[\eps_{i_1t}\eps_{i_2t}\big]\Big\vert + 3 \sum_{i_1,i_2\in\mathcal I_1} \sum_{s=1}^t \Big\vert\E\big[\eps_{i_1s}\eps_{i_2s}\big]\Big\vert
\Big\vert \bigg\}\nn\\
&\le \frac{C^2 r^2}{n^2 T^4}T \sum_{t=1}^T\E\big[F_{jt}^2\big] K_2^2 M_3 (n+n^\delta t ) = O\l(\frac{1}{nT}\r)+O\l(\frac{1}{n^{2-\delta}}\r),\nn
\end{align}
where we used Assumption \ref{asm:factor}(b) of uniformly bounded loadings, Assumption \ref{ASS:idio}(a) and (e) of independent idiosyncratic shocks also independent of the common shocks, Assumptions \ref{ASS:idio} and \ref{asm:rates} which bound the cross-sectional dependence of idiosyncratic components, square summability of the coefficients, with $K_2$ defined in \eqref{eq:sqsumMM2}, Cauchy-Schwarz inequality, and Lemma \ref{lem:main}(ii). This proves part (i).\smallskip

For part (ii) we have:
\begin{align}
\E&\bigg[\bigg\Vert \frac 1{\sqrt nT^{2}} \sum_{t=1}^T \mbf F_t\bm\xi_t'\bigg\Vert^2\bigg] = \frac1{nT^4}\sum_{j=1}^r \sum_{i=1}^n\E\bigg[\bigg(\sum_{t=1}^T F_{jt}\xi_{it}\bigg)^2\bigg]\le\frac{1}{nT^4} \sum_{t,s=1}^T\sum_{j=1}^r \sum_{i_1,i_2=1}^n \Big\vert\E\big[F_{jt} F_{js} \xi_{i_1t}\xi_{i_2s}\big]\Big\vert\nn\\
&\le\frac{1}{nT^4}\sum_{t,s=1}^T\sum_{j=1}^r \E\big[F_{jt}^2\big] \bigg\{
\sum_{i_1,i_2\in\mathcal I_1^c}\Big\vert\E\big[\xi_{i_1t}\xi_{i_2s}\big]\Big\vert + 3 \sum_{i_1,i_2\in\mathcal I_1}\Big\vert\E\big[\xi_{i_1t}\xi_{i_2s}\big]\Big\vert \bigg\}\nn\\
&\le \frac{1}{n T^4} \sum_{t,s=1}^T\sum_{j=1}^r\E\big[F_{jt}^2\big] K_2^2\Big\{  \sum_{i_1,i_2\in\mathcal I_1^c}\Big\vert\E\big[\eps_{i_1t}\eps_{i_2t}\big]\Big\vert + 3 \sum_{i_1,i_2\in\mathcal I_1} \sum_{s=1}^t \Big\vert\E\big[\eps_{i_1s}\eps_{i_2s}\big]\Big\vert
\Big\vert \bigg\}\nn\\
&\le \frac{r}{n T^4}T \sum_{t=1}^T\E\big[F_{jt}^2\big] K_2^2 M_3 (n+n^\delta t ) 
= O\l(\frac{1}{T}\r)+O\l(\frac{1}{n^{1-\delta}}\r),\nn
\end{align}
using the same arguments used for proving part (i). This proves part (ii).\smallskip

Turning to part (iii):
\begin{align}
\E&\bigg[\bigg\Vert \frac 1{n^2 T^2} \sum_{t=1}^T \bm\Lambda'\bm\xi_t\bm\xi_t'\bm\Lambda\bigg\Vert^2\bigg]
=\frac 1{n^4T^4}\sum_{j_1,j_2=1}^r\E\bigg[\bigg(\sum_{t=1}^T \bigg(\sum_{i_1=1}^n \lambda_{i_1j_1}\xi_{i_1t}\bigg)\bigg(\sum_{i_2=1}^n \lambda_{i_2j_2}\xi_{i_2t}\bigg)\bigg)^2\bigg]\nn\\
&\le \frac {C^4r^2}{n^4T^4} \sum_{t,s=1}^T\sum_{i_1,i_1'=1}^n\sum_{i_2,i_2'=1}^n\Big\vert\E\big[\xi_{i_1t}\xi_{i_1't}\xi_{i_2s}\xi_{i_2's}\big]\Big|\nn\\
&\le \frac {C^4r^2K_2^4}{n^4T^4} \sum_{t,s=1}^T \bigg\{ \! \sum_{i_1,i_1'\in\mathcal I_1^c}\sum_{i_2,i_2'\in\mathcal I_1^c}\Big\vert\E\big[\eps_{i_1t}\eps_{i_1't}\eps_{i_2s}\eps_{i_2's}\big]\Big\vert+ 15\! \!\!\sum_{i_1,i_1'\in\mathcal I_1}\sum_{i_2,i_2'\in\mathcal I_1}\sum_{t_1',t_2'=1}^t\sum_{s_1',s_2'=1}^s \Big\vert\E\big[\eps_{i_1t_1'}\eps_{i_1't_2'}\eps_{i_2s_1'}\eps_{i_2's_2'}\big]\Big\vert
 \bigg\}
\nn\\
&\le \frac {C^4r^2K_2^4}{n^4T^4} \sum_{t,s=1}^T\bigg\{\! \sum_{i_1,i_1'\in\mathcal I_1}\Big\vert\E\big[\eps_{i_1t}\eps_{i_1't}\big]\Big\vert\sum_{i_2,i_2'\in\mathcal I_1}\Big\vert\E\big[\eps_{i_2s}\eps_{i_2's}\big]\Big\vert+
15\!\!\!\!\sum_{i_1,i_1'\in\mathcal I_1}\sum_{t'=1}^t\Big\vert\E\big[\eps_{i_1t'}\eps_{i_1't'}\big]\Big\vert\sum_{i_2,i_2'\in\mathcal I_1}\sum_{s'=1}^s\Big\vert\E\big[\eps_{i_2s'}\eps_{i_2's'}\big]\Big\vert\bigg\}\nn\\
&\le \frac {C^4r^2K_2^4}{n^4T^4}\sum_{t,s=1}^T \bigg\{\bigg(\sum_{i_1,i_2\in\mathcal I_1^c}\Big\vert\E\big[\eps_{i_1t}\eps_{i_2t}\big]\Big\vert\bigg)^2
+15\bigg(\sum_{i_1,i_2\in\mathcal I_1}\sum_{s=1}^t\Big\vert\E\big[\eps_{i_1s}\eps_{i_2s}\big]\Big\vert\bigg)^2
\bigg\}\nn\\
&\le \frac {C^4r^2K_2^4M_3^4}{n^4T^4} T^2  (n^2+15n^{2\delta} t^2)= O\l( \frac 1{n^2T^2}\r)+O\l( \frac 1{n^{2(2-\delta)}}\r),\nn
\end{align}
using the same arguments used for proving part (i). This proves part (iii).\smallskip

For part (iv) we have:
\begin{align}
\E&\bigg[\bigg\Vert \frac 1{ n T^2} \sum_{t=1}^T \bm\xi_t\bm\xi_t'\bigg\Vert^2\bigg]=\frac 1{n^2T^4}\sum_{i,j=1}^n\E\bigg[\bigg(\sum_{t=1}^T \xi_{it}\xi_{jt}\bigg)^2\bigg]\le \frac 1 {n^2T^4}\sum_{t,s=1}^T\Big\vert\E\big[\xi_{it}\xi_{is}\xi_{jt}\xi_{js}\big]\Big\vert\nn\\
&\le\frac{K_2^4}{n^2 T^4}\sum_{t,s=1}^T \bigg\{\bigg(\sum_{i_1,i_2\in\mathcal I_1^c}\Big\vert\E\big[\eps_{i_1t}\eps_{i_2t}\big]\Big\vert\bigg)^2
+15\bigg(\sum_{i_1,i_2\in\mathcal I_1}\sum_{s=1}^t\Big\vert\E\big[\eps_{i_1s}\eps_{i_2s}\big]\Big\vert\bigg)^2\bigg\}\nn\\
&\le \frac{K_2^4M_3^4}{n^2 T^4} T^2 (n^2 + 15 n^{2\delta}t^2)= O\l( \frac 1{T^2}\r)+O\l( \frac 1{n^{2(1-\delta)}}\r),\nn
\end{align}
using the same arguments used for proving part (i). This proves part (iv). Parts (v) and (vi) follow from parts (i) and (ii) respectively. Parts (vii) and (ix)  follow from part (iii), while part (viii) follows from part (iv). This completes the proof. \hfill $\Box$\\
\begin{lem}\label{lem:trend}
Under Assumptions \ref{ASS:common} through \ref{ASS:trend}, as $n,T\to\infty$, \smallskip
\begin{compactenum}
\item [(i)] $\E[\Vert(nT^2)^{-1}\sum_{t=1}^T \mbf F_t(\wh{\mbf x}_t-\mbf x_t)'\bm\Lambda\Vert^2]=O(n^{-2(1-\eta)})$;
\item [(ii)] $\E[\Vert(\sqrt n T^2)^{-1}\sum_{t=1}^T \mbf F_t(\wh{\mbf x}_t-\mbf x_t)'\Vert^2]=O(n^{-(1-\eta)})$;
\item [(iii)] $\E[\Vert(n^2T^2)^{-1}\sum_{t=1}^T \bm\Lambda'(\wh{\mbf x}_t-\mbf x_t)(\wh{\mbf x}_t-\mbf x_t)'\bm\Lambda\Vert^2]=O(n^{-4(1-\eta)})$;
\item [(iv)] $\E[\Vert(nT^2)^{-1}\sum_{t=1}^T (\wh{\mbf x}_t-\mbf x_t)(\wh{\mbf x}_t-\mbf x_t)'\Vert^2]=O(n^{-2(1-\eta)})$;
\item [(v)] $\E[\Vert(nT)^{-1}\sum_{t=1}^T \Delta \mbf F_t (\wh{\mbf x}_t-\mbf x_t)'\bm\Lambda\Vert^2]=O(T n^{-2(1-\eta)})$;
\item [(vi)] $\E[\Vert(\sqrt nT)^{-1}\sum_{t=1}^T \Delta \mbf F_t (\wh{\mbf x}_t-\mbf x_t)'\Vert^2]=O(T n^{-(1-\eta)})$;
\item [(vii)] $\E[\Vert (n^2T)^{-1} \sum_{t=1}^T \bm\Lambda'(\Delta\wh{\mbf x}_t-\Delta\mbf x_t)(\wh{\mbf x}_t-\mbf x_t)'\bm\Lambda\Vert^2]=O(T n^{-4(1-\eta)})$;
\item [(viii)] $\E[\Vert (nT)^{-1} \sum_{t=1}^T (\Delta\wh{\mbf x}_t-\Delta\mbf x_t)(\wh{\mbf x}_t-\mbf x_t)'\Vert^2]=O(T n^{-2(1-\eta)})$.
\end{compactenum}
\end{lem}
 
\noindent{\textbf{Proof.}} We start with two preliminary results. First, note that for all $j=1,\ldots, r$ and all $t,s=1,\ldots, T$ we have
\begin{align}
\E&\big[F_{jt}^2F_{js}^2\big]\le q^4 K_1^4\E\bigg[\bigg(\sum_{t'=1}^t u_{jt'}\bigg)^2\bigg(\sum_{s'=1}^s u_{js'}\bigg)^2\bigg]\le q^4 K_1^4 \sum_{t,t'=1}^T \sum_{s,s'=1}^T \Big\vert\E\big[u_{jt} u_{jt'} u_{js} u_{js'} \big] \Big\vert\nn\\
&\le q^4 K_1^4\bigg\{\sum_{t=1}^T \E[u_{jt}^4] + \sum_{t,s=1}^T \E[u_{jt}^2u_{js}^2]\bigg\}\le q^4 K_1^4 M_1 T^2,\label{F4u4}
\end{align}
where we used square summability of the coefficients, with $K_1$ defined in \eqref{eq:sqsumMM1}, and Assumption \ref{ASS:common}(a) of independence of the common shocks and finite fourth moments. Second, by using the same reasoning as in \eqref{eq:blem2} and \eqref{eq:blem3} in the proof of Lemma \ref{lem:load}, we have that $\E[(\sum_{t=1}^T x_{it})^4]= O(T^6)$ and $\E[(\sum_{t=1}^T tx_{it})^4]= O(T^{10})$ for all $i=1,\ldots, n$. Therefore, 
\beq
\E[(\wh b_i-b_i)^4] =\frac{\E\l[\l(\sum_{t=1}^T t x_{it}-\frac {T+1}2\sum_{t=1}^T x_{it}\r)^4\r]}{\l(\frac 1{12}T(T^2-1)\r)^4}= \frac{C_1}{T^2}.\label{trendb4}
\eeq
for some positive real $C_1$ independent of $i$.\smallskip

Now let us consider part (i):
\begin{align}
\E&\bigg[\bigg\Vert\frac 1{nT^2}\sum_{t=1}^T \mbf F_t(\wh{\mbf x}_t-\mbf x_t)'\bm\Lambda\bigg\Vert^2\bigg]=
\frac1{n^2T^4}\sum_{j_1,j_2=1}^r \E\bigg[\bigg(\sum_{t=1}^T F_{j_1t}\sum_{i=1}^n\lambda_{ij_2} ( b_i-\wh b_i)t\bigg)^2\bigg]\nn\\
&\le \frac{C^2r}{n^2T^4}\sum_{t,s=1}^T ts\sum_{j=1}^r\sum_{i_1,i_2\in\mathcal I_b} \Big\vert\E\big[F_{jt}F_{js} ( b_{i_1}-\wh b_{i_1})( b_{i_2}-\wh b_{i_2})\big]\Big\vert\nn\\
&\le \frac{C^2r}{n^2T^4}\sum_{t,s=1}^T ts\sum_{j=1}^r\sqrt{\E\big[F_{jt}^2F_{js}^2\big]}\sum_{i_1,i_2\in\mathcal I_b}\sqrt{\E[(b_{i_1}-\wh b_{i_1})^2(b_{i_2}-\wh b_{i_2})^2]}\nn\\
&\le \frac{C^2r}{n^2T^4}\sum_{t,s=1}^T ts\sum_{j=1}^r\sqrt{\E\big[F_{jt}^2F_{js}^2\big]}\, n^{\eta}\sum_{i\in\mathcal I_b}\sqrt{\E[(b_i-\wh b_i)^4]}\nn\\
&\le \frac{C^2r^2}{n^2T^4}\l(\frac 1{12} T(T+1)(T+2)(3T+1)\r) q^2 K_1^2 \sqrt {M_1} T n^{2\eta} \frac {\sqrt{C_1}} T = O\l(\frac 1{n^{2(1-\eta)}}\r),\nn
\end{align}
where we Assumption \ref{asm:factor}(b) of uniformly bounded loadings, Cauchy-Schwarz inequality, Assumption \ref{ASS:trend} (a) which bounds the number of deterministic linear trends, \eqref{F4u4}, and \eqref{trendb4}. This proves part (i).\smallskip

For part (ii) we have
\begin{align}
\E&\bigg[\bigg\Vert\frac 1{\sqrt nT^2}\sum_{t=1}^T \mbf F_t(\wh{\mbf x}_t-\mbf x_t)'\bigg\Vert^2\bigg]= \frac1{nT^4}\sum_{j=1}^r \sum_{i=1}^n\E\bigg[\bigg(\sum_{t=1}^T F_{jt}(b_i-\wh b_i)t\bigg)^2\bigg]\nn\\
&\le \frac{1}{nT^4}\sum_{t,s=1}^T ts \sum_{j=1}^r \sum_{i\in\mathcal I_b}\Big\vert \E\big[F_{jt}F_{js} (b_i-\wh b_i)^2\big]\Big\vert\le \frac{1}{nT^4}\sum_{t,s=1}^T ts \sum_{j=1}^r\sqrt{\E\big[F_{jt}^2F_{js}^2\big]}\sum_{i\in\mathcal I_b}\sqrt{\E[(b_i-\wh b_i)^4]}\nn\\
&\le \frac{r}{nT^4}\l(\frac 1{12} T(T+1)(T+2)(3T+1)\r) q^2 K_1^2 \sqrt {M_1} T n^{\eta} \frac {\sqrt{C_1}} T = O\l(\frac 1{n^{1-\eta}}\r),\nn
\end{align}
using the same arguments used for proving part (i). This proves part (ii).\smallskip

Turning to part (iii):
\begin{align}
\E&\bigg[\bigg\Vert\frac 1{n^2T^2}\sum_{t=1}^T \bm\Lambda'(\wh{\mbf x}_t-\mbf x_t)(\wh{\mbf x}_t-\mbf x_t)'\bm\Lambda\bigg\Vert^2\bigg]=\frac 1{n^4T^4}\!\!\!\sum_{j_1,j_2=1}^r\!\!\! \E\bigg[\bigg(\sum_{t=1}^T\bigg(\sum_{i_1=1}^n\lambda_{i_1j_1}(b_{i_1}-\wh b_{i_1})t\bigg)\bigg(\sum_{i_2=1}^n\lambda_{i_2j_2}(b_{i_2}-\wh b_{i_2})t\bigg)\bigg)^2\bigg]\nn\\
&\le \frac{C^4r^2}{n^4T^4}\sum_{t,s=1}^T t^2s^2n^{2\eta}\sum_{i,j\in\mathcal I_b}\E[(b_{i}-\wh b_{i})^2(b_{j}-\wh b_{j})^2]\le \frac{C^4r^2}{n^4T^4}\sum_{t,s=1}^T t^2s^2 n^{3\eta}\sum_{i\in\mathcal I_b}\E[(b_{i}-\wh b_{i})^4]\nn\\
&\le \frac{C^4r^2}{n^4T^4}\bigg(\frac 1{30} T(T+1)(T+2)(2T+1)(2T+3)(5T-1)\bigg) n^{4\eta}\frac{C_1}{T^2}=O\l(\frac 1{n^{4(1-\eta)}}\r),\nn
\end{align}
using the same arguments used for proving part (i). This proves part (iii).\smallskip

For part (iv) we have:
\begin{align}
\E&\bigg[\bigg\Vert\frac 1{nT^2}\sum_{t=1}^T (\wh{\mbf x}_t-\mbf x_t)(\wh{\mbf x}_t-\mbf x_t)'\bigg\Vert^2\bigg]
=\frac 1{n^2T^4}\sum_{i,j=1}^n\E\bigg[\bigg(\sum_{t=1}^T(b_{i}-\wh b_{i})(b_{j}-\wh b_{j})t^2\bigg)\bigg)^2\bigg]\nn\\
&\le \frac 1{n^2T^4}\sum_{t,s=1}^T t^2 s^2 \sum_{i,j=1}^n\E[(b_{i}-\wh b_{i})^2(b_{j}-\wh b_{j})^2]\le \frac 1{n^2T^4}\sum_{t,s=1}^T t^2 s^2 n^{\eta}\sum_{i=1}^n\E[(b_{i}-\wh b_{i})^4]\nn\\
&\le \frac{1}{n^2T^4}\bigg(\frac 1{30} T(T+1)(T+2)(2T+1)(2T+3)(5T-1)\bigg) n^{2\eta}\frac{C_1}{T^2}=O\l(\frac 1{n^{2(1-\eta)}}\r),\nn
\end{align}
using the same arguments used for proving part (i). This proves part (iv). Parts (v) and (vi) follow from parts (i) and (ii) respectively. Part (vii) follows from part (iii), while part (viii) follows from part (iv). This completes the proof. \hfill $\Box$\\
\begin{lem}\label{lem:trendxi}
Under Assumptions \ref{ASS:common} through \ref{asm:rates}, as $n,T\to\infty$, \smallskip
\begin{compactenum}
\item [(i)] $\E[\Vert(n^2T^2)^{-1}\sum_{t=1}^T \bm\Lambda'\bm\xi_t(\wh{\mbf x}_t-\mbf x_t)'\bm\Lambda\Vert^2]=O(n^{-2(2-\delta-\eta}))$;
\item [(ii)] $\E[\Vert(nT^2)^{-1}\sum_{t=1}^T \bm\xi_t(\wh{\mbf x}_t-\mbf x_t)'\Vert^2]=O(n^{-(2-\delta-\eta)})$.
\end{compactenum}
 \end{lem}

\noindent{\textbf{Proof.}} First, note that for all $i,j\in\mathcal I_i$ and all $t,s=1,\ldots, T$ we have
\begin{align}
\E&\big[\xi_{it}^2\xi_{js}^2\big]\le K_2^4\E\bigg[\bigg(\sum_{t'=1}^t \eps_{it'}\bigg)^2\bigg(\sum_{s'=1}^s \eps_{js'}\bigg)^2\bigg]\le  K_2^4 \sum_{t,t'=1}^T \sum_{s,s'=1}^T \Big\vert\E\big[\eps_{it} \eps_{it'} \eps_{js} \eps_{js'} \big] \Big\vert\nn\\
&\le K_2^4\bigg\{\sum_{t=1}^T \E[\eps_{it}^2\eps_{jt}^2] + \sum_{t,s=1}^T \E[\eps_{it}^2\eps_{js}^2]\bigg\}\le  K_2^4 M_2 T^2,\label{xi4e4}
\end{align}
where we used square summability of the coefficients, with $K_2$ defined in \eqref{eq:sqsumMM2}, and Assumption \ref{ASS:idio}(a) of independence of the idiosyncratic shocks and finite fourth moments.\smallskip

Then, consider part (i):
\begin{align}
\E&\bigg[\bigg\Vert\frac 1{n^2T^2}\sum_{t=1}^T \bm\Lambda'\bm\xi_t(\wh{\mbf x}_t-\mbf x_t)'\bm\Lambda\bigg\Vert^2\bigg]=
\frac1{n^4T^4}\sum_{j_1,j_2=1}^r \E\bigg[\bigg(\sum_{t=1}^T \bigg(\sum_{i_1=1}^n\lambda_{i_1 j_1}\xi_{i_1t}\bigg)\bigg(\sum_{i_2=1}^n\lambda_{i_2j_2} ( b_{i_2}-\wh b_{i_2})t\bigg)\bigg)^2\bigg]\nn\\
&\le \frac{C^4r^2}{n^4T^4}\sum_{t,s=1}^T ts\sum_{i_1,i_1'=1}^n\sum_{i_2,i_2'\in\mathcal I_b} \Big\vert\E\big[\xi_{i_1t}\xi_{i_1's} ( b_{i_2}-\wh b_{i_2})( b_{i_2'}-\wh b_{i_2'})\big]\Big\vert\nn\\
&\le \frac{C^4r^2}{n^4T^4}\sum_{t,s=1}^T ts \bigg\{\sum_{i_1,i_1'\in\mathcal I_1^c}\sqrt{\E\big[\xi_{i_1t}^2\xi_{i_1's}^2\big]}+3\sum_{i_1,i_1'\in\mathcal I_1}\sqrt{\E\big[\xi_{i_1t}^2\xi_{i_1's}^2\big]}\bigg\}
n^{\eta}\sum_{i_2\in\mathcal I_b}\sqrt{\E[(b_{i_2}-\wh b_{i_2})^4]}
\nn\\
&\le \frac{C^4r^2}{n^4T^4}\l(\frac 1{12} T(T+1)(T+2)(3T+1)\r)K_2^2 \sqrt{M_2}(n^2+n^{2\delta}  T) n^{2\eta}\frac{\sqrt{C_1}}{T}
=O\l(\frac 1{n^{2(1-\eta)}T}\r)+ O\l(\frac 1{n^{2(2-\delta-\eta)}}\r),\nn
\end{align}
where we Assumption \ref{asm:factor}(b) of uniformly bounded loadings, Cauchy-Schwarz inequality, Assumption \ref{ASS:trend} (a) which bounds the number of deterministic linear trends, Assumption \ref{asm:rates} which bounds the number of $I(1)$ idiosyncratic components, \eqref{xi4e4}, and \eqref{trendb4} in the proof of Lemma \ref{lem:trend}. This proves part (i).\smallskip

For part (ii) we have:
\begin{align}
\E&\bigg[\bigg\Vert\frac 1{nT^2}\sum_{t=1}^T \bm\xi_t(\wh{\mbf x}_t-\mbf x_t)'\bigg\Vert^2\bigg]=
\frac1{n^2T^4}\sum_{i,j=1}^n \E\bigg[\bigg(\sum_{t=1}^T \xi_{it}( b_{j}-\wh b_{j})t\bigg)^2\bigg]\nn\\
&\le \frac 1{n^2T^4}\sum_{t,s=1}^T ts \sum_{i,j=1}^n \Big\vert\E\big[\xi_{it}\xi_{is}( b_{j}-\wh b_{j})^2\big]\Big\vert\nn\\
&\le \frac 1{n^2T^4}\sum_{t,s=1}^T ts \bigg\{\sum_{i\in\mathcal I_1^c}\sqrt{\E\big[\xi_{it}^2\xi_{is}^2\big]}+3\sum_{i\in\mathcal I_1}\sqrt{\E\big[\xi_{it}^2\xi_{is}^2\big]}\bigg\}\sum_{j\in\mathcal I_b}\sqrt{\E[(b_{j}-\wh b_{j})^4]}\nn\\
&\le \frac 1{n^2T^4}\l(\frac 1{12} T(T+1)(T+2)(3T+1)\r)K_2^2 \sqrt{M_2}(n+n^{\delta}  T) n^{\eta}\frac{\sqrt{C_1}}{T}=O\l(\frac 1{n^{(1-\eta)}T}\r)+ O\l(\frac 1{n^{2-\delta-\eta}}\r),\nn
\end{align}
using the same arguments used for proving part (i). This proves part (ii). \hfill$\Box$\\
\begin{lem}\label{lem:mhat} Define the matrices 
\bea
&&\wh{\mbf M}_{00}=\frac 1 T\sum_{t=1}^T\Delta\wh{\mbf F}_t\Delta\wh{\mbf F}_t',\quad\wh{\mbf M}_{01}=\frac 1 T\sum_{t=1}^T\Delta\wh{\mbf F}_t\wh{\mbf F}_{t-1}',\quad\wh{\mbf M}_{02}=\frac 1 T\sum_{t=1}^T\Delta\wh{\mbf F}_t\Delta\wh{\mbf F}_{t-1}',\nn\\
&&\wh{\mbf M}_{11}=\frac 1 T\sum_{t=1}^T\wh{\mbf F}_t\wh{\mbf F}_t',\quad \wh{\mbf M}_{21}=\frac 1 T\sum_{t=1}^T\Delta\wh{\mbf F}_{t-1}'\wh{\mbf F}_{t-1},\quad \wh{\mbf M}_{22}=\frac 1 T\sum_{t=1}^T\Delta\wh{\mbf F}_{t-1}\Delta\wh{\mbf F}_{t-1}',\nn
\eea
and denote by $\mbf M_{ij}$, for $i,j=0,1,2$, the analogous ones but computed by using $\wt{\mbf F}_t=\mbf J\mbf F_t$. Define also $\wt{\bm\beta}=\mbf J\bm\beta$. Under Assumptions \ref{ASS:common} through \ref{ASS:trend}, as $n,T\to\infty$,  \smallskip
\begin{compactenum}[(i)]
\item $\Vert T^{-1}\wh{\mbf M}_{11}-T^{-1}\mbf M_{11}\Vert =O_p(\max(n^{-1/2},T^{-1/2}, n^{-(1-\eta)}))$; 
\item $\Vert \wh{\mbf M}_{00}-\mbf M_{00}\Vert =O_p(\max(n^{-1/2},T^{-1/2}, n^{-(1-\eta)}))$;
\item $\Vert \wh{\mbf M}_{02}-\mbf M_{02}\Vert =O_p(\max(n^{-1/2},T^{-1/2}, n^{-(1-\eta)}))$;
\item $\Vert \wh{\mbf M}_{22}-\mbf M_{22}\Vert =O_p(\max(n^{-1/2},T^{-1/2}, n^{-(1-\eta)}))$.
\end{compactenum}
If also Assumption \ref{asm:rates} holds, then, \smallskip
\begin{compactenum}
\item [(v)] $\Vert \wh{\mbf M}_{01}\wt{\bm\beta}-\mbf M_{01}\wt{\bm\beta}\Vert =O_p(\vartheta_{nT,\delta,\eta})$; 
\item [(vi)] $\Vert \wh{\mbf M}_{21}\wt{\bm\beta}-\mbf M_{21}\wt{\bm\beta}\Vert =O_p(\vartheta_{nT,\delta,\eta})$;  
\item [(vii)] $\Vert T^{-1/2}\wh{\mbf M}_{01}-T^{-1/2}\mbf M_{01}\Vert =O_p(\vartheta_{nT,\delta,\eta})$;  
\item [(viii)] $\Vert T^{-1/2}\wh{\mbf M}_{21}-T^{-1/2}\mbf M_{21}\Vert =O_p(\vartheta_{nT,\delta,\eta})$; 
\item [(ix)] $\Vert \wt{\bm\beta}'\wh{\mbf M}_{11}\wt{\bm\beta}-\wt{\bm\beta}'\mbf M_{11}\wt{\bm\beta}\Vert =O_p(\vartheta_{nT,\delta,\eta})$.
\end{compactenum}
\end{lem}

\noindent{\textbf{Proof.}} Throughout, we use $\Vert \bm\beta\Vert =O(1)$ and obviously $\Vert \mbf J\Vert =1$ and  the fact that, since $\sqrt T/n\to 0$, as $n,T\to\infty$ we have (see also \eqref{eq:what2} and \eqref{eq:whatlemma3} in the proof of Lemma \ref{lem:load})
\beq\label{UNNAMED_L}
\bigg\Vert \frac{\widehat{\bm\Lambda} - \bm\Lambda\mbf J}{\sqrt n}\bigg\Vert =O_p\l(\frac 1 {\sqrt T}\r)\; \mbox{ and }\;
\bigg\Vert \frac{\widehat{\bm\Lambda}'\bm\Lambda}n - \mbf J\bigg\Vert=O_p\l(\frac 1 {\sqrt T}\r).
\eeq
and therefore $\Vert n^{-1}\wh{\bm\Lambda}'\bm\Lambda\Vert =O_p(1)$. \smallskip

Start with part (i). By adding and subtracting $\mbf J\mbf F_t$ from $\wh{\mbf F}_t$, we have
\begin{align}
\bigg\Vert \frac 1 {T^2} \sum_{t=1}^T \wh{\mbf F}_t\wh{\mbf F}_t'-\frac 1{T^2}\sum_{t=1}^T\wt{\mbf F}_t\wt{\mbf F}_t' \bigg\Vert  &\leq 
2\bigg\Vert \frac 1 {T^2}\sum_{t=1}^T\Big(\wh{\mbf F}_t-\mbf J\mbf F_t\Big)\Big(\mbf J\mbf F_t\Big)' \bigg\Vert
\nn\\
&+\bigg\Vert \frac 1 {T^2}\sum_{t=1}^T \Big(\wh{\mbf F}_t-\mbf J\mbf F_t\Big) \Big(\wh{\mbf F}_t-\mbf J\mbf F_t\Big)'\bigg\Vert.\label{eq:mhata}
\end{align}
Using \eqref{macron} and \eqref{parlacondudu}, the first term on the rhs of \eqref{eq:mhata} is such that
\begin{align}
\bigg\Vert \frac 1 {T^2}&\sum_{t=1}^T \Big(\wh{\mbf F}_t-\mbf J\mbf F_t\Big) \Big(\mbf J\mbf F_t\Big)'\bigg\Vert  =\bigg\Vert \frac 1 {T^2}\sum_{t=1}^T \Big(\frac{\wh{\bm\Lambda}'\wh{\mbf x}_t }{n}-\mbf J\mbf F_t\Big) \Big(\mbf J\mbf F_t\Big)'\bigg\Vert \nn\\
&=\bigg\Vert\frac 1 {T^2}\sum_{t=1}^T\bigg(\frac{\wh{\bm\Lambda}'\bm\Lambda\mbf F_t}{n}-\mbf J\mbf F_t+\frac{\wh{\bm\Lambda}'\bm\xi_t}{n}+\frac{\wh{\bm\Lambda}'(\wh{\mbf x}_t-\mbf x_t)}{n}\bigg)\Big(\mbf J\mbf F_t\Big)'\bigg\Vert\label{eq:mhata3}\\
&\le \underbrace{\bigg\Vert\frac 1 {T^2}\sum_{t=1}^T\bigg(\frac{\wh{\bm\Lambda}'\bm\Lambda\mbf F_t}{n}-\mbf J\mbf F_t\bigg)
\Big(\mbf J\mbf F_t\Big)'\bigg\Vert}_{\mathcal A_1}
+\underbrace{\bigg\Vert\frac 1 {T^2}\sum_{t=1}^T\frac{\wh{\bm\Lambda}'\bm\xi_t\mbf F_t'\mbf J}{n}\bigg\Vert}_{\mathcal B_1}
+\underbrace{\bigg\Vert\frac 1 {T^2}\sum_{t=1}^T\frac{\wh{\bm\Lambda}'(\wh{\mbf x}_t-\mbf x_t)\mbf F_t'\mbf J}{n}\bigg\Vert}_{\mathcal C_1}\nn
\end{align}
Now, consider each of the three terms in \eqref{eq:mhata3} separately:
\begin{align}
\mathcal A_1 \le \bigg\Vert\frac{\wh{\bm\Lambda}'\bm\Lambda}{n}-\mbf J\bigg\Vert \; \bigg\Vert\frac 1 {T^2}\sum_{t=1}^T \mbf F_t\mbf F_t' \bigg\Vert =O_p\l(\frac 1 {\sqrt T}\r),\nn
\end{align}
because of \eqref{UNNAMED_L} and Lemma \ref{rem:DFF}(ii). Then, considering the worst case, i.e. $\delta=1$, we have
\begin{align}
\mathcal B_1 \le \bigg\Vert\frac{\wh{\bm\Lambda}-\bm\Lambda\mbf J}{\sqrt n}\bigg\Vert\;  
\bigg\Vert\frac 1{T^2}\sum_{t=1}^T\frac{\bm\xi_t\mbf F_t'}{\sqrt n}\bigg\Vert
+  \bigg\Vert\frac 1{T^2}\sum_{t=1}^T\frac{\bm\Lambda'\bm\xi_t\mbf F_t'}{n}\bigg\Vert=O_p\l(\max\l(\frac 1{\sqrt T},\frac 1{\sqrt n}\r)\r),\nn
\end{align}
because of \eqref{UNNAMED_L} and Lemma \ref{lem:fidio}(i) and \ref{lem:fidio}(ii). Last,
\begin{align}
\mathcal C_1  \le \bigg\Vert\frac{\wh{\bm\Lambda}-\bm\Lambda\mbf J}{\sqrt n}\bigg\Vert\;  \bigg\Vert\frac 1{T^2}\sum_{t=1}^T\frac{(\wh{\mbf x}_t-\mbf x_t)\mbf F_t'}{\sqrt n}\bigg\Vert
+\bigg\Vert\frac 1{T^2}\sum_{t=1}^T\frac{\bm\Lambda'(\wh{\mbf x}_t-\mbf x_t)\mbf F_t'}{n}\bigg\Vert=O_p\l(\frac 1{n^{(1-\eta)/2}\sqrt T}\r)+O_p\l(\frac 1{n^{1-\eta}}\r),\nn
\end{align}
because of \eqref{UNNAMED_L} and Lemma \ref{lem:trend}(i) and  \ref{lem:trend}(ii). 

Consider the second term on the rhs of \eqref{eq:mhata3}
\begin{align}
\bigg\Vert \frac 1 {T^2}&\sum_{t=1}^T \Big(\wh{\mbf F}_t-\mbf J\mbf F_t\Big) \Big(\wh{\mbf F}_t-\mbf J\mbf F_t\Big)'\bigg\Vert = \bigg\Vert \frac 1 {T^2} \sum_{t=1}^T \bigg(\frac{\wh{\bm\Lambda}'\wh{\mbf x}_t}{n}-\mbf J\mbf F_t\bigg) \bigg(\frac{\wh{\bm\Lambda}'\wh{\mbf x}_t}{n}-\mbf J\mbf F_t\bigg)'\bigg\Vert \nn\\
=&\bigg\Vert \frac 1 {T^2} \sum_{t=1}^T \bigg(\frac{\wh{\bm\Lambda}'\bm\Lambda\mbf F_t}{n}-\mbf J\mbf F_t+\frac{\wh{\bm\Lambda}'\bm\xi_t}{n}+\frac{\wh{\bm\Lambda}'(\wh{\mbf x}_t-\mbf x_t)}{n}\bigg) \bigg(\frac{\wh{\bm\Lambda}'\bm\Lambda\mbf F_t}{n}-\mbf J\mbf F_t+\frac{\wh{\bm\Lambda}'\bm\xi_t}{n}+\frac{\wh{\bm\Lambda}'(\wh{\mbf x}_t-\mbf x_t)}{n}\bigg)'\bigg\Vert \nn\\
\leq& 
\underbrace{\bigg\Vert \frac1 {T^2}\sum_{t=1}^T\frac{\wh{\bm\Lambda}'\bm\Lambda\mbf F_t\mbf F_t'}{n}\bigg(\frac{\bm\Lambda'\wh{\bm\Lambda}}{n}-\mbf J\bigg)+\mbf J\mbf F_t\mbf F_t'\bigg(\mbf J-\frac{\bm\Lambda'\wh{\bm\Lambda}}{n}\bigg)\bigg\Vert }_{\mathcal D_1} +\underbrace{2\bigg\Vert \frac 1{T^2}\sum_{t=1}^T\frac{\wh{\bm\Lambda}'\bm\Lambda\mbf F_t\bm\xi_t'\wh{\bm\Lambda}}{n^2}\bigg\Vert }_{\mathcal E_1}\nn\\
&+\underbrace{2\bigg\Vert \frac 1{T^2}\sum_{t=1}^T\frac{\wh{\bm\Lambda}'\bm\xi_t\mbf F_t'\mbf J}{n}\bigg\Vert }_{\mathcal F_1} + 
\underbrace{\bigg\Vert \frac 1 {T^2}\sum_{t=1}^T\frac{\wh{\bm\Lambda}'\bm\xi_t\bm\xi_t'\wh{\bm\Lambda}}{n^2}\bigg\Vert }_{\mathcal G_1} 
+\underbrace{2\bigg\Vert \frac 1 {T^2}\sum_{t=1}^T\bigg(\frac{\wh{\bm\Lambda}'\bm\Lambda}{n}-\mbf J\bigg)\frac{\mbf F_t(\wh{\mbf x}_t-\mbf x_t)'\wh{\bm\Lambda}}{n}
\bigg\Vert}_{\mathcal H_1}\nn\\
&+\underbrace{2\bigg\Vert \frac 1 {T^2}\sum_{t=1}^T\frac{\wh{\bm\Lambda}'\bm\xi_t(\wh{\mbf x}_t-\mbf x_t)'\wh{\bm\Lambda}}{n^2}
\bigg\Vert}_{\mathcal J_1}
+\underbrace{\bigg\Vert \frac 1 {T^2}\sum_{t=1}^T\frac{\wh{\bm\Lambda}'(\wh{\mbf x}_t-\mbf x_t)(\wh{\mbf x}_t-\mbf x_t)'\wh{\bm\Lambda}}{n^2}\bigg\Vert}_{\mathcal K_1}
.\label{eq:mhata2}
\end{align}
Now, consider each of the terms in \eqref{eq:mhata2} separately. Term $\mathcal D_1$ behaves like $\mathcal A_1$, $\mathcal E_1$ and $\mathcal F_1$ behave like $\mathcal B_1$. Then term $\mathcal H_1$ is dominated by $\mathcal C_1$. Moreover, by Lemma \ref{lem:trendxi}(i) and \ref{lem:trendxi}(ii) term $\mathcal J_1$ is dominated by $\mathcal H_1$ and by Lemma \ref{lem:trend}(iii) and \ref{lem:trend}(iv) term $\mathcal K_1$ is also dominated by $\mathcal H_1$. We are left with $\mathcal G_1$, which, considering the worst case, i.e. $\delta=1$, is such that
\begin{align}
\mathcal G_1\le&\bigg\Vert\frac{\wh{\bm\Lambda}-\bm\Lambda\mbf J}{\sqrt n}\bigg\Vert^2 \;\bigg\Vert\frac 1 {T^2}\sum_{t=1}^T\frac{\bm\xi_t\bm\xi_t'}{n} \bigg\Vert + \bigg\Vert\frac 1 {T^2}\sum_{t=1}^T\frac{\bm\Lambda'\bm\xi_t\bm\xi_t'\bm\Lambda}{n^2} \bigg\Vert\nn\\ 
&+ 2\bigg\Vert\frac{\wh{\bm\Lambda}-\bm\Lambda\mbf J}{\sqrt n}\bigg\Vert \;\bigg\Vert\frac 1 {T^2}\sum_{t=1}^T\frac{\bm\xi_t\bm\xi_t'}{n}\bigg\Vert\;\bigg\Vert\frac{\bm\Lambda}{\sqrt n}\bigg\Vert=O_p\l(\max\l(\frac 1 {\sqrt T},\frac 1n\r)\r),\nn
\end{align}
because of \eqref{UNNAMED_L} and Lemma \ref{lem:fidio}(iii) and \ref{lem:fidio}(iv). By substituting \eqref{eq:mhata3} and \eqref{eq:mhata2} into \eqref{eq:mhata}, we prove part (i). Part (ii), (iii), (iv) are proved analogously by noting that since in these cases we deal with differenced data the terms due to the de-trending are all $O_p(T^{-1/2})$ (this can be proved by simple modifications in the proof of Lemma \ref{lem:trend}).\smallskip

Now, consider part (v):
\begin{align}
\bigg\Vert \frac 1 {T} &\sum_{t=1}^T \Delta\wh{\mbf F}_t\wh{\mbf F}_{t-1}'\wt{\bm\beta}-\frac 1{T}\sum_{t=1}^T\Delta\wt{\mbf F}_t\wt{\mbf F}_{t-1}'\wt{\bm\beta}\bigg\Vert  
\leq 
\bigg\Vert \frac 1 {T}\sum_{t=1}^T\Big(\mbf J\Delta\mbf F_t\Big)\Big(\wh{\mbf F}_{t-1}-\mbf J\mbf F_{t-1}\Big)'\wt{\bm\beta}\bigg\Vert \nn\\
&+\bigg\Vert \frac 1 {T}\sum_{t=1}^T\Big(\Delta\wh{\mbf F}_t-\mbf J\Delta\mbf F_t\Big)\Big(\check{\bm\beta}'\mbf J\mbf F_{t-1}\Big)' \bigg\Vert 
+\bigg\Vert \frac 1 {T}\sum_{t=1}^T \Big(\Delta\wh{\mbf F}_t-\mbf J\Delta\mbf F_t\Big) \Big(\wh{\mbf F}_{t-1}-\mbf J\mbf F_{t-1}\Big)'\wt{\bm\beta}\bigg\Vert .\label{eq:mhatb}
\end{align}
Consider the first term on the rhs of \eqref{eq:mhatb}
\begin{align}
\bigg\Vert \frac 1 {T}&\sum_{t=1}^T\Big(\mbf J\Delta\mbf F_t\Big)\Big(\wh{\mbf F}_{t-1}-\mbf J\mbf F_{t-1}\Big)'\wt{\bm\beta}\bigg\Vert=
\bigg\Vert \frac 1 {T}\sum_{t=1}^T\Big(\mbf J\Delta\mbf F_t\Big)\bigg(\frac{\wh{\bm\Lambda}'\wh{\mbf x}_{t-1}}{n}-\mbf J\mbf F_{t-1}\bigg)'\wt{\bm\beta}\bigg\Vert\nn\\
=&\bigg\Vert \frac 1 {T}\sum_{t=1}^T\Big(\mbf J\Delta\mbf F_t\Big)\bigg(\frac{\wh{\bm\Lambda}'\bm\Lambda\mbf F_{t-1}}{n}-\mbf J\mbf F_{t-1}+\frac{\wh{\bm\Lambda}'\bm\xi_{t-1}}{n}+\frac{\wh{\bm\Lambda}'(\wh{\mbf x}_{t-1}-\mbf x_{t-1})}{n}\bigg)'\wt{\bm\beta}\bigg\Vert\label{eq:mhatb3}\\
=&\underbrace{
\bigg\Vert \frac 1 {T}\sum_{t=1}^T\Big(\mbf J\Delta\mbf F_t\Big)\bigg(\frac{\wh{\bm\Lambda}'\bm\Lambda\mbf F_{t-1}}{n}-\mbf J\mbf F_{t-1}\bigg)'\wt{\bm\beta}\bigg\Vert}_{\mathcal A_2}
+\underbrace{\bigg\Vert \frac 1 {T}\sum_{t=1}^T\frac{\mbf J\Delta\mbf F_t\bm\xi_{t-1}'\wh{\bm\Lambda}\wt{\bm\beta}}n\bigg\Vert
}_{\mathcal B_2}\nn\\
&+\,\underbrace{\bigg\Vert \frac 1 {T}\sum_{t=1}^T\frac{\mbf J\Delta\mbf F_t(\wh{\mbf x}_{t-1}-\mbf x_{t-1})'\wh{\bm\Lambda}\wt{\bm\beta}}n\bigg\Vert
}_{\mathcal C_2}.\nn
\end{align}
Now, consider each of the three terms in \eqref{eq:mhatb3} separately:
\begin{align}
\mathcal A_2 \le \bigg\Vert\frac{\wh{\bm\Lambda}'\bm\Lambda}{n}-\mbf J\bigg\Vert \; \bigg\Vert\frac 1 {T}\sum_{t=1}^T \Delta\mbf F_t\mbf F_{t-1}' \bigg\Vert \;\big\Vert \wt{\bm\beta}\big\Vert=O_p\l(\frac 1 {\sqrt T}\r),\nn
\end{align}
because of \eqref{UNNAMED_L} and Lemma \ref{rem:DFF}(iii). Then,
\begin{align}
\mathcal B_2 \le \bigg\Vert\frac{\wh{\bm\Lambda}-\bm\Lambda\mbf J}{\sqrt n}\bigg\Vert\;  
\bigg\Vert\frac 1{T}\sum_{t=1}^T\frac{\Delta\mbf F_t\bm\xi_{t-1}'}{\sqrt n}\bigg\Vert
+  \bigg\Vert\frac 1{T}\sum_{t=1}^T\frac{\Delta\mbf F_t\bm\Lambda'\bm\xi_{t-1}}{n}\bigg\Vert
\;\big\Vert \wt{\bm\beta}\big\Vert
=O_p\l(\max\l(\frac 1{n^{(1-\delta)/2}},\frac {\sqrt T}{n^{(2-\delta)/2}}\r)\r),\nn
\end{align}
because of \eqref{UNNAMED_L} and Lemma \ref{lem:fidio}(v) and \ref{lem:fidio}(vi). Last,
\begin{align}
\mathcal C_2  \le \bigg\Vert\frac{\wh{\bm\Lambda}-\bm\Lambda\mbf J}{\sqrt n}\bigg\Vert\;  \bigg\Vert\frac 1{T}\sum_{t=1}^T\frac{\Delta \mbf F_t(\wh{\mbf x}_{t-1}-\mbf x_{t-1})'}{\sqrt n}\bigg\Vert
+\bigg\Vert\frac 1{T}\sum_{t=1}^T\frac{\Delta\mbf F_t(\wh{\mbf x}_{t-1}-\mbf x_{t-1})'\bm\Lambda}{n}\bigg\Vert=O_p\l(\frac {1}{n^{(1-\eta)/2}},\frac{\sqrt T}{n^{1-\eta}}\r),\nn
\end{align}
because of \eqref{UNNAMED_L} and Lemma \ref{lem:trend}(v) and  \ref{lem:trend}(vi). The second term on the rhs of \eqref{eq:mhatb} contains only stationary terms, thus is dominated by the first one.

Then, consider the third term on the rhs of \eqref{eq:mhatb}
\begin{align}
\bigg\Vert \frac 1 {T}&\sum_{t=1}^T \Big(\Delta\wh{\mbf F}_t-\mbf J\Delta\mbf F_{t}\Big) \Big(\wt{\bm\beta}'\wh{\mbf F}_{t-1}-\wt{\bm\beta}'\mbf J\mbf F_{t-1}\Big)'\bigg\Vert = \bigg\Vert \frac 1 {T} \sum_{t=1}^T \bigg(\frac{\wh{\bm\Lambda}'\Delta\wh{\mbf x}_t}{n}-\mbf J\Delta\mbf F_t\bigg) \bigg(\frac{\wh{\bm\Lambda}'\wh{\mbf x}_{t-1}}{n}-\mbf J \mbf F_{t-1}\bigg)'\wt{\bm\beta}\bigg\Vert \nn\\
\leq& \underbrace{\bigg\Vert \frac1 {T}\sum_{t=1}^T\frac{\wh{\bm\Lambda}'\bm\Lambda\Delta\mbf F_t\mbf F_{t-1}'}{n}\bigg(\frac{\bm\Lambda'\wh{\bm\Lambda}}{n}-\mbf J\bigg)\wt{\bm\beta}+\mbf J\Delta\mbf F_t\mbf F_{t-1}'\bigg(\mbf J-\frac{\bm\Lambda'\wh{\bm\Lambda}}{n}\bigg)\wt{\bm\beta}\bigg\Vert }_{\mathcal D_2} +
\underbrace{\bigg\Vert \frac 1{T}\sum_{t=1}^T\frac{\wh{\bm\Lambda}'\bm\Lambda\Delta\mbf F_t\bm\xi_{t-1}'\wh{\bm\Lambda}\wt{\bm\beta}}{n^2}\bigg\Vert }_{\mathcal E_2}\nn\\
&+\underbrace{\bigg\Vert \frac 1 T \sum_{t=1}^T \frac{\wh{\bm\Lambda}'\Delta\bm\xi_t\mbf F_{t-1}'\bm\Lambda'\wh{\bm\Lambda}\wt{\bm\beta}}{n^2}\bigg\Vert }_{\mathcal F_2}
+\underbrace{\bigg\Vert \frac 1 T \sum_{t=1}^T \frac{\mbf J\Delta\mbf F_t\bm\xi_{t-1}'\wh{\bm\Lambda}\wt{\bm\beta}}{n}\bigg\Vert }_{\mathcal G_2}
+\underbrace{\bigg\Vert \frac 1 T \sum_{t=1}^T \frac{\wh{\bm\Lambda}'\Delta\bm\xi_t\mbf F_{t-1}'\mbf J\wt{\bm\beta}}{n}\bigg\Vert }_{\mathcal H_2}\nn\\
&+\underbrace{\bigg\Vert \frac 1 T \sum_{t=1}^T \frac{\wh{\bm\Lambda}'\Delta\bm\xi_t\bm\xi_{t-1}'\wh{\bm\Lambda}\wt{\bm\beta}}{n^2}\bigg\Vert }_{\mathcal J_2}
+\underbrace{\bigg\Vert \frac 1 {T}\sum_{t=1}^T\bigg(\frac{\wh{\bm\Lambda}'\bm\Lambda}{n}-\mbf J\bigg)\frac{\Delta\mbf F_t(\wh{\mbf x}_{t-1}-\mbf x_{t-1})'\wh{\bm\Lambda}\wt{\bm\beta}}{n}
\bigg\Vert}_{\mathcal K_2}\nn\\
&+\underbrace{\bigg\Vert \frac 1 {T}\sum_{t=1}^T\frac{\mbf F_t(\Delta\wh{\mbf x}_{t-1}-\Delta\mbf x_{t-1})'\wh{\bm\Lambda}\wt{\bm\beta}}{n}
\bigg(\frac{\wh{\bm\Lambda}'\bm\Lambda}{n}-\mbf J\bigg)'
\bigg\Vert}_{\mathcal H_2}
+\underbrace{\bigg\Vert \frac 1 {T}\sum_{t=1}^T\frac{\wh{\bm\Lambda}'\Delta\bm\xi_t(\wh{\mbf x}_{t-1}-\mbf x_{t-1})'\wh{\bm\Lambda}\wt{\bm\beta}}{n^2}
\bigg\Vert}_{\mathcal L_2}\nn\\
&+\underbrace{\bigg\Vert \frac 1 {T}\sum_{t=1}^T\frac{\wh{\bm\Lambda}'\bm\xi_t(\Delta\wh{\mbf x}_{t-1}-\Delta\mbf x_{t-1})'\wh{\bm\Lambda}\wt{\bm\beta}}{n^2}\bigg\Vert}_{\mathcal M_2}
+\underbrace{\bigg\Vert \frac 1 {T}\sum_{t=1}^T\frac{\wh{\bm\Lambda}'(\Delta\wh{\mbf x}_{t}-\Delta\mbf x_{t})(\wh{\mbf x}_{t-1}-\mbf x_{t-1})'\wh{\bm\Lambda}\wt{\bm\beta}}{n^2}\bigg\Vert}_{\mathcal N_2}.\label{eq:mhatb2}
\end{align}
Term $\mathcal D_2$ behaves like term $\mathcal A_2$, $\mathcal E_2$ and $\mathcal G_2$ behave like term $\mathcal B_2$, then  since $\wt{\bm\beta}\mbf J\mbf F_t'=\bm\beta'\mbf F_t$ and therefore it is stationary, and because of because of \eqref{UNNAMED_L}, $\mathcal F_2$ is $O_p(\max(T^{-1/2},n^{-1/2}))$ (this can be proved by simple modifications in the proof of Lemma \ref{lem:fidio}). Terms $\mathcal H_2$, $\mathcal K_2$, and $\mathcal N_2$ are dominated by $\mathcal C_2$. Terms $\mathcal L_2$ and $\mathcal M_2$ behave as $\mathcal C_2$. We are left with term $\mathcal J_2$, which is such that
\begin{align}
\mathcal J_2\le& \bigg\Vert\frac{\wh{\bm\Lambda}-\bm\Lambda\mbf J}{\sqrt n}\bigg\Vert^2 \;\bigg\Vert\frac 1 {T}\sum_{t=1}^T\frac{\Delta\bm\xi_t\bm\xi_{t-1}'}{n} \bigg\Vert \;\big\Vert\wt{\bm\beta}\big\Vert + 
\bigg\Vert\frac 1 {T}\sum_{t=1}^T\frac{\bm\Lambda'\Delta\bm\xi_t\bm\xi_{t-1}'{\bm\Lambda}}{n^{2}} \bigg\Vert\; \big\Vert\wt{\bm\beta}\big\Vert\nn\\
&+2\bigg\Vert\frac{\wh{\bm\Lambda}-\bm\Lambda\mbf J}{\sqrt n}\bigg\Vert\;
\bigg\Vert\frac 1 {T}\sum_{t=1}^T\frac{\Delta\bm\xi_t\bm\xi_{t-1}'}{n} \bigg\Vert\;\bigg\Vert\frac{\bm\Lambda}{\sqrt n}\bigg\Vert\; \big\Vert\wt{\bm\beta}\big\Vert=O_p\l( \frac {\sqrt T} {n^{2-\delta}}\r)+O_p\l(\frac 1{n^{1-\delta}}\r),\nn
\end{align}
because of Lemma \ref{lem:fidio}(vii) and \ref{lem:fidio}(viii). Therefore, $\mathcal J_2$ is dominated by $\mathcal B_2$. By substituting \eqref{eq:mhatb3} and \eqref{eq:mhatb2} we have that \eqref{eq:mhatb} is $O_p(\max(T^{1/2}n^{-(1-\delta/2)}, T^{1/2}n^{-(1-\eta)}, n^{(1-\delta)/2},n^{(1-\eta)/2}, T^{-1/2} ))$ and since $T^{1/2}n^{-(1-\delta/2)}<T^{1/2}n^{-(1-(\delta+\eta)/2)}$, then \eqref{eq:mhatb} is also $O_p(\vartheta_{nT,\delta,\eta})$. 
Parts (vi), (vii), and (viii) are proved in the same way.\smallskip

Last consider part (ix)
\begin{align}
\bigg\Vert \frac 1 {T} \sum_{t=1}^T \wt{\bm\beta}'\wh{\mbf F}_t\wh{\mbf F}_{t}'\wt{\bm\beta}-\frac 1{T}\sum_{t=1}^T\wt{\bm\beta}'\wt{\mbf F}_t\wt{\mbf F}_{t}'\wt{\bm\beta}\bigg\Vert  &\leq 2\bigg\Vert \frac 1 {T}\sum_{t=1}^T\wt{\bm\beta}'\Big(\wh{\mbf F}_t-\mbf J\mbf F_t\Big)\Big(\check{\bm\beta}'\mbf J\mbf F_t\Big)' \bigg\Vert\nn\\
& +\bigg\Vert \frac 1 {T}\sum_{t=1}^T\wt{\bm\beta}' \Big(\wh{\mbf F}_t-\mbf J\mbf F_t\Big) \Big(\wh{\mbf F}_{t}-\mbf J\mbf F_{t}\Big)'\wt{\bm\beta}\bigg\Vert.\label{eq:mhatc}
\end{align}
The first term on the rhs of \eqref{eq:mhatc} behaves exactly as the first term on the rhs of \eqref{eq:mhatb}, so we just have to consider the second term on the rhs of \eqref{eq:mhatc} 
\begin{align}
\bigg\Vert \frac 1 {T}&\sum_{t=1}^T\wt{\bm\beta}' \Big(\wh{\mbf F}_t-\mbf J\mbf F_t\Big) \Big(\wh{\mbf F}_{t}-\mbf J\mbf F_{t}\Big)'\wt{\bm\beta}\bigg\Vert 
= \bigg\Vert \frac 1 {T}\sum_{t=1}^T\wt{\bm\beta}' \bigg(\frac{\wh{\bm\Lambda}'\wh{\mbf x}_t}n-\mbf J\mbf F_t\bigg)\bigg(\frac{\wh{\bm\Lambda}'\wh{\mbf x}_t}n-\mbf J\mbf F_t\bigg)'\wt{\bm\beta}\bigg\Vert \nn\\
\leq &\underbrace{\bigg\Vert \frac 1 {T}\sum_{t=1}^T\frac{\wt{\bm\beta}'\wh{\bm\Lambda}'\bm\Lambda\mbf F_t\mbf F_t'}{n}\bigg(\frac{\bm\Lambda'\wh{\bm\Lambda}}{n}-\mbf J\bigg)\wt{\bm\beta}+\wt{\bm\beta}'\mbf J\mbf F_t\mbf F_t'\bigg(\mbf J-\frac{\bm\Lambda'\wh{\bm\Lambda}}{n}\bigg)\wt{\bm\beta}\bigg\Vert }_{\mathcal A_3} + \underbrace{2\bigg\Vert \frac 1 {T}\sum_{t=1}^T\frac{\wt{\bm\beta}'\wh{\bm\Lambda}'\bm\Lambda\mbf F_t\bm\xi_t'\wh{\bm\Lambda}\wt{\bm\beta}}{n^2}\bigg\Vert }_{\mathcal B_3}\nn\\
&+\underbrace{2\bigg\Vert \frac 1 {T}\sum_{t=1}^T\frac{\wt{\bm\beta}'\mbf J\mbf F_t\bm\xi_t'\wh{\bm\Lambda}\wt{\bm\beta}}{n}\bigg\Vert }_{\mathcal C_3}
+\underbrace{\bigg\Vert \frac 1 {T}\sum_{t=1}^T\frac{\wt{\bm\beta}'\wh{\bm\Lambda}'\bm\xi_t\bm\xi_t'\wh{\bm\Lambda}\wt{\bm\beta}}{n^2}\bigg\Vert }_{\mathcal D_3}
+\underbrace{2\bigg\Vert\frac 1 T\sum_{t=1}^T\wt{\bm\beta}'\bigg(\frac{\wh{\bm\Lambda}'\bm\Lambda}{n}-\mbf J\bigg)\frac{\mbf F_t(\wh{\mbf x}_t-\mbf x_t)'\wh{\bm\Lambda}\wt{\bm\beta}}{n}\bigg\Vert}_{\mathcal E_3}\nn\\
&+\underbrace{2\bigg\Vert\frac 1 T\sum_{t=1}^T\frac{\wt{\bm\beta}'\wh{\bm\Lambda}'\bm\xi_t(\wh{\mbf x}_t-\mbf x_t)'\wh{\bm\Lambda}\wt{\bm\beta}}{n^2}\bigg\Vert}_{\mathcal F_3}
+\underbrace{\bigg\Vert\frac 1 T\sum_{t=1}^T\frac{\wt{\bm\beta}'\wh{\bm\Lambda}'(\wh{\mbf x}_t-\mbf x_t)(\wh{\mbf x}_t-\mbf x_t)'\wh{\bm\Lambda}\wt{\bm\beta}}{n^2}\bigg\Vert}_{\mathcal G_3}.
\label{eq:mhatc2}
\end{align}
Now term $\mathcal A_3$ is $O_p(T^{-1/2})$, because of \eqref{UNNAMED_L} and Lemma \ref{rem:DFF}(v), terms $\mathcal B_3$ and $\mathcal C_3$ behave like term $\mathcal B_2$ in \eqref{eq:mhatb3}, while term $\mathcal E_3$ is dominated by $\mathcal C_2$ in \eqref{eq:mhatb3}. Then,
\begin{align}
\mathcal D_3\le& \bigg\Vert\frac{\wh{\bm\Lambda}-\bm\Lambda \mbf J}{\sqrt n}\bigg\Vert^2\; \bigg\Vert\frac 1 T\sum_{t=1}^T\frac{\bm\xi_t\bm\xi_t'}{n} \bigg\Vert\;\big\Vert\wt{\bm\beta}\big\Vert^2 + 
\bigg\Vert\frac 1 {T}\sum_{t=1}^T\frac{\bm\Lambda'\bm\xi_t\bm\xi_{t}'{\bm\Lambda}}{n^{2}} \bigg\Vert\; \big\Vert\wt{\bm\beta}\big\Vert^2\nn\\
&+2\bigg\Vert\frac{\wh{\bm\Lambda}-\bm\Lambda\mbf J}{\sqrt n}\bigg\Vert\;
\bigg\Vert\frac 1 {T}\sum_{t=1}^T\frac{\bm\xi_t\bm\xi_{t}'\bm\Lambda}{n^{3/2}} \bigg\Vert\; \big\Vert\wt{\bm\beta}\big\Vert^2
=O_p \l( \frac T {n^{2-\delta}}\r)+O_p \l( \frac {\sqrt T} {n^{(3-2\delta)/2}}\r)=O_p\l(\frac{\sqrt T}{n^{(2-\delta)/2}}\r),\nn
\end{align}
because of Lemma \ref{lem:fidio}(iii), \ref{lem:fidio}(iv), and \ref{lem:fidio}(ix) (multiplying the statements by $T^2$). 
Moreover,
\begin{align}
\mathcal F_3\le & \bigg\Vert\frac{\wh{\bm\Lambda}-\bm\Lambda \mbf J}{\sqrt n}\bigg\Vert^2\; \bigg\Vert\frac 1 T\sum_{t=1}^T\frac{\bm\xi_t(\wh{\mbf x}_t-\mbf x_t)'}{n} \bigg\Vert\;\big\Vert\wt{\bm\beta}\big\Vert^2 + \bigg\Vert\frac 1 {T}\sum_{t=1}^T\frac{\bm\Lambda'\bm\xi_t(\wh{\mbf x}_t-\mbf x_t)'{\bm\Lambda}}{n^{2}} \bigg\Vert\; \big\Vert\wt{\bm\beta}\big\Vert^2\nn\\
&+2\bigg\Vert\frac{\wh{\bm\Lambda}-\bm\Lambda\mbf J}{\sqrt n}\bigg\Vert\;
\bigg\Vert\frac 1 {T}\sum_{t=1}^T\frac{\bm\xi_t(\wh{\mbf x}_t-\mbf x_t)'}{n} \bigg\Vert\;\bigg\Vert\frac{\bm\Lambda}{\sqrt n}\bigg\Vert\; \big\Vert\wt{\bm\beta}\big\Vert^2 = O_p\l(\frac{T}{n^{(2-\eta-\delta)}}\r)+O_p\l(\frac{\sqrt T}{n^{(2-\eta-\delta)/2}}\r),\nn
\end{align}
because of Lemma \ref{lem:trendxi}(i) and \ref{lem:trendxi}(ii) (multiplying the statements by $T^2$). Last,
\begin{align}
\mathcal G_3\le & \bigg\Vert\frac{\wh{\bm\Lambda}-\bm\Lambda \mbf J}{\sqrt n}\bigg\Vert^2\; \bigg\Vert\frac 1 T\sum_{t=1}^T\frac{\bm(\wh{\mbf x}_t-\mbf x_t)(\wh{\mbf x}_t-\mbf x_t)'}{n} \bigg\Vert\;\big\Vert\wt{\bm\beta}\big\Vert^2 + \bigg\Vert\frac 1 {T}\sum_{t=1}^T\frac{\bm\Lambda'(\wh{\mbf x}_t-\mbf x_t)(\wh{\mbf x}_t-\mbf x_t)'{\bm\Lambda}}{n^{2}} \bigg\Vert\; \big\Vert\wt{\bm\beta}\big\Vert^2\nn\\
&+2\bigg\Vert\frac{\wh{\bm\Lambda}-\bm\Lambda\mbf J}{\sqrt n}\bigg\Vert\;
\bigg\Vert\frac 1 {T}\sum_{t=1}^T\frac{(\wh{\mbf x}_t-\mbf x_t)(\wh{\mbf x}_t-\mbf x_t)'}{n} \bigg\Vert\;\bigg\Vert\frac{\bm\Lambda}{\sqrt n}\bigg\Vert\; \big\Vert\wt{\bm\beta}\big\Vert^2 = O_p\l(\frac{T}{n^{2(1-\eta)}}\r)+O_p\l(\frac{\sqrt T}{n^{1-\eta}}\r),\nn
\end{align}
because of Lemma \ref{lem:trend}(iii) and \ref{lem:trend}(iv) (multiplying the statements by $T^2$). 

By noticing that as $n,T\to\infty$, we have $\sqrt T n^{-(2-\eta-\delta)/2}\to 0$  (in $\mathcal F_3$) and $\sqrt T n^{-(1-\eta)}\to 0$ (in $\mathcal G_3$), we have
\begin{align}
\mathcal D_3+\mathcal F_3+\mathcal G_3&=O_p\l(\frac{\sqrt T}{n^{(2-\delta)/2}}\r)+O_p\l(\frac{\sqrt T}{n^{(2-\eta-\delta)/2}}\r)+O_p\l(\frac{\sqrt T}{n^{1-\eta}}\r).\label{DDFFGG}
\end{align}
By substituting \eqref{DDFFGG} into \eqref{eq:mhatc2} and then \eqref{eq:mhatc2} into the second term on the rhs of \eqref{eq:mhatc} and the results of part (v) for the  second term on the rhs of \eqref{eq:mhatc}, we prove part (ix). This completes the proof.\hfill $\Box$\\
\begin{lem}\label{lem:shat} Define the matrices 
\begin{align}
\wh{\mbf S}_{00}=\wh{\mbf M}_{00}-\wh{\mbf M}_{02}\wh{\mbf M}_{22}^{-1}\wh{\mbf M}_{20},&&\wh{\mbf S}_{01}=\wh{\mbf M}_{01}-\wh{\mbf M}_{02}\wh{\mbf M}_{22}^{-1}\wh{\mbf M}_{21},&&\wh{\mbf S}_{11}=\wh{\mbf M}_{11}-\wh{\mbf M}_{12}\wh{\mbf M}_{22}^{-1}\wh{\mbf M}_{21},\nn
\end{align}
where $\wh{\mbf M}_{10}=\wh{\mbf M}_{01}'$, $\wh{\mbf M}_{20}=\wh{\mbf M}_{02}'$, and $\wh{\mbf M}_{12}=\wh{\mbf M}_{21}'$. 
Denote by $\mbf S_{ij}$, for $i,j=0,1$, the analogous ones but computed by using $\wt{\mbf F}_t=\mbf J\mbf F_t$. Define also $\wt{\bm\beta}=\mbf J\bm\beta$ and $\wt{\bm\beta}_{\perp*}=\wt{\bm\beta}_{\perp}^{}(\wt{\bm\beta}_{\perp}'\wt{\bm\beta}_{\perp}^{})^{-1}$, where $\wt{\bm\beta}_{\perp}= \mbf J\bm\beta_{\perp}$ such that $\wt{\bm\beta}_{\perp}'\wt{\bm\beta}=\mbf 0_{r-c\times r}$. Under Assumptions \ref{ASS:common} through \ref{ASS:trend}, as $n,T\to\infty$,   \smallskip
\begin{compactenum}[(i)]
\item $\Vert  \wh{\mbf S}_{00} - \mbf S_{00}\Vert =O_p(\max(n^{-1/2},T^{-1/2},n^{-(1-\eta)}))$.
\end{compactenum}
If also Assumption \ref{asm:rates} holds, then, \smallskip
\begin{compactenum}
\item [(ii)] $\Vert \wt{\bm\beta}' \wh{\mbf S}_{11}\wt{\bm\beta} - \wt{\bm\beta}'\mbf S_{11}\wt{\bm\beta}\Vert =O_p(\vartheta_{nT,\delta,\eta})$;
\item [(iii)] $\Vert T^{-1/2}\wt{\bm\beta}'\wh{\mbf S}_{11}\wt{\bm\beta}_{\perp*}-T^{-1/2}\wt{\bm\beta}'{\mbf S}_{11}\wt{\bm\beta}_{\perp*}\Vert =O_p(\vartheta_{nT,\delta,\eta})$;
\item [(iv)] $\Vert T^{-1/2}\wt{\bm\beta}'\wh{\mbf S}_{10}\wh{\mbf S}_{00}^{-1}\wh{\mbf S}_{01} \wt{\bm\beta}_{\perp*}-T^{-1/2}\wt{\bm\beta}'{\mbf S}_{10}{\mbf S}_{00}^{-1}{\mbf S}_{01} \wt{\bm\beta}_{\perp*}\Vert =O_p(\vartheta_{nT,\delta,\eta})$;
\item [(v)] $\Vert T^{-1}\wt{\bm\beta}_{\perp*}'\wh{\mbf S}_{10}\wh{\mbf S}_{00}^{-1}\wh{\mbf S}_{01} \wt{\bm\beta}_{\perp*}^{}-T^{-1}\wt{\bm\beta}_{\perp*}'{\mbf S}_{10}{\mbf S}_{00}^{-1}{\mbf S}_{01} \wt{\bm\beta}_{\perp*}^{}\Vert =O_p(\vartheta_{nT,\delta,\eta})$;
\item [(vi)] $\Vert T^{-1}\wt{\bm\beta}_{\perp*}' \wh{\mbf S}_{11}\wt{\bm\beta}_{\perp*}-T^{-1}\wt{\bm\beta}_{\perp*}' {\mbf S}_{11}\wt{\bm\beta}_{\perp*}\Vert =O_p(\vartheta_{nT,\delta,\eta})$.
\end{compactenum}
\end{lem}

\noindent{\textbf{Proof.}} Throughout we use the fact that $\Vert \check{\bm\beta}_{\perp*}\Vert =O(1)$. Part (i) is proved using Lemma \ref{lem:mhat}(ii), \ref{lem:mhat}(iii) and \ref{lem:mhat}(iv). For proving part (ii) we use Lemma \ref{lem:mhat}(iv), \ref{lem:mhat}(vi) and \ref{lem:mhat}(ix). Part (iii) is proved by combining part (ii) with Lemma \ref{lem:mhat}(v) and \ref{lem:mhat}(ix), and by noticing that $\Vert T^{-1/2}\mbf F_t\Vert =O_p(1)$ from Lemma \ref{lem:main}(ii). For proving part (iv) we combine part (i) with Lemma \ref{lem:mhat}(v), \ref{lem:mhat}(vii) and \ref{lem:mhat}(viii). Part (v) is proved by combining part (i) with Lemma \ref{lem:mhat}(vii) and \ref{lem:mhat}(viii). Finally, part (vi) follows from Lemma \ref{lem:mhat}(i) and \ref{lem:mhat}(viii). This completes the proof. \hfill$\Box$\\
\begin{lem}\label{lem:shat2} Consider the matrices $\mbf S_{ij}$ defined in Lemma \ref{lem:shat}, with $i,j=0,1$. Define $\wt{\mbf F}_t=\mbf J\mbf F_t$, $\wt{\bm\beta}=\mbf J\bm\beta$ and the conditional covariance matrices
\[
\wt{\bm\Omega}_{00}=\E[\Delta\wt{\mbf F}_t\Delta\wt{\mbf F}_t'|\Delta\wt{\mbf F}_{t-1}],\quad \wt{\bm\Omega}_{\wt{\beta}\wt{\beta}}=\E[\wt{\bm\beta}'\wt{\mbf F}_{t-1}\wt{\mbf F}_{t-1}'\wt{\bm\beta}|\Delta\wt{\mbf F}_{t-1}],\quad
\check{\bm\Omega}_{0\wt{\beta}}=\E[\Delta\wt{\mbf F}_t\wt{\mbf F}_{t-1}'\wt{\bm\beta}|\Delta\wt{\mbf F}_{t-1}].\nn
\]
Under Assumptions \ref{ASS:common} and \ref{initcond}, as $T\to\infty$,\smallskip
\begin{compactenum}[(i)]
\item $\Vert  {\mbf S}_{00} - \wt{\bm\Omega}_{00}\Vert =O_p(T^{-1/2})$;
\item $\Vert  \wt{\bm\beta}'\mbf S_{11}\wt{\bm\beta}-\wt{\bm\Omega}_{\wt{\beta}\wt{\beta}}\Vert =O_p(T^{-1/2})$;
\item $\Vert {\mbf S}_{01} \wt{\bm\beta}-\check{\bm\Omega}_{0\wt{\beta}}\Vert =O_p(T^{-1/2})$.
\end{compactenum}
\end{lem}

\noindent{\textbf{Proof.}} For part (i), notice that
\begin{align}
\wt{\bm\Omega}_{00} &= \E[\Delta\wt{\mbf F}_t\Delta\wt{\mbf F}_t'] - \E[\Delta\wt{\mbf F}_t\Delta\wt{\mbf F}_{t-1}']\Big(\E[\Delta\wt{\mbf F}_{t-1}\Delta\wt{\mbf F}_{t-1}']\Big)^{-1}\E[\Delta\wt{\mbf F}_{t-1}\Delta\wt{\mbf F}_{t}']=\bm\Gamma_0^{\Delta F}-\bm\Gamma_1^{\Delta F}\Big(\bm\Gamma_0^{\Delta F}\Big)^{-1}\bm\Gamma_1^{\Delta F},\nn
\end{align}
and
\begin{align}
\mbf S_{00} &=\frac 1 T\sum_{t=1}^T \Delta\wt{\mbf F}_t\Delta\wt{\mbf F}_t'-\bigg(\frac 1 T\sum_{t=2}^T \Delta\wt{\mbf F}_t\Delta\wt{\mbf F}_{t-1}'\bigg)\bigg(\frac 1 T\sum_{t=2}^T \Delta\wt{\mbf F}_{t-1}\Delta\wt{\mbf F}_{t-1}'\bigg)^{-1}\frac 1 T\sum_{t=2}^T \Delta\wt{\mbf F}_{t-1}\Delta\wt{\mbf F}_t'\nn\\
&= \mbf M_{00}-\mbf M_{02}\mbf M_{22}^{-1}\mbf M_{20}.\nn
\end{align}
Using Lemma \ref{rem:DFF}(i), we have the result. Parts (ii) and (iii) are proved in the same way, but using Lemma \ref{rem:DFF}(v) and \ref{rem:DFF}(vi), respectively. This completes the proof. \hfill$\Box$\\
\begin{lem} \label{rem:idiospect} 
Under Assumptions \ref{ASS:common} through \ref{ASS:idio},  there exist positive reals $\underline{M}_9$, $\overline{M}_9$, $M_{10}$, $\underline{M}_{11}$, $\overline{M}_{11}$ and an integer $\bar n$ such that
\begin{compactenum}[(i)]
\item $\underline M_9\leq n^{-1}\nu_{j}^{\Delta\chi}(\theta)\leq \overline M_9$ a.e. in $[-\pi,\pi]$, and for any $j=1,\ldots,q$ and $n>\bar n$;
\item $\sup_{\theta\in[-\pi,\pi]}\nu_1^{\Delta\xi}(\theta)\leq M_{10}$, for any $n\in\mathbb N$;
\item $\underline M_{11}^{\vphantom{\Delta\chi}}\leq n^{-1}\nu_{j}^{\Delta x}(\theta)\leq \overline M_{11}$ a.e. in $[-\pi,\pi]$, and for any $j=1,\ldots,q$ and $n>\bar n$;
\item $\sup_{\theta\in[-\pi,\pi]}\nu_{q+1}^{\Delta x}(\theta)\leq M_{10}$, for any $n\in\mathbb N$;
\item $\underline M_{12}^{\vphantom{\Delta\chi}}\leq n^{-1}\nu_{j}^{\Delta x}(0)\leq \overline M_{12}$, for any $j=1,\ldots,\tau$ and $n>\bar n$; 
\item $\nu_{\tau+1}^{\Delta x}(0)\leq M_{10}$, for any $n\in\mathbb N$.
\end{compactenum}
\end{lem}

\noindent{\textbf{Proof.}} For part (i) we can follow a reasoning similar to Lemma \ref{lem:evalcov}(i). The spectral density matrix of the first difference of the common factors can be written as $\bm\Sigma^{\Delta F}(\theta)=(2\pi)^{-1} \mbf C(e^{-i\theta})\overline{\mbf C'(e^{-i\theta})}$ and, since $\mbox{rk}(\mbf C(e^{-i\theta}))=q$ a.e.~in $[-\pi,\pi]$, then it has $q$ non-zero real eigenvalues and $r-q$ zero eigenvalues. Notice also that we have $\mbox{rk}(\mbf C(e^{-i\theta}))\leq q$ for any $\theta\in[-\pi,\pi]$. Moreover, given square summability of the coefficients of $\mbf C(L)$ as a consequence of Assumption \ref{ASS:common}(b), the non-zero eigenvalues are also finite for any $\theta\in[-\pi,\pi]$. Thus, by denoting as $\nu_j^{\Delta F}(\theta)$ such eigenvalues, there exist positive reals $\underline M_{10}$ and $\overline M_{10}$ such that a.e. in $[-\pi,\pi]$
\beq\label{eq:M9}
\underline M_{10}\leq \nu_j^{\Delta F}(\theta)\leq \overline M_{10},\quad j=1,\ldots, q.
\eeq
Therefore, we can write $\bm\Sigma^{\Delta F}(\theta)=\mbf W^{\Delta F}(\theta) \mbf M^{\Delta F}(\theta)\overline{\mbf W^{\Delta F'}(\theta)}$, where $\mbf W^{\Delta F}(\theta)$ is the $r\times q$ matrix of normalised  eigenvectors, i.e. such that $\overline{\mbf W^{\Delta F'}(\theta)}\mbf W^{\Delta F}(\theta)=\mbf I_q$ for any $\theta\in[-\pi,\pi]$, and $\mbf M^{\Delta F}(\theta)$ is the corresponding $q\times q$ diagonal matrix of  eigenvalues. \smallskip

Define $\bm L(\theta)= \bm\Lambda\mbf W^{\Delta F}(\theta) (\mbf M^{\Delta F}(\theta))^{1/2}$ for any $\theta\in[-\pi,\pi]$. Then the spectral density matrix of the first differences of the common component is given by
\beq\nn
\frac{\bm\Sigma^{\Delta\chi}(\theta)}{n} = \frac 1 n \bm\Lambda \bm\Sigma^{\Delta F}(\theta) \bm\Lambda' =\frac 1 n \bm\Lambda\mbf W^{\Delta F}(\theta) \mbf M^{\Delta F}(\theta)\overline{\mbf W^{\Delta F'}(\theta)}\bm\Lambda'=\frac{\bm L(\theta)\overline{\bm L'(\theta)}}{n}, \quad \theta\in[-\pi,\pi].
\eeq
Moreover, since because of Assumption \ref{asm:factor}(a), there exists an integer $\bar n$ such that $n^{-1}\bm\Lambda'\bm\Lambda = \mbf I_r$, for any $n>\bar n$, then
\beq\label{eq:LLMFnspect}
\frac{\overline{\bm L'(\theta)}\bm L(\theta)}{n} = \mbf M^{\Delta F}(\theta), \quad \theta\in[-\pi,\pi].
\eeq
Therefore, a.e.~in $[-\pi,\pi]$ the non-zero dynamic eigenvalues of $\bm\Sigma^{\Delta\chi}(\theta)$ are the same as those of ${\overline{\bm L'(\theta)}}\bm L(\theta)$, and from \eqref{eq:LLMFnspect}, we have for any $n>\bar n$ and a.e.~in $[-\pi,\pi]$, $n^{-1}\nu_j^{\Delta\chi}(\theta) = \nu_j^{\Delta F}(\theta)$, for any $j=1,\ldots,r$. Part (i) then follows from \eqref{eq:M9}.\smallskip

As for part (ii), from Assumption \ref{ASS:idio}(c), for any $\theta\in [-\pi, \pi]$, there exists a positive real $M_4$ such that 
\beq\label{eq:k3}
\sup_{i\in\mathbb N}\big|\check{d}_i(e^{-i\theta})\big|\leq \sup_{i\in\mathbb N}\bigg|\sum_{k=0}^{\infty} \check d_{ik} e^{-ik\theta}\bigg|\leq \sup_{i\in\mathbb N}\sum_{k=0}^{\infty} \big|\check d_{ik}\big| \leq M_4.
\eeq
Define as $\sigma_{ij}(\theta)$ the generic $(i,j)$-th entry of $\bm\Sigma^{\Delta\xi}(\theta)$. Then, for any $n>\bar n$, 
\begin{align}
\sup_{\theta\in[-\pi,\pi]}&\big\Vert\bm\Sigma^{\Delta\xi}(\theta)\big\Vert_1=\sup_{\theta\in[-\pi,\pi]}\max_{i=1,\ldots, n}\sum_{j=1}^{n} |\sigma_{ij}(\theta)|= \sup_{\theta\in[-\pi,\pi]}\max_{i=1,\ldots, n}\frac 1 {2\pi}\sum_{j=1}^{n}\big|\check{d}_i(e^{-i\theta})\E[\eps_{it}\eps_{jt}]\;\check{d}_j(e^{i\theta})\big|\nn\\
&\leq \frac {M_4^2} {2\pi}\max_{i=1,\ldots, n}\sum_{j=1}^{n}|\E[\eps_{it}\eps_{jt}]| \leq\frac {M_4^2M_3} {2\pi},\label{eq:sigmaij}
\end{align}
where we used \eqref{eq:k3} and Assumption \ref{ASS:idio}(b). From \eqref{eq:12inf} and \eqref{eq:sigmaij}, we have, for any $n>\bar n$,
\begin{align}
\sup_{\theta\in[-\pi,\pi]}\nu_1^{\Delta\xi}(\theta) = \sup_{\theta\in[-\pi,\pi]}\big\Vert\bm\Sigma^{\Delta\xi}(\theta)\big\Vert \leq\sup_{\theta\in[-\pi,\pi]} \big\Vert\bm\Sigma^{\Delta\xi}(\theta)\big\Vert_1\leq \frac {M_4^2M_3} {2\pi},
\end{align}
and part (ii) is proved by defining $M_{11}= {M_4^2M_3} (2\pi)^{-1}$.\smallskip 

Finally, parts (iii) and (iv), are immediate consequences of Assumption \ref{ASS:idio}(e), which implies that $\bm\Sigma^{\Delta x}(\theta)=\bm\Sigma^{\Delta\chi}(\theta)+\bm\Sigma^{\Delta\xi}(\theta)$, for any $\theta\in[-\pi,\pi]$, and of Weyl's inequality \eqref{eq:weyl}. So, for $j=1,\ldots, q$, and for any $n>\bar n$ and a.e. in $[-\pi,\pi]$, there exist positive reals $\underline M_{12}$ and $\overline M_{12}$ such that
\begin{align}
\frac{\nu_j^{\Delta x}(\theta)}n &\leq \frac{\nu_j^{\Delta\chi}(\theta)}n+ \frac{\nu_1^{\Delta \xi}(\theta)}n \leq \overline M_{10} + \sup_{\theta\in[-\pi,\pi]}\frac{\nu_1^{\Delta \xi}(\theta)}n \leq \overline M_{10}+\frac {M_{11}} n= \overline M_{12},\nn\\
\frac{\nu_j^{\Delta x}(\theta)}n &\geq \frac{\nu_j^{\Delta\chi}(\theta)}n+ \frac{\nu_n^{\Delta \xi}(\theta)}n \geq \underline M_{10} + \inf_{\theta\in[-\pi,\pi]}\frac{\nu_n^{\Delta \xi}(\theta)}n= \underline M_{12} .\nn
\end{align}
because of parts (i) and (ii). This proves part (iii). When $j=q+1$, using parts (i) and (ii), and since $\mbox{rk}(\bm\Sigma^{\Delta\chi}(\theta))\leq q$, for any $\theta\in[-\pi,\pi]$, we have $\nu_{q+1}^{\Delta x}(\theta) \leq \nu_{q+1}^{\Delta\chi}(\theta) + \nu_1^{\Delta \xi(\theta)}=\nu_1^{\Delta \xi(\theta)}\leq M_{11}$, thus proving part (iv). \smallskip

Finally, for parts (v) and (vi) consider parts (iii) and (iv) but when $\theta=0$. Then, $\mbox{rk}(\bm\Sigma^{\Delta\chi}(0))=\tau\leq q$ which implies $\underline M_{10}\leq n^{-1}\nu_{\tau}^{\Delta\chi}(0)\leq \overline{M}_{10}$, but $\nu_{\tau+1}^{\Delta\chi}(0)=0$. Using again parts (i) and (ii) and Weyl's inequality \eqref{eq:weyl}, we prove parts (v) and (vi).  This completes the proof.\hfill$\Box$



\clearpage
\setcounter{subsection}{-1}
\setcounter{equation}{0}
\setcounter{lem}{0}
\renewcommand{\thesection}{E}
\renewcommand{\thesubsection}{E\arabic{subsection}}
\renewcommand{\thelem}{E\arabic{lem}}
\renewcommand{\theequation}{E\arabic{equation}}

\section{Details on identification of IRFs and their confidence bands}\label{app:ident}

\subsection{Identification}\label{app:ident1}
As we discuss in Section \ref{sec:vecmvar}, the IRFs in \eqref{eq:IRF1_NOTID} are in general not identified unless we also estimate the orthogonal $q\times q$ transformation $\mbf R$. Economic theory tells us that the choice of the identifying transformation can be determined by the economic meaning attached to the common shocks, $\mbf u_t$. In general, for a given set of identifying restrictions, ${\mbf R}$ depends on the other parameters of the model, that is, it is determined by a mapping ${\mbf R}\equiv {\mbf R}({\bm\Lambda},{\mbf A}(L),{\mbf K})$. In the typical case of just- or under-identifying restrictions, to estimate $\mbf R$ we just have to consider the $q$ rows of the raw estimated IRFs, denoted as $\widetilde{\bm\Phi}_{[q]}(L)$, corresponding to the economic variables which are relevant for identification of the shocks. Therefore, we define the estimator $\wh{\mbf R}$ such that $\widetilde{\bm\Phi}_{[q]}(L)\wh{\mbf R}$ satisfies our desired restrictions. In this case, due to orthogonality, an estimator $\wh{\mbf R}$ is obtained by solving a linear system of  $q(q - 1)/2$ equations with $q(q - 1)/2$ unknowns, which depends on $\widetilde{\bm\Phi}_{[q]}(L)$ and therefore on $\wh{\bm\Lambda}$, $\wh{\mbf A}^{\mbox{\tiny{VECM}}}(L)$, and $\wh{\mbf K}$. Once we have computed $\wh{\mbf R}$, the $n\times q$ matrix of identified IRFs is $\wh{\bm \Phi}(L)=\widetilde{\bm \Phi}(L)\wh{\mbf R}$. Finally, if we denote the raw shocks as $\widetilde{\mbf u}_t$, the identified shocks are given by $\wh{\mbf u}_t=\wh{\mbf R}'\widetilde{\mbf u}_t$. Details on the two identification schemes adopted in Section \ref{sec:emp} are given below.

\ben[wide, labelwidth=!, labelindent=0pt]
\item [\textbf{Application 1: Oil price shock.}] To identify the oil price shock, \citet{stockwatson16} use a standard recursive identification scheme such that an oil price shock is the only shock having contemporaneous effect on the oil price. Specifically, when $q=3$, let $x_{1t}$ be the oil price, $x_{2t}$ be GDP, and $x_{3t}$ be consumption; then, $\wh{\mbf R}$ must be such that $\wh{\bm \Phi}_{[3]}(0) = \widetilde{\bm \Phi}_{[3]}(0) \wh{\mbf R}$ is lower triangular, i.e. such that the identified IRFs are given by
\[
\wh{\bm \Phi}_{[3]}(0)=\begin{bmatrix}
\wh{\phi}_{11}(0) & 0 & 0 \\
\wh{\phi}_{21}(0) & \wh{\phi}_{22}(0) & 0 \\
\wh{\phi}_{31}(0) & \wh{\phi}_{32}(0) & \wh{\phi}_{33}(0) \\
\end{bmatrix}= \widetilde{\bm \Phi}_{[3]}(0) \wh{\mbf R}.
\]
Therefore, we can choose $\wh{\mbf R}=[\widetilde{\bm \Phi}_{[3]}(0)]^{-1}\widetilde{\mbf R}$, where $\widetilde{\mbf R}$ is  the lower triangular Choleski factor such that $\widetilde{\bm \Phi}_{[3]}(0)\widetilde{\bm \Phi}_{[3]}(0)'=\widetilde{\mbf R}\widetilde{\mbf R}'$. The oil price shock is then obtained as $\wh{u}_{1t}=\wh{\mbf r}_1'\widetilde{\mbf u}_{t}$, where $\wh{\mbf r}_1$ is the first column of $\wh{\mbf R}$. The identified IRFs, reported in Figure \ref{fig:IRFoil}, are given by the entries of the first column of $\wh{\bm \Phi}(L)$, corresponding to the variables considered.

\item [\textbf{Application 2: News shock.}] To identify the news shock, \citet{fornigambettisala} proceed as follows: first, they identify what they call a ``surprise technology shock'' as the only shock having a contemporaneous effect on TFP; next, they identify the news shock by imposing that out of the remaining four shocks, the news shock is the one with maximal impact on TFP at lag 60. In practice, this identification is obtained as follows---recall that the considered FAVAR is composed of two variables (TFP and stock prices) and three estimated factors so that $q=5$: Let $x_{1t}$ and $x_{2t}$ be TFP and stock prices, respectively, and let $x_{3t}$, $x_{4t}$, $x_{5t}$ be GDP, consumption, and investment. 

\begin{compactenum}
	\item The surprise technology shock is identified by setting $\wh{\mbf R}$ such that $\wh{\bm \Phi}_{[5]}(0) = \widetilde{\bm \Phi}_{[5]}(0) \wh{\mbf R}$ is lower triangular, i.e. such that the identified IRFs are given by
	\[
	\wh{\bm \Phi}_{[5]}(0)=\begin{bmatrix}
	\wh{\phi}_{11}(0) & 0 & 0 & 0 & 0\\
	\wh{\phi}_{21}(0) & \wh{\phi}_{22}(0) & 0 & 0 & 0 \\
	\wh{\phi}_{31}(0) & \wh{\phi}_{32}(0) & \wh{\phi}_{33}(0)& 0 & 0 \\
	\wh{\phi}_{41}(0) & \wh{\phi}_{42}(0) & \wh{\phi}_{43}(0)& \wh{\phi}_{44}(0) & 0 \\
	\wh{\phi}_{51}(0) & \wh{\phi}_{52}(0) & \wh{\phi}_{53}(0)& \wh{\phi}_{54}(0) & \wh{\phi}_{55}(0)  \\
	\end{bmatrix}=\widetilde{\bm \Phi}_{[5]}(0) \wh{\mbf R}.
	\]
	Therefore, we can choose $\wh{\mbf R}=[\widetilde{\bm \Phi}_{[5]}(0)]^{-1}\widetilde{\mbf R}$, where $\widetilde{\mbf R}$ is  the lower triangular Choleski factor such that $\widetilde{\bm \Phi}_{[5]}(0)\widetilde{\bm \Phi}_{[5]}(0)'=\widetilde{\mbf R}\widetilde{\mbf R}'$.  
	\item The news shock is then identified by choosing the $4\times 1$ vector ${\wh{\mbf r}_2}=(0\, \wh r_{22} \, \wh r_{32} \,\wh r_{42} \,\wh r_{52})'$ such that ${\wh{\mbf r}_2}'{\wh{\mbf r}_2}=1$ and it maximizes 
	the element $(1,1)$ of $\wh{\bm \Phi}_{[5]}(60) = \widetilde{\bm \Phi}_{[5]}(60) \wh{\mbf r}_2$, which is the effect of the news shock on TFP at lag 60. The news shock is then obtained as $\wh{u}_{2t}=\wh{\mbf r}_2'\widetilde{\mbf u}_{t}$. The identified IRFs to a news shock, reported in Figure \ref{fig:IRFnews}, are given by the entries of the second column of $\wh{\bm \Phi}(L)$, corresponding to the variables considered.
\end{compactenum}
\een

\subsection{Bootstrap confidence bands in practice}

In order to build confidence intervals for the estimated IRFs, we use a bootstrap algorithm. In detail, at each iteration $d=1,\ldots, 1000$, we generate bootstrap shocks $\mbf u^d_t$ by drawing randomly with replacement from the estimated shocks $\wh{\mbf u}_t$  and we generate bootstrap common factors $\mbf F_t^d$. Then, we estimate $\widehat{\mbf A}(L)^d$, $\wh{\mbf K}^d$, and $\wh{\mbf R}^d$ in \eqref{eq:IRF1} or \eqref{eq:IRF1var}, thus obtaining a bootstrap IRF $\wh{\bm \Phi}(L)^d=\wh{\bm \Lambda} [\wh{\mbf A}(L)^d]^{-1}\wh{\mbf K}^d\wh{\mbf R}^d$. Repeating this procedure several times gives, for each $i,j$ and lag $k$, a bootstrap distribution of the IRF: $\{\widehat{\phi}_{ij,k}^d,\, d=1,\ldots 1000\}$ (for simplicity below we omit the dependence on $i$ and $j$ of the IRF). 

In order to compute the $(1-\alpha)$ confidence interval, at each lag $k$ we compute the sample variance of $\{\widehat{\phi}_{k}^d\}$, which we denote as  $\sigma_{k}^2$, and then we construct the $(1-\alpha)$ confidence interval is given by $[\widehat{\phi}_{k} + z_{\alpha/2}\, \sigma_{k},\ \widehat{\phi}_{k} + z_{1-\alpha/2} \,\sigma_{k}]$, where $z_{\alpha/2}=-z_{1-\alpha/2}$ is the $\alpha/2$ quantile of a standard normal, see also Chapter 12 in \citet{KL17}. By proceeding in this way we obtain symmetric confidence bands around the estimated IRF. 

\citet{stockwatson16} adopt a procedure very similar to the one described above. By contrast, \citet{fornigambettisala} compute the confidence bands as the percentiles of $\{\widehat{\phi}_{k}^d\}$ over the replication $d$. This is also a a possible strategy, which yields confidence bands that are not symmetrical by construction, but does not ensure that the estimated IRF is within the confidence bands.

\subsection{Estimated identified shocks}
In Section \ref{sec:emp}, we show and discuss the estimated IRFs, which are our main object of interest. In contrast, we said nothing about the identified shocks, which, although they are not the object of interest in the empirical application, they are intimately intertwined with the IRFs, as we explain in Section \ref{app:ident1}. 

\begin{figure}[h!]\caption{Estimated identified shocks}\label{fig:shocks}
\centering

\begin{tabular}{@{}cc@{}}
\footnotesize Oil price shock & \footnotesize News shock\\
\includegraphics[width=.495\textwidth]{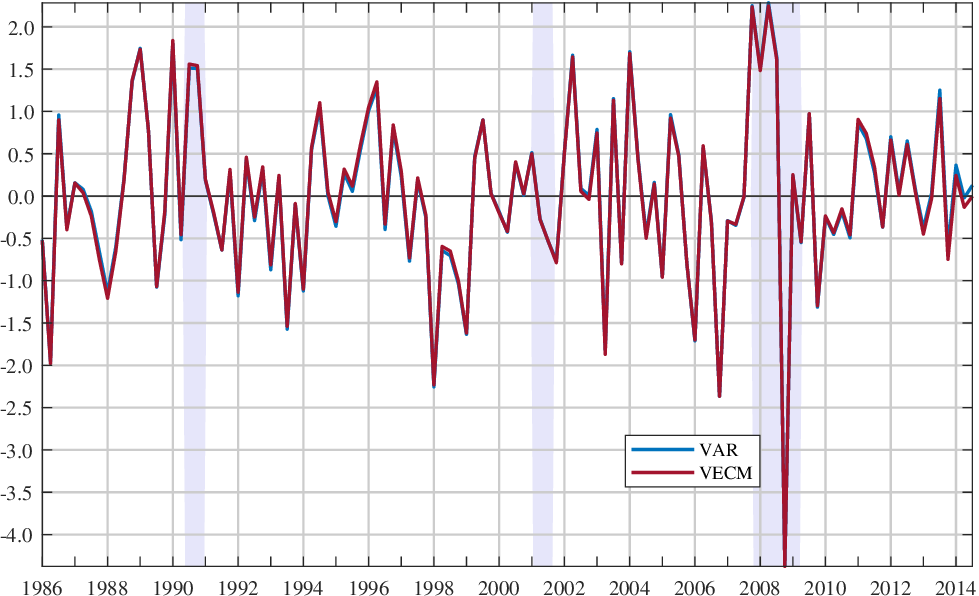}&
\includegraphics[width=.495\textwidth]{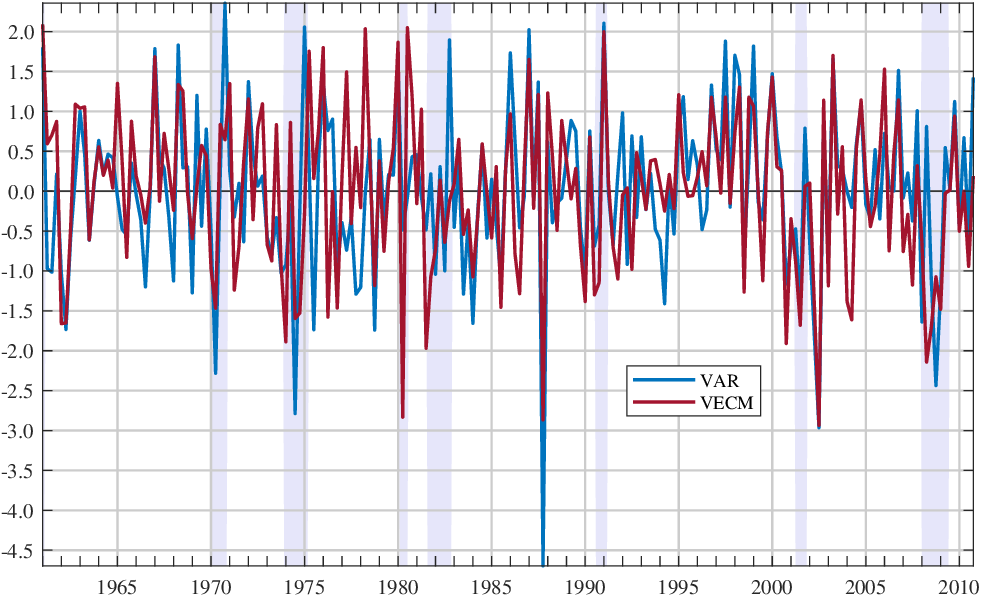} \\
\end{tabular}
\begin{tabular}{@{}p{\textwidth}}
\end{tabular}
\end{figure}

Figure \ref{fig:shocks} shows the estimated shocks. The left plot reports the oil price shock identified as in \citet{stockwatson16}, while the right plot reports the news shock identified as in \citet{fornigambettisala}. The figure shows both the estimate obtained by estimating an unrestricted VAR on $\wh{\mbf F}_t$ or a VECM on $\Delta \wh{\mbf F}_t$. As we can see, the two estimates of the oil price shock are nearly indistinguishable, which dovetail with the estimated IRFs shown in Figure \ref{fig:IRFoil} in the paper. By contrast, the news shock differs depending on which law of motion is estimated for the common factors, which, as we explained in Section \ref{sec:est} in the paper, depends on the fact that the restriction is imposed at lag 60, and therefore it depends on the estimated of the long-run IRFs.

\setcounter{subsection}{-1}
\setcounter{equation}{0}
\setcounter{lem}{0}
\renewcommand{\thesection}{F}
\renewcommand{\thesubsection}{F\arabic{subsection}}
\renewcommand{\thelem}{F\arabic{lem}}
\renewcommand{\theequation}{F\arabic{equation}}

\section{Factor Augment VAR models}

\subsection{On the relation between FAVAR and DFM}\label{app:favar}

Consider the FAVAR model proposed by \citet{BBE05}:
\begin{align}
\mbf w_t = \bm L^f \mbf f_t +\bm L^z \mbf z_t+\mbf e_t, \qquad
\bm\Psi(L) 
\begin{bmatrix}
\mbf f_t \\ \mbf z_t
\end{bmatrix}
=   \mbf v_t,
\label{FAVAR}
\end{align}
where $\mbf z_t$ is an $m$-dimensional vector of observable economic variables of interest, $\mbf f_t$ is a $k$-dimensional vector of latent factors summarising additional information contained in the $N$-dimensional vector $\mbf w_t$. In this setting $\mbf e_t$ is the idiosyncratic component of $\mbf w_t$ and $\mbf v_t$ is a white noise process containing the structural shocks that we are interested in and it is of dimension $k+m\ll N$. 

Following \citet[Section 5.2]{stockwatson16}, let 
\[
\mbf x_t=
\begin{bmatrix}
\mbf w_t\\
\mbf z_t
\end{bmatrix} \qquad \mathrm{and} \qquad 
\bar{\mbf F}_t=
\begin{bmatrix}
\mbf f_t\\
\mbf z_t
\end{bmatrix},
\]
where ${\mbf x}_t$ is the vector of all observed time series of dimension $n=N+m$ and $\bar{\mbf F}_t$ is $(m+k)$-dimensional.
Then, we can rewrite \eqref{FAVAR} as:
\begin{align}
{\mbf x}_t = \bar{\bm \Lambda} \bar{\mbf F}_t + \bar{\bm \xi}_t, \qquad \bm\Psi(L)\bar{\mbf F}_t =  \mbf v_t, \label{equiv}
\end{align}
where:
$$
\bar{\bm \Lambda} =
\begin{bmatrix}
\bm L^f &  \bm L^z \\ \mbf 0_{m\times r} & \mbf I_m
\end{bmatrix}
\qquad \mathrm{and} \qquad 
\bar{\bm\xi}_t =
\begin{bmatrix} \mbf e_t \\ \mbf 0_{m\times 1} \end{bmatrix}.
$$
On the other hand the DFM reads
\begin{align}
\mbf x_t = \bm \Lambda \mbf F_t + \bm\xi_t, \qquad \mbf A(L) \mbf F_t =\mbf K  \mbf u_t.\label{DFM}
\end{align}
Therefore, the FAVAR \eqref{equiv} is a restricted version of the DFM \eqref{DFM}, where the variables $\mbf z_t$ have unit factor loadings and zero idiosyncratic component and the number of factors is $r=k+m$, which is equal to the number of common shocks, i.e. in \eqref{DFM} we also impose $r=q$ and thus $\mbf K=\mbf I_r$. In other words in a FAVAR the variables of interest $\mbf z_t$ are considered as ``observable'' factors.  
Although the FAVAR has been mainly studied in a stationary setting, the same reasoning applies if we have non-stationary data. Note that deterministic linear trends can also be included in the FAVAR as we discuss in the next section.

\subsection{FAVAR estimation}\label{app:FAVARapp}
Let $y_{it}$ be the observed data, then in our framework the FAVAR is written as
\begin{align}
y_{it} &=  a_i +  b_i t +  x_{it},\nn\\
x_{it} & = (\bm l_{i}^{f'}  \quad \bm l_{i}^{z'})  (\mbf f_t' \quad \mathbf  z_t')'+ \xi_{it},\nn\\
\bm\Psi(L) (\mbf f_t' \quad \mathbf  z_t')'
&=   \mbf v_t,\nn
\end{align}
where $\mbf z_t$ are the ``observed'' common factors, and $\mbf f_t$ are the ``unobserved'' common factors.
The model is estimated as follows:
\ben
	\item estimate the unobserved common factors $\wh{\mbf f}_t$ from $\mbf y_t=(y_{1t}\cdots y_{nt})'$ as explained in Section \ref{sec:est}, thus de-trending series first (if needed);
	\item estimate $\wh{\bm\Psi}(L)$ by  fitting either a VECM on $(\Delta\wh{\mbf f}_t' \quad \Delta{\mbf z}_t')'$ or an unrestricted VAR on $(\wh{\mbf f}_t' \quad {\mbf z}_t')'$ as explained in Section \ref{sec:est};
	\item estimate  $(\wh{\bm l}_{i}^{f'}  \quad \wh{\bm l}_{i}^{z'})$  by regressing $\Delta y_{it}$ onto a constant and the vector $(\Delta \mbf f_t' \quad \Delta \wh{\mbf z}_t')'$; 
	\item estimate IRFs as $(\wh{\bm l}_{i}^{f'}  \quad \wh{\bm l}_{i}^{z'}) [\wh{\bm\Psi}(L)]^{-1}$.
%
	
\een
In contrast, in the approach by \citet{fornigambettisala} the factors are extracted directly from the observed data $y_{it}$, without controlling for the presence of possible deterministic linear trends. Therefore, the FAVAR is written as
\beq
y_{it} = (\bm l_{i}^{f'}  \quad \bm l_{i}^{z'})  (\mbf f_t' \quad \mathbf  z_t')'+ \xi_{it}.\nn
\eeq
The model is estimated as follows:
\ben
	\item estimate the unobserved common factors from PC analysis of $\mbf y_t=(y_{1t}\cdots y_{nt})'$ as in \citet{bai04};
	\item estimate an unrestricted VAR on $(\wh{\mbf f}_t' \quad {\mbf z}_t')'$ as explained in Section \ref{sec:est} to get $\wh{\bm\Psi}(L)$;
	\item estimate  $(\wh{\bm l}_{i}^{f'}  \quad \wh{\bm l}_{i}^{z'})$  by regressing $\Delta y_{it}$ onto a constant and the vector $(\Delta \mbf f_t' \quad \Delta \wh{\mbf z}_t')'$; 
	\item estimate IRFs as $(\wh{\bm l}_{i}^{f'}  \quad \wh{\bm l}_{i}^{z'}) [\wh{\bm\Psi}(L)]^{-1}$.
%
	
\een



\end{document}